\documentclass[11pt]{article}
\pdfoutput=1
\usepackage{epsfig}
\usepackage{psfrag}
\usepackage{latexsym}
\usepackage{indentfirst}
\usepackage{fancyhdr}
\usepackage{dsfont}
\usepackage{amsmath}
\usepackage{amssymb}
\usepackage{amsfonts}
\usepackage{mathrsfs}
\usepackage{amsthm}
\usepackage{pifont}
\usepackage{dsfont}
\usepackage{multirow}
\usepackage{array}
\usepackage{chngpage}
\usepackage{longtable}
\usepackage{cite}
\usepackage{bbold}
\usepackage{color}
\usepackage{colordvi}
\usepackage{fancybox}
\usepackage[footnotesize]{caption}
\usepackage{graphicx}
\usepackage[center,footnotesize,hang]{subfigure}
\usepackage{bbm}
\usepackage[colorlinks, linkcolor=red, anchorcolor=black, citecolor=green]{hyperref}
\usepackage{url}
\usepackage{multirow}
\usepackage{array}
\usepackage{fancybox}
\usepackage{enumitem}
\allowdisplaybreaks
\LTcapwidth=\textwidth

\theoremstyle{remark}

\makeatletter
\@addtoreset{equation}{section}
\makeatother

\begin{document}

\title{\hfill ~\\[-30mm] \hfill\mbox{\small USTC-ICTS-16-06}\\[10mm]\textbf{CP Symmetry and Lepton Mixing from a Scan of Finite Discrete Groups}}

\date{}

\author{\\[1mm] Chang-Yuan Yao\footnote{E-mail: {\tt phyman@mail.ustc.edu.cn}}~,~~Gui-Jun Ding\footnote{E-mail: {\tt dinggj@ustc.edu.cn}}\\ \\
\it{\small Interdisciplinary Center for Theoretical Study and  Department of Modern Physics, }\\
  \it{\small University of Science and
    Technology of China, Hefei, Anhui 230026, China}\\[4mm] }

\maketitle

\begin{abstract}

Including the generalized CP symmetry, we have performed a comprehensive scan of leptonic mixing patterns which can be obtained from finite discrete groups with order less than 2000. Both the semidirect approach and its variant are considered. The lepton mixing matrices which can admit a good agreement with experimental data can be organized into eight different categories up to possible row and column permutations. These viable mixing patterns can be completely obtained from the discrete flavor groups $\Delta(6n^2)$, $D^{(1)}_{9n,3n}$, $A_5$ and $\Sigma(168)$ combined with CP symmetry. We perform a detailed analytical and numerical analysis for each possible mixing patterns. The resulting predictions for lepton mixing parameter, neutrinoless double decay and flavored leptogenesis are studied.

\end{abstract}
\thispagestyle{empty}
\vfill

\newpage
\setcounter{page}{1}

\section{\label{sec:introduction}Introduction}

The origin of fermion mass and flavor mixing is one of longstanding open questions beyond the Standard Model physics. The discovery of neutrino oscillations and the precise measurements of the three lepton mixing angles $\theta_{12}$, $\theta_{23}$ and $\theta_{13}$ shed light on the flavor puzzle and help to establish underlying physics principle. One most popular approach is to invoke a discrete flavor symmetry to explain the observed patterns. In this paradigm, a given mixing pattern is related to certain residual symmetry of the leptonic mass matrices, and the residual symmetry  may arise from the breaking of the complete flavor symmetry group $G_{f}$ of some unknown extension of the Standard Model. The residual symmetry groups and their embedding in $G_f$ is sufficient to predict the values of the mixing angles, and the detailed dynamics of symmetry breaking is not necessary. Many different discrete flavor symmetry groups and their application in model building have been studied in the literature, please see Refs.~\cite{Altarelli:2010gt,Ishimori:2010au,King:2013eh}
for review.

In recent years, the flavor symmetry is extended to include the generalized CP symmetry in order to understand the observed values of the mixing angles and simultaneously predict the unknown CP violating phases~\cite{Feruglio:2012cw,Holthausen:2012dk}. Note that low significance hints for a maximal Dirac CP phase $\delta_{CP}\simeq-\pi/2$ has been reported~\cite{Abe:2015awa}, and the measurement of the Dirac CP phase is an important physical motivation of forthcoming neutrino oscillation experiments. From the bottom-up view, the neutrino and
the charged lepton mass matrices admit both residual flavor symmetry and residual CP symmetry, and the residual flavor symmetry can be generated by the residual CP transformations~\cite{Chen:2014wxa,Chen:2015nha}. One generally presumes that these residual symmetries originate from a large symmetry group (a flavor symmetry $G_{f}$ and the generalized CP) at high energy scale whose breaking leads to the symmetries of the mass matrices. Imposing a flavor symmetry as well as generalized CP symmetry, one can constrain the CP violation phases besides mixing angles. This can lead to very predictive scenarios in which the mixing angles and CP phases are determined in terms of few input parameters~\cite{Feruglio:2012cw,Chen:2014wxa,Chen:2015nha}. Discrete flavor symmetry combined with CP symmetry turns out to be a rather powerful framework. A variety of flavor symmetry groups and their interplay with the CP symmetry have been studied such as  $A_{4}$~\cite{Ding:2013bpa}, $S_{4}$~\cite{Feruglio:2012cw,Ding:2013hpa,Li:2014eia,Feruglio:2013hia, Luhn:2013vna, Li:2013jya},
$\Delta(27)$~\cite{Branco:2015gna}, $\Delta(48)$~\cite{Ding:2013nsa}, $A_{5}$~\cite{Li:2015jxa,DiIura:2015kfa,Ballett:2015wia}, $\Delta(96)$~\cite{Ding:2014ssa} and $\Sigma(36\times 3)$~\cite{Rong:2016cpk}. In particular the lepton mixing patterns arising from flavor symmetry group series  $\Delta(3n^{2})$~\cite{Hagedorn:2014wha,Ding:2015rwa}, $\Delta(6n^{2})$~\cite{Hagedorn:2014wha,King:2014rwa,Ding:2014ora} and $D^{(1)}_{9n, 3n}$~\cite{Li:2016ppt} in combination with a CP symmetry have been analyzed for an arbitrary index $n$. Some models with flavor and CP symmetry have been constructed~\cite{Ding:2013bpa,Ding:2013hpa,Li:2014eia,Feruglio:2013hia, Luhn:2013vna,Li:2013jya,Ding:2013nsa,Li:2015jxa}, where the required vacuum alignment needed to achieve the remnant symmetries is dynamically realized. Moreover, the phenomenological implications of residual flavor and CP symmetry in neutrinoless double beta ($0\nu\beta\beta$) decay~\cite{Ding:2013hpa,Li:2014eia,Li:2015jxa,Ding:2014ora,Li:2016ppt,Hagedorn:2016lva} and leptogenesis~\cite{Hagedorn:2016lva,Chen:2016ptr} have been studied. It is remarkable that the residual CP transformation could be systematically classified according to the number of its zero elements~\cite{Chen:2016ica}.

The powerful computer algebra software \texttt{GAP}~\cite{GAP} has been frequently used to investigate the lepton mixing matrices achievable from finite discrete groups~\cite{Lam:2012ga,Holthausen:2012wt,Holthausen:2013vba,Lavoura:2014kwa,Joshipura:2014pqa,Joshipura:2014qaa,Talbert:2014bda,Yao:2015dwa,Varzielas:2016zuo,King:2016pgv}. In this paper, we shall include the generalized CP symmetry and performed a comprehensive scan of all finite subgroups up to order 2000 with the help of \texttt{GAP}. The CP transformations are assumed to correspond to class-inverting automorphisms of the flavor symmetry group. All the possible residual flavor symmetries would be considered. We shall find out all the admissible lepton  lepton mixing patterns which can be compatible with the experimental data for certain values of the free parameter $\theta$. To our surprise, these viable lepton mixing matrices can be categorized into eight cases up to permutations of rows and columns, and they can be completely reproduced from the $\Delta(6n^2)$, $D^{(1)}_{9n,3n}$, $A_5$ and $\Sigma(168)$ flavor symmetry groups and CP symmetry. We give the analytic formulas of mixing angles and CP invariants in each of these cases. Moreover, we present the analytic expressions for the effective Majorana neutrino mass $|m_{ee}|$ in neutrinoless double beta decay and the lepton asymmetry parameters $\epsilon_{\alpha}$ ($\alpha=e$, $\mu$, $\tau$) relevant to leptogenesis. Furthermore, the allowed values of $|m_{ee}|$ and the baryon asymmetry $Y_{B}$ are analyzed numerically for the smallest values of the index $n$ that admit a good agreement with the experimental data on the mixing angles.

This paper is structured as follows: we shall elaborate the method to obtain the lepton mixing PMNS matrix from any given residual symmetry in the semidirect approach and the variant of the semidirect approach in section~\ref{sec:approach}. The mixing matrix can be determined from the representation matrices of the residual symmetry without reconstructing the lepton mass matrices. We outline the procedure of group scanning in section~\ref{sec:scan}. The resulting mixing patterns which can accommodate the experimental data, and the predictions for mixing angles and CP invariants are presented. Moreover the phenomenological predictions for $0\nu\beta\beta$ decay and flavored thermal leptogenesis are studied. Finally we conclude in section~\ref{sec:conclusion}. In appendix~\ref{sec:equivalence}, we derive the criteria to determine whether two residual symmetries leads to the same mixing pattern, if the redefinition of the free parameter $\theta$ is used.

\section{\label{sec:approach}Framework}

Both family symmetry and CP symmetry acts on the flavor space in a non-trivial way, and the interplay between them should be treated carefully. In order to consistently combine the generalized CP symmetry with a flavor symmetry group $G_{f}$, the CP transformation should be related to an automorphism $u: G_{f}\rightarrow G_{f}$, and the so called consistency condition has to be fulfilled~\cite{Feruglio:2012cw,Holthausen:2012dk,Ecker:1981wv},
\begin{equation}
\label{eq:consistency_condition}X_{\mathbf{r}}\rho^{*}_{\mathbf{r}}(g)X^{\dagger}_{\mathbf{r}}=\rho_{\mathbf{r}}(u(g)),\qquad \forall g\in G_{f}\,,
\end{equation}
where the subscript ``$\mathbf{r}$'' refers to the representation space acted on, $\rho_{\mathbf{r}}(g)$ is the representation matrix of the element $g$, and $X_{\mathbf{r}}$ is the generalized CP transformation. For a given CP transformation $X_{\mathbf{r}}$, $\rho_{\textbf{r}}(h)X_{\textbf{r}}$ with $h\in G_f$ also satisfies the consistency equation of Eq.~\eqref{eq:consistency_condition}, and consequently it is an admissible CP transformation as well. Obviously $\rho_{\textbf{r}}(h)X_{\textbf{r}}$ corresponds to performing a flavor symmetry transformation $\rho_{\textbf{r}}(h)$ followed by a CP transformation $X_{\textbf{r}}$. It is easy to check that the generalized CP transformation $\rho_{\textbf{r}}(h)X_{\textbf{r}}$ maps the group element $g$ into $hu(g)h^{-1}$. Hence the automorphism related to $\rho_{\textbf{r}}(h)X_{\textbf{r}}$ is an composition of $u$ and an inner automorphism $\mu_h: g\to hgh^{-1}$ with $h,g\in G_f$. This implies that the effect of the inner automorphism $\mu_h$ amounts to a flavor symmetry transformation $\rho_{\mathbf{r}}(h)$. As a result, one could focus on the outer automorphism of $G_f$ when searching for the most general CP transformations compatible with $G_{f}$. Furthermore, it has been shown that that the physically well-defined CP transformations should be given by class-inverting automorphism of $G_f$~\cite{Chen:2014tpa}. In other words, the automorphism $u$ should map each class of $G_f$ into its inverse class. In the present work, we shall be concerned with the CP transformations corresponding to the class-inverting automorphisms.

Let us now consider a theory with both flavor symmetry $G_{f}$ and CP symmetry $H_{CP}$ which denotes the CP transformations consistent with $G_{f}$. Thus the original symmetry at high energy scale is generically $G_{f}\rtimes H_{CP}$. Notice that the mathematical structure of the group comprising $G_f$ and $H_{CP}$ is a semi-direct product~\cite{Feruglio:2012cw}, because the flavor symmetry and CP transformations are not commutable in general. The experimental data clearly shows that all lepton masses are unequal and there is flavor
mixing among the three mass eigenstates. Therefore the parent symmetry $G_{f}\rtimes H_{CP}$ should be broken down to different residual subgroups $G_{l}\rtimes H^{l}_{CP}$ and $G_{\nu}\times H^{\nu}_{CP}$ in the charged lepton and neutrino sectors, respectively. It is remarkable that the lepton flavor mixing is fully fixed by the group structure of $G_{f}\rtimes H_{CP}$ and the residual symmetries~\cite{Chen:2014wxa,Chen:2015nha}. The details of the breaking mechanisms realizing the assumed residual symmetries are irrelevant. Assuming that neutrinos are Majorana particles, the mass terms of leptons obtained through flavor and CP symmetry breaking take the following form:
\begin{equation}
\label{eq:mass_Lag}\mathcal{L}_{m}=-\bar{l}_{R}m_{l}l_{L}-\frac{1}{2}\nu_{L}^{T}Cm_{\nu}\nu_{L}+h.c.\,,
\end{equation}
where $C$ is the charge conjugation matrix, $l_L \equiv (e_L,\mu_L,\tau_L)^T$ and $l_R \equiv (e_R,\mu_R,\tau_R)^T$ denote the three left-handed (LH) and right-handed (RH) charged lepton fields, respectively, and $\nu_L \equiv (\nu_{eL}, \nu_{\mu L}, \nu_{\tau L})^T$ contains the three LH neutrino fields. Both the charged lepton and neutrino mass matrices $m_{l}$ and $m_{\nu}$ are subject to the constraints of the remnant symmetries, such that the lepton mixing matrix can be fixed. Bottom-up analysis shows that the residual flavor symmetry $G_{l}$ can be any Abelian subgroup of $G_{f}$ while $G_{\nu}$ is either a $K_4\cong Z_2\times Z_2$ Klein subgroup or a $Z_2$ subgroup for Majorana neutrinos~\cite{Chen:2014wxa,Chen:2015nha}.
If the remnant flavor symmetry $G_{\nu}$ is restricted to be a Klein subgroup of $G_{f}$ and the left-handed leptons $l_{L}$ transform as three unequivalent one dimensional representations under $G_{l}$, both the lepton mixing angles and Dirac CP violating phase would be fully determined by residual symmetries. This scenario has been studied comprehensively in the literature~\cite{Holthausen:2012wt,Yao:2015dwa,Fonseca:2014koa}. The Majorana CP phase $\alpha_{31}$ would be predicted to be trivial and another Majorana phase $\alpha_{21}$ can only be a rational multiple of $\pi$ after the CP symmetry is taken into account~\cite{Chen:2015nha}.

In this work, we shall discuss two different types of remnant symmetries dubbed as ``semidirect'' and ''variant of semidirect'' approaches. In the semidirect approach, the residual symmetry in the neutrino sector is $Z_2\times H^{\nu}_{CP}$ while $G_{l}$ is able to distinguish among the three generations of charged lepton fields. As a result, one column of the PMNS matrix is completely fixed by the residual symmetries in this case. In the variant of semidirect approach, the remnant symmetries in the charged lepton and neutrino sectors are assumed to be $Z_2\times H^{l}_{CP}$ and $K_4\times H^{\nu}_{CP}$ respectively, and one row of the PMNS matrix can be fixed. It turns out that the lepton mixing matrix depends on a single real parameter $\theta$ in both approaches. Consequently the mixing angles and CP violating phases are strongly correlated with each other. In the following, the master formula of the prediction for lepton flavor mixing would be derived. As usual the three generations of left-handed leptons are asigned to a faithful irreducible three-diemensional representation of $G_{f}$ which is denoted as $\mathbf{3}$ henceforth.

\subsection{\label{subsec:semi_app}Semidirect approach}

We first analyze the residual symmetry constraints in the charged lepton sector. The requirement that $G_{l}\rtimes H^{l}_{CP}$ is a symmetry of the charged lepton mass matrix $m_{l}$ entails that the hermitian combination $m^{\dagger}_{l}m_{l}$ should be invariant under the action of $G_{l}\rtimes H^{l}_{CP}$, i.e.,
\begin{align}
\label{eq:flavor_cons_Charg}&\rho^{\dagger}_{\mathbf{3}}(g_{l})m^{\dagger}_{l}m_{l}\rho_{\mathbf{3}}(g_{l})=m^{\dagger}_{l}m_{l},  \qquad g_{l}\in G_{l}\,, \\
\label{eq:CP_cons_Charg}&X^{\dagger}_{l\mathbf{3}}m^{\dagger}_{l}m_{l}X_{l\mathbf{3}}=(m^{\dagger}_{l}m_{l})^{*},  \qquad X_{l\mathbf{3}} \in H^{l}_{CP}\,.
\end{align}
The residual flavor symmetry $G_{l}$ and the residual CP symmetry $H^{l}_{CP}$ has to be compatible with each other such that the following restricted consistency equation must be satisfied~\cite{Chen:2014wxa,Chen:2015nha,Li:2014eia},
\begin{equation}
\label{eq:GlXl2}X_{l\mathbf{r}}\rho^*_{\mathbf{r}}(g_l)X_{l\mathbf{r}}^{-1}=\rho_{\mathbf{r}}(g_l^{-1}),\quad
g_{l}\in G_{l},~~X_{l\mathbf{r}}\in H^{l}_{CP} \,.
\end{equation}
The hermitian matrix $m^{\dagger}_{l}m_{l}$ is diagonalized by the unitary transformation $U_{l}$ with $U^{\dagger}_{l}m^{\dagger}_{l}m_{l}U_{l}=\text{diag}(m^2_{e}, m^2_{\mu}, m^2_{\tau})$. The explicit form of $m^{\dagger}_{l}m_{l}$ could be constructed from Eqs.~(\ref{eq:flavor_cons_Charg},\ref{eq:CP_cons_Charg}), and thus $U_{l}$ can be determined. In fact, one can directly extract the constraints on $U_{l}$ from Eqs.~(\ref{eq:flavor_cons_Charg},\ref{eq:CP_cons_Charg}) without resorting to mass matrix $m^{\dagger}_{l}m_{l}$ as follows
\begin{align}
\label{eq:GlUl2}U_l^{\dagger}\rho_{\mathbf{3}}(g_l)U_l&=\rho^{diag}_{\mathbf{3}}(g_l)\,,\\
\label{eq:XlUl}U_l^{\dagger}X_{l\mathbf{3}}U_l^*&=X^{diag}_{l\mathbf{3}}\,,
\end{align}
where $\rho^{diag}_{\mathbf{3}}(g_l)$ and $X^{diag}_{l\mathbf{3}}$ are diagonal phase matrices. We see that the residual CP transformation $X_{l\mathbf{3}}$ should be a symmetric unitary matrix, and $\rho_{\mathbf{3}}(g_l)$ and $m^{\dagger}_{l}m_{l}$ can be diagonalized by the same unitary matrix $U_{l}$. Given a specific residual symmetry group $G_{l}$ and the three-dimensional representation of $G_{f}$, the three normalized and mutually
orthogonal eigenvectors of $\rho_{\mathbf{3}}(g_l)$ can be easily found and they constitute a unitary matrix $\Sigma_{l}$ fulfilling $\Sigma_l^{\dagger}\rho_{\mathbf{3}}(g_l)\Sigma_l=\rho^{diag}_{\mathbf{3}}(g_l)$. Since we consider a scenario in which the three generations of left-handed leptons can be distinguished by $G_{l}$, and no further assumption or prediction is made about the charged lepton masses.
Therefore $U_{l}$ is uniquely fixed up to permutations and phases of its column vectors, i.e.
\begin{equation}
\label{eq:Ul1}U_l=\Sigma_lP_lQ_l\,,
\end{equation}
where $Q_l$ is an arbitrary diagonal phase matrix, and $P_l$ is a permutation matrix. Moreover, it is straightforward to check that the constraint of Eq.~\eqref{eq:XlUl} arising from remnant CP is automatically fulfilled for the admissible CP transformation $X_{l\mathbf{r}}$ satisfying the restricted consistency condition in Eq.~\eqref{eq:GlXl2}. That is to say, the mixing matrix $U_{l}$ of charged leptons is fully determined by the residual flavor symmetry $G_{l}$, and the residual CP symmetry $H^{l}_{CP}$ doesn't lead to additional new constraint in the semidirect approach.

Then we proceed to the neutrino sector. The invariance of the neutrino mass matrix $m_{\nu}$ under the action of the residual symmetry $Z_2\times H^{\nu}_{CP}$ gives rise to
\begin{align}
\label{eq:mnuGnu2}&\rho^T_{\mathbf{3}}(g_\nu)m_\nu\rho_{\mathbf{3}}(g_\nu)=m_\nu,\qquad g_{\nu}\in G_{\nu}\,,\\
\label{eq:mnuXnu2}&X_{\nu\mathbf{3}}^Tm_\nu X_{\nu\mathbf{3}}=m_\nu^*,\qquad X_{\nu\mathbf{3}} \in H^{\nu}_{CP}\,,
\end{align}
where $g_\nu$ is the generator of the residual flavor symmetry $G_\nu=Z_2$ such that the equality $g^2_{\nu}=1$ is satisfied. The restricted consistency condition reads as
\begin{equation}
\label{eq:GnuXnu2}X_{\nu\mathbf{r}}\rho^*_{\mathbf{r}}(g_\nu)X_{\nu\mathbf{r}}^{-1}=\rho_{\mathbf{r}}(g_\nu),\quad
g_{\nu}\in G_{\nu},~~X_{\nu\mathbf{r}}\in H^{\nu}_{CP}\,.
\end{equation}
We denote the diagonalization matrix of $m_{\nu}$ as $U_{\nu}$ which fulfills $U_\nu^Tm_\nu U_\nu=\text{diag}(m_1, m_2, m_3)$. Neutrino oscillation experiments reveal that three light neutrino masses $m_{1,2,3}$ are not degenerate. Inserting $U_\nu^Tm_\nu U_\nu=\text{diag}(m_1, m_2, m_3)$ into Eqs.~(\ref{eq:mnuGnu2}, \ref{eq:mnuXnu2}), we can derive the following constraints on the unitary transformation $U_{\nu}$,
\begin{align}
\label{eq:GnuUnu2}&U_{\nu}^{\dagger}\rho_{\mathbf{3}}(g_\nu)U_\nu=\text{diag}(\pm 1,\pm 1,\pm 1)\,,\\
\label{eq:XnuUnu}&U_\nu^{\dagger}X_{\nu\mathbf{3}}U_{\nu}^*=\text{diag}(\pm 1,\pm 1,\pm 1)\equiv Q^2_{\nu}\,,
\end{align}
where the ``$\pm$'' signs can be chosen independently. The unitary matrix $Q_{\nu}=\text{diag}(\sqrt{\pm1},\sqrt{\pm1},\sqrt{\pm1})$ is diagonal, and its non-vanishing entries are $\pm1$ or $\pm i$. Obviously the residual CP transformation $X_{\nu\mathbf{3}}$ is a unitary symmetric matrix as well. Since $g_{\nu}$ is an element of order two and its representation matrix $\rho_{\mathbf{3}}(g_{\nu})$ satisfies $\rho^2_{\mathbf{3}}(g_{\nu})=1$, the eigenvalues of $\rho_{\mathbf{3}}(g_{\nu})$ can only be $+1$ or $-1$. Without loss of generality, we choose the three eigenvalues of $\rho_{\mathbf{3}}(g_{\nu})$ to be $+1$, $-1$ and $-1$ respectively. In the following, we shall list the procedures of how to extract the prediction for $U_{\nu}$.

Firstly $\rho_{\mathbf{3}}(g_{\nu})$ can be diagonalized by a unitary matrix $\Sigma_{\nu 1}$ with
\begin{equation}
\Sigma_{\nu 1}^{\dagger}\rho_{\mathbf{3}}(g_\nu)\Sigma_{\nu 1}=\text{diag}(1,-1,-1)\,.
\end{equation}
Note that $\Sigma_{\nu 1}$ is determined up to a unitary rotation of the second and third column vectors because $\rho_{\mathbf{3}}(g_{\nu})$ has two degenerate eigenvalues $-1$. Subsequently plugging the expression $\rho_{\mathbf{3}}(g_\nu)=\Sigma_{\nu1}\text{diag}(1,-1,-1)\Sigma^{\dagger}_{\nu 1}$ into the the consistency condition of Eq.~\eqref{eq:GnuXnu2}, we obtain
\begin{equation}
\Sigma_{\nu 1}^{\dagger}X_{\nu\mathbf{3}}\Sigma_{\nu 1}^*\text{diag}(1, -1, -1)=\text{diag}(1, -1, -1)\Sigma_{\nu 1}^{\dagger}X_{\nu\mathbf{3}}\Sigma_{\nu 1}^*\,,
\end{equation}
which implies that $\Sigma_{\nu 1}^{\dagger}X_{\nu}\Sigma_{\nu 1}^*$ is a block-diagonal matrix, and it is of the form
\begin{equation}
\Sigma_{\nu 1}^{\dagger}X_{\nu\mathbf{3}}\Sigma_{\nu 1}^*=
\left(\begin{array}{cc}
 e^{i\gamma}&0\\
 0&u_{2\times 2}
  \end{array}
\right)\,,
\end{equation}
where $u_{2\times2}$ is a symmetric unitary matrix, and it can be written as $u_{2\times 2}=\sigma_{2\times 2}\sigma_{2\times 2}^T$ by performing the Takagi factorization. As a consequence, the residual CP transformation $X_{\nu\mathbf{3}}$ can be factorized as
\begin{equation}
\label{eq:XnuSigma}
X_{\nu\mathbf{3}}=\Sigma_{\nu}\Sigma_{\nu}^T\,,
\end{equation}
where $\Sigma_{\nu}=\Sigma_{\nu 1}\Sigma_{\nu2}$ with
\begin{equation}
\Sigma_{\nu2}=\left(\begin{array}{cc}
e^{i\gamma/2}&0\\
 0&\sigma_{2\times 2}
  \end{array}
\right)\,.
\end{equation}
It is easy to check that the residual flavor symmetry transformation $\rho_{\mathbf{3}}(g_{\nu})$ can be diagonalized by $\Sigma_{\nu}$ as well,
\begin{equation}
\Sigma_{\nu}^{\dagger}\rho_{\mathbf{3}}(g_{\nu})\Sigma_{\nu}=\text{diag}(1, -1, -1)\,.
\end{equation}
Then we discuss the constraint on $U_{\nu}$ from the remnant CP. Substituting the relation $X_{\nu\mathbf{3}}=\Sigma_{\nu}\Sigma_{\nu}^T$ of Eq.~\eqref{eq:XnuSigma} into Eq.~\eqref{eq:XnuUnu}, we have
\begin{equation}
\left(Q^{\dagger}_{\nu}U_\nu^{\dagger}\Sigma_{\nu}\right)\left(Q^{\dagger}_{\nu}U_\nu^{\dagger}\Sigma_{\nu}\right)^{T}=\mathbb{1}\,.
\end{equation}
This implies that the combination $Q^{\dagger}_{\nu}U_\nu^{\dagger}\Sigma_{\nu}$ is a orthogonal matrix, and it is also a unitary matrix. Therefore $Q^{\dagger}_{\nu}U_\nu^{\dagger}\Sigma_{\nu}$ is a real orthogonal matrix denoted by $O_{3\times3}$. Then the unitary transformation $U_{\nu}$ takes the following form
\begin{equation}
\label{eq:UnuSigma}
U_{\nu}=\Sigma_{\nu} O_{3\times 3}^TQ^{\dagger}_{\nu}\,.
\end{equation}
This indicated that $U_{\nu}$ is fixed up to a real orthogonal matrix $O_{3\times3}$ by the remnant CP transformation $X_{\nu\mathbf{3}}$~\cite{Chen:2014wxa}. Furthermore, $U_{\nu}$ is subject to the constraint of residual $Z_2$ flavor symmetry shown in Eq.~\eqref{eq:GnuUnu2}, i.e.
\begin{equation}
\label{eq:GnuUnu2_Pnu}
U_{\nu}^{\dagger}\rho_{\mathbf{3}}(g_\nu)U_\nu=P^T_\nu\text{diag}(1,-1,-1)P_\nu\,,
\end{equation}
where $P_{\nu}$ is a permutation matrix, because the neutrino masses can not be pinned down in this approach and the neutrino mass spectrum can be either normal ordering (NO) or inverted ordering (IO). One finds from Eq.~\eqref{eq:GnuUnu2_Pnu} that
\begin{equation}
P_{\nu}Q_{\nu}O_{3\times3}\text{diag}(1,-1,-1)=\text{diag}(1,-1,-1)P_{\nu}Q_{\nu}O_{3\times 3}\,,
\end{equation}
which leads to
\begin{equation}
O_{3\times3}=P^T_{\nu}S^T_{23}(\theta)\,,
\end{equation}
where $S_{23}(\theta)$ is a rotation matrix, it is given by
\begin{equation}
S_{23}(\theta)=\left(
\begin{array}{ccc}
 1 & 0 & 0 \\
 0 & \cos \theta  & -\sin \theta  \\
 0 & \sin \theta  & \cos \theta
\end{array}
\right)\,.
\end{equation}
As a result, the residual symmetry $Z_2\times CP$ of the neutrino mass matrix enforces the unitary diagonalization matrix $U_{\nu}$ of the following form
\begin{equation}
\label{eq:Unu1}U_{\nu}=\Sigma_{\nu} S_{23}(\theta)P_{\nu}Q^{\dagger}_{\nu}\,.
\end{equation}
Thus we summarize the lepton mixing matrix is determined to be
\begin{equation}
\label{eq:Upmns1}
U=U_l^{\dagger}U_{\nu}=Q_l^{\dagger}P_l^T\Sigma_l^{\dagger}\Sigma_{\nu} S_{23}(\theta)P_{\nu}Q^{\dagger}_{\nu}\,.
\end{equation}
Note that PMNS matrix only depend on one free parameter $\theta$, the phase matrix $Q_{l}$ can be absorbed into the charged lepton fields, and the same result has been obtained by using various methods~\cite{Feruglio:2012cw,Chen:2014wxa}. This is our master formula to extract the mixing matrix from the postulated residual symmetry in semidirect approach. It would be frequently exploited when we scan the finite groups in section~\ref{sec:scan}.

\subsection{\label{subsec:variant_semi_app}Variant of semidirect approach}

In this scenario, the original symmetry $G_{f}\rtimes H_{CP}$ is broken down to $Z_2\times H^{l}_{CP}$ in the charged lepton sector. The generator of the residual $Z_2$ flavor symmetry group is called $g_{l}$ with $g^2_{l}=1$. For the symmetry $Z_2\times H^{l}_{CP}$ to hold, the charged lepton mass matrix has to fulfill
\begin{align}
\label{eq:mlGl3}\rho^{\dagger}_{\mathbf{3}}(g_l)m_l^{\dagger}m_l\rho_{\mathbf{3}}(g_l)&=m_l^{\dagger}m_l\,,\\
\label{eq:mlXl3}X_{l\mathbf{3}}^{\dagger}m_l^{\dagger}m_lX_{l\mathbf{3}}&=(m_l^{\dagger}m_l)^*,\qquad X_{l\mathbf{3}} \in H^{l}_{CP}\,,
\end{align}
The remnant symmetry $Z_2\times H^{l}_{CP}$ is well defined only if the restricted consistency condition is satisfied,
\begin{equation}
\label{eq:GlXl3}X_{l\mathbf{r}}\rho^*_{\mathbf{r}}(g_l)X_{l\mathbf{r}}^{-1}=\rho_{\mathbf{r}}(g_l),\qquad X_{l\mathbf{r}} \in H^{l}_{CP}\,.
\end{equation}
From Eqs.~(\ref{eq:mlGl3}, \ref{eq:mlXl3}), we find that the residual symmetry $Z_2\times H^{l}_{CP}$ leads to the following constraints on the unitary transformation $U_{l}$,
\begin{align}
\label{eq:GlUl3}&U_l^{\dagger}\rho_{\mathbf{3}}(g_l)U_l=\text{diag}(\pm 1,\pm 1,\pm 1)\,,\\
\label{eq:XlUl3}&
U_l^{\dagger}X_{l\mathbf{3}}U_l^*=\text{diag}\left(e^{i\alpha_e},e^{i\alpha_\mu},e^{i\alpha_\tau}\right)\equiv Q^2_l\,,
\end{align}
where $Q_l=\text{diag}\left(e^{i\alpha_e/2},e^{i\alpha_\mu/2},e^{i\alpha_\tau/2}\right)$ and $\alpha_{e, \mu, \tau}$ are real parameters. Note that $X_{l\mathbf{3}}$ should be symmetric, and the entries of the diagonal matrix is $\pm1$ in Eq.~\eqref{eq:GlUl3} because $g_{l}$ is of order two here. We assume that the eigenvalues of $\rho_{\mathbf{3}}(g_l)$ are $+1$, $-1$ and $-1$ without loss of generality. In the same fashion as we analyze the neutrino sector in the semidirect approach, a proper Takagi factorization of $X_{l\mathbf{3}}$ can be found to satisfy
\begin{equation}
\label{eq:Sigma_l1}X_{l\mathbf{3}}=\Sigma_{l}\Sigma^{T}_{l},\qquad \Sigma_{l}^{\dagger}\rho_{\mathbf{3}}(g_l)\Sigma_{l}=\text{diag}(1,-1,-1)\,,
\end{equation}
where $\Sigma_{l}$ is a unitary matrix. Substituting $X_{l\mathbf{3}}$ from this equation in Eq.~\eqref{eq:XlUl3} we obtain
\begin{equation}
\label{eq:CP_cons_simp}(Q^{\dagger}_lU_l^{\dagger}\Sigma_l)(Q^{\dagger}_lU_l^{\dagger}\Sigma_l)^T=\mathbb{1}\,.
\end{equation}
Hence $Q^{\dagger}_lU_l^{\dagger}\Sigma_l$ is a real orthogonal matrix denoted as $O_{3\times3}$, and thus $U_{l}$ can be expressed as
\begin{equation}
\label{eq:UlSigma3}U_l=\Sigma_{l}O_{3\times 3}^TQ^{\dagger}_l\,.
\end{equation}
Furthermore, we take into account the constraint of the residual $Z_2$ flavor symmetry,
\begin{equation}
\label{eq:GlUl3p}U_l^{\dagger}\rho_{\mathbf{3}}(g_l)U_l=P^T_l\text{diag}(1,-1,-1)P_l\,,
\end{equation}
where $P_{l}$ is a permutation matrix since no prediction can be made for the charged lepton masses. Inserting Eq.~\eqref{eq:UlSigma3} into Eq.~\eqref{eq:GlUl3p}, we obtain
\begin{equation}
\left(P_{l}Q_lO_{3\times3}\right)\text{diag}(1, -1, -1)=\text{diag}(1,-1,-1)\left(P_lQ_lO_{3\times 3}\right)\,.
\end{equation}
As a consequence, $O_{3\times3}$ can only be a block diagonal rotation matrix
\begin{equation}
\label{eq:O3x3_Char}O_{3\times3}=P^T_lS^T_{23}(\theta)\,.
\end{equation}
Hence the charged lepton mass matrix $m^{\dagger}_{l}m_{l}$ can be diagonalized by
\begin{equation}
\label{eq:Ul_charg_Z2}U_l=\Sigma_lS_{23}(\theta)P_lQ^{\dagger}_l\,.
\end{equation}
In the neutrino sector, the residual flavor symmetry $G_{\nu}$ is identified with a Klein group,
\begin{equation}
G_{\nu}=\left\{1, g_{\nu1}, g_{\nu2}, g_{\nu3}\right\}
\end{equation}
with the properties
\begin{equation}
g^2_{\nu i}=1,\quad g_{\nu i}g_{\nu j}=g_{\nu j}g_{\nu i}=g_{\nu k},~~\text{for}~~i\neq j\neq k\,.
\end{equation}
The residual CP symmetry $H^{\nu}_{CP}$ arises from the breaking of $H_{CP}$, and it has to be compatible with residual flavor symmetry $G_{\nu}$,
\begin{equation}
\label{eq:GnuXnu4}X_{\nu\mathbf{r}}\rho^*_{\mathbf{r}}(g_{\nu i})X_{\nu\mathbf{r}}^{-1}=\rho_{\mathbf{r}}(g_{\nu i}),\qquad X_{\nu\mathbf{r}} \in H^{\nu}_{CP},~~i=1, 2, 3\,.
\end{equation}
The $G_{\nu}\times H^{\nu}_{CP}$ transformation on $\nu_{L}$ leaves the Majorana neutrino mass term in Eq.~\eqref{eq:mass_Lag} invariant. This implies that
\begin{align}
\label{eq:mnuGnu4}&\rho^T_{\mathbf{3}}(g_{\nu i})m_\nu\rho_{\mathbf{3}}(g_{\nu i})=m_\nu,\quad i=1,2,3\,,\\
\label{eq:mnuXnu4}&X_{\nu\mathbf{3}}^Tm_\nu X_{\nu\mathbf{3}}=m_\nu^*,\quad X_{\nu\mathbf{3}} \in H^{\nu}_{CP}\,,
\end{align}
Equivalently the neutrino diagonalization matrix $U_{\nu}$ should satisfy
\begin{align}
\label{eq:GlUl4}&U_{\nu}^{\dagger}\rho_{\mathbf{3}}(g_{\nu i})U_{\nu}=\text{diag}\left(\pm 1, \pm 1,\pm 1\right),\\
\label{eq:XnuUnu4}&U_\nu^{\dagger}X_{\nu\mathbf{3}}U_{\nu}^*=\text{diag}(\pm 1,\pm 1,\pm 1)\equiv Q^2_{\nu}\,,
\end{align}
where $Q_{\nu}=\text{diag}\left(\sqrt{\pm1}, \sqrt{\pm1}, \sqrt{\pm1}\right)$. As $g_{\nu i}$ is of order two, we have $\text{det}\left(\rho_{\mathbf{3}}(g_{\nu i})\right)=\pm 1$. Thus each residual flavor symmetry transformation $\rho_{\mathbf{3}}(g_{\nu i})$ has a unique normalized eigenvector $v_i$ with eigenvalue equal to $\text{det}\left(\rho_{\mathbf{3}}(g_{\nu i})\right)$. These three unique eigenvectors $v_i$ ($i=1, 2, 3$, one for each non-trivial Klein group element) constitute a unitary matrix $\Sigma'_{\nu}\equiv\left(v_1, v_2, v_3\right)$. It is easy to see that $\Sigma'_{\nu}$ simultaneously diagonalizes all the three representation matrices $\rho_{\mathbf{3}}(g_{\nu i})$. Therefore $U_{\nu}$ coincides with $\Sigma'_{\nu}$ up to an arbitrary diagonal phase matrix $Q'_{\nu}$ and permutation matrix $P_{\nu}$ multiplied from the right-handed side,
\begin{equation}
\label{eq:Unu4v1}U_{\nu}=\Sigma'_{\nu}P_{\nu}Q'_{\nu}\,.
\end{equation}
From the consistency condition of Eq.~\eqref{eq:GnuXnu4}, we can straightforwardly derive that the remnant CP transformation $X_{\nu\mathbf{3}}$ would be diagonalized by $\Sigma'_{\nu}$ as follow:
\begin{equation}
\label{eq:XSigma4}\Sigma'^{\dagger}_{\nu}X_{\nu\mathbf{3}}\Sigma'^*_{\nu}=\text{diag}(e^{i\beta_e},e^{i\beta_{\mu}},e^{i\beta_{\tau}})\equiv D^2_{\nu}\,,
\end{equation}
where $D_{\nu}=\text{diag}(e^{i\beta_e/2},e^{i\beta_{\mu}/2},e^{i\beta_{\tau}/2})$, and $\beta_{e, \mu, \tau}$ are real. The diagonal matrix $Q'_{\nu}$ would contribute to the Majorana CP phases. Considering the constraint of the remnant CP transformation in Eq.~\eqref{eq:XnuUnu4} and using the relation of Eq.~\eqref{eq:XSigma4}, we find
\begin{equation}
Q'_{\nu}=P^{T}_{\nu}D_{\nu}P_{\nu}Q^{\dagger}_{\nu}\,.
\end{equation}
Therefore the unitary matrix $U_{\nu}$ is uniquely determined (up to permutations and phases of the column vectors)
\begin{equation}
U_{\nu}=\Sigma'_{\nu}D_{\nu}P_{\nu}Q^{\dagger}_{\nu}\equiv\Sigma_{\nu}P_{\nu}Q^{\dagger}_{\nu}\,,
\end{equation}
where we have denoted $\Sigma_{\nu}=\Sigma'_{\nu}D_{\nu}$.
Hence in this approach, the master formula for constructing the PMNS matrix is given by
\begin{equation}
\label{eq:Upmns2}U=U^{\dagger}_{l}U_{\nu}=Q_{l}P^{T}_{l}S^{T}_{23}(\theta)\Sigma^{\dagger}_{l}\Sigma_{\nu}P_{\nu}Q^{\dagger}_{\nu}\,,
\end{equation}
where $Q_{l}$ is unphysical as it can be absorbed by redefinition of the charged lepton fields. In contrast with the semidirect approach, one row instead of one column is fixed by the remnant symmetries while the PMNS matrix depends on a single free parameter $\theta$ in both cases.

Notice that if another pair of remnant subgroups $\{G'_l\rtimes H^{l'}_{CP},G'_{\nu}\times H^{\nu'}_{CP}\}$ are conjugate to $\{G_l\rtimes H^{l}_{CP}, G_{\nu}\times H^{\nu}_{CP}\}$ under a group element of $G_f$, i.e.,
\begin{align}
&G'_l=hG_lh^{-1},\quad G'_{\nu}=hG_{\nu}h^{-1},\quad h\in G_f\,,\\
&H^{l^{\prime}}_{CP}=\rho_{\mathbf{r}}(h)H^{l}_{CP}\rho^{T}_{\mathbf{r}}(h),\quad H^{\nu^{\prime}}_{CP}=\rho_{\mathbf{r}}(h)H^{\nu}_{CP}\rho^{T}_{\mathbf{r}}(h)\,,
\end{align}
The unitary diagonalization matrices of the charged lepton and neutrino would be related by $U'_{l}=\rho_{\mathbf{3}}(h)U_{l}$ and $U'_{\nu}=\rho_{\mathbf{3}}(h)U_{\nu}$. As a consequence, the same result for the PMNS matrix would be obtained. In Appendix~\ref{sec:equivalence}, we present the most general criteria to determine whether the predicted PMNS for different residual symmetries are equivalent.

\section{\label{sec:scan}Lepton mixing from scan of finite groups and phenomenology}

In this section, we shall perform an exhaustive scan over the discrete groups of order less than 2000 with the help of the computer algebra  program \texttt{GAP}~\cite{GAP}, and all the possible lepton mixing patterns achievable from the semidirect approach and the variant of the semidirect approach would be studied. In order to avoid duplicating subgroups which have been scanned, we shall only consider the groups with faithful three-dimensional irreducible representations. In our previous work, the possible lepton flavor mixing from flavor symmetry breaking (without generalized CP) have been systematically analyzed~\cite{Yao:2015dwa}, and all discrete groups of size smaller than 2000 are considered by using \texttt{GAP}. The CP symmetry would be taken into account further in the present work.

As a proper generalized CP symmetry corresponds to a class-inverting automorphism of the flavor symmetry group~\cite{Chen:2014tpa}, we should firstly determine whether a finite group have a class-inverting automorphism. The \texttt{GAP} command \texttt{AutomorphismGroup(.)} can be exploited to obtain all the automorphisms of a given group $G_f$, then we can search for the existence of class-inverting automorphisms which map the classes of $G_f$ into their inverse. However, this might be a tough job for groups of large order, since there are generically large amount of  automorphisms. We notice that all the automorphisms of $G_{f}$
constitute a group called automorphism group $\mathrm{Aut}(G_f)$.
The inner automorphism group $\mathrm{Inn}(G_f)$ is generated by the group conjugation $\mu_h:g\to hgh^{-1}$ with $h,g\in G_f$. $\mathrm{Inn}(G_f)$ is a normal subgroup of $\mathrm{Aut}(G_f)$, and it can be easily obtained by using the command \texttt{InnerAutomorphismsAutomorphismGroup(.)}. Obviously the inner automorphism maps each conjugacy class into itself. As a result, if $\mathfrak{u}$ is a class-inverting automorphism, so will be the composition $\mu_{h}\circ\mathfrak{u}$. The search for class-inverting automorphism can be greatly simplified by considering the quotient group $\mathrm{Out}(G_f)\equiv\mathrm{Aut}(G_{f})/\mathrm{Inn}(G_f)$ which is called outer automorphism group. $\mathrm{Out}(G_f)$ can be obtained by the \texttt{GAP} command \texttt{NaturalHomomorphismByNormalSubgroup(.)}. If there exists a class-inverting outer automorphism, a generalized CP transformation consistent with $G_{f}$ can be imposed for a generic field content. For a class-inverting outer automorphism $\mathfrak{u}$, the corresponding CP transformation $X_{0\mathbf{r}}$ can be
fixed by solving the consistency equation
\begin{equation}
\label{eq:cons_Gf}X_{0\mathbf{r}}\rho^*_{\mathbf{r}}(g)X_{0\mathbf{r}}^{-1}=\rho_{\mathbf{r}}(\mathfrak{u}(g)),\qquad g\in G_{f}\,.
\end{equation}
Note that it is sufficient to impose this consistency equation on the generators of $G_{f}$. Including the contribution of the inner automorphism, the most general CP transformation compatible with the flavor symmetry $G_{f}$ takes the form
\begin{equation}
X_{\mathbf{r}}=\rho_{\mathbf{r}}(h)X_{0\mathbf{r}},\qquad h\in G_{f}\,.
\end{equation}
On the other hand, if $G_{f}$ doesn't possess a class-inverting automorphism, CP symmetry can only be introduced in the case that a special subset of irreducible representations is present in a model. We shall not consider such flavor symmetry since the generalized CP symmetry and the resulting predictions for lepton mixing are model dependent.

The residual flavor symmetries $G_{l}$ and $G_{\nu}$ are Abelian subgroups of the flavor symmetry $G_{f}$~\cite{Yao:2015dwa,Chen:2014wxa,Chen:2015nha}. Hence we find all the Abelian subgroups of $G_{f}$ with \texttt{GAP}, and the corresponding group structures and generators are extracted. For a generic residual flavor symmetry group $G_{R}$ which can be either $G_{l}$ or $G_{\nu}$, the residual CP transformation $X_{R\mathbf{r}}=\rho_{\mathbf{r}}(f_{R})X_{0\mathbf{r}}$ with $f_{R}\in G_{f}$ should be a symmetric unitary matrix and it satisfies the consistency condition
\begin{equation}
X_{R\mathbf{r}}\rho^*_{\mathbf{r}}(h_{R})X_{R\mathbf{r}}^{-1}=\rho_{\mathbf{r}}(h^{-1}_{R}),\qquad h_{R}\in G_{R}\,,
\end{equation}
which gives rise to
\begin{equation}
\label{eq:res_X_gap}f^{-1}_{R}h^{-1}_{R}f_{R}=\mathfrak{u}(h_{R})\,.
\end{equation}
The permissible solutions to $f_{R}$ can be straightforwardly found by \texttt{GAP}. Notice that $G_{R}$ is an Abelian group, therefore all the elements in the right coset $G_{R}f_{R}$ also satisfy Eq.~\eqref{eq:res_X_gap} for a given solution $f_{R}$. In other words,  $\rho_{\mathbf{r}}(h_{R})X_{R\mathbf{r}}$ with $h_{R}\in G_{R}$ is also an admissible residual CP transformation, and it imposes the same constraints on the lepton mass matrices as $X_{R\mathbf{r}}$ because of the remnant flavor symmetry invariance. In this manner, we can find out all the possible remnant CP symmetries $H^{l}_{CP}$ and $H^{\nu}_{CP}$ which are compatible with the postulated remnant flavor symmetry groups $G_{l}$ and $G_{\nu}$ respectively.

Our comprehensive scan over the discrete finite group up to order 2000 reveals that there are 574 groups which possess both faithful three-dimensional irreducible representation and class-inverting automorphism. For each of the 574 groups, the class-inverting automorphism and the corresponding CP transformation $X_{0\mathbf{r}}$ in the triplet representation, its Abelian subgroups as well as the residual CP transformations are calculated. Furthermore, we investigate the possible lepton mixing patterns achievable from the semidirect approach and the variant of the semidirect approach by considering all the admitted residual symmetries. The predictions for the PMNS matrix are obtained by using the master formulas in Eqs.~(\ref{eq:Upmns1}, \ref{eq:Upmns2}). In order to measure quantitatively how well the obtained mixing patterns can explain the current experimental data, we perform a conventional $\chi^2$ analysis. The $\chi^2$ function is defined in the usual way
\begin{equation}
\chi^2=\sum_{ij=12,13,23}\frac{\left(\sin^2\theta_{ij}-\left(\sin^2\theta_{ij}\right)^{\mathrm{bf}}\right)^2}{\sigma^2_{ij}}\,,
\end{equation}
where $\sin^2\theta_{ij}$ are the mixing angles predicted for different remnant symmetries, and they depend on the free parameter $\theta$. $\left(\sin^2\theta_{ij}\right)^{\mathrm{bf}}$ denote the best fit values of the lepton mixing angles and $\sigma_{ij}$ their corresponding $1\sigma$ errors. We use the current global fit of neutrino oscillation data in Ref.~\cite{Gonzalez-Garcia:2014bfa}. The results of our analysis are available at the website~\cite{webdata}.
It is remarkable that we find many interesting mixing patterns which can accommodate the experimental data on lepton mixing for certain values of $\theta$. Moreover, these phenomenologically viable mixing patterns can be categorized into several cases, as will be shown below.

\subsection{\label{subsec:semidirect_scan}Mixing patterns derived from semidirect approach}

In this section we shall report the lepton mixing patterns which can be obtained in the semidirect approach. The contributions of the permutations of the rows and columns would be considered. We shall give the analytical expressions for mixing angles and CP invariants $J_{CP}$, $I_1$ and $I_2$. Moreover, the resulting phenomenological implications in neutrinoless double decay and leptogenesis will be discussed. In the following, three rotation matrices $S_{12}(\theta)$, $S_{13}(\theta)$ and $S_{23}(\theta)$ would be used with the convention
\begin{equation}
  \label{eq:RotationMatrix}
\begin{aligned}
S_{12}(\theta)&=\left(
\begin{array}{ccc}
 \cos\theta  & -\sin\theta  & 0 \\
 \sin\theta  & \cos\theta  & 0 \\
 0 & 0 & 1
\end{array}
\right)\,,\\
S_{13}(\theta)&=\left(
\begin{array}{ccc}
 \cos\theta  & 0 & \sin\theta  \\
 0 & 1 & 0 \\
 -\sin\theta  & 0 & \cos\theta
\end{array}
\right)\,,\\
  S_{23}(\theta)&=\left(
\begin{array}{ccc}
 1 & 0 & 0 \\
 0 & \cos \theta & -\sin \theta \\
 0 & \sin \theta  & \cos \theta
\end{array}
\right)\,.
\end{aligned}
\end{equation}
The permutation matrices $P_{l}$ and $P_{\nu}$ in Eq.~\eqref{eq:Upmns1} can take the following six forms:
\begin{equation}
\begin{aligned}
\label{eq:per_matrix}
P_{123}=&\left(\begin{array}{ccc}
1  & 0  &  0 \\
0  & 1  &  0\\
0  & 0  &  1
\end{array}\right),\quad
P_{231}=\left(\begin{array}{ccc}
0&  1&  0 \\
0&  0&  1  \\
1&  0&  0
\end{array}\right),\quad
P_{312}=\left(\begin{array}{ccc}
0&  0  &1  \\
1&  0  &0 \\
0&  1  &  0
\end{array}\right)\,,\\
P_{132}=&\left(\begin{array}{ccc}
1  &  0 &  0 \\
0  &  0 &  1 \\
0  &  1 &  0
\end{array}\right),\quad
P_{213}=\left(\begin{array}{ccc}
0  &  1  &  0 \\
1  &  0  &  0 \\
0  &  0  &  1
\end{array}\right),\quad
P_{321}=\left(\begin{array}{ccc}
0 &0 &1  \\
0 &1 &0  \\
1 &0 &0
\end{array}\right)\,.
\end{aligned}
\end{equation}
It is known that if the second and third rows of the PMNS matrix are exchanged, the atmosphere mixing angle $\theta_{23}$ becomes $\pi/2-\theta_{23}$, the Dirac $CP$ phase $\delta_{CP}$ becomes $\pi+\delta_{CP}$, and other mixing parameters are invariant. Therefore generically the two permutations of a certain pattern related through the exchange of the second and third rows of the PMNS matrix can (or can't) accommodate the experimental data on mixing angles simultaneously, as will be shown in the following.

\begin{description}[labelindent=-0.8em, leftmargin=0.3em]
\item[Case \uppercase\expandafter{\romannumeral1}(a)]

\begin{equation}
\label{eq:pmnsIa}
U^{I(a)}=\frac{1}{\sqrt{3}}
\left(\begin{array}{ccc}
\sqrt{2} \sin \varphi _1 & e^{i \varphi _2} & \sqrt{2}\cos\varphi_1 \\
\sqrt{2}\cos\left(\varphi _1-\frac{\pi }{6}\right) & -e^{i\varphi_2} & -\sqrt{2} \sin \left(\varphi_1-\frac{\pi}{6}\right) \\
\sqrt{2}\cos\left(\varphi_1+\frac{\pi}{6}\right)   &  e^{i\varphi_2} & -\sqrt{2}\sin\left(\varphi_1+\frac{\pi}{6}\right)
\end{array}
\right)S_{23}(\theta)Q^{\dagger}_{\nu}\,,
\end{equation}
where $\varphi_1$ and $\varphi_2$ are rational angles, and they are determined by the residual symmetries. The mixing patterns originating from the permutations of rows are related to this matrix through a redefinition of the parameters $\varphi_1$ and $\theta$. The viable values of $\varphi_1$ and $\varphi_2$ and the corresponding representative flavor symmetry groups are collected in table~\ref{tab:Ia}. Note that the mixing patterns with the signs of $\varphi_1$ and $\varphi_2$ reversed can also can be produced, and the same predictions for the mixing angles are obtained except all the $CP$ phases become their opposite. However, these viable values are not shown in table~\ref{tab:Ia} in order not to appear too lengthy. From this table, we can see that most of the groups can predict more than one mixing patterns, and some groups predict the same mixing patterns. We only show one or two representative flavor symmetry groups in table~\ref{tab:Ia}, and a full summary of the results is available at our website~\cite{webdata}. The subscripts $\Delta$ and $\Delta^{\prime}$ of the group identity denote that the corresponding groups belong to the type D group series $D_{n,n}^{(0)}\cong\Delta(6n^2)$ and $D_{9n^{\prime},3n^{\prime}}^{(1)}\cong(Z_{9n^{\prime}}\times Z_{3n^{\prime}})\rtimes S_3$, respectively. It is notable that all these interesting mixing patterns can be obtained from the $\Delta(6n^2)$ or $D_{9n^{\prime},3n^{\prime}}^{(1)}$ flavor symmetry groups combined with CP symmetry. In particular, widely studied smaller groups $S_4\cong[24,12]$ and $\Delta(96)\cong [96,64]$ can admit a reasonably good fit to the experimental data. This is compatible with the known results in the literature~\cite{Feruglio:2012cw,Ding:2013hpa,Li:2014eia,Li:2013jya,Ding:2014ssa}.
From the PMNS matrix $U^{I(a)}_{PMNS}$ in Eq.~\eqref{eq:pmnsIa}, we can read out the lepton mixing angles as follow
\begin{equation}
\begin{aligned}
\sin^2\theta_{13}=&\frac{1}{3}\left(1+\cos^2\theta\cos2\varphi_1-\sqrt{2}\sin2\theta\cos\varphi_1\cos\varphi_2\right),\\
\sin^2\theta_{12}=&\frac{1+\sin^2\theta\cos2\varphi_1+\sqrt{2}\sin2\theta\cos\varphi_1\cos\varphi_2}
{2-\cos^2\theta\cos2\varphi_1+\sqrt{2}\sin2\theta\cos\varphi_1\cos\varphi_2},\\
\sin^2\theta_{23}=&\frac{1-\cos^2\theta\sin\left(\pi/6+2\varphi_1\right)+\sqrt{2}\sin2\theta\cos\varphi_2\sin\left(\pi/6-\varphi_1\right)}
{2-\cos^2\theta\cos2\varphi_1+\sqrt{2}\sin2\theta\cos\varphi_1\cos\varphi_2}\,.
\end{aligned}
\end{equation}
We see that the solar and reactor mixing angles are correlated as,
\begin{equation}
3\cos^2\theta_{12}\cos^2\theta_{13}=2\sin^2\varphi_1\,,
\end{equation}
For the experimentally measured values $0.270\leq \sin^2\theta_{12} \leq 0.344$ and $0.0188 \leq \sin^2\theta_{13} \leq 0.0251$ at $3\sigma$ level~\cite{Gonzalez-Garcia:2014bfa},  we find the allowed intervals of the parameter $\varphi_1$ is
\begin{equation}
\label{eq:varphi_range} \varphi _1\in[0.435\pi, 0.565\pi]\cup [1.435\pi, 1.565\pi]
\end{equation}
Obviously $\varphi_1$ should be around $\pi/2$ or $3\pi/2$. Moreover, the three CP rephasing invariants $J_{CP}$ , $I_1$ and $I_2$ are predicted to be
\begin{equation}
\begin{aligned}
&\left|J_{CP}\right|=\frac{1}{6\sqrt{6}}\left|\sin2\theta\sin\varphi_2\sin3\varphi_1\right|\,,\\
&|I_1|=\frac{4}{9} \left|\cos \theta  \sin ^2\varphi _1 \sin \varphi _2 \left(\cos \theta  \cos \varphi _2+\sqrt{2} \sin \theta  \cos \varphi _1\right)\right|\,,\\
&|I_2|=\frac{4}{9} \left|\sin \theta  \sin ^2\varphi _1 \sin \varphi _2 \left(\sin \theta  \cos \varphi _2-\sqrt{2} \cos \theta  \cos \varphi _1\right)\right|\,.
\end{aligned}
\end{equation}
The above three CP invariants are conventionally defined as~\cite{Jarlskog:1985ht,Branco:2011zb,Branco:1986gr,Jenkins:2007ip}
\begin{equation}
\begin{split}
J_{CP}&\equiv\Im\left(U_{11}U_{33}U^{*}_{13}U^{*}_{31}\right)=\frac{1}{8}\sin2\theta_{12}\sin2\theta_{13}\sin2\theta_{23}\cos\theta_{13}\sin\delta_{CP}\,,\\
I_1&\equiv\Im\left(U^{*2}_{11}U^2_{12}\right)=\frac{1}{4}\sin^22\theta_{12}\cos^4\theta_{13}\sin\alpha_{21}\,,\\
I_2&\equiv\Im\left(U^{\ast2}_{11}U^2_{13}\right)=\frac{1}{4}\sin^22\theta_{13}\cos^2\theta_{12}\sin(\alpha_{31}-2\delta_{CP})\,,
\end{split}
\end{equation}
where $\delta_{CP}$ is the Dirac CP violation phase, $\alpha_{21}$ and $\alpha_{31}$ are the Majorana CP phases in the standard parameterization of the lepton mixing matrix~\cite{Agashe:2014kda}. In this work, we shall present the absolute values of $J_{CP}$, $I_1$ and $I_2$ because the signs of $I_1$ and $I_2$ depend on the CP parity of the neutrino states which is encoded in the matrix $Q_{\nu}$ and the overall signs of all the three CP invariant would be changed if the left-handed lepton doublets are assigned to conjugate triplet $\bar{\mathbf{3}}$ instead of $\mathbf{3}$.

Furthermore, we can derive the following exact sum rule among the mixing angles and Dirac CP phase,
\begin{equation}
\label{eq:angle_phse_caseIa}\cos\delta_{CP}=\frac{\cos2\theta_{23}\left(3\cos2\theta_{12}-2\sin^2\varphi_1\right)+\sqrt{3}\sin2\varphi_1}{3\sin2\theta_{12}\sin\theta_{13}\sin2\theta_{23}}\,.
\end{equation}
This sum rule can also be obtained from $|U_{\mu1}|^2=2\cos^2(\varphi_1-\pi/6)/3$ and $|U_{\tau1}|^2=2\cos^2(\varphi_1+\pi/6)/3$. Because the parameter $\varphi_1$ should be around $\pi/2$ or $3\pi/2$ as shown in Eq.~\eqref{eq:varphi_range}, the sum rule of Eq.~\eqref{eq:angle_phse_caseIa} is approximately
\begin{equation}
\label{eq:sum_app_caseIa}\cos\delta_{CP}\simeq\frac{(3\cos2\theta_{12}-2)\cot2\theta_{23}}{3\sin2\theta_{12}\sin\theta_{13}}\,.
\end{equation}
This implies that $\delta_{CP}$ would be nearly maximal if the atmospheric angle $\theta_{23}$ takes the maximal value $\theta_{23}=\pi/4$. We allow the three mixing angles to freely vary in the experimentally preferred $3\sigma$ ranges~\cite{Gonzalez-Garcia:2014bfa}, then the sum rule Eq.~\eqref{eq:sum_app_caseIa} leads to
\begin{equation}
-0.643\leq\cos\delta_{CP}\leq0.819\,.
\end{equation} 
Needless to say, the improved measurement of the mixing angles particularly $\theta_{12}$ and $\theta_{23}$ could help to make more precise prediction for $\delta_{CP}$ in our framework.

If the light neutrinos with definite mass $\nu_i$ are Majorana fermions, their exchange can trigger the neutrinoless double beta ($0\nu\beta\beta$) decay processes $(A, Z)\rightarrow (A, Z+2)+e^{-}+e^{-}$ in which the total lepton number changes by two units. Most importantly, the experimental detection of this lepton number violating decay will proof the Majorana nature of neutrinos. In addition, the lifetime of the $0\nu\beta\beta$ decay is related to the neutrino masses so that its measurement will also probe the unknown absolute neutrino mass and hierarchy. The $0\nu\beta\beta$ decay amplitude has the form $\mathcal{A}^{0\nu\beta\beta}=G^2_{F}m_{ee}\mathcal{M}^{0\nu\beta\beta}$, where where $G_F$ is the Fermi constant, $m_{ee}$ is the $0\nu\beta\beta$ decay effective Majorana mass and $\mathcal{M}^{0\nu\beta\beta}$ is the nuclear matrix element of the process. The effective mass $m_{ee}$ contains all the dependence of $\mathcal{A}^{0\nu\beta\beta}$ on the neutrino mixing parameters with~\cite{Agashe:2014kda}
\begin{eqnarray}
\nonumber\hskip-0.2in|m_{ee}|&=&\left|\sum^{3}_{i=1}m_iU^2_{1i}\right|\\
\hskip-0.2in&=&\left|m_1\cos^2\theta_{12}\cos^2\theta_{13}+m_2\sin^2\theta_{12}\cos^2\theta_{13}e^{i\alpha_{21}}+m_3\sin^2\theta_{13}e^{i(\alpha_{31}-2\delta_{CP})}\right|\,,
\end{eqnarray}
where $m_{1,2,3}$ are the light Majorana neutrino masses. One can see that $m_{ee}$ depends on the values of the Majorana phase $\alpha_{21}$ and the Majorana-Dirac phase difference $\alpha_{31}'\equiv\alpha_{31}-2\delta_{CP}$. We recall that the two heavier neutrino masses can be expressed in
terms of the lightest neutrino mass and the two neutrino mass-squared differences measured in neutrino oscillation experiments. For the NO spectrum, one gets
\begin{equation}
m_1=m_{\mathrm{lightest}},\quad m_2=\sqrt{m^2_{\mathrm{lightest}}+\Delta m^2_{21}},\quad m_3=\sqrt{m^2_{\mathrm{lightest}}+\Delta m^2_{31}}\,,
\end{equation}
while for the IO spectrum:
\begin{equation}
m_1=\sqrt{m^2_{\mathrm{lightest}}-\Delta m^2_{32}-\Delta m^2_{21}},\quad m_2=\sqrt{m^2_{\mathrm{lightest}}-\Delta m^2_{32}},\quad m_3=m_{\mathrm{lightest}}\,,
\end{equation}
where $\Delta m^2_{ij}=m^2_{i}-m^2_{j}$. In our numerical analysis, we shall use the best fit values of $\Delta m^2_{21}$ and $\Delta m^2_{31(32)}$ obtained in the global analysis~\cite{Gonzalez-Garcia:2014bfa},
\begin{equation}
\Delta m^2_{21}=7.50\times10^{-5}\text{eV}^2,\quad \Delta m^2_{31}=2.457\times10^{-3}\text{eV}^2,\quad  \Delta m^2_{32}=-2.449\times10^{-3}\text{eV}^2\,,
\end{equation}
where the quoted values of $\Delta m^2_{31}$ and $\Delta m^2_{32}$ correspond to the NO and IO spectrums, respectively. The numerical results would only change a little bit if the experimental uncertainties of the neutrino mass squared splittings are considered. For the mixing pattern $U^{I(a)}$, the effective Majorana mass $|m_{ee}|$ is given by
\begin{equation}
\begin{split}
|m_{ee}|=\frac{1}{3} \left|2 m_1\sin ^2\varphi _1+q_1m_2\left(e^{i \varphi _2} \cos \theta +\sqrt{2} \cos \varphi _1\sin \theta \right)^2 \right.\\\left.+  q_2 m_3\left(\sqrt{2} \cos \theta \cos\varphi _1-e^{i \varphi _2} \sin \theta \right)^2\right|\,,
\end{split}
\end{equation}
where $q_1,q_2=\pm 1$ originates from the ambiguity of the CP parity matrix $Q_{\nu}$. We show $|m_{ee}|$ versus the lightest neutrino mass $m_{\text{lightest}}$ in Fig.~\ref{fig:mee_CaseI}, where the three
mixing angles are required to lie in the $3\sigma$ regions. We display the allowed ranges of the effective mass $|m_{ee}|$ under the assumption of $\varphi_1$ and $\varphi_2$ as free continuous continuous parameters and for the specific value of $(\varphi_1, \varphi_2)=(\pi/2, \pi/2)$. The case of $(\varphi_1, \varphi_2)=(\pi/2, \pi/2)$ can be naturally reproduced from the $S_4$ flavor symmetry combined with CP symmetry. Accordingly $|m_{ee}|$ is predicted to close to 0.017eV or around the upper bound 0.048eV for IO neutrino mass spectrum, which are within the future sensitivity of forthcoming $0\nu\beta\beta$ decay experiments. However, for NO spectrum, $|m_{ee}|$ strongly depends on the lightest neutrino mass $m_{\mathrm{lightest}}$, and it can even be approximately vanishing for particular value of $m_{\mathrm{lightest}}$. Although exploring the NO region experimentally is beyond the
reach of any planned experiment, if $0\nu\beta\beta$ decays are not
observed and neutrino oscillation experiments establish that the neutrino masses are NO, it would be important to test $|m_{ee}|$ values in the NO region by combining the information on the absolute mass scale from cosmology.

It is recently found that lepton flavor mixing as well as leptogenesis is strongly constrained by the residual discrete flavor and CP symmetries of the neutrino and charged lepton sectors~\cite{Chen:2016ptr}. For the widely studied scenario of leptogenesis in type-I seesaw model with a hierarchical heavy neutrinos mass spectrum $M_{2,3}\gg M_1$, the CP asymmetry generated by the $N_1$ decay process $N_1\to l_\alpha+H$, $\alpha=e,\mu,\tau$ process is approximately given by~\cite{Covi:1996wh,Endoh:2003mz,Abada:2006ea,Abada:2006fw,Fong:2013wr}
\begin{equation}
\label{eq:epsilon_R}
\begin{split}
\epsilon_\alpha&\equiv\frac{\Gamma(N_1\rightarrow H l_{\alpha})-\Gamma(
N_1\rightarrow \overline{H} \overline{l}_{\alpha})}{
\sum_{\alpha}[\Gamma(N_1\rightarrow H
l_{\alpha})+\Gamma(N_1\rightarrow \overline{H} \overline{l}_{\alpha})]}\\
&=-\frac{3M_1}{16\pi v^2}\frac{\Im\left(\sum_{ij}\sqrt{m_im_j}\,m_jR_{1i}R_{1j}U^{\ast}_{\alpha i}U_{\alpha j}\right)}{\sum_j m_j|R_{1j}|^2}\,,
\end{split}
\end{equation}
where $v$ is the Higgs vacuum expectation value given by $v=174$ GeV, $U$ is the PMNS matrix, and $R$ is the Casas-Ibarra parametrization of the neutrino Yukawa matrix $\lambda$~\cite{Casas:2001sr}:
\begin{equation}
R=vM^{-\frac{1}{2}}\lambda Um^{-\frac{1}{2}}\,,
\end{equation}
where $M\equiv\text{diag}(M_1, M_2, M_3)$ and $m\equiv\text{diag}(m_1, m_2, m_3)$. One sees that $R$ is a generic complex orthogonal matrix fulfilling $RR^{T}=R^{T}R=1$. Besides the CP asymmetry parameter $\epsilon_{\alpha}$, the final baryon asymmetry depends on
washout mass parameter $\widetilde{m}_\alpha$ for each flavor $\alpha$ with
\begin{equation}
\label{eq:mtilde_alpha}\widetilde{m}_\alpha=\Big|\sum_j m_j^{1/2} R_{1j} U_{\alpha j}^*\Big|^2\,.
\end{equation}
In the present work we will be concerned with temperature window $10^{9}\,\text{GeV}\leq T\sim M_1\leq 10^{12}$ GeV. In this range only the interactions mediated by the $\tau$ Yukawa coupling are in equilibrium, and the final baryon asymmetry is well approximated by
\begin{equation}
\label{eq:Yb2}Y_{B}\simeq -\frac{12}{37\,g^*}\left[\epsilon_{2}\eta\left(\frac{417}{589}{\widetilde{m}_2}\right)\,+\,\epsilon_{\tau}\eta\left(\frac{390}{589}{\widetilde{m}_{\tau}}\right)\right]\,,
\end{equation}
where $g_{\ast}$ is the effective number of spin-degrees of
freedom in thermal equilibrium with $g_{\ast}=106.75$ in the standard model, $\epsilon_2=\epsilon_{e}+\epsilon_{\mu}$, $\widetilde{m}_2=\widetilde{m}_{e}+\widetilde{m}_{\mu}$ and
\begin{equation}
\label{eq:efficency_factor}\eta(\widetilde{m}_{\alpha})\simeq\left[\left(\frac{\widetilde{m}_{\alpha}}{8.25\times 10^{-3}\,{\rm eV}}\right)^{-1}+\left(\frac{0.2\times 10^{-3}\,{\rm eV}}{\widetilde{m}_{\alpha}}\right)^{-1.16}\ \right]^{-1}\,.
\end{equation}
Then we recapitulate the main results for leptogenesis predicted by residual flavor and CP symmetries in Ref.~\cite{Chen:2016ptr}. If both the neutrino Yukawa coupling and the RH neutrino mass matrix (after the electroweak and flavor symmetries breaking) are invariant under two set of residual CP transformation $X_{\nu1}$, $X_{\nu2}$ of the LH neutrino fields $\nu_{L}$ and $X_{N1}$, $X_{N2}$ of the RH neutrino fields, or equivalently a $Z_2$ flavor symmetry and a CP symmetry are preserved in the neutrino sector, the $R$-matrix would be constrained to be block diagonal~\cite{Chen:2016ptr},
\begin{equation}
\label{eq:R2cp}P_{N}RP^{T}_{\nu}=
\begin{pmatrix}
\times &\; 0 & \;0\\
0 &\;\times  &\; \times\\
0 &\;\times  &\;\times
\end{pmatrix}\,,
\end{equation}
where the notation ``$\times$'' denotes a nonzero matrix element, $P_N$ and $P_{\nu}$ are the permutation matrices. In order to generate a nonvanishing lepton asymmetry, there cannot be two zero elements in the first row of the $R-$matrix. As a consequence, depending on the values of $P_{\nu}$, we have three possible cases named $C_{12}, C_{13}$ and $C_{23}$~\cite{Chen:2016ptr},
\begin{equation}
\label{eq:R_1st_row}C_{12}: R=\begin{pmatrix}\times&\times&0\\... \end{pmatrix}\,,\quad
C_{13}: R=\begin{pmatrix}\times&0&\times\\... \end{pmatrix}\,,\quad
C_{23}: R=\begin{pmatrix}0&\times&\times\\... \end{pmatrix}\,.
\end{equation}
Furthermore, each element of $R$-matrix is either real or purely imaginary because of the residual CP invariance. To facilitate the discussion, we introduce the notations
\begin{equation}
U^{\prime}=UQ_{\nu1},\qquad R^{\prime}=Q_{N1}R Q_{\nu1}\,,
\end{equation}
where $Q_{N1}$ and $Q_{\nu1}$ are the CP parity matrices of the RH and LH neutrino fields respectively, they are diagonal matrices with entries $\pm1$ and $\pm i$, and their values are not constrained by residual symmetries. Thus $R^{\prime}$ would be a block diagonal real matrix, and it satisfies
\begin{equation}
\sum^{3}_{i=1}R'^{2}_{1i}K_i=1\,,
\end{equation}
where $K_{i}$ is equal to $+1$ or $-1$ with
\begin{equation}
K_i= (Q^2_{N1})_{11}(Q^2_{\nu1})_{ii}\,.
\end{equation}
Moreover, for each case $C_{ab}$ with $ab=12$, 13 and 23 listed in Eq.~\eqref{eq:R_1st_row}, the lepton asymmetry $\epsilon_{\alpha}$ and washout mass $\widetilde{m}_\alpha$ can be written into a quite simple form
\begin{eqnarray}
\label{eq:epsilon_alpha_2CP}\epsilon_\alpha&=&-\frac{3M_1}{16\pi v^2}W_{ab}\,I^{\alpha}_{ab}\,,\\
\label{eq:m_tilde_2CP}\widetilde{m}_\alpha&=&\left|m^{1/2}_aR^{\prime}_{1a}U^{\prime}_{\alpha a}+m^{1/2}_{b}R^{\prime}_{1b}U^{\prime}_{\alpha b}\right|^2\,,
\end{eqnarray}
where
\begin{equation}
\label{eq:W_2CP}W_{ab}=\frac{\sqrt{m_am_b}\,R^{\prime}_{1a}R^{\prime}_{1b}(m_aK_a-m_bK_b)}{m_a(R_{1a}^{\prime})^2+m_b(R_{1b}^{\prime})^2},\qquad I^{\alpha}_{ab}=\mathrm{Im}\big(U^{\prime}_{\alpha a}U^{\prime*}_{\alpha b}\big)\,.
\end{equation}
We would like to remind the readers that the repeated indices are not summed over in Eqs.~(\ref{eq:epsilon_alpha_2CP}, \ref{eq:m_tilde_2CP}, \ref{eq:W_2CP}). We notice that the lepton asymmetry $\epsilon_{\alpha}$ are closely related to the lower energy CP phases in this framework. The observation of CP violation in future neutrino oscillation and neutrinoless double decay experiments would imply the existence of a baryon asymmetry.
We give the most general parametrization of the first column of $R^{\prime}$ and corresponding expressions of $W_{12}$, $W_{13}$ and $W_{23}$ in table~\ref{tab:R_W_para}. For the predicted mixing pattern $U^{I(a)}$ in Eq.~\eqref{eq:pmnsIa}, the rephasing invariants $I^{\alpha}_{23}$ are of the form
\begin{equation}
\begin{split}
I^e_{23}=&\frac{\sqrt{2}}{3}\cos\varphi_1\sin\varphi_2\,,\\
I^{\mu}_{23}=&-\frac{\sqrt{2}}{3}\sin\left(\frac{\pi}{6}-\varphi_1\right)\sin\varphi_2\,,\\
I^{\tau}_{23}=&-\frac{\sqrt{2}}{3}\sin\left(\frac{\pi}{6}+\varphi_1\right)\sin\varphi_2\,.
\end{split}
\end{equation}
As shown in table~\ref{tab:Ia}, the parameter values $(\varphi_1,\varphi_2)=(\pi/2, \pi/2)$ can be obtained when the flavor symmetry group $G_{f}$ is $S_4$. Accordingly both atmospheric mixing angle and Dirac CP phase are predicted to be maximal. We find that the best fit value of the parameter $\theta$ is $\theta_{\mathrm{bf}}=\pm 0.082\pi (\pm 0.083\pi)$, and the global minimum of the $\chi^2$ function is $\chi^2_{\text{min}}=2.089 (5.783)$ for NO (IO) spectrum. The predictions for $Y_{B}$ as a function of the parameter $\eta$ are plotted in figure~\ref{fig:leptogenesis_CaseIa}. We see that the realistic value of $Y_{B}$ can be reproduced for appropriate values of $\eta$ except in the case of NO with $(K_1, K_2, K_3)=(\pm,-,+)$, while for IO spectrum the correct value of $Y_B$ can be achieved when  $(K_1, K_2, K_3)=(\pm,+,-)$ for $\theta_{\mathrm{bf}}=0.083\pi$ or $(K_1, K_2, K_3)=(\pm,+,-)$, $(\pm,-,+)$ for $\theta_{\mathrm{bf}}=-0.083\pi$.

\begin{table}[hptb]
\begin{center}
\begin{tabular}{|m{0.25\columnwidth}<{\centering}|m{0.7\columnwidth}<{\centering}|}
\hline\hline
\texttt{Group Id} & $(\varphi_1,\varphi_2)$ \\
\hline
$[24, 12]_{\vartriangle}$, $[48, 48]$ & $\left(\frac{\pi }{2},\frac{\pi }{2}\right)$\\ \hline $[150, 5]_{\vartriangle}$, $[300, 26]$ & $\left(\frac{7 \pi }{15},-\frac{\pi }{5}\right)$, $\left(\frac{7 \pi }{15},0\right)$, $\left(\frac{7 \pi }{15},\frac{2 \pi }{5}\right)$, $\left(\frac{8 \pi }{15},-\frac{\pi }{5}\right)$, $\left(\frac{8 \pi }{15},0\right)$, $\left(\frac{8 \pi }{15},\frac{2 \pi }{5}\right)$\\ \hline $[162, 10]$, $[162, 12]$ & $\left(\frac{5 \pi }{9},0\right)$, $\left(\frac{5 \pi }{9},\frac{\pi }{3}\right)$\\ \hline $[294, 7]_{\vartriangle}$, $[588, 39]$ & $\left(\frac{10 \pi }{21},-\frac{3 \pi }{7}\right)$, $\left(\frac{10 \pi }{21},-\frac{2 \pi }{7}\right)$, $\left(\frac{10 \pi }{21},-\frac{\pi }{7}\right)$, $\left(\frac{10 \pi }{21},0\right)$, $\left(\frac{11 \pi }{21},-\frac{3 \pi }{7}\right)$, $\left(\frac{11 \pi }{21},-\frac{2 \pi }{7}\right)$, $\left(\frac{11 \pi }{21},-\frac{\pi }{7}\right)$, $\left(\frac{11 \pi }{21},0\right)$\\ \hline $[384, 568]_{\vartriangle}$, $[768, 1085727]$ & $\left(\frac{11 \pi }{24},-\frac{\pi }{4}\right)$, $\left(\frac{11 \pi }{24},0\right)$, $\left(\frac{11 \pi }{24},\frac{\pi }{8}\right)$, $\left(\frac{11 \pi }{24},\frac{3 \pi }{8}\right)$, $\left(\frac{11 \pi }{24},\frac{\pi }{2}\right)$, $\left(\frac{\pi }{2},-\frac{3 \pi }{8}\right)$, $\left(\frac{13 \pi }{24},-\frac{\pi }{4}\right)$, $\left(\frac{13 \pi }{24},0\right)$, $\left(\frac{13 \pi }{24},\frac{\pi }{8}\right)$, $\left(\frac{13 \pi }{24},\frac{3 \pi }{8}\right)$, $\left(\frac{13 \pi }{24},\frac{\pi }{2}\right)$\\ \hline $[600, 179]_{\vartriangle}$, $[1200, 1011]$ & $\left(\frac{7 \pi }{15},\frac{\pi }{10}\right)$, $\left(\frac{7 \pi }{15},\frac{3 \pi }{10}\right)$, $\left(\frac{7 \pi }{15},\frac{\pi }{2}\right)$, $\left(\frac{\pi }{2},-\frac{2 \pi }{5}\right)$, $\left(\frac{\pi }{2},-\frac{3 \pi }{10}\right)$, $\left(\frac{8 \pi }{15},\frac{\pi }{10}\right)$, $\left(\frac{8 \pi }{15},\frac{3 \pi }{10}\right)$, $\left(\frac{8 \pi }{15},\frac{\pi }{2}\right)$\\ \hline $[648, 259]_{\vartriangle'}$, $[648, 260]$ & $\left(\frac{\pi }{2},\frac{\pi }{3}\right)$, $\left(\frac{5 \pi }{9},-\frac{\pi }{6}\right)$, $\left(\frac{5 \pi }{9},\frac{\pi }{2}\right)$\\ \hline $[726, 5]_{\vartriangle}$, $[1452, 23]$ & $\left(\frac{5 \pi }{11},-\frac{2 \pi }{11}\right)$, $\left(\frac{5 \pi }{11},0\right)$, $\left(\frac{5 \pi }{11},\frac{\pi }{11}\right)$, $\left(\frac{5 \pi }{11},\frac{3 \pi }{11}\right)$, $\left(\frac{5 \pi }{11},\frac{4 \pi }{11}\right)$, $\left(\frac{5 \pi }{11},\frac{5 \pi }{11}\right)$, $\left(\frac{16 \pi }{33},-\frac{5 \pi }{11}\right)$, $\left(\frac{16 \pi }{33},-\frac{3 \pi }{11}\right)$, $\left(\frac{16 \pi }{33},-\frac{2 \pi }{11}\right)$, $\left(\frac{16 \pi }{33},-\frac{\pi }{11}\right)$, $\left(\frac{16 \pi }{33},0\right)$, $\left(\frac{16 \pi }{33},\frac{4 \pi }{11}\right)$, $\left(\frac{17 \pi }{33},-\frac{5 \pi }{11}\right)$, $\left(\frac{17 \pi }{33},-\frac{3 \pi }{11}\right)$, $\left(\frac{17 \pi }{33},-\frac{2 \pi }{11}\right)$, $\left(\frac{17 \pi }{33},\frac{4 \pi }{11}\right)$, $\left(\frac{6 \pi }{11},-\frac{2 \pi }{11}\right)$, $\left(\frac{6 \pi }{11},0\right)$, $\left(\frac{6 \pi }{11},\frac{\pi }{11}\right)$, $\left(\frac{6 \pi }{11},\frac{3 \pi }{11}\right)$, $\left(\frac{6 \pi }{11},\frac{4 \pi }{11}\right)$, $\left(\frac{6 \pi }{11},\frac{5 \pi }{11}\right)$\\ \hline $[1014, 7]_{\vartriangle}$ & $\left(\frac{6 \pi }{13},-\frac{5 \pi }{13}\right)$, $\left(\frac{6 \pi }{13},-\frac{3 \pi }{13}\right)$, $\left(\frac{6 \pi }{13},0\right)$, $\left(\frac{6 \pi }{13},\frac{\pi }{13}\right)$, $\left(\frac{6 \pi }{13},\frac{2 \pi }{13}\right)$, $\left(\frac{6 \pi }{13},\frac{4 \pi }{13}\right)$, $\left(\frac{6 \pi }{13},\frac{6 \pi }{13}\right)$, $\left(\frac{19 \pi }{39},-\frac{5 \pi }{13}\right)$, $\left(\frac{19 \pi }{39},-\frac{3 \pi }{13}\right)$, $\left(\frac{19 \pi }{39},0\right)$, $\left(\frac{19 \pi }{39},\frac{\pi }{13}\right)$, $\left(\frac{19 \pi }{39},\frac{2 \pi }{13}\right)$, $\left(\frac{19 \pi }{39},\frac{4 \pi }{13}\right)$, $\left(\frac{19 \pi }{39},\frac{6 \pi }{13}\right)$, $\left(\frac{20 \pi }{39},-\frac{5 \pi }{13}\right)$, $\left(\frac{20 \pi }{39},-\frac{3 \pi }{13}\right)$, $\left(\frac{20 \pi }{39},\frac{4 \pi }{13}\right)$, $\left(\frac{20 \pi }{39},\frac{6 \pi }{13}\right)$, $\left(\frac{7 \pi }{13},-\frac{5 \pi }{13}\right)$, $\left(\frac{7 \pi }{13},-\frac{3 \pi }{13}\right)$, $\left(\frac{7 \pi }{13},0\right)$, $\left(\frac{7 \pi }{13},\frac{\pi }{13}\right)$, $\left(\frac{7 \pi }{13},\frac{2 \pi }{13}\right)$, $\left(\frac{7 \pi }{13},\frac{4 \pi }{13}\right)$, $\left(\frac{7 \pi }{13},\frac{6 \pi }{13}\right)$\\ \hline $[1176, 243]_{\vartriangle}$ & $\left(\frac{19 \pi }{42},-\frac{3 \pi }{7}\right)$, $\left(\frac{19 \pi }{42},-\frac{2 \pi }{7}\right)$, $\left(\frac{19 \pi }{42},-\frac{\pi }{7}\right)$, $\left(\frac{19 \pi }{42},0\right)$, $\left(\frac{19 \pi }{42},\frac{\pi }{14}\right)$, $\left(\frac{19 \pi }{42},\frac{3 \pi }{14}\right)$, $\left(\frac{19 \pi }{42},\frac{5 \pi }{14}\right)$, $\left(\frac{19 \pi }{42},\frac{\pi }{2}\right)$, $\left(\frac{10 \pi }{21},\frac{\pi }{14}\right)$, $\left(\frac{10 \pi }{21},\frac{3 \pi }{14}\right)$, $\left(\frac{10 \pi }{21},\frac{5 \pi }{14}\right)$, $\left(\frac{10 \pi }{21},\frac{\pi }{2}\right)$, $\left(\frac{\pi }{2},-\frac{3 \pi }{7}\right)$, $\left(\frac{\pi }{2},\frac{2 \pi }{7}\right)$, $\left(\frac{\pi }{2},\frac{5 \pi }{14}\right)$, $\left(\frac{11 \pi }{21},\frac{\pi }{14}\right)$, $\left(\frac{11 \pi }{21},\frac{3 \pi }{14}\right)$, $\left(\frac{11 \pi }{21},\frac{5 \pi }{14}\right)$, $\left(\frac{11 \pi }{21},\frac{\pi }{2}\right)$, $\left(\frac{23 \pi }{42},-\frac{3 \pi }{7}\right)$, $\left(\frac{23 \pi }{42},-\frac{2 \pi }{7}\right)$, $\left(\frac{23 \pi }{42},-\frac{\pi }{7}\right)$, $\left(\frac{23 \pi }{42},0\right)$, $\left(\frac{23 \pi }{42},\frac{\pi }{14}\right)$, $\left(\frac{23 \pi }{42},\frac{3 \pi }{14}\right)$, $\left(\frac{23 \pi }{42},\frac{5 \pi }{14}\right)$, $\left(\frac{23 \pi }{42},\frac{\pi }{2}\right)$\\ \hline $[1458, 659]_{\vartriangle'}$, $[1458, 663]$ & $\left(\frac{13 \pi }{27},-\frac{2 \pi }{9}\right)$, $\left(\frac{13 \pi }{27},-\frac{\pi }{9}\right)$, $\left(\frac{13 \pi }{27},0\right)$, $\left(\frac{13 \pi }{27},\frac{\pi }{3}\right)$, $\left(\frac{13 \pi }{27},\frac{4 \pi }{9}\right)$, $\left(\frac{14 \pi }{27},-\frac{2 \pi }{9}\right)$, $\left(\frac{14 \pi }{27},-\frac{\pi }{9}\right)$, $\left(\frac{14 \pi }{27},0\right)$, $\left(\frac{14 \pi }{27},\frac{\pi }{3}\right)$, $\left(\frac{14 \pi }{27},\frac{4 \pi }{9}\right)$, $\left(\frac{5 \pi }{9},-\frac{2 \pi }{9}\right)$, $\left(\frac{5 \pi }{9},\frac{\pi }{9}\right)$, $\left(\frac{5 \pi }{9},\frac{4 \pi }{9}\right)$\\ \hline $[1536, 408544632]_{\vartriangle}$ & $\left(\frac{11 \pi }{24},-\frac{7 \pi }{16}\right)$, $\left(\frac{11 \pi }{24},-\frac{5 \pi }{16}\right)$, $\left(\frac{11 \pi }{24},-\frac{\pi }{16}\right)$, $\left(\frac{11 \pi }{24},\frac{3 \pi }{16}\right)$, $\left(\frac{23 \pi }{48},-\frac{5 \pi }{16}\right)$, $\left(\frac{23 \pi }{48},-\frac{3 \pi }{16}\right)$, $\left(\frac{23 \pi }{48},0\right)$, $\left(\frac{23 \pi }{48},\frac{\pi }{16}\right)$, $\left(\frac{23 \pi }{48},\frac{\pi }{8}\right)$, $\left(\frac{23 \pi }{48},\frac{\pi }{4}\right)$, $\left(\frac{23 \pi }{48},\frac{3 \pi }{8}\right)$, $\left(\frac{23 \pi }{48},\frac{7 \pi }{16}\right)$, $\left(\frac{23 \pi }{48},\frac{\pi }{2}\right)$, $\left(\frac{\pi }{2},\frac{5 \pi }{16}\right)$, $\left(\frac{\pi }{2},\frac{7 \pi }{16}\right)$, $\left(\frac{25 \pi }{48},-\frac{5 \pi }{16}\right)$, $\left(\frac{25 \pi }{48},-\frac{3 \pi }{16}\right)$, $\left(\frac{25 \pi }{48},0\right)$, $\left(\frac{25 \pi }{48},\frac{\pi }{16}\right)$, $\left(\frac{25 \pi }{48},\frac{\pi }{8}\right)$, $\left(\frac{25 \pi }{48},\frac{\pi }{4}\right)$, $\left(\frac{25 \pi }{48},\frac{3 \pi }{8}\right)$, $\left(\frac{25 \pi }{48},\frac{7 \pi }{16}\right)$, $\left(\frac{25 \pi }{48},\frac{\pi }{2}\right)$, $\left(\frac{13 \pi }{24},-\frac{7 \pi }{16}\right)$, $\left(\frac{13 \pi }{24},-\frac{5 \pi }{16}\right)$, $\left(\frac{13 \pi }{24},-\frac{\pi }{16}\right)$, $\left(\frac{13 \pi }{24},\frac{3 \pi }{16}\right)$\\ \hline $[1734, 5]_{\vartriangle}$ & $\left(\frac{23 \pi }{51},-\frac{8 \pi }{17}\right)$, $\left(\frac{23 \pi }{51},-\frac{6 \pi }{17}\right)$, $\left(\frac{23 \pi }{51},-\frac{4 \pi }{17}\right)$, $\left(\frac{23 \pi }{51},-\frac{3 \pi }{17}\right)$, $\left(\frac{23 \pi }{51},-\frac{2 \pi }{17}\right)$, $\left(\frac{23 \pi }{51},-\frac{\pi }{17}\right)$, $\left(\frac{23 \pi }{51},0\right)$, $\left(\frac{23 \pi }{51},\frac{5 \pi }{17}\right)$, $\left(\frac{23 \pi }{51},\frac{7 \pi }{17}\right)$, $\left(\frac{8 \pi }{17},-\frac{8 \pi }{17}\right)$, $\left(\frac{8 \pi }{17},-\frac{7 \pi }{17}\right)$, $\left(\frac{8 \pi }{17},-\frac{6 \pi }{17}\right)$, $\left(\frac{8 \pi }{17},-\frac{5 \pi }{17}\right)$, $\left(\frac{8 \pi }{17},-\frac{4 \pi }{17}\right)$, $\left(\frac{8 \pi }{17},-\frac{3 \pi }{17}\right)$, $\left(\frac{8 \pi }{17},-\frac{2 \pi }{17}\right)$, $\left(\frac{8 \pi }{17},-\frac{\pi }{17}\right)$, $\left(\frac{8 \pi }{17},0\right)$, $\left(\frac{25 \pi }{51},-\frac{8 \pi }{17}\right)$, $\left(\frac{25 \pi }{51},-\frac{6 \pi }{17}\right)$, $\left(\frac{25 \pi }{51},-\frac{5 \pi }{17}\right)$, $\left(\frac{25 \pi }{51},-\frac{4 \pi }{17}\right)$, $\left(\frac{25 \pi }{51},-\frac{3 \pi }{17}\right)$, $\left(\frac{25 \pi }{51},\frac{7 \pi }{17}\right)$, $\left(\frac{26 \pi }{51},-\frac{8 \pi }{17}\right)$, $\left(\frac{26 \pi }{51},-\frac{6 \pi }{17}\right)$, $\left(\frac{26 \pi }{51},-\frac{5 \pi }{17}\right)$, $\left(\frac{26 \pi }{51},-\frac{4 \pi }{17}\right)$, $\left(\frac{26 \pi }{51},\frac{7 \pi }{17}\right)$, $\left(\frac{9 \pi }{17},-\frac{8 \pi }{17}\right)$, $\left(\frac{9 \pi }{17},-\frac{7 \pi }{17}\right)$, $\left(\frac{9 \pi }{17},-\frac{6 \pi }{17}\right)$, $\left(\frac{9 \pi }{17},-\frac{5 \pi }{17}\right)$, $\left(\frac{9 \pi }{17},-\frac{4 \pi }{17}\right)$, $\left(\frac{9 \pi }{17},-\frac{3 \pi }{17}\right)$, $\left(\frac{9 \pi }{17},-\frac{2 \pi }{17}\right)$, $\left(\frac{9 \pi }{17},-\frac{\pi }{17}\right)$, $\left(\frac{9 \pi }{17},0\right)$, $\left(\frac{28 \pi }{51},-\frac{8 \pi }{17}\right)$, $\left(\frac{28 \pi }{51},-\frac{6 \pi }{17}\right)$, $\left(\frac{28 \pi }{51},-\frac{4 \pi }{17}\right)$, $\left(\frac{28 \pi }{51},-\frac{3 \pi }{17}\right)$, $\left(\frac{28 \pi }{51},-\frac{2 \pi }{17}\right)$, $\left(\frac{28 \pi }{51},-\frac{\pi }{17}\right)$, $\left(\frac{28 \pi }{51},0\right)$, $\left(\frac{28 \pi }{51},\frac{5 \pi }{17}\right)$, $\left(\frac{28 \pi }{51},\frac{7 \pi }{17}\right)$\\
\hline \hline
\end{tabular}
\caption{\label{tab:Ia}The predictions for PMNS matrix of the form $U^{I(a)}$, where the first column shows the group identification in \texttt{GAP} system, and the second column displays the achievable values of the parameters $\varphi_1$ and $\varphi_2$. We have shown at most two representatives flavor symmetry groups in the first column. If there is only one group predicting the corresponding values of $\varphi_1$ and $\varphi_2$ in the second column, this unique group would be listed. The full results of our analysis are provided at the website~\cite{webdata}. The subscripts $\Delta$ and $\Delta^{\prime}$ indicate that the corresponding groups belong to the type D group series $D_{n,n}^{(0)}\cong\Delta(6n^2)$ and $D_{9n^{\prime},3n^{\prime}}^{(1)}\cong(Z_{9n^{\prime}}\times Z_{3n^{\prime}})\rtimes S_3$, respectively. }
\end{center}
\end{table}

\begin{table}[ht!]
\begin{center}
\begin{tabular}{|c|c|c|c|}\hline\hline
\texttt{Case $C_{ab}$} & $(K_1,K_2,K_3)$ &$(R^{\prime}_{11}, R^{\prime}_{12}, R^{\prime}_{13})$&  $W_{ab}$ \\\hline
& & & \\ [-0.16in]
\multirow{4}{*}[-5pt]{$a=1, b=2$}
&$(+,\,+,\,\pm)$&$(\cos\eta,\sin\eta,0)$ & $\frac{\sqrt{m_1m_2}\left(m_1-m_2\right)\sin\eta\cos\eta}{m_1\cos^2\eta+m_2\sin^2\eta}$\\[0.08in]\cline{2-4}

& & & \\ [-0.16in]

&$(+,\,-,\,\pm)$&$(\cosh\eta, \sinh\eta,0)$  & $\frac{\sqrt{m_1m_2}\left(m_1+m_2\right)\sinh\eta\cosh\eta}{m_1\cosh^2\eta+m_2\sinh^2\eta}$\\[0.08in]\cline{2-4}

&& & \\ [-0.16in]

&$(-,\,+,\,\pm)$&$( \sinh\eta,\cosh\eta,0)$ & $-\frac{\sqrt{m_1m_2}\left(m_1+m_2\right)\sinh\eta\cosh\eta}{m_1\sinh^2\eta+m_2\cosh^2\eta}$ \\[0.08in]\hline\hline

& & & \\ [-0.16in]

\multirow{4}{*}[-5pt]{$a=1, b=3$}
&$(+,\,\pm\,,+)$&$(\cos\eta,0,\sin\eta)$ & $\frac{\sqrt{m_1m_3}\left(m_1-m_3\right)\sin\eta\cos\eta}{m_1\cos^2\eta+m_3\sin^2\eta}$ \\[0.08in]\cline{2-4}

& & & \\ [-0.16in]

&$(+,\,\pm\,,-)$&$(\cosh\eta,0, \sinh\eta)$  & $\frac{\sqrt{m_1m_3}\left(m_1+m_3\right)\sinh\eta\cosh\eta}{m_1\cosh^2\eta+m_3\sinh^2\eta}$ \\[0.08in]\cline{2-4}

& & & \\ [-0.16in]

&$(-,\,\pm\,,+)$&$(\sinh\eta,0,\cosh\eta)$  &  $-\frac{\sqrt{m_1m_3}\left(m_1+m_3\right)\sinh\eta\cosh\eta}{m_1\sinh^2\eta+m_3\cosh^2\eta}$\\[0.08in]\hline\hline

&  & & \\ [-0.16in]

\multirow{4}{*}[-5pt]{$a=2, b=3$}
&$(\pm\,,+,\,+)$&$(0,\cos\eta,\sin\eta)$ & $\frac{\sqrt{m_2m_3}\left(m_2-m_3\right)\sin\eta\cos\eta}{m_2\cos^2\eta+m_3\sin^2\eta}$ \\[0.08in]\cline{2-4}

&  & & \\ [-0.16in]

&$(\pm\,,+,\,-)$&$(0,\cosh\eta,\sinh\eta)$ & $\frac{\sqrt{m_2m_3}\left(m_2+m_3\right)\sinh\eta\cosh\eta}{m_2\cosh^2\eta+m_3\sinh^2\eta}$\\[0.08in]\cline{2-4}

&  & & \\ [-0.16in]

&$(\pm\,,-,\,+)$&$(0, \sinh\eta, \cosh\eta)$ & $-\frac{\sqrt{m_2m_3}\left(m_2+m_3\right)\sinh\eta\cosh\eta}{m_2\sinh^2\eta+m_3\cosh^2\eta}$\\[0.08in]\hline\hline
\end{tabular}
\end{center}
\caption{\label{tab:R_W_para}The parametrization of the first column of $R^{\prime}$-matrix and the corresponding expressions of $W_{12}$, $W_{13}$ and $W_{23}$ in the three interesting cases $C_{12}$, $C_{13}$ and $C_{23}$.}
\end{table}

\item[Case \uppercase\expandafter{\romannumeral1}(b)]
\begin{equation}
\label{eq:pmnsIb}
U^{I(b)}=\frac{1}{\sqrt{3}}
\left(
\begin{array}{ccc}
 \sqrt{2}\cos\varphi_1 & e^{i \varphi _2} & \sqrt{2} \sin \varphi _1 \\
 -\sqrt{2}\sin \left(\varphi _1-\frac{\pi }{6}\right) & -e^{i \varphi _2} & \sqrt{2}\cos \left(\varphi _1-\frac{\pi }{6}\right) \\
 -\sqrt{2}\sin\left(\varphi _1+\frac{\pi }{6}\right) & e^{i \varphi _2} & \sqrt{2}\cos\left(\varphi _1+\frac{\pi }{6}\right)
\end{array}
\right)S_{12}(\theta)Q^{\dagger}_{\nu}\,,
\end{equation}
where the admissible values of $\varphi_1$ and $\varphi_2$ and the corresponding representative flavor symmetry groups are listed in table~\ref{tab:Ib}. One can refer to the full results at the website~\cite{webdata}. It is remarkable that all these phenomenological viable mixing patterns can be achieved from the type D group series $\Delta(6n^2)$ or $D^{(1)}_{9n,3n}$ combined with CP symmetry. The smallest group which can admit a good fit to the experimental data is $[649, 259]\cong D^{(1)}_{9\times2, 3\times2}$ in this case. The PMNS matrix $U^{I(b)}$ is related to $U^{I(a)}$ by column permutations, and the constant column vector $\left(\sqrt{2}\sin\varphi_1, \sqrt{2}\cos \left(\varphi _1-\frac{\pi }{6}\right),  \sqrt{2}\cos\left(\varphi _1+\frac{\pi }{6}\right)\right)^{T}/\sqrt{3}$ enforce by residual symmetries is arranged at the third column in this case. The patterns originating from the six possible row permutations of  $U^{I(b)}$ can be obtained through redefinitions of $\varphi_1$ and $\theta$. We can extract the mixing angles from Eq.~\eqref{eq:pmnsIb} in the usual way and find
\begin{equation}
\begin{split}
&\sin^2\theta_{13}=\frac{2}{3}\sin^2\varphi_1,\qquad\sin^2\theta_{23}=\frac{1+\sin\left(\pi/6+2\varphi_1\right)}{2+\cos2\varphi_1},\\
&\sin^2\theta_{12}=\frac{1+\sin^2\theta\cos2\varphi_1-\sqrt{2}\sin2\theta\cos\varphi_2\cos\varphi_1}{2+\cos2\varphi_1}\,.
\end{split}
\end{equation}
Notice that both the reactor angle $\theta_{13}$ and the atmospheric mixing angle $\theta_{23}$ only depend on the discrete parameter $\varphi_{1}$ while all the three parameters $\theta$, $\varphi_1$ and $\varphi_2$ are involved in the solar mixing angle $\theta_{12}$. Moreover, we easily see that the mixing angles fulfill the following sum rule
\begin{equation}
  2\sin^2\theta_{23}=1\pm\tan\theta_{13}\sqrt{2-\tan^2\theta_{13}}\,.
\end{equation}
Using the best fit value $\sin^2\theta_{13}=0.0218$~\cite{Gonzalez-Garcia:2014bfa}, we obtain
\begin{equation}
\sin^2\theta_{23}\simeq0.395,\quad\text{or}\quad \sin^2\theta_{23}\simeq0.605\,.
\end{equation}
Consequently $\theta_{23}$ deviates from maximal mixing but it is in the experimentally preferred $3\sigma$ range~\cite{Gonzalez-Garcia:2014bfa}. As regards the CP invariants, we find
\begin{equation}
\begin{aligned}
\left|J_{CP}\right|&=\frac{1}{6\sqrt{6}}\left|\sin2\theta\sin3\varphi_1\sin\varphi_2\right|\,,\\
\left|I_1\right|&=\frac{1}{9}\left|\cos\varphi_1\sin\varphi_2\left(4\cos2\theta\cos\varphi_1\cos \varphi_2-\sqrt{2}\sin2\theta\cos2\varphi_1\right)\right|\,, \\
\left|I_2\right|&=\frac{2\sqrt{2}}{9} \left|\sin^2\varphi_1\sin\varphi_2 \left(\sqrt{2}\sin^2\theta\cos\varphi_2+\sin 2\theta\cos\varphi _1\right)\right| \,.
\end{aligned}
\end{equation}
For this mixing pattern $U^{I(b)}$, the effective Majorana mass $|m_{ee}|$ in $0\nu\beta\beta$ is given by
\begin{equation}
\begin{split}
|m_{ee}|=\frac{1}{3} \left|2 m_3\sin ^2\varphi _1+q_1m_2\left(e^{i \varphi _2} \cos \theta -\sqrt{2} \cos \varphi _1\sin \theta \right)^2 \right.\\\left.+  q_2 m_1\left(\sqrt{2} \cos \theta \cos\varphi _1+e^{i \varphi _2} \sin \theta \right)^2\right|\,,
\end{split}
\end{equation}
where $q_1,q_2=\pm 1$ appears due to the undetermined CP parity of the neutrino states encoded in the matrix $Q_{\nu}$. In the limit of $|G_{f}|\rightarrow\infty$, where $|G_{f}|$ represents the order of $G_{f}$, $\varphi_1$ and $\varphi_2$ tends to be continuous parameters. Then one can almost reproduce the whole regions of $|m_{ee}|$ obtained by varying the oscillation parameters over their current $3\sigma$ global ranges, as shown in figure~\ref{fig:mee_CaseI}. For the smallest group $G_{f}=[649, 259]$, the admissible values of $\varphi_1$ and $\varphi_2$ are $(\varphi_1,\varphi_2)=(\frac{\pi}{18},-\frac{\pi}{6})$, $(\frac{\pi}{18}, 0)$, $(\frac{\pi}{18},\frac{\pi}{3})$, $(\frac{\pi}{18},\frac{\pi}{2})$, $(\frac{17\pi}{18},-\frac{\pi}{6})$, $(\frac{17\pi}{18}, 0)$, $(\frac{17\pi}{18},\frac{\pi}{3})$ and $(\frac{17\pi}{18},\frac{\pi}{2})$.
The corresponding predictions for the $0\nu\beta\beta$ decay effective mass $|m_{ee}|$ versus the lightest neutrino mass $m_{\text{lightest}}$ are plotted in figure~\ref{fig:mee_CaseI}. We see that $|m_{ee}|$ is close to 0.029eV or 0.042eV for IO neutrino mass spectrum, which are within the future sensitivity of planned $0\nu\beta\beta$ decay experiments. On the other hand, $|m_{ee}|$ is always bigger than $0.7\times10^{-4}$ eV in the case of NO spectrum.

Now we proceed to discuss the predictions for leptogenesis. The bilinear invariant $I^{\alpha}_{12}$ can be read out as follow
\begin{equation}
\begin{split}
I^e_{12}=&-\frac{\sqrt{2}}{3}\cos\varphi_1\sin\varphi_2\,,\\
I^{\mu}_{12}=&\frac{\sqrt{2}}{3}\sin\left(\frac{\pi}{6}-\varphi_1\right)\sin\varphi_2\,,\\
I^{\tau}_{12}=&\frac{\sqrt{2}}{3}\sin\left(\frac{\pi}{6}+\varphi_1\right)\sin\varphi_2\,,
\end{split}
\end{equation}
which are generally nonzero except $\varphi_2=0, \pi$. The value of baryon asymmetry can be straightforwardly calculated from any given values of $\varphi_1$ and $\varphi_2$. We shall study the smallest viable flavor symmetry $[649, 259]$ for illustration.
The results of the $\chi^2$ analysis are summarized in table~\ref{tab:CaseIbparas}. We display the values of the mixing angles and CP phases at $\theta_{\text{bf}}$, the best fit points for which the $\chi^2$ function has a global minimum $\chi^2_{\text{min}}$. Obviously the mixing angles can be in accordance with the experimental data for particular values of $\theta$. The leptogenesis asymmetries $\epsilon_{\alpha}$ are vanishing for $(\varphi_1,\varphi_2)=(\pi/18, 0)$, $(17\pi/18, 0)$. For the remaining six admissible values of $\varphi_1$ and $\varphi_2$, the variations of $Y_{B}$ as a function of $\eta$ are plotted in
figures~(\ref{fig:leptogenesis_CaseIb1}-\ref{fig:leptogenesis_CaseIb6}).
We see that the correct value of $Y_B$ can be reproduced for certain values of $\eta$ and $K_{1,2,3}$.

\begin{table}[t!]\tabcolsep=0.11cm
\begin{center}
\scalebox{0.82}{\begin{tabular}{|c|c|c|c|c|c|c|c|c|c|}
\hline\hline
\multirow{2}{*}{$(\varphi_1,\varphi_2)$} & \multirow{2}{*}{$\theta_{\text{bf}}/\pi$} & \multirow{2}{*}{$\chi^2_{\text{min}}$} & \multirow{2}{*}{$\sin^2\theta_{13}$} & \multirow{2}{*}{$\sin^2\theta_{12}$} & \multirow{2}{*}{$\sin^2\theta_{23}$} & \multirow{2}{*}{$\delta_{CP}/\pi$} &$\alpha_{21}/\pi$ & $\alpha'_{31}/\pi$ & \multirow{2}{*}{$(K_1,K_2,K_3)$} \\
&  &  &  &  &  &  & (mod 1) & (mod 1) &\\
\hline
\multirow{4}{*}{$(\frac{\pi}{18},-\frac{\pi}{6})$} &  \multirow{2}{*}{0.014} &  & \multirow{4}{*}{0.0201}  &  \multirow{4}{*}{0.304}  &  \multirow{4}{*}{0.601}  &  \multirow{2}{*}{0.984}  &  \multirow{2}{*}{0.656}  &  \multirow{2}{*}{0.010} & $(-,+,\pm)$\\
&  & 11.065 &  &  &  &  &  &  & [$(-,+,\pm)$] \\
\cline{2-2}\cline{7-10}
& \multirow{2}{*}{0.367}  & [3.989] &  &  &  &  \multirow{2}{*}{0.132}  &  \multirow{2}{*}{0.344}  &  \multirow{2}{*}{0.207} &$(+,-,\pm)$\\
&  &  &  &  &  & &  &  &[$(+,+,\pm)$, $(+,-,\pm)$]\\
\hline
\multirow{4}{*}{$(\frac{\pi}{18},0)$} &  \multirow{2}{*}{0.012} &  & \multirow{4}{*}{0.0201}  &  \multirow{4}{*}{0.304}  &  \multirow{4}{*}{0.601}  &  \multirow{2}{*}{1}  &  \multirow{2}{*}{0}  &  \multirow{2}{*}{0} &  \multirow{2}{*}{\rule{0.06\textwidth}{.4pt}}\\
&  & 11.065 &  &  &  &  &  &  &  \\
\cline{2-2}\cline{7-10}
& \multirow{2}{*}{0.384}  & [3.989] &  &  &  &  \multirow{2}{*}{0}  &  \multirow{2}{*}{0}  &  \multirow{2}{*}{0} &  \multirow{2}{*}{\rule{0.06\textwidth}{.4pt}}\\
&  &  &  &  &  & &  &  &\\
\hline
\multirow{4}{*}{$(\frac{\pi}{18},\frac{\pi}{3})$} &  \multirow{2}{*}{0.026} &  & \multirow{4}{*}{  0.0201  } &  \multirow{4}{*}{  0.304  } &  \multirow{4}{*}{  0.601  } &  \multirow{2}{*}{1.049}  &  \multirow{2}{*}{0.701}  &  \multirow{2}{*}{0.969}  & $(+,-,\pm)$ \\
&  & 11.065 &  &  &  &  &  &  & [$(+,+,\pm)$, $(+,-,\pm)$, $(-,+,\pm)$] \\
\cline{2-2}\cline{7-10}
& \multirow{2}{*}{0.285} & [3.989] &  &  &  &  \multirow{2}{*}{1.629}  &  \multirow{2}{*}{0.299}  &  \multirow{2}{*}{0.686}  &$(+,-,\pm)$\\
&  &  &  &  &  & &  &  &[$(+,+,\pm)$, $(+,-,\pm)$, $(-,+,\pm)$]\\
\hline
\multirow{2}{*}{$(\frac{\pi}{18},\frac{\pi}{2})$} &  \multirow{2}{*}{0} & 18.807 & \multirow{2}{*}{ 0.0201}  & \multirow{2}{*}{0.340} &  \multirow{2}{*}{0.601} &  \multirow{2}{*}{1}  &  \multirow{2}{*}{0}  & \multirow{2}{*}{0} &$(+,-,\pm)$, $(-,+,\pm)$\\
&  & [11.731] &  &  &  &  &  &  & [$(+,+,\pm)$, $(-,+,\pm)$]\\
\hline
\multirow{4}{*}{$(\frac{17\pi}{18},-\frac{\pi}{6})$} &  \multirow{2}{*}{0.633} & & \multirow{4}{*}{  0.0201  } &  \multirow{4}{*}{  0.304  } &  \multirow{4}{*}{  0.399  } &  \multirow{2}{*}{1.132}  &  \multirow{2}{*}{0.344}  &  \multirow{2}{*}{0.207}  & $(+,-,\pm)$, $(-,+,\pm)$ \\
&  & 6.432  &  &  &  & &  &  &  [$(+,+,\pm)$, $(-,+,\pm)$]\\
\cline{2-2}\cline{7-10}
&\multirow{2}{*}{0.986}  & [26.835] &  &  &  &\multirow{2}{*}{1.984}  &  \multirow{2}{*}{0.656}  & \multirow{2}{*}{0.010} &$(-,+,\pm)$\\
&  &  &  &  &  & &  &  &[$(-,+,\pm)$]\\
\hline
\multirow{4}{*}{$(\frac{17\pi}{18},0)$} &  \multirow{2}{*}{0.616} &  & \multirow{4}{*}{0.0201}  &  \multirow{4}{*}{0.304}  &  \multirow{4}{*}{0.399}  &  \multirow{2}{*}{1}  &  \multirow{2}{*}{0}  &  \multirow{2}{*}{0} &  \multirow{2}{*}{\rule{0.06\textwidth}{.4pt}}\\
&  & 6.432 &  &  &  &  &  &  &  \\
\cline{2-2}\cline{7-10}
& \multirow{2}{*}{0.988}  & [26.835] &  &  &  &  \multirow{2}{*}{0}  &  \multirow{2}{*}{0}  &  \multirow{2}{*}{0} & \multirow{2}{*}{\rule{0.06\textwidth}{.4pt}}\\
&  &  &  &  &  & &  &  &\\
\hline
\multirow{4}{*}{$(\frac{17\pi}{18},\frac{\pi}{3})$} &  \multirow{2}{*}{0.715} &  & \multirow{4}{*}{  0.0201  } &  \multirow{4}{*}{  0.304  } &  \multirow{4}{*}{  0.399  } &  \multirow{2}{*}{0.629}  &  \multirow{2}{*}{0.299}  &  \multirow{2}{*}{0.686}  & $(+,-,\pm)$, $(-,+,\pm)$ \\
& & 6.432 &  &  &  &  &  &  &  [$(+,+,\pm)$, $(-,+,\pm)$]\\
\cline{2-2}\cline{7-10}
& \multirow{2}{*}{0.974}  & [26.835] &  &  &  &\multirow{2}{*}{0.049}  &  \multirow{2}{*}{0.701}  & \multirow{2}{*}{0.969}  &$(+,-,\pm)$\\
&  &  &  &  &  & &  &  &[$(+,+,\pm)$, $(+,-,\pm)$, $(-,+,\pm)$] \\
\hline
\multirow{2}{*}{$(\frac{17\pi}{18},\frac{\pi}{2})$} &  \multirow{2}{*}{0} & 14.174 & \multirow{2}{*}{ 0.0201}  & \multirow{2}{*}{0.340} &  \multirow{2}{*}{0.399} &  \multirow{2}{*}{0}  &  \multirow{2}{*}{0}  & \multirow{2}{*}{0} & $(+,-,\pm)$ \\
&  & [34.576] &  &  &  &  &  &  & [$(+,+,\pm)$, $(+,-,\pm)$, $(-,+,\pm)$]\\
\hline\hline
\end{tabular}}
\caption {Results of the $\chi^2$ analysis for case I(b) with the flavor symmetry $G_{f}=[649, 259]$. As shown in table~\ref{tab:Ib}, the experimentally measured values of the mixing angles can be accommodated in the case of $(\varphi_1,\varphi_2)=(\frac{\pi}{18},-\frac{\pi}{6})$, $(\frac{\pi}{18},0)$, $(\frac{\pi}{18},\frac{\pi}{3})$, $(\frac{\pi}{18},\frac{\pi}{2})$, $(\frac{17\pi}{18},-\frac{\pi}{6})$, $(\frac{17\pi}{18},0)$, $(\frac{17\pi}{18},\frac{\pi}{3})$ and $(\frac{17\pi}{18},\frac{\pi}{2})$. We display the best fit value $\theta_{\text{bf}}$ for $\theta$, and $\chi^2_{\text{min}}$ is the smallest value of $\chi^2$ that can be obtained at the best fit value $\theta_{\text{bf}}$. The mixing angles and the CP violating phases for $\theta=\theta_{\text{bf}}$ are presented as well. Note that the CP parity matrix $Q_{\nu}$ can shift the Majorana phases $\alpha_{21}$ and $\alpha'_{31}$ by $\pi$. In the last column we give the values of $K_{1,2,3}$ for which the observed baryon asymmetry can be generated via leptogenesis. The values in the square brackets are the corresponding results for the case of IO mass spectrum. The net baryon asymmetry can not be generated for $\varphi_2=0, \pi$.\label{tab:CaseIbparas}}
\end{center}
\end{table}

\begin{table}[hptb]
\begin{center}
\begin{tabular}{|m{0.25\columnwidth}<{\centering}|m{0.7\columnwidth}<{\centering}|}
\hline\hline
\texttt{Group Id} & $(\varphi_1,\varphi_2)$ \\
\hline

$[648, 259]_{\vartriangle'}$, $[648, 260]$ & $\left(\frac{\pi }{18},-\frac{\pi }{6}\right)$, $\left(\frac{\pi }{18},0\right)$, $\left(\frac{\pi }{18},\frac{\pi }{3}\right)$, $\left(\frac{\pi }{18},\frac{\pi }{2}\right)$, $\left(\frac{17 \pi }{18},-\frac{\pi }{6}\right)$, $\left(\frac{17 \pi }{18},0\right)$, $\left(\frac{17 \pi }{18},\frac{\pi }{3}\right)$, $\left(\frac{17 \pi }{18},\frac{\pi }{2}\right)$\\ \hline $[726, 5]_{\vartriangle}$, $[1452, 23]$ & $\left(\frac{2 \pi }{33},-\frac{2 \pi }{11}\right)$, $\left(\frac{2 \pi }{33},0\right)$, $\left(\frac{2 \pi }{33},\frac{\pi }{11}\right)$, $\left(\frac{2 \pi }{33},\frac{3 \pi }{11}\right)$, $\left(\frac{2 \pi }{33},\frac{4 \pi }{11}\right)$, $\left(\frac{2 \pi }{33},\frac{5 \pi }{11}\right)$, $\left(\frac{31 \pi }{33},-\frac{2 \pi }{11}\right)$, $\left(\frac{31 \pi }{33},0\right)$, $\left(\frac{31 \pi }{33},\frac{\pi }{11}\right)$, $\left(\frac{31 \pi }{33},\frac{3 \pi }{11}\right)$, $\left(\frac{31 \pi }{33},\frac{4 \pi }{11}\right)$, $\left(\frac{31 \pi }{33},\frac{5 \pi }{11}\right)$\\ \hline $[1734, 5]_{\vartriangle}$ & $\left(\frac{\pi }{17},-\frac{8 \pi }{17}\right)$, $\left(\frac{\pi }{17},-\frac{6 \pi }{17}\right)$, $\left(\frac{\pi }{17},0\right)$, $\left(\frac{\pi }{17},\frac{\pi }{17}\right)$, $\left(\frac{\pi }{17},\frac{2 \pi }{17}\right)$, $\left(\frac{\pi }{17},\frac{3 \pi }{17}\right)$, $\left(\frac{\pi }{17},\frac{4 \pi }{17}\right)$, $\left(\frac{\pi }{17},\frac{5 \pi }{17}\right)$, $\left(\frac{\pi }{17},\frac{7 \pi }{17}\right)$, $\left(\frac{16 \pi }{17},-\frac{8 \pi }{17}\right)$, $\left(\frac{16 \pi }{17},-\frac{6 \pi }{17}\right)$, $\left(\frac{16 \pi }{17},0\right)$, $\left(\frac{16 \pi }{17},\frac{\pi }{17}\right)$, $\left(\frac{16 \pi }{17},\frac{2 \pi }{17}\right)$, $\left(\frac{16 \pi }{17},\frac{3 \pi }{17}\right)$, $\left(\frac{16 \pi }{17},\frac{4 \pi }{17}\right)$, $\left(\frac{16 \pi }{17},\frac{5 \pi }{17}\right)$, $\left(\frac{16 \pi }{17},\frac{7 \pi }{17}\right)$\\ \hline \hline
\end{tabular}
\caption{\label{tab:Ib}The predictions for PMNS matrix of the form $U^{I(b)}$, where the first column shows the group identification in \texttt{GAP} system, and the second column displays the achievable values of the parameters $\varphi_1$ and $\varphi_2$. We have shown at most two representatives flavor symmetry groups in the first column. If there is only one group predicting the corresponding values of $\varphi_1$ and $\varphi_2$ in the second column, this unique group would be listed. The full results of our analysis are provided at the website~\cite{webdata}. The subscripts $\Delta$ and $\Delta^{\prime}$ indicate that the corresponding groups belong to the type D group series $D_{n,n}^{(0)}\cong\Delta(6n^2)$ and $D_{9n^{\prime},3n^{\prime}}^{(1)}\cong(Z_{9n^{\prime}}\times Z_{3n^{\prime}})\rtimes S_3$, respectively. }
\end{center}
\end{table}

\begin{figure}[ht!]
\begin{center}
\includegraphics[width=0.45\linewidth]{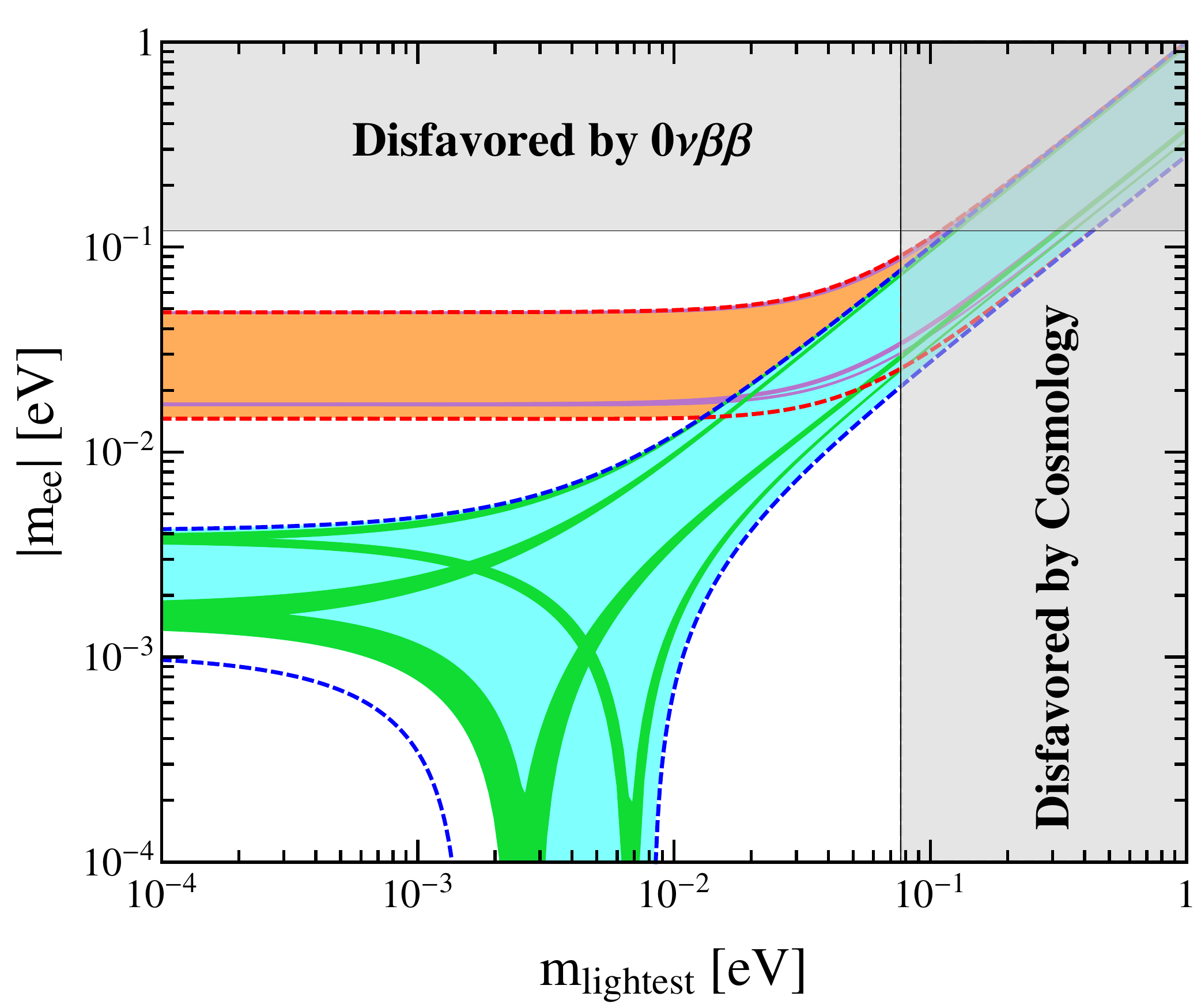}~~~
\includegraphics[width=0.45\linewidth]{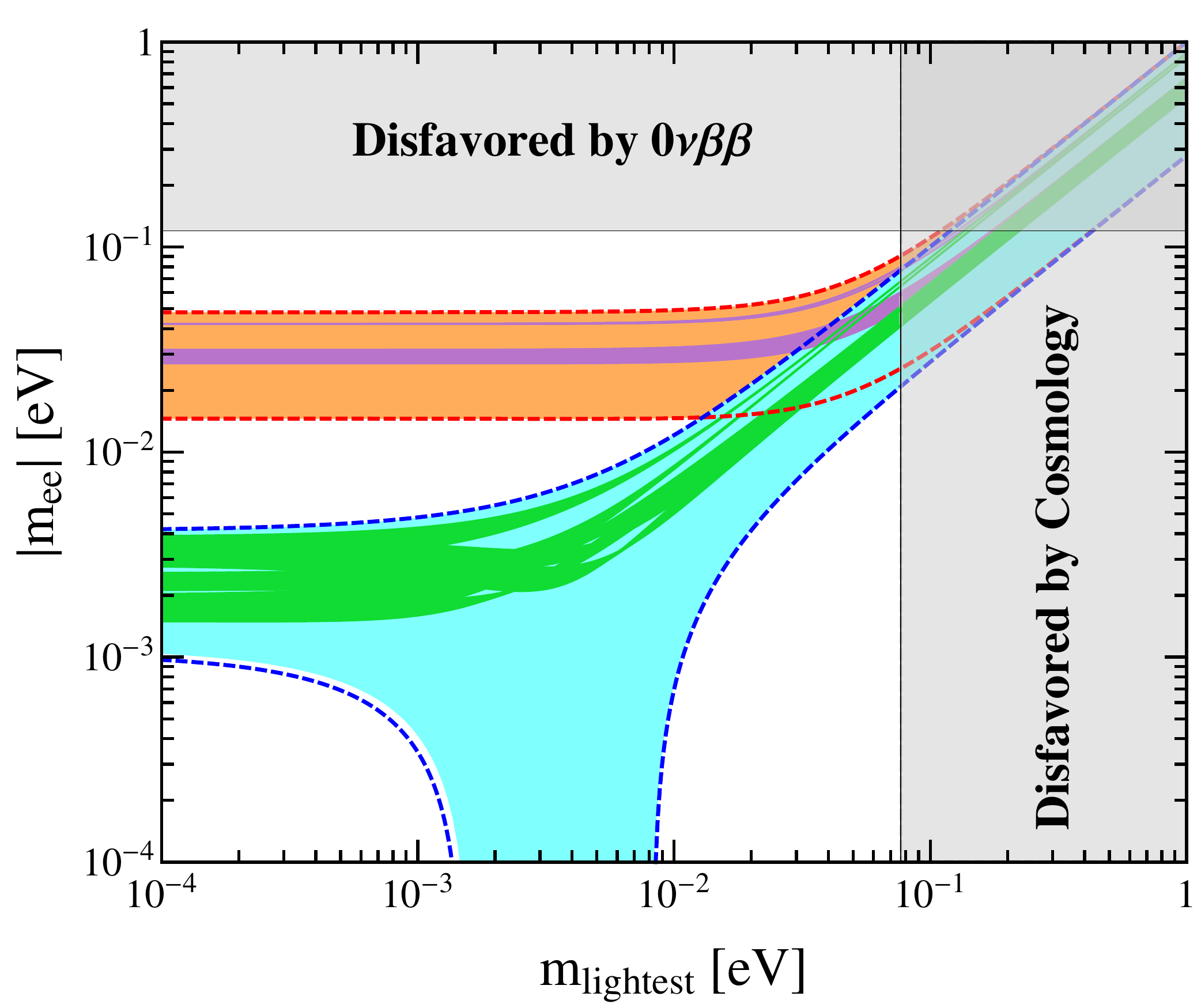}
\caption{\label{fig:mee_CaseI} Predictions for the $0\nu\beta\beta$ decay effective mass $|m_{ee}|$ with respect to the lightest neutrino mass $m_{\text{lightest}}$ in the case I. The left and right panels are for the mixing patterns $U^{I(a)}$ and $U^{I(b)}$ respectively. The red (blue) dashed lines indicate the most general allowed regions for IO (NO) spectrum obtained by varying the mixing parameters within their $3\sigma$ ranges~\cite{Gonzalez-Garcia:2014bfa}. The orange (cyan) areas denote the achievable values of $|m_{ee}|$ when $\varphi_1$ and $\varphi_2$ are taken to be free continuous parameters in the case of IO (NO). The purple and green regions are the theoretical predictions of the smallest flavor symmetry group which can generate these two mixing patterns. Note that the purple (green) region overlaps the orange (cyan) one. The present most stringent upper limits $|m_{ee}|<0.120$ eV from EXO-200~\cite{Auger:2012ar, Albert:2014awa} and KamLAND-ZEN~\cite{Gando:2012zm} is shown by horizontal grey band. The vertical grey exclusion band is the current limit on $m_{\text{lightest}}$ from the cosmological data of $\sum m_i<0.230$ eV by the Planck collaboration~\cite{Ade:2013zuv}.}
\end{center}
\end{figure}

\begin{figure}[ht!]
\begin{center}
\includegraphics[width=0.42\linewidth]{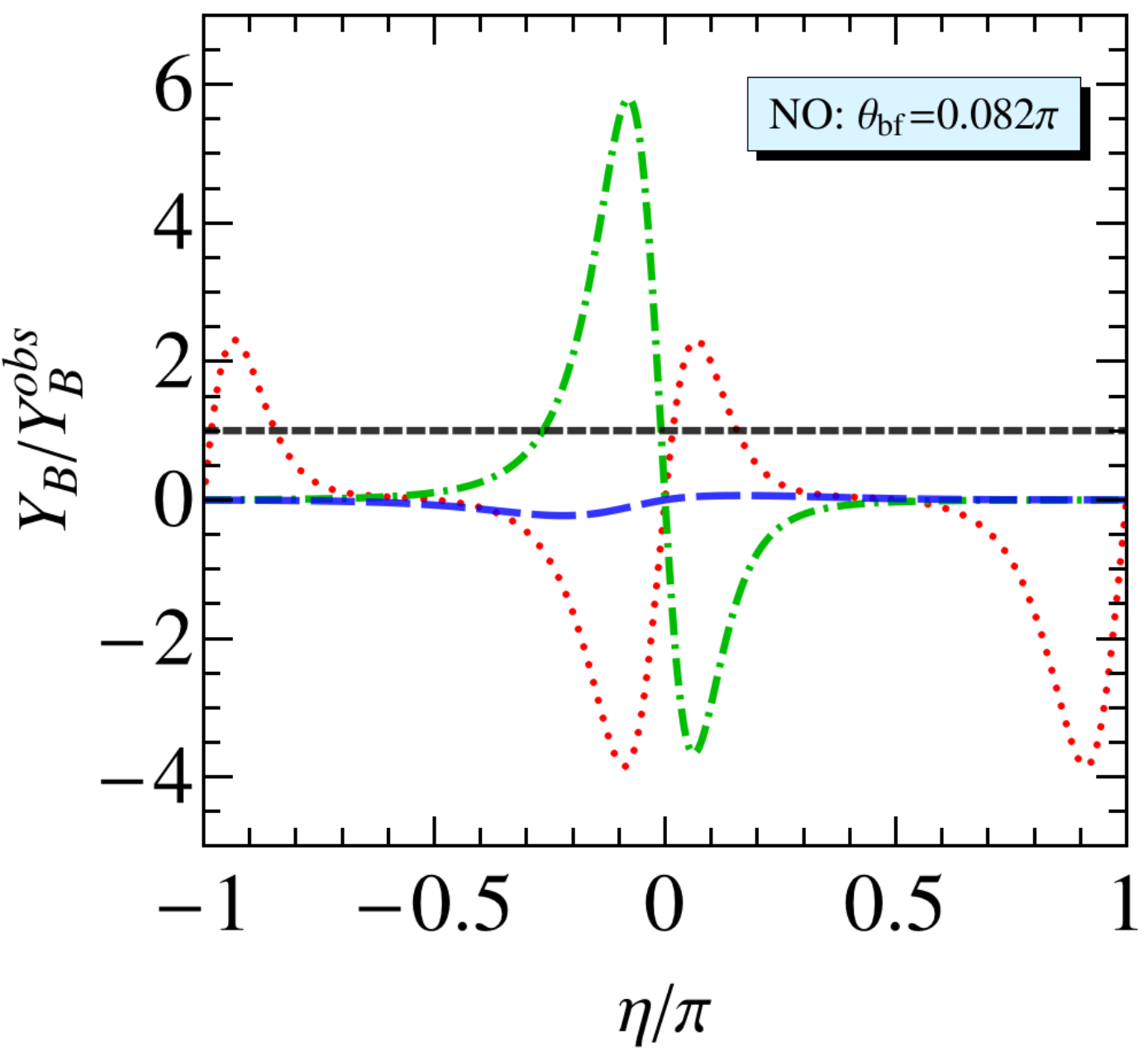}~~~
\includegraphics[width=0.445\linewidth]{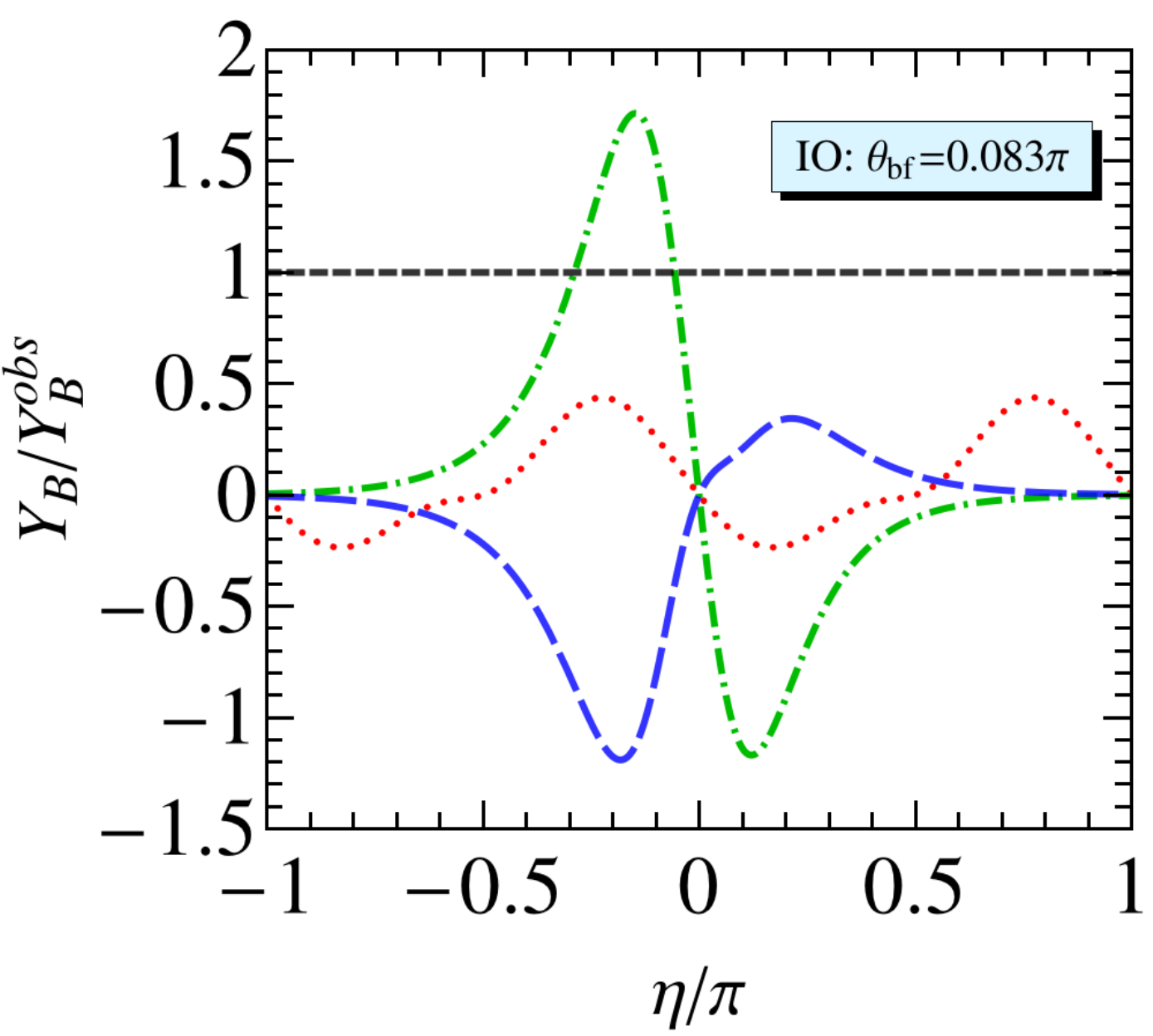}
\includegraphics[width=0.42\linewidth]{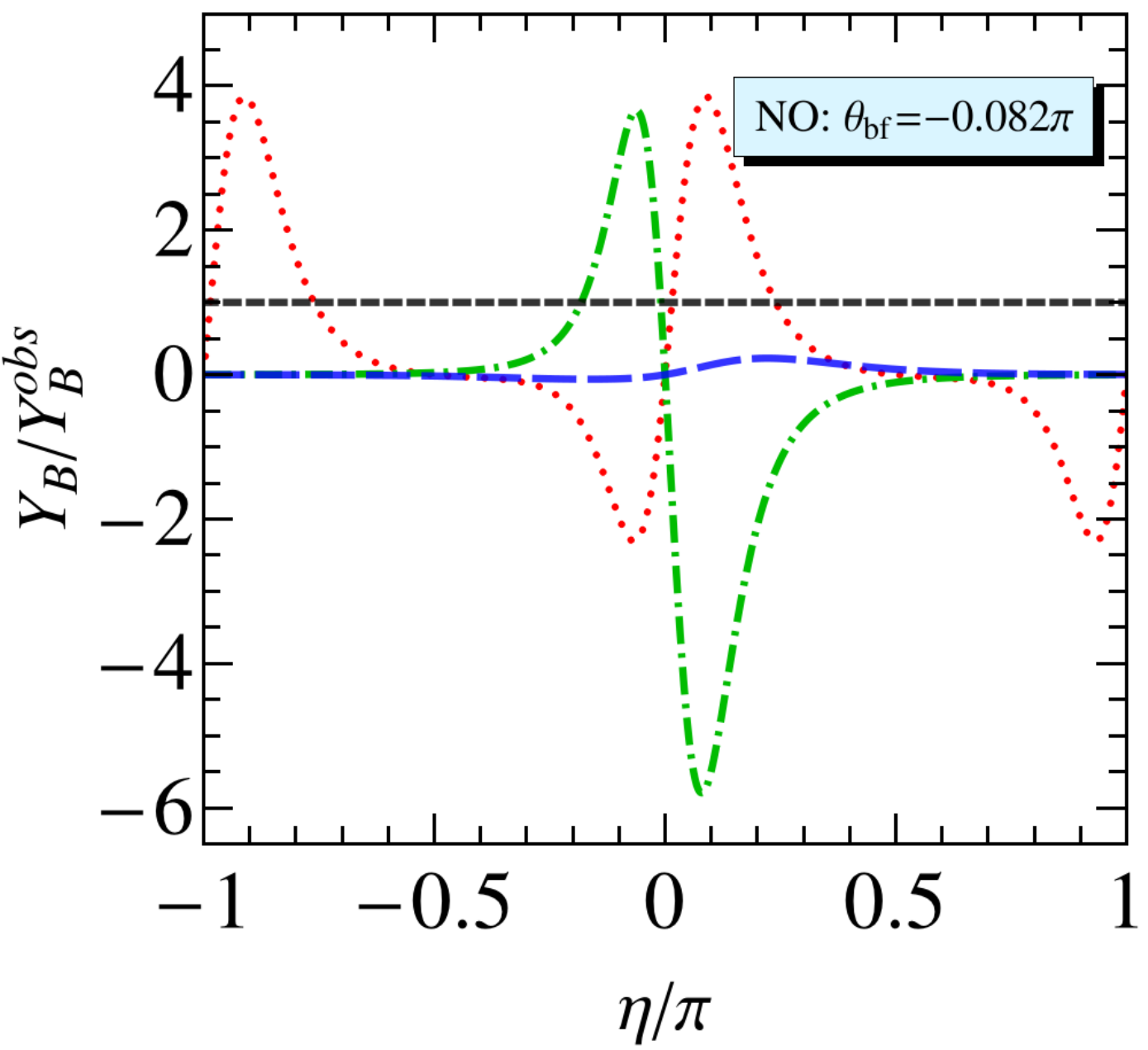}~~~
\includegraphics[width=0.445\linewidth]{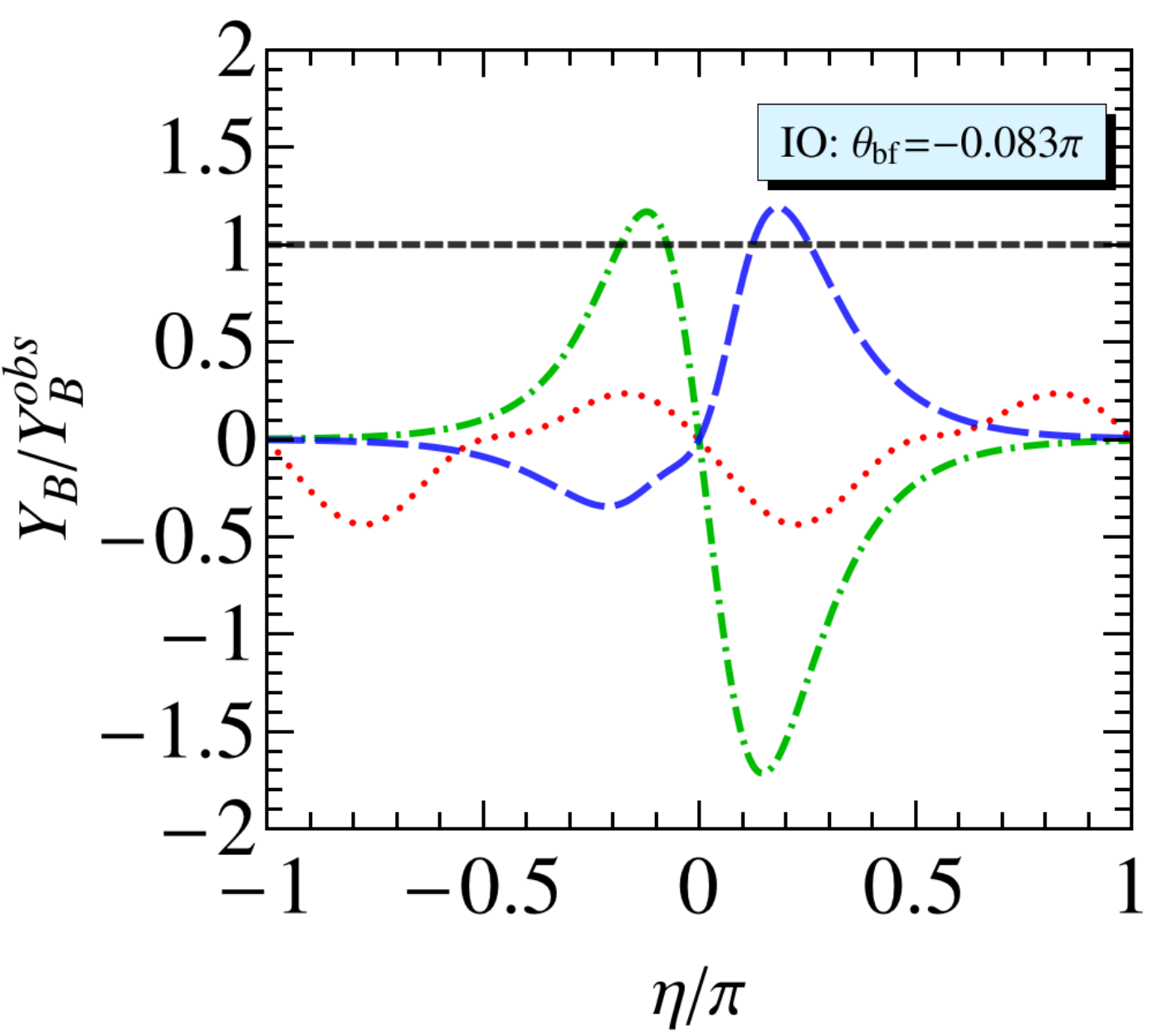}
\caption{\label{fig:leptogenesis_CaseIa}
The prediction for $Y_B/Y_B^{obs}$ as a function of $\eta$ in case I(a) with $(\varphi_1,\varphi_2)=(\frac{\pi}{2},\frac{\pi}{2})$, where $\theta_{\text{bf}}$ is the best fit value of $\theta$. Note that minor difference in $\theta_{\text{bf}}$ is obtained for NO and IO spectrums, because the best fit value as well as $1\sigma$ error of $\sin^2\theta_{13}$ and $\sin^2\theta_{23}$ slightly depend on the mass ordering~\cite{Gonzalez-Garcia:2014bfa}. We choose $M_1=5\times 10^{11}$ GeV and the lightest neutrino mass $m_1$ (or $m_3$) = 0.01eV. The red dotted, green dot-dashed, blue dashed lines correspond to $(K_1,K_2,K_3)=(\pm,+,+),(\pm,+,-)$ and $(\pm,-,+)$ respectively. The experimentally observed value $Y_B^{obs}$ is represented by the horizontal black dashed line.}
\end{center}
\end{figure}

\begin{figure}[hptb]
\begin{center}
\includegraphics[width=0.42\linewidth]{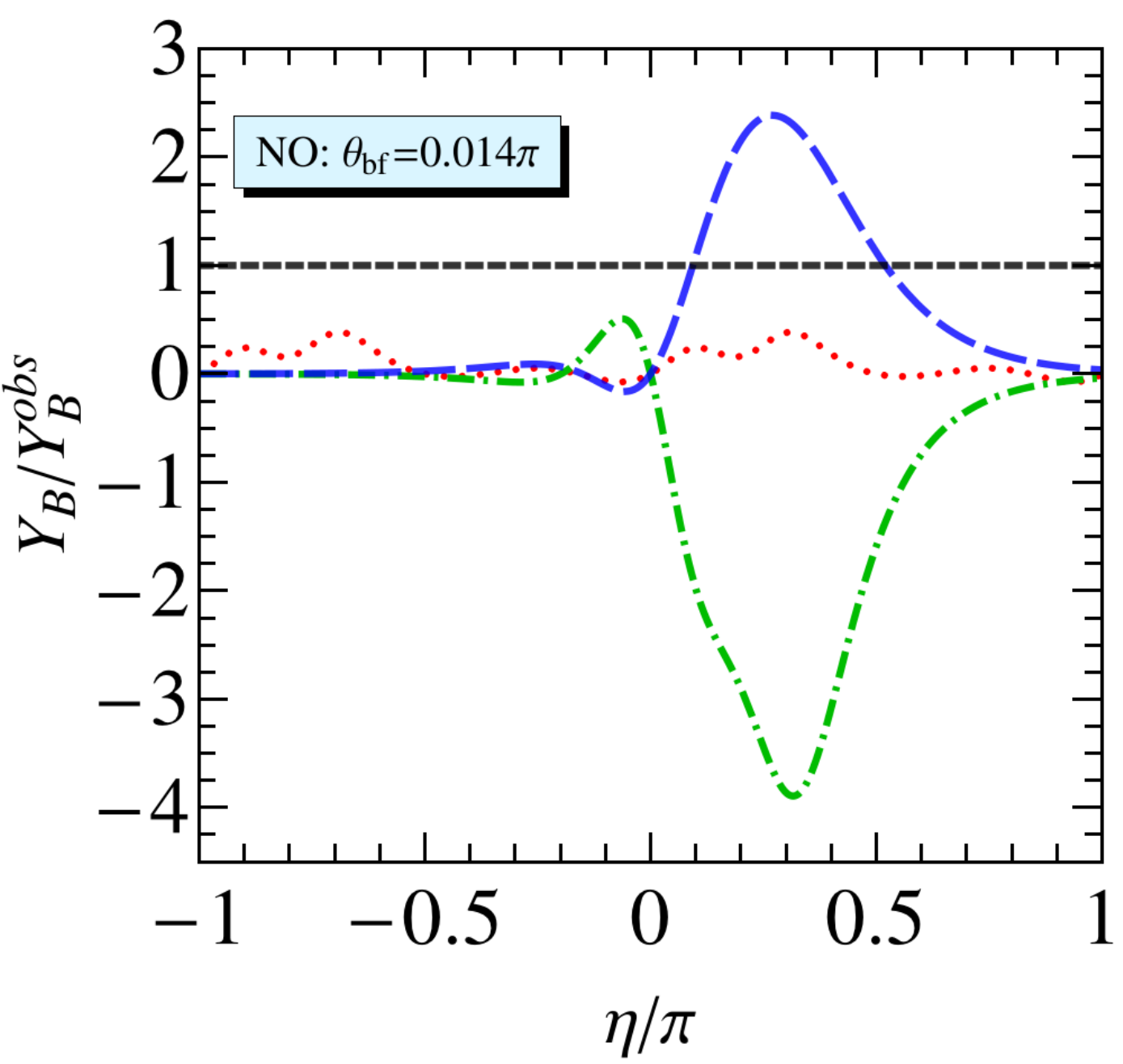}~~~
\includegraphics[width=0.42\linewidth]{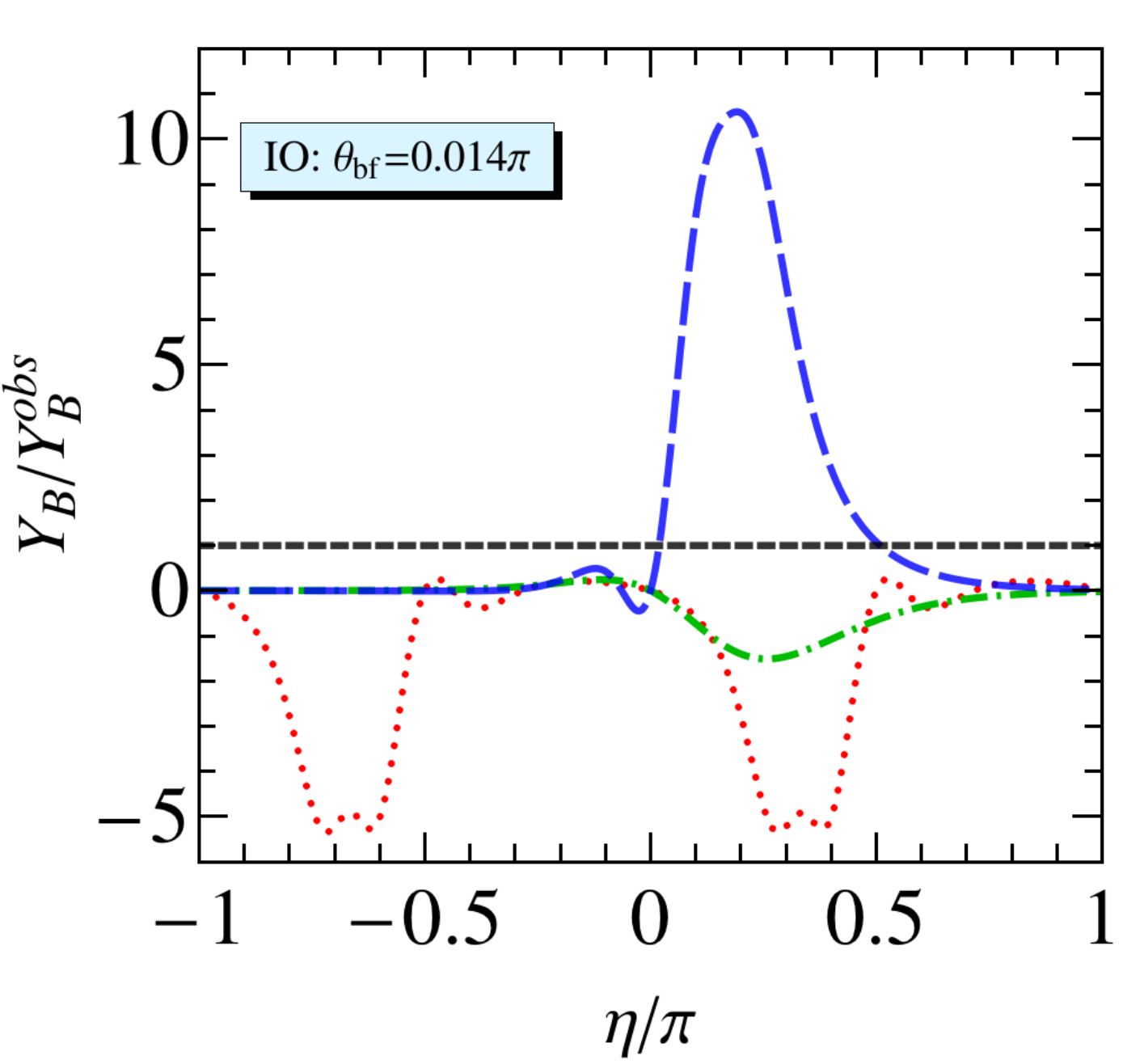}
\includegraphics[width=0.42\linewidth]{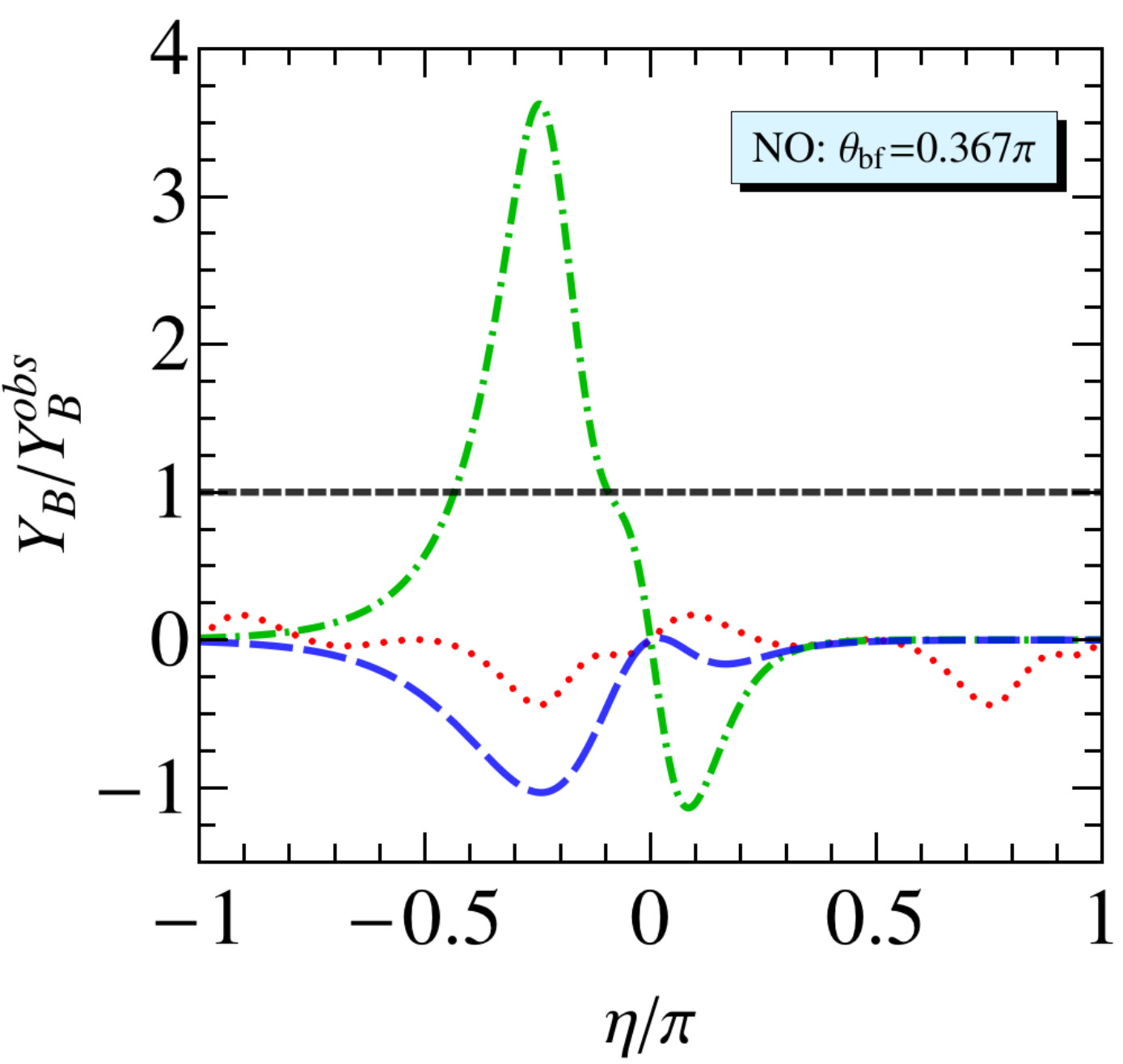}~~~
\includegraphics[width=0.42\linewidth]{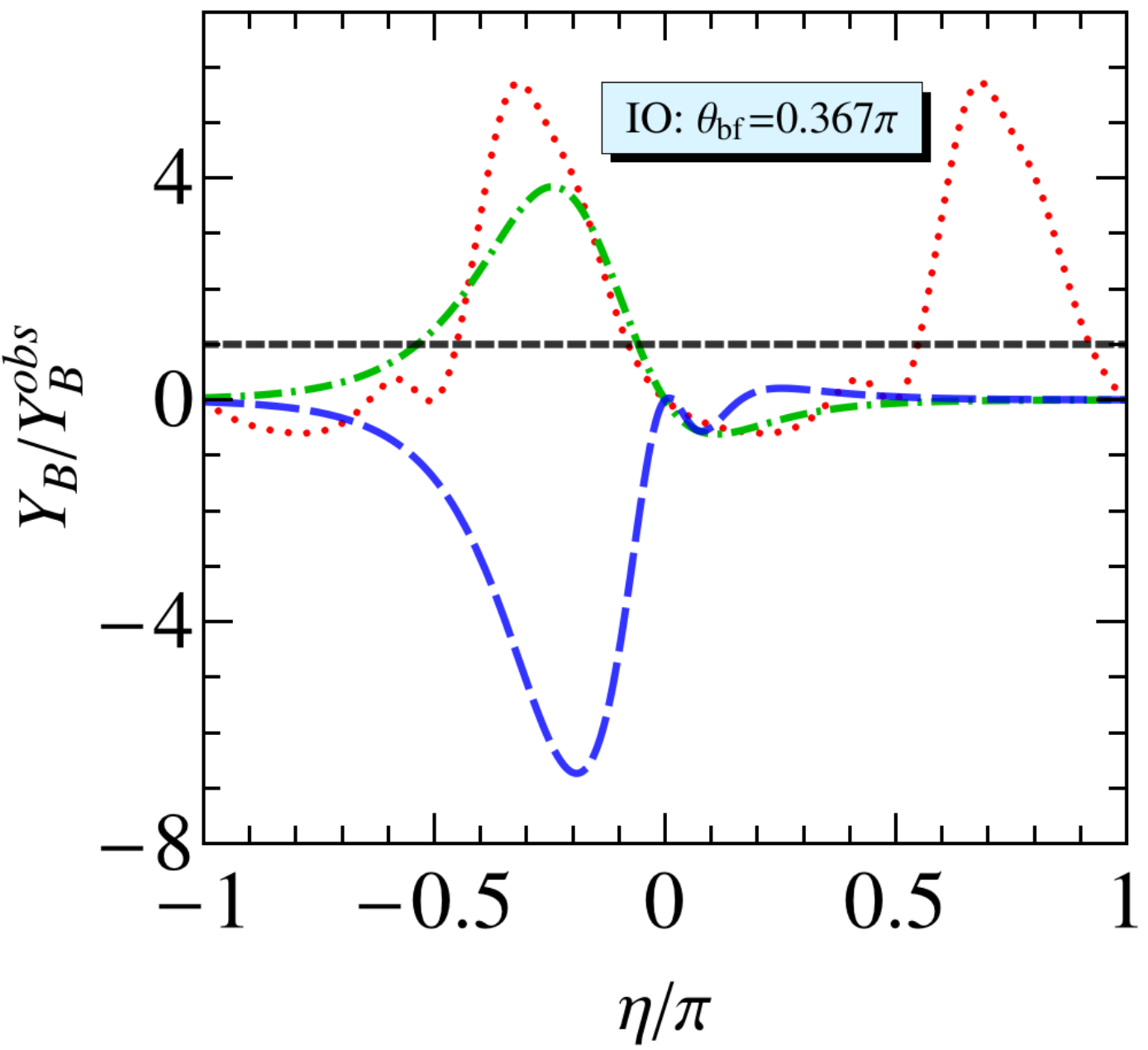}
\caption{\label{fig:leptogenesis_CaseIb1}
The prediction for $Y_B/Y_B^{obs}$ as a function of $\eta$ in case I(b) with $(\varphi_1,\varphi_2)=(\frac{\pi}{18},-\frac{\pi}{6})$, where $\theta_{\text{bf}}$ is the best fit value of $\theta$. We choose $M_1=5\times 10^{11}$ GeV and the lightest neutrino mass $m_1$ (or $m_3$) = 0.01eV. The red dotted,  green dot-dashed, blue dashed lines correspond to $(K_1,K_2,K_3)=(+,+,\pm),(+,-,\pm)$ and $(-,+,\pm)$ respectively. The experimentally observed value $Y_B^{obs}$ is represented by the horizontal black dashed line.}
\end{center}
\end{figure}

\begin{figure}[hptb]
\begin{center}
\includegraphics[width=0.42\linewidth]{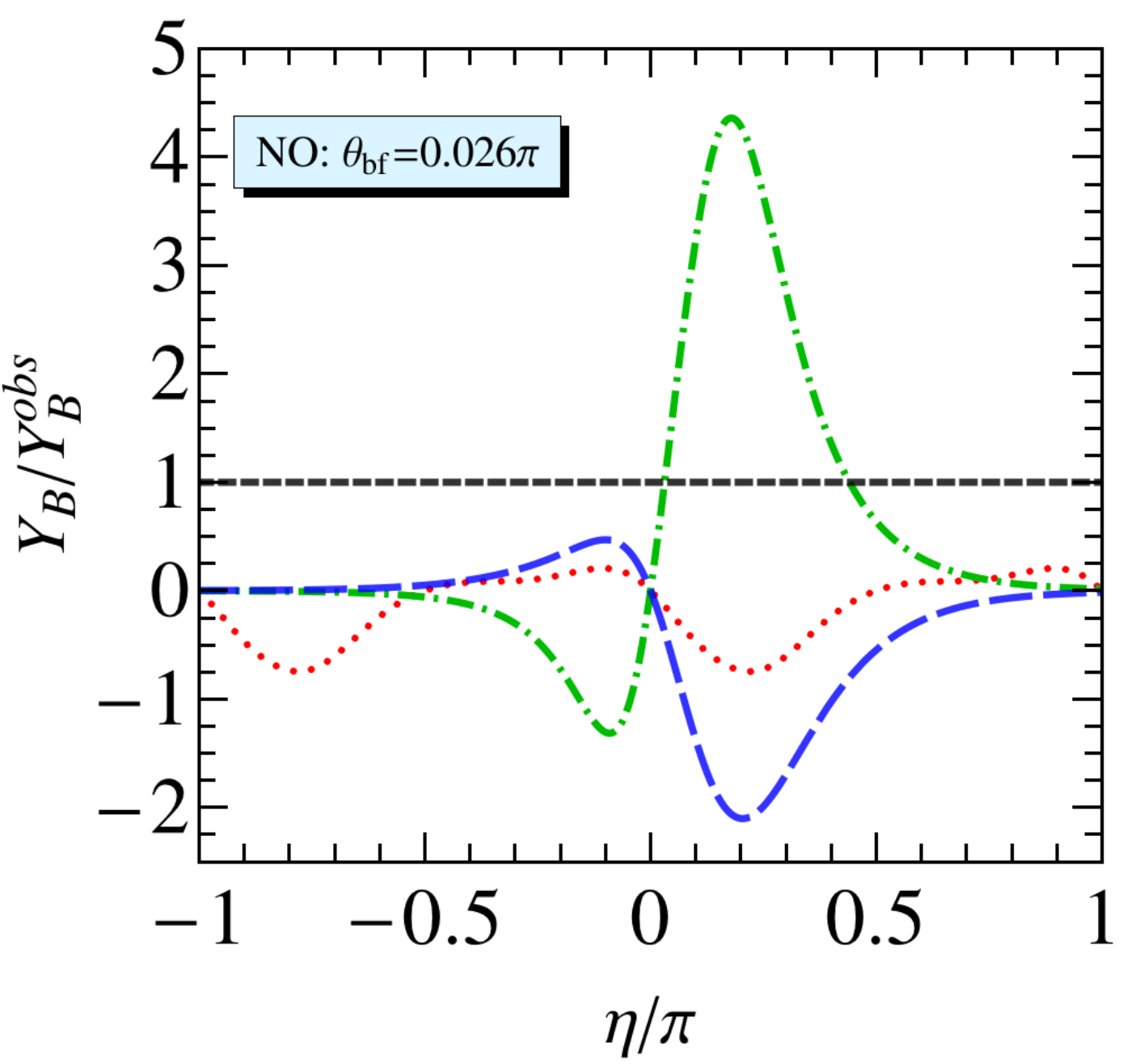}~~~
\includegraphics[width=0.435\linewidth]{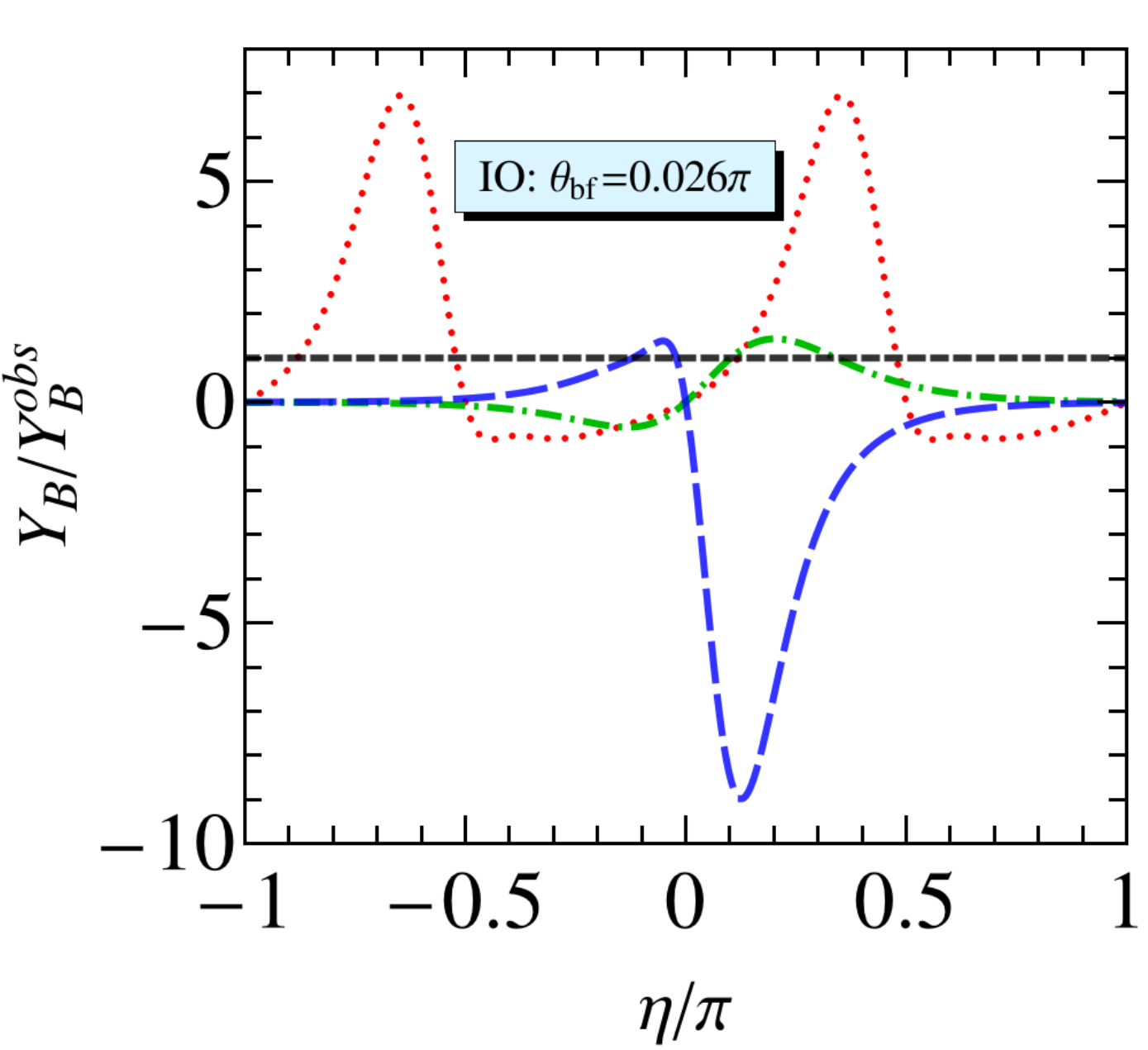}
\includegraphics[width=0.42\linewidth]{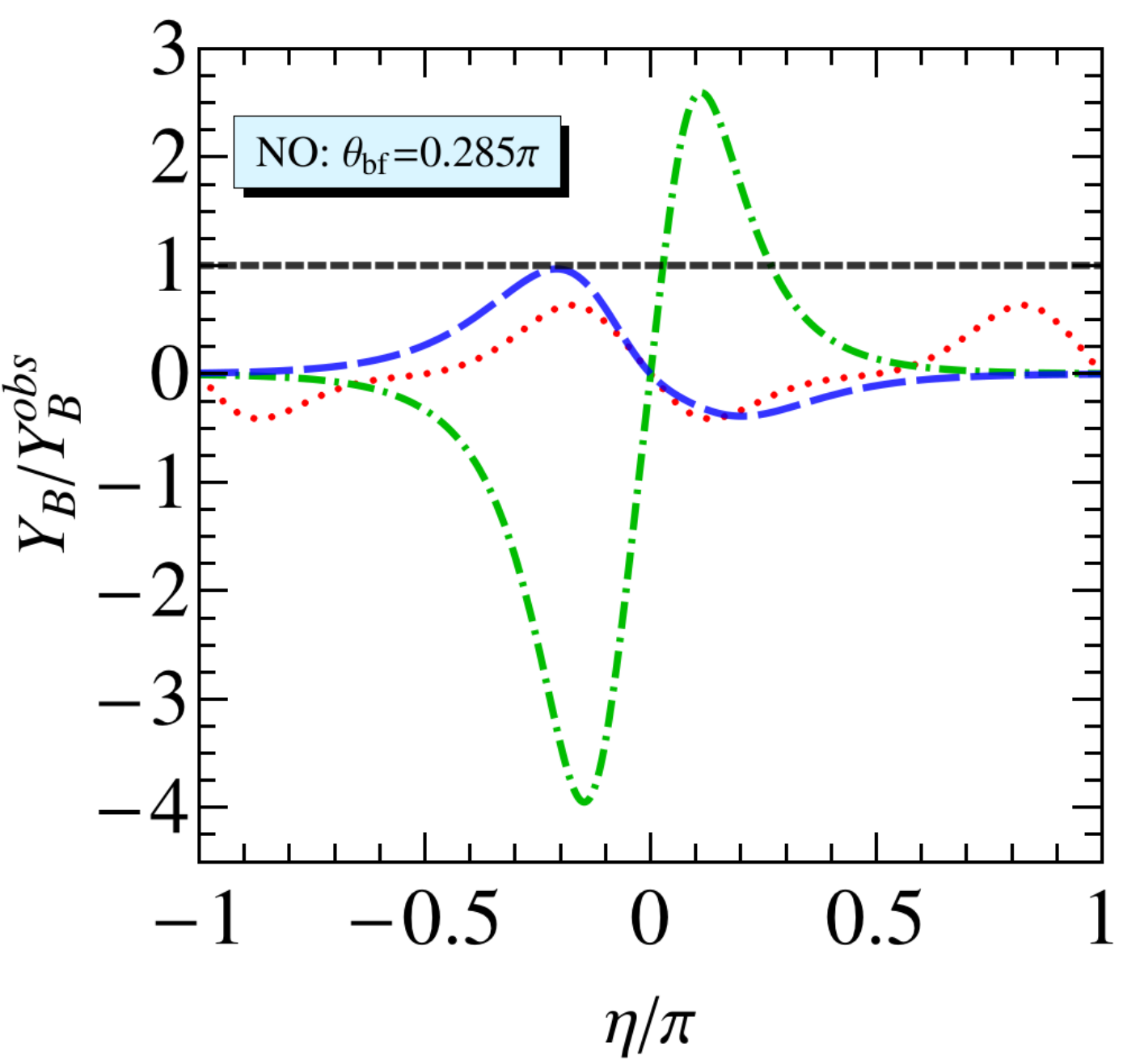}~~~~~
\includegraphics[width=0.42\linewidth]{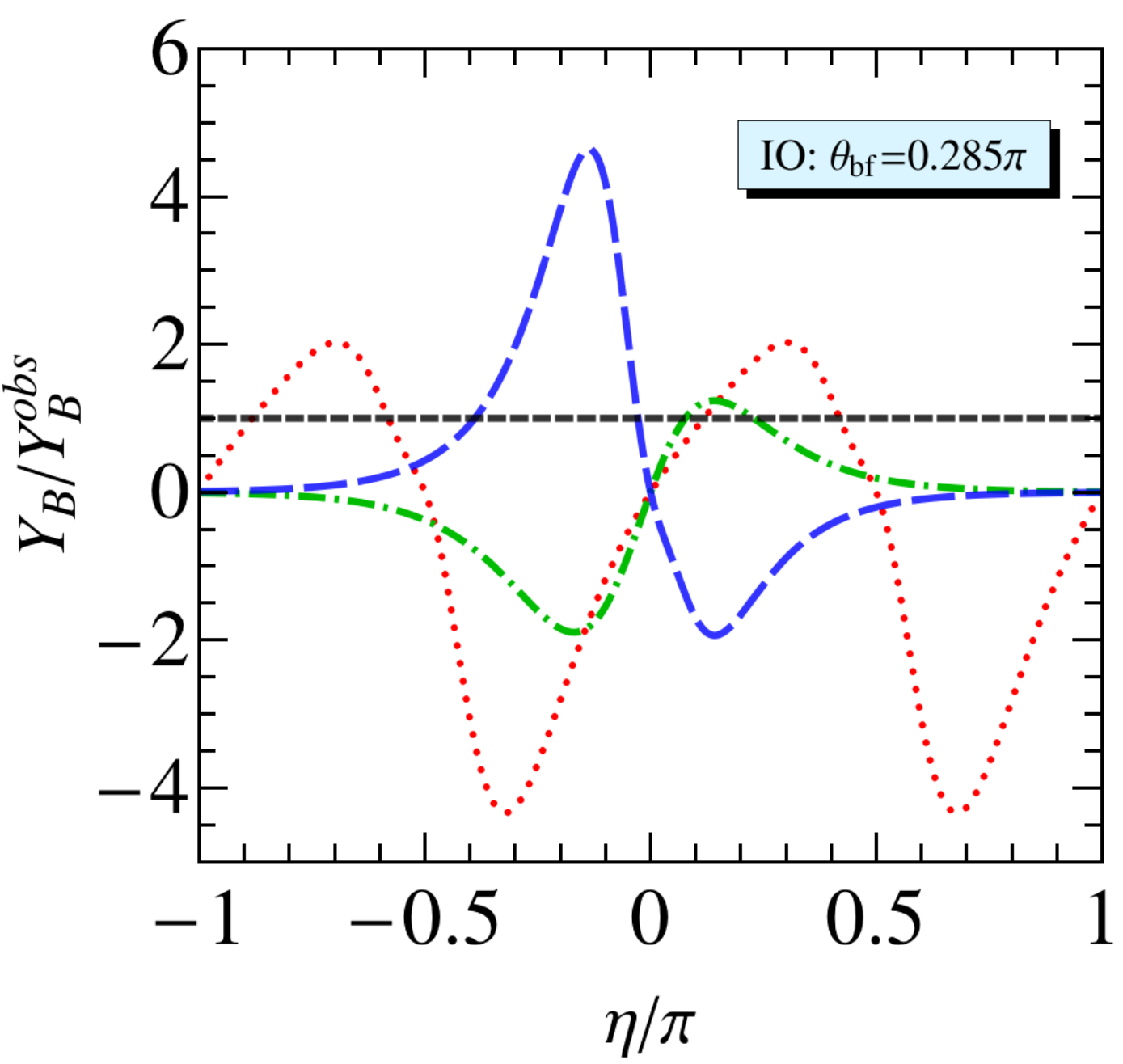}
\caption{\label{fig:leptogenesis_CaseIb2}
The prediction for $Y_B/Y_B^{obs}$ as a function of $\eta$ in case I(b) with $(\varphi_1,\varphi_2)=(\frac{\pi}{18},\frac{\pi}{3})$, where $\theta_{\text{bf}}$ is the best fit value of $\theta$. We choose $M_1=5\times 10^{11}$ GeV and the lightest neutrino mass $m_1$ (or $m_3$) = 0.01eV. The red dotted, green dot-dashed, blue dashed lines correspond to $(K_1,K_2,K_3)=(+,+,\pm),(+,-,\pm)$ and $(-,+,\pm)$ respectively. The experimentally observed value $Y_B^{obs}$ is represented by the horizontal black dashed line. }
\end{center}
\end{figure}

\begin{figure}[hptb]
\begin{center}
\includegraphics[width=0.42\linewidth]{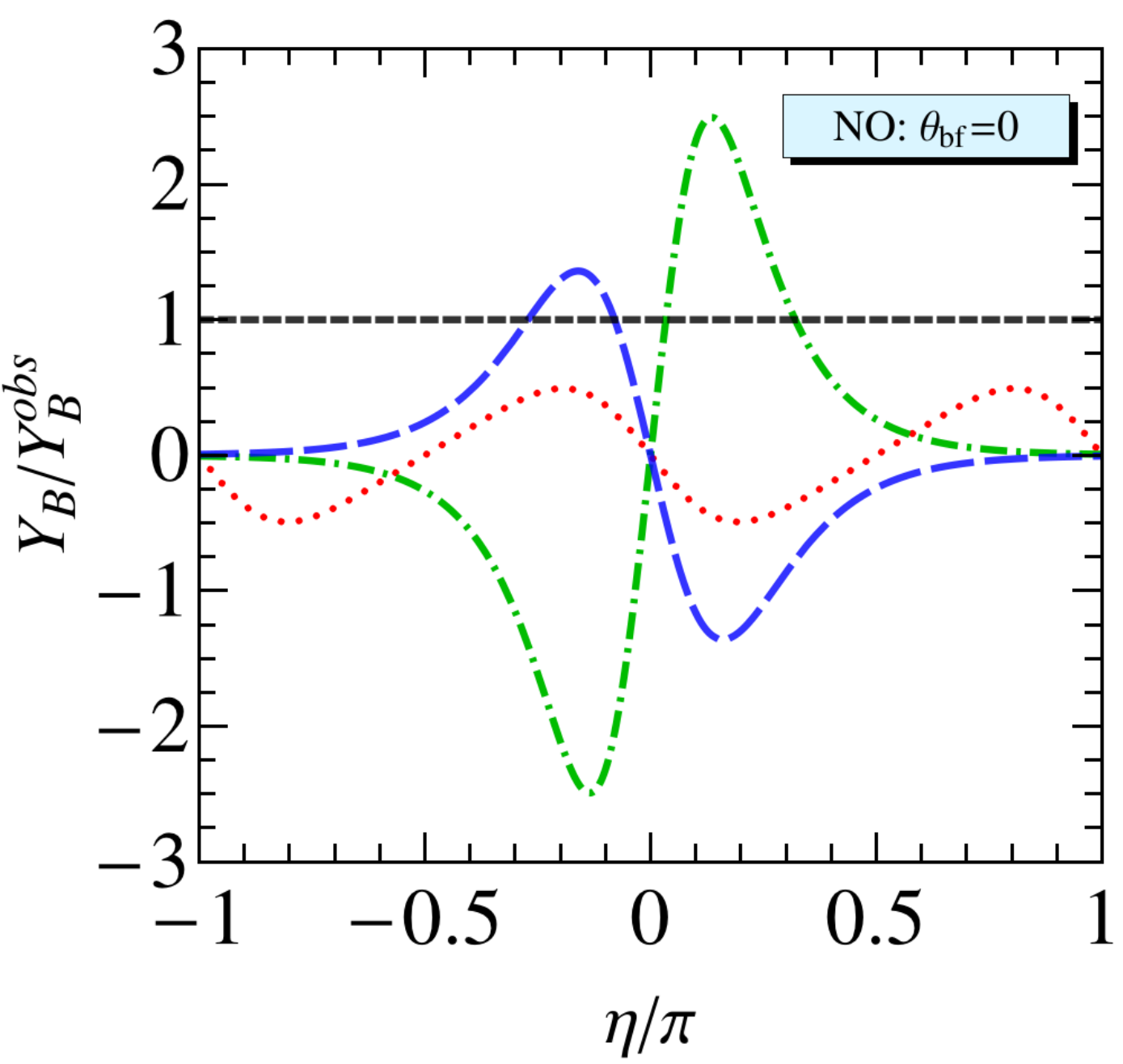}
\includegraphics[width=0.42\linewidth]{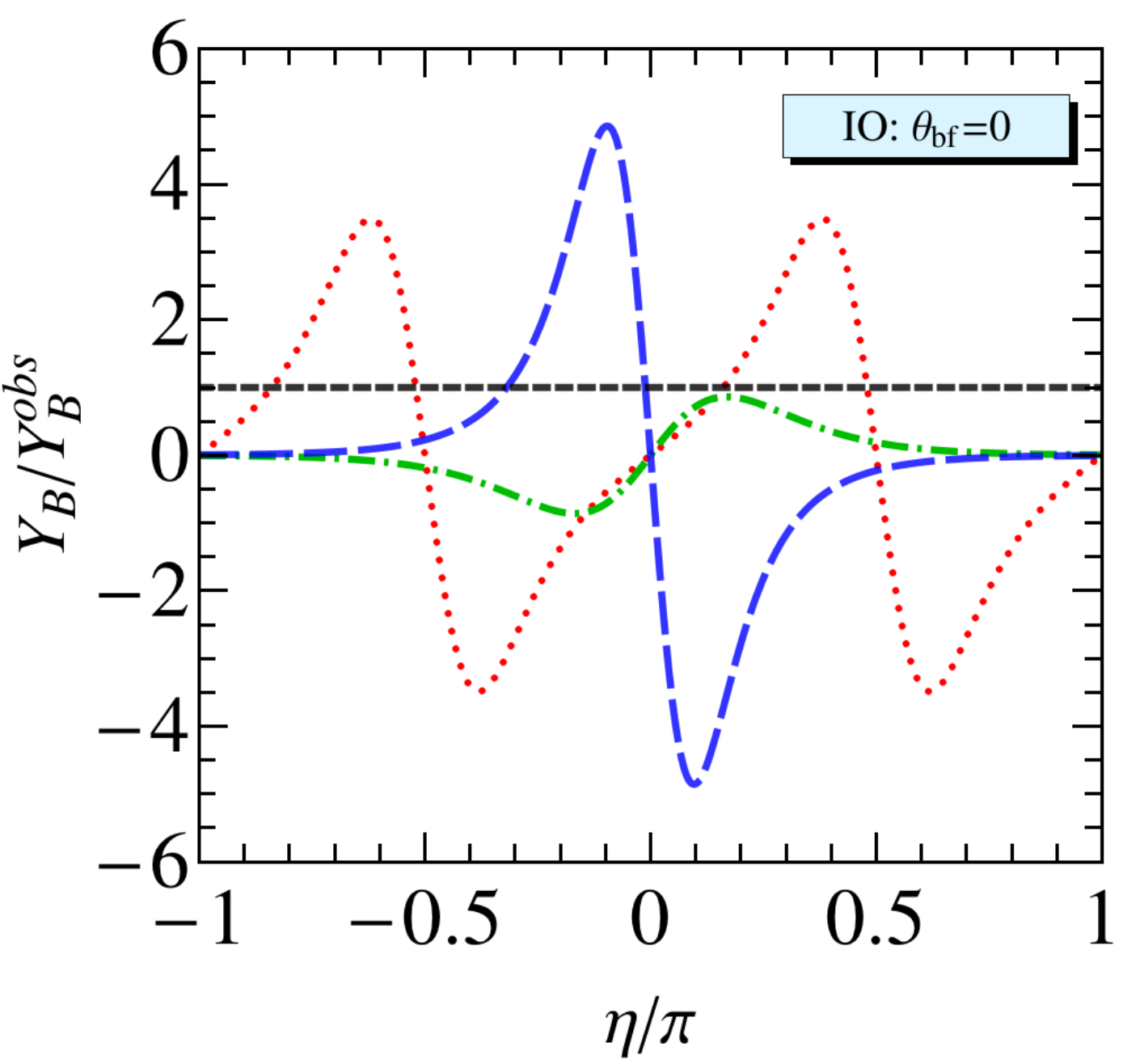}
\caption{\label{fig:leptogenesis_CaseIb3}
The prediction for $Y_B/Y_B^{obs}$ as a function of $\eta$ in case I(b) with $(\varphi_1,\varphi_2)=(\frac{\pi}{18},\frac{\pi}{2})$, where $\theta_{\text{bf}}$ is the best fit value of $\theta$. We choose $M_1=5\times 10^{11}$ GeV and the lightest neutrino mass $m_1$ (or $m_3$) = 0.01eV. The red dotted, green dot-dashed, blue dashed lines correspond to $(K_1,K_2,K_3)=(+,+,\pm),(+,-,\pm)$ and $(-,+,\pm)$ respectively. The experimentally observed value $Y_B^{obs}$ is represented by the horizontal black dashed line. }
\end{center}
\end{figure}

\begin{figure}[hptb]
\begin{center}
\includegraphics[width=0.42\linewidth]{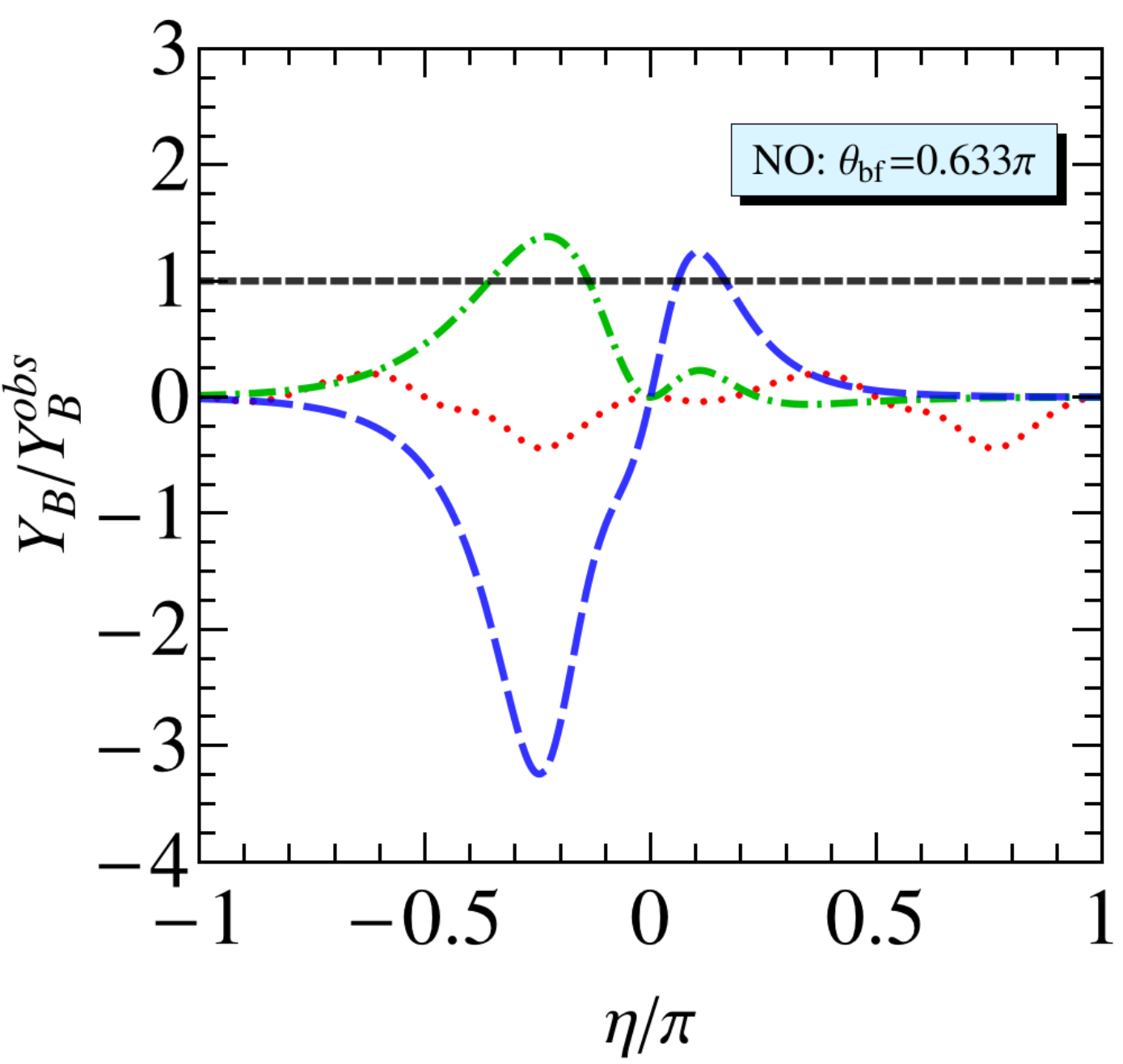}~~~
\includegraphics[width=0.435\linewidth]{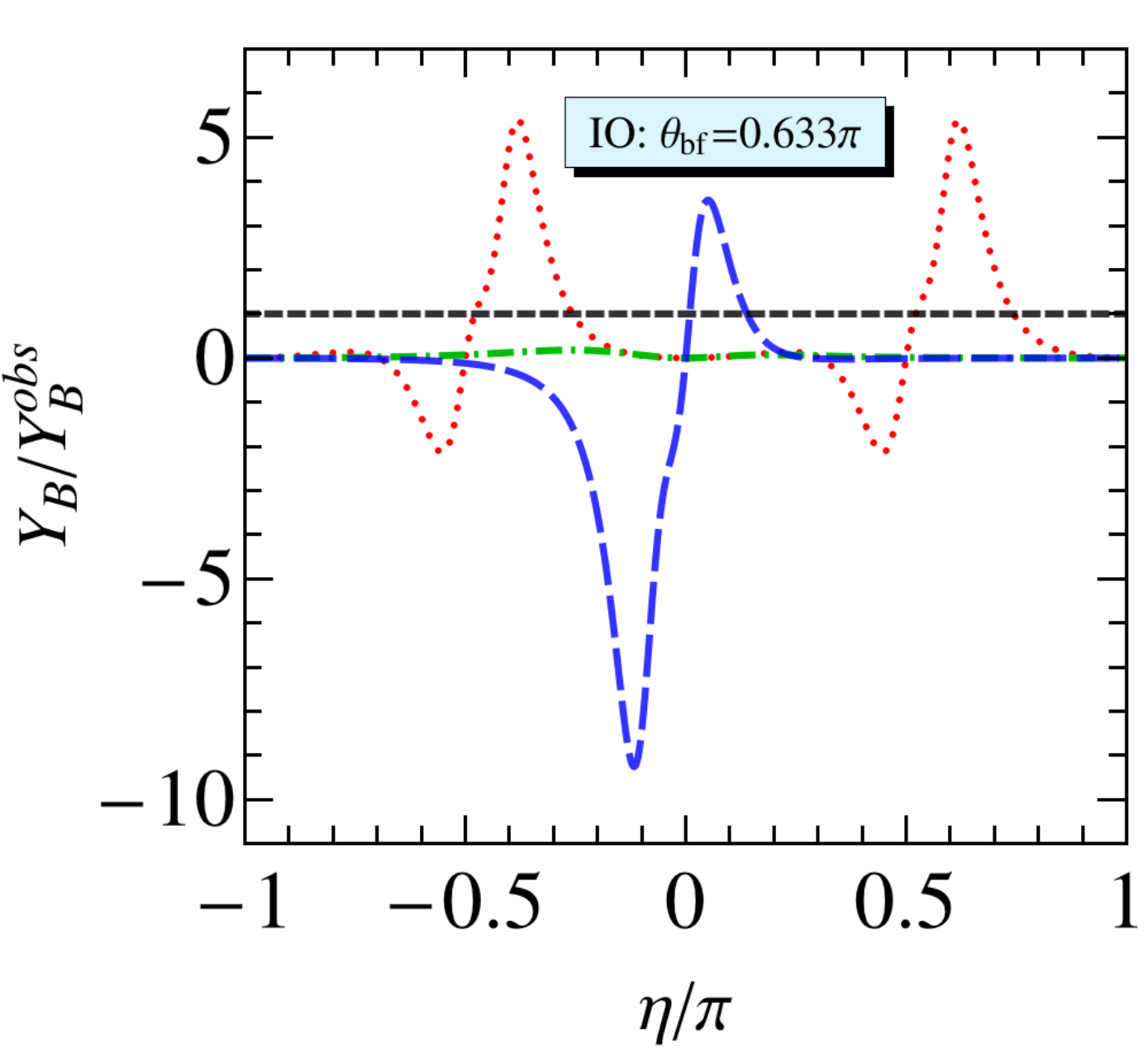}
\includegraphics[width=0.42\linewidth]{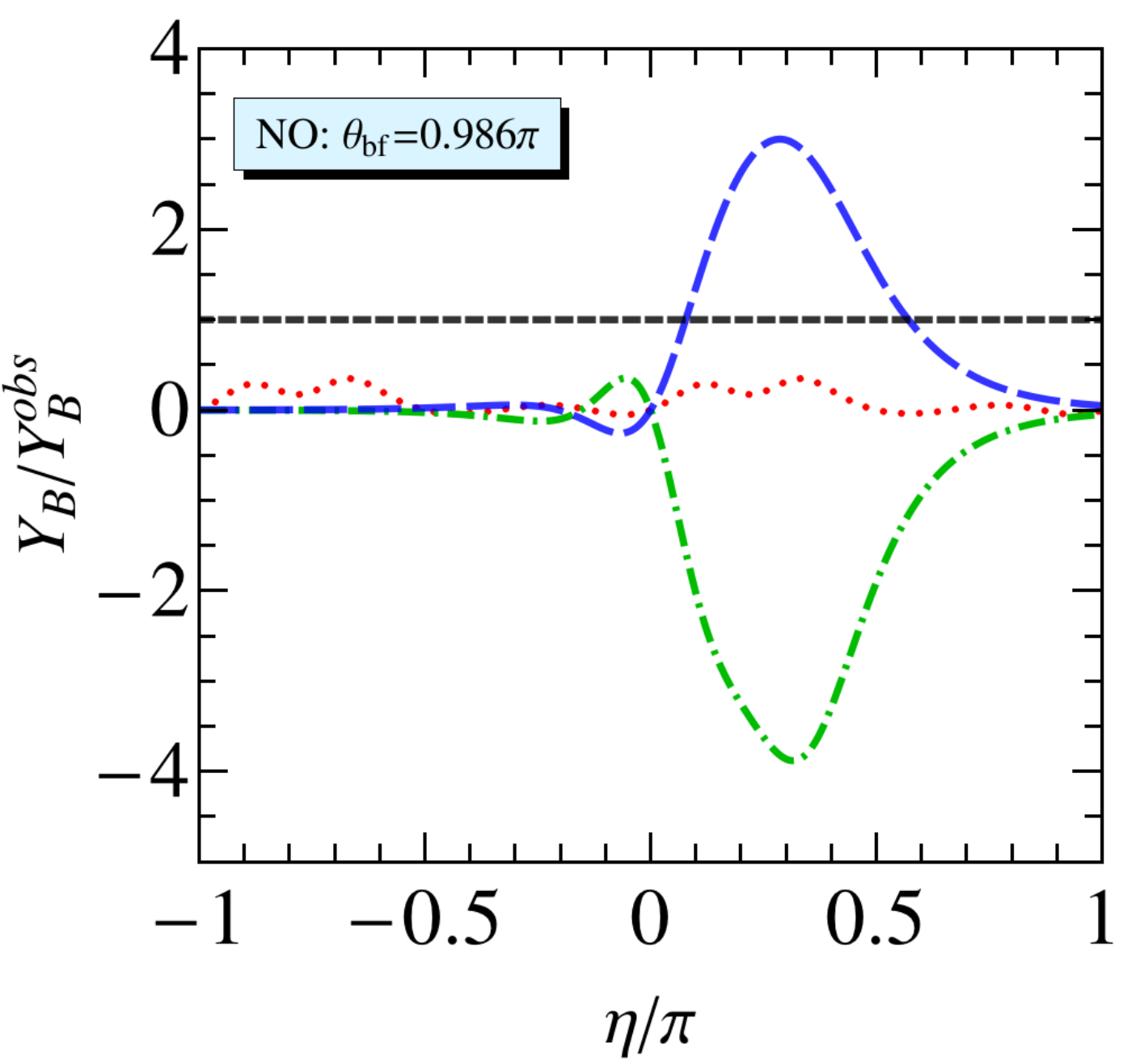}~~~~
\includegraphics[width=0.42\linewidth]{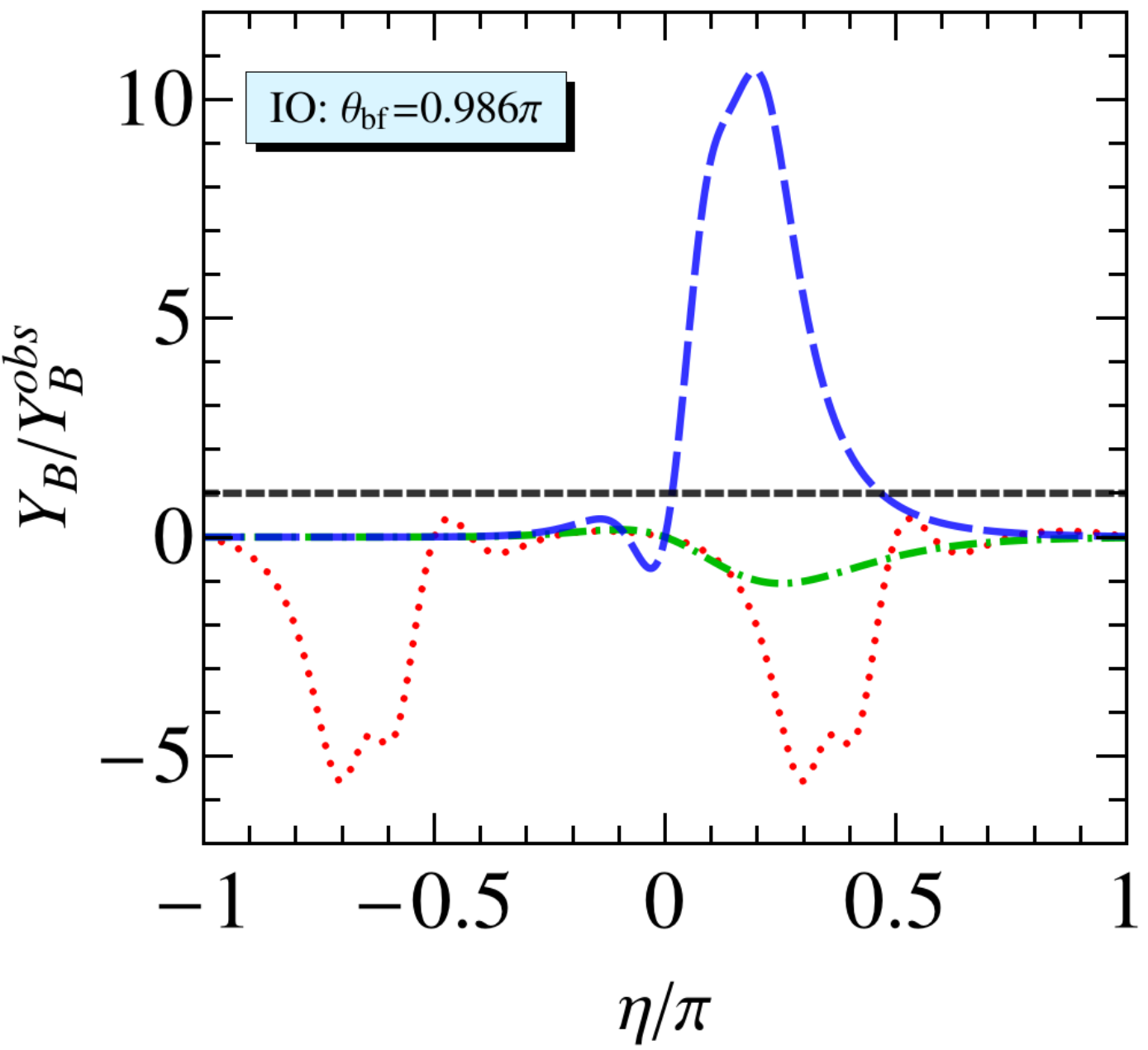}
\caption{\label{fig:leptogenesis_CaseIb4}
The prediction for $Y_B/Y_B^{obs}$ as a function of $\eta$ in case I(b) with  $(\varphi_1,\varphi_2)=(\frac{17\pi}{18},-\frac{\pi}{6})$, where $\theta_{\text{bf}}$ is the best fit value of $\theta$. We choose $M_1=5\times 10^{11}$ GeV and the lightest neutrino mass $m_1$ (or $m_3$) = 0.01eV. The red dotted,  green dot-dashed, blue dashed lines correspond to $(K_1,K_2,K_3)=(+,+,\pm),(+,-,\pm)$ and $(-,+,\pm)$ respectively. The experimentally observed value $Y_B^{obs}$ is represented by the horizontal black dashed line.}
\end{center}
\end{figure}

\begin{figure}[tbp]
\begin{center}
\includegraphics[width=0.42\linewidth]{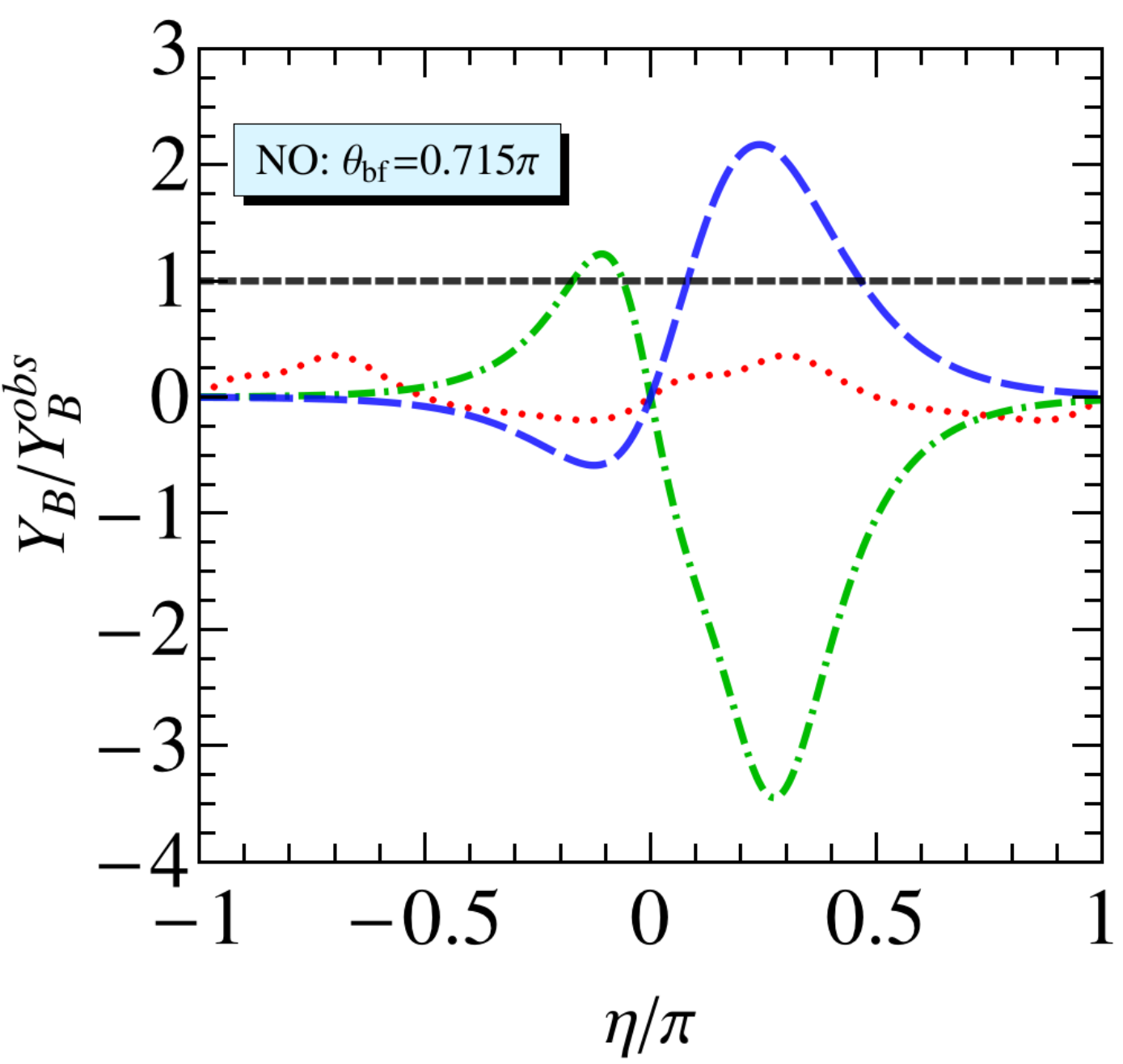}~~~
\includegraphics[width=0.42\linewidth]{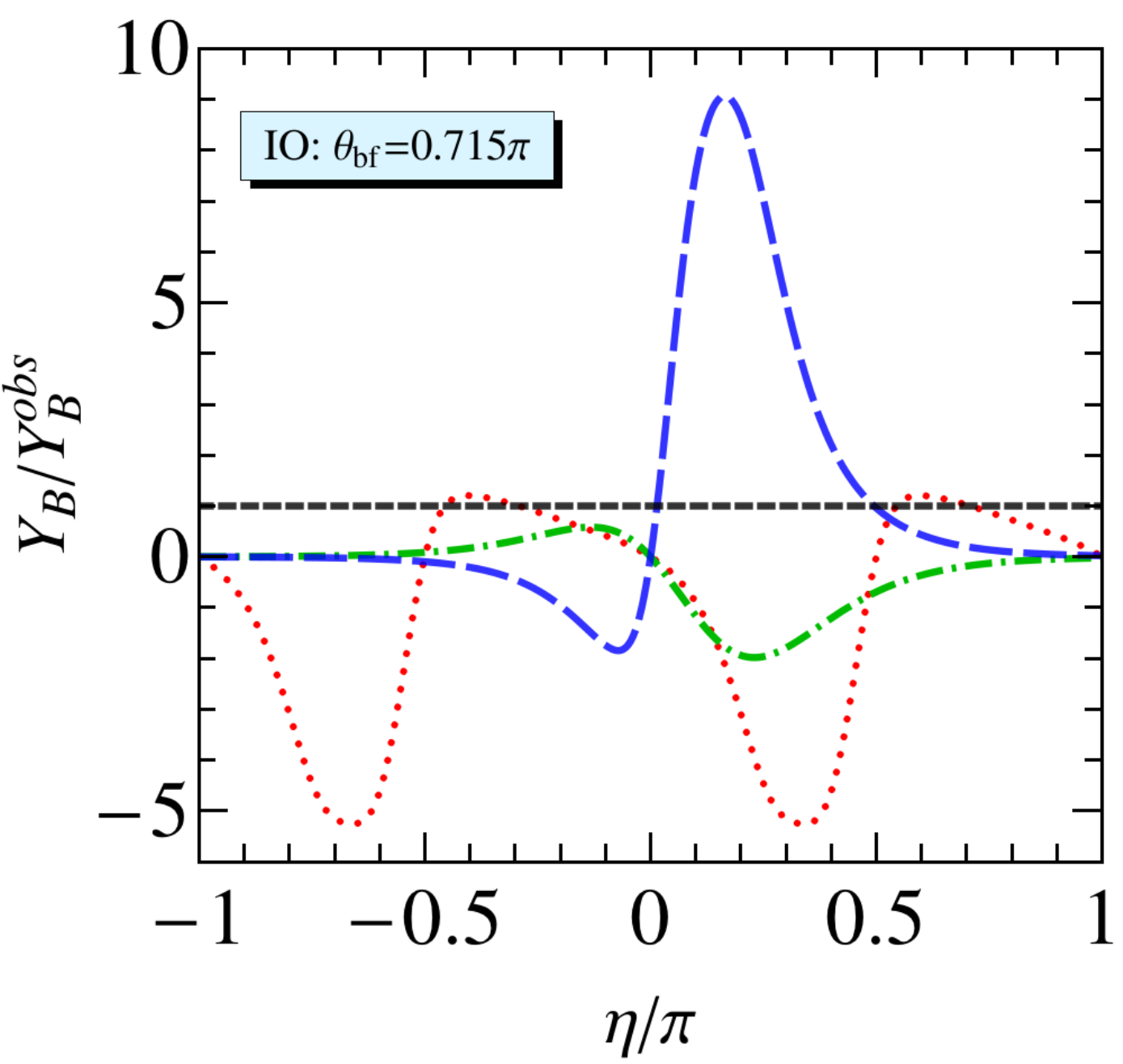}
\includegraphics[width=0.42\linewidth]{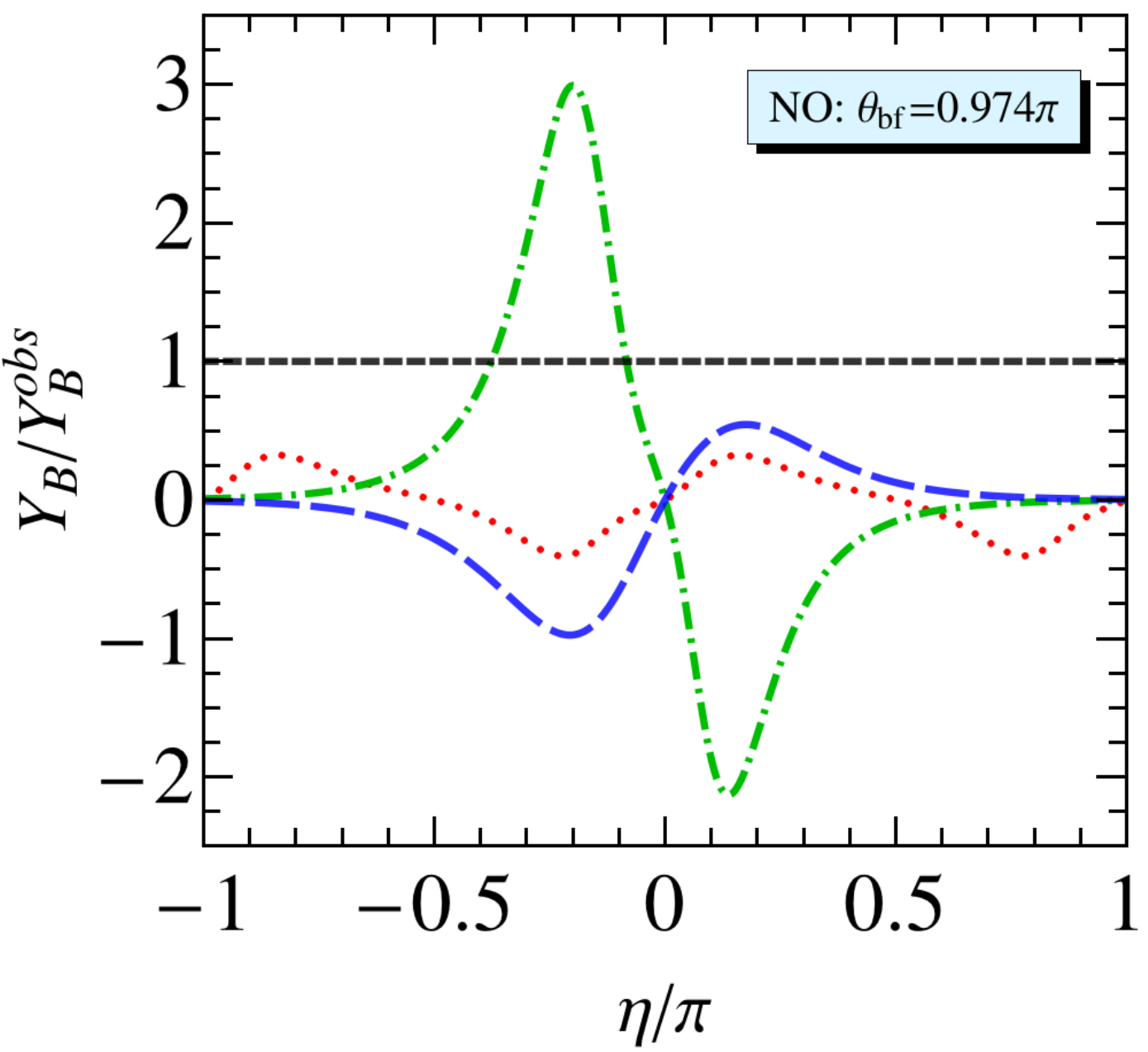}~~~
\includegraphics[width=0.42\linewidth]{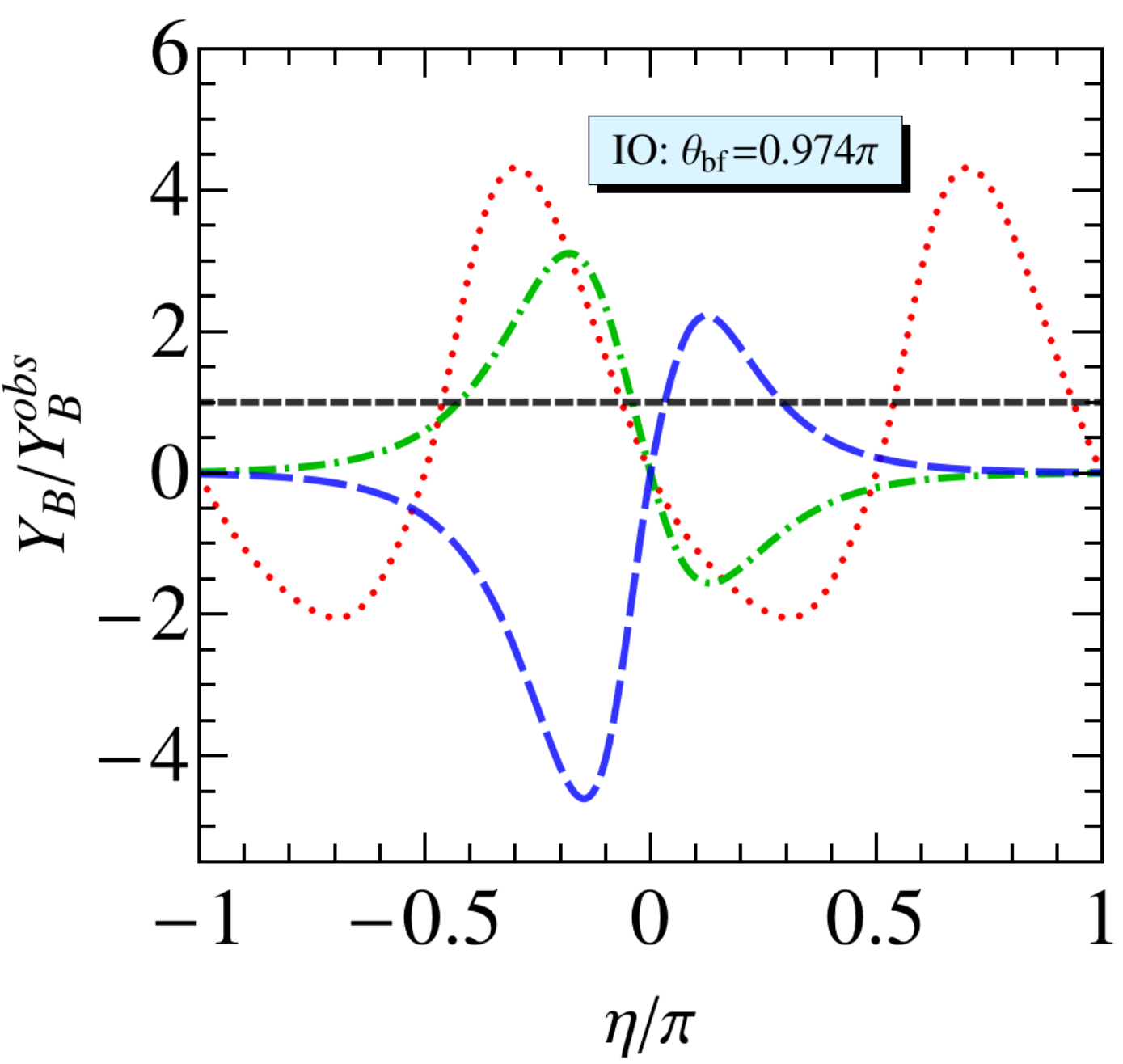}
\caption{\label{fig:leptogenesis_CaseIb5}
The prediction for $Y_B/Y_B^{obs}$ as a function of $\eta$ in case I(b) with $(\varphi_1,\varphi_2)=(\frac{17\pi}{18},\frac{\pi}{3})$, where $\theta_{\text{bf}}$ is the best fit value of $\theta$. We choose $M_1=5\times 10^{11}$ GeV and the lightest neutrino mass $m_1$ (or $m_3$) = 0.01eV. The red dotted, green dot-dashed, blue dashed lines correspond to $(K_1,K_2,K_3)=(+,+,\pm),(+,-,\pm)$ and $(-,+,\pm)$ respectively. The experimentally observed value $Y_B^{obs}$ is represented by the horizontal black dashed line.}
\end{center}
\end{figure}

\begin{figure}[tbp]
\begin{center}
\includegraphics[width=0.42\linewidth]{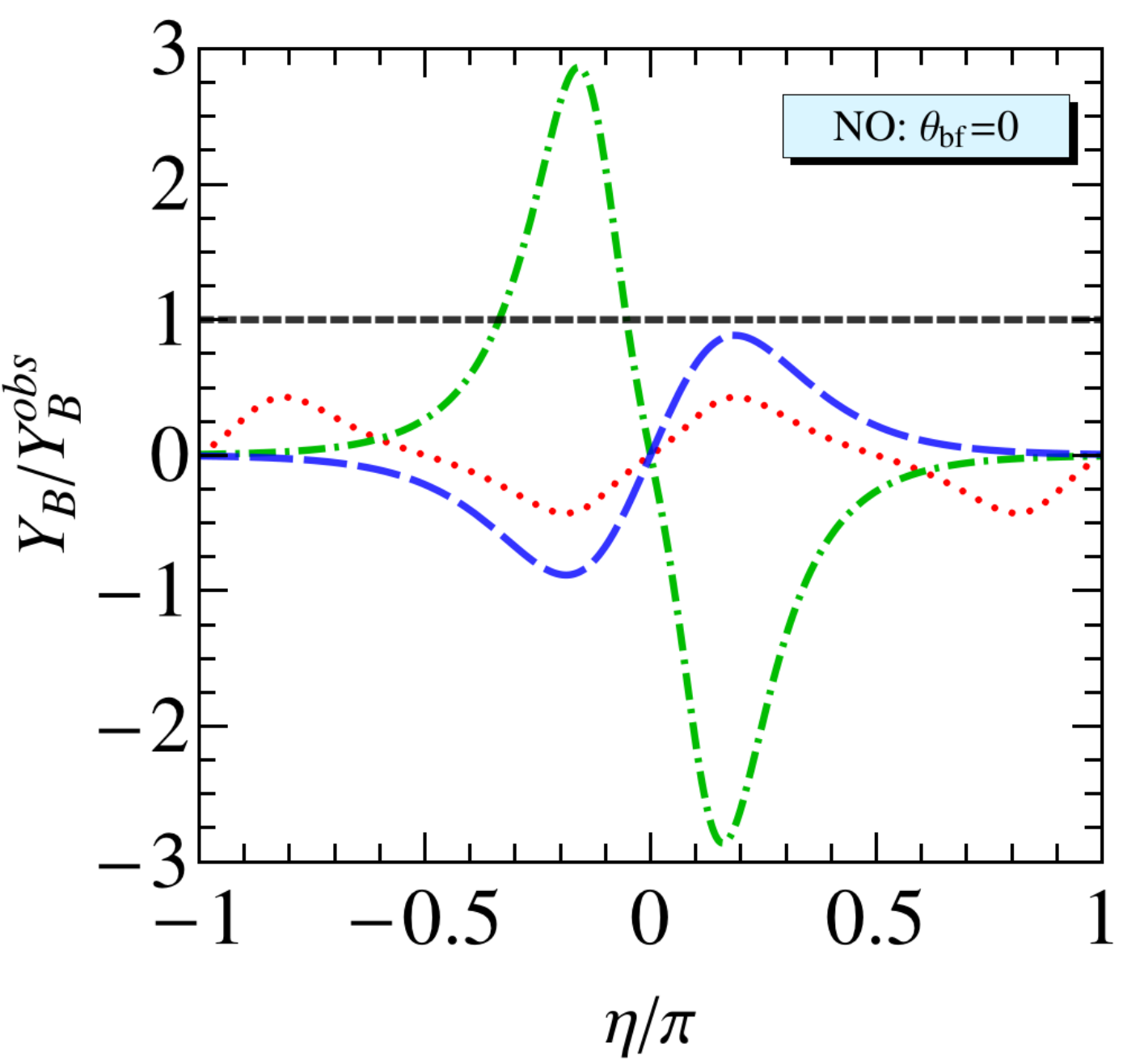}~~~
\includegraphics[width=0.42\linewidth]{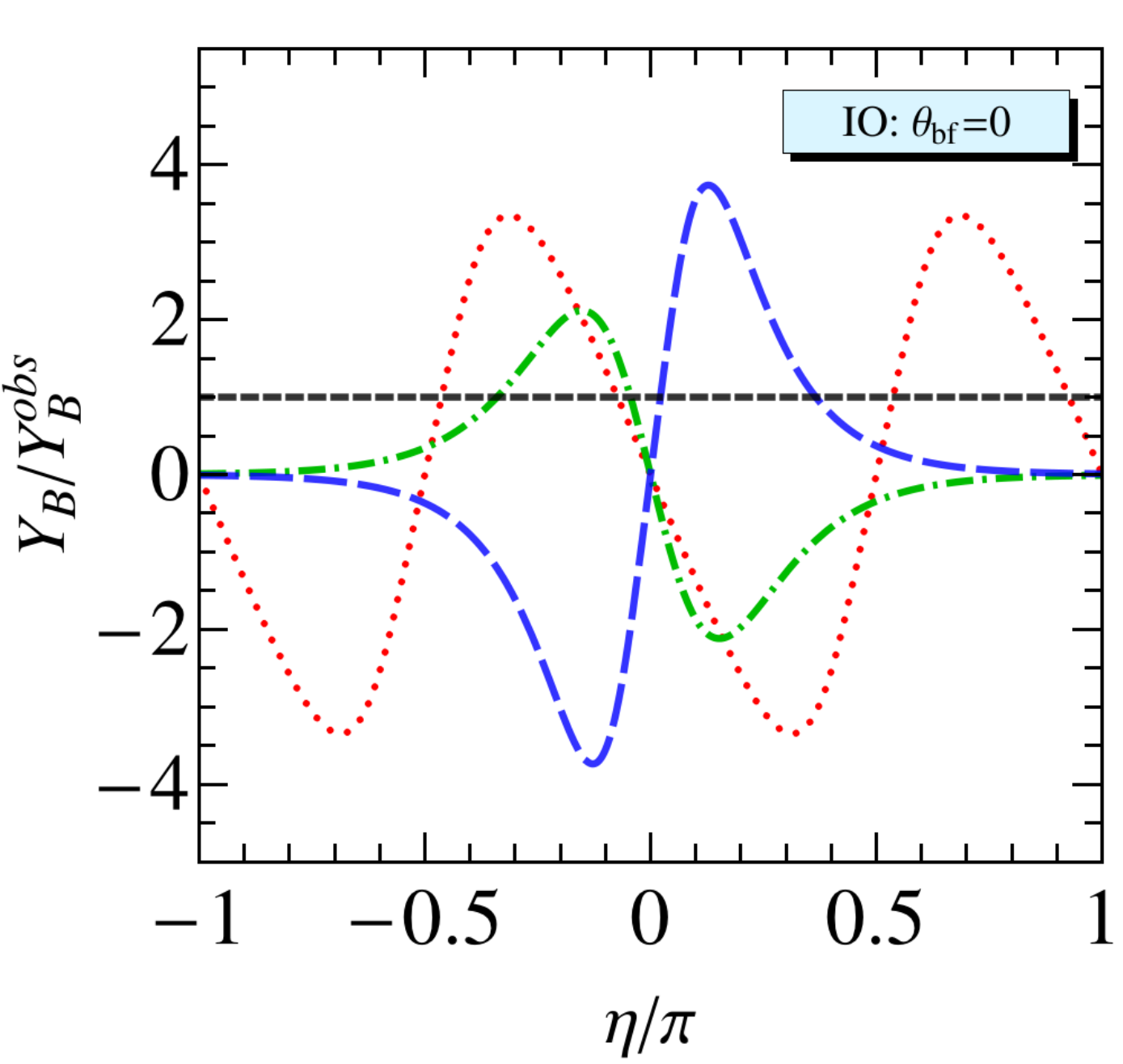}
\caption{\label{fig:leptogenesis_CaseIb6}
The prediction for $Y_B/Y_B^{obs}$ as a function of $\eta$ in case I(b) with $(\varphi_1,\varphi_2)=(\frac{17\pi}{18},\frac{\pi}{2})$, where $\theta_{\text{bf}}$ is the best fit value of $\theta$. We choose $M_1=5\times 10^{11}$ GeV and the lightest neutrino mass $m_1$ (or $m_3$)=0.01eV. The red dotted, green dot-dashed, blue dashed lines correspond to $(K_1,K_2,K_3)=(+,+,\pm),(+,-,\pm)$ and $(-,+,\pm)$ respectively. The experimentally observed value $Y_B^{obs}$ is represented by the horizontal black dashed line.}
\end{center}
\end{figure}

\item[Case \uppercase\expandafter{\romannumeral2}]

\begin{eqnarray}
\label{eq:pmnsIIa}
U^{II(a)}&=&\frac{1}{\sqrt{3}}
 \left(
\begin{array}{ccc}
 e^{i \varphi _1} & 1 & e^{i \varphi _2} \\
 \omega  e^{i \varphi _1} & 1 & \omega ^2 e^{i \varphi _2} \\
 \omega ^2 e^{i \varphi _1} & 1 & \omega  e^{i \varphi _2} \\
\end{array}
\right)S_{13}(\theta)Q^{\dagger}_{\nu}\,,\\
\label{eq:pmnsIIb}
U^{II(b)}&=&\frac{1}{\sqrt{3}}
 \left(
\begin{array}{ccc}
 e^{i \varphi _1} & 1 & e^{i \varphi _2} \\
 \omega ^2 e^{i \varphi _1} & 1 & \omega  e^{i \varphi _2} \\
 \omega  e^{i \varphi _1} & 1 & \omega ^2 e^{i \varphi _2}
\end{array}
\right)S_{13}(\theta)Q^{\dagger}_{\nu}\,,
\end{eqnarray}
where $\omega=e^{i2\pi/3}$. The viable values of $\varphi_1$ and $\varphi_2$ and corresponding representative flavor symmetry groups are listed in table~\ref{tab:II}. Please see the website~\cite{webdata} for the full results. The smallest group which can describe the experimentally measured values of the mixing angles for certain values of $\theta$ is $S_4$. The mixing pattern in Eq.~\eqref{eq:pmnsIIb} results from the permutation of the second and third rows of the PMNS mixing matrix in Eq.~\eqref{eq:pmnsIIa}. The second column of $U^{II(a)}$ and $U^{II(b)}$ are $(1, 1, 1)^{T}/\sqrt{3}$, and consequently they are the trimaximal pattern. We can extract the following results for the lepton mixing angles
\begin{equation}
\label{eq:mixing_angles_case_II}
\begin{split}
\sin^2\theta_{13}&=\frac{1}{3}\left[1+\sin2\theta\cos(\varphi_2-\varphi_1)\right],\\
\sin^2\theta_{12}&=\frac{1}{2-\sin2\theta\cos(\varphi_2-\varphi_1)}\,,\\
\sin^2\theta_{23}&=\frac{1-\sin2\theta\sin\left(\varphi_2-\varphi_1+\pi/6\right)}{2-\sin2\theta\cos(\varphi_2-\varphi_1)}~~~\mathrm{for}~~~U^{II(a)}\,,\\
\sin^2\theta_{23}&=\frac{1+\sin2\theta\sin\left(\varphi_2-\varphi_1-\pi/6\right)}{2-\sin2\theta\cos(\varphi_2-\varphi_1)}~~~\mathrm{for}~~~U^{II(b)}\,,
\end{split}
\end{equation}
Therefore the solar and the reactor mixing angles fulfill the well known sum rule
\begin{equation}
3\cos^2\theta_{13}\sin^2\theta_{12}=1\,.
\end{equation}
Hence the solar mixing angle admits a lower bound $\sin^2\theta_{12}>1/3$. Using for $\sin^2\theta_{13}$ its $3\sigma$ range $0.0188\leq\sin^2\theta_{13}\leq0.0251$~\cite{Gonzalez-Garcia:2014bfa}, we find $0.340\leq\sin^2\theta_{12}\leq0.342$. The JUNO experiment will be capable of reducing the error of $\sin^2\theta_{12}$ to about $0.1^{\circ}$ or around $0.3\%$~\cite{An:2015jdp}. Future long baseline experiments such as DUNE~\cite{Acciarri:2016crz} and Hyper-Kamiokande~\cite{Kearns:2013lea} can also make very precise measurements of the solar mixing angle. If significant deviations from $1/3$ of $\sin^2\theta_{12}$ were detected, this mixing pattern would be ruled out. Moreover, the reactor mixing angle and the atmospheric mixing angle are related as follow
\begin{equation}
\begin{split}
&\frac{3\cos^2\theta_{13}\sin^2\theta_{23}-1}{1-3\sin^2\theta_{13}}=\frac{1}{2}+\frac{\sqrt{3}}{2}\tan\left(\varphi_2-\varphi_1\right),~~~\text{for}~~~U^{II(a)}\,,\\
&\frac{3\cos^2\theta_{13}\sin^2\theta_{23}-1}{1-3\sin^2\theta_{13}}=\frac{1}{2}-\frac{\sqrt{3}}{2}\tan\left(\varphi_2-\varphi_1\right),~~~\text{for}~~~U^{II(b)}\,.
\end{split}
\end{equation}
For the mixing matrices $U^{II(a)}$ and $U^{II(b)}$, the CP invariants take the form
\begin{equation}
\begin{split}
\left|J_{CP}\right|&=\frac{1}{6\sqrt{3}}\left|\cos2\theta\right|,\\ |I_1|&=\frac{2}{9}\left|\left(\cos\theta\cos\varphi_1-\sin\theta\cos\varphi_2\right) \left(\cos\theta\sin\varphi_1-\sin\theta\sin\varphi_2\right)\right|\,,\\
|I_2|&=\frac{1}{9}\left|\cos2\theta\sin\left(2\varphi_1-2\varphi_2\right)\right|\,.
\end{split}
\end{equation}
We find that the mixing angles and Dirac CP violating phase fulfill the following sum rule
\begin{equation}
\label{eq:angle_phase_caseII}\cos\delta_{CP}=\frac{\cos2\theta_{13}\cot2\theta_{23}}{\sqrt{3\cos^2\theta_{13}-1}\,\sin\theta_{13}}\simeq\frac{\sqrt{2}\left(\pi/4-\theta_{23}\right)}{\theta_{13}}\,.
\end{equation}
Therefore the value of $\delta_{CP}$ is closely related with the deviation of $\theta_{23}$ from maximal mixing. Inputting the $3\sigma$ regions $0.0188\leq\sin^2\theta_{13}\leq0.0251$ and $0.385\leq\sin^2\theta_{23}\leq0.644$ from the global fit~\cite{Gonzalez-Garcia:2014bfa}, we see $\cos\delta_{CP}$ can be any value in the interval of $[-1,1]$. Hence no definite prediction can be made for $\delta_{CP}$ at present. However, if the uncertainly of the atmospheric mixing angle $\theta_{23}$ is reduced considerably by future neutrino experiments, the above sum rule in Eq.~\eqref{eq:angle_phase_caseII} could impose a strong constraint on the value of $\delta_{CP}$.

As shown in table~\ref{tab:II}, the group $G_{f}=S_4$ can give rise to the mixing patterns $U^{II(a)}$ and $U^{II(b)}$ with $(\varphi_1, \varphi_2)=(\pi, 0)$. Then the atmospheric angle $\theta_{23}$ as well as the Dirac CP phase $\delta_{CP}$ are predicted to be maximal while both Majorana phases are 0 or $\pi$. In fact, $U^{II(a)}$ and $U^{II(b)}$ are essentially the same mixing pattern in this case, since they are related by the redefinition of $\theta$ and $Q_{\nu}$
\begin{equation}
U^{II(b)}(\theta,\varphi_1=\pi, \varphi_2=0)=U^{II(a)}(\frac{\pi}{2}-\theta,\varphi_1=\pi, \varphi_2=0)\text{diag}(1,1,-1)\,.
\end{equation}
Furthermore we find there are two best fit solutions $\theta_{\text{bf}}=0.192\pi, 0.308\pi$ $(0.192\pi, 0.308\pi)$ for $U^{II(a)}$ in case of NO (IO) spectrum, and the minimal value of the $\chi^2$ function is $\chi^2_{\text{min}}=8.843$ (12.565).

Regarding the $0\nu\beta\beta$ decay, the effective mass $|m_{ee}|$ is given by
\begin{equation}
|m_{ee}|=\frac{1}{3} \left|m_1(e^{i\varphi_1}\cos\theta-e^{i\varphi_2}\sin\theta)^2+q_1m_2+q_2m_3(e^{i\varphi_2}\cos\theta+e^{i\varphi_1}\sin\theta)^2\right|\,,
\end{equation}
where $q_1, q_2=\pm1$. The predicted values of $|m_{ee}|$ are displayed in figure~\ref{fig:mee_CaseII}, where we require the three lepton mixing angles are within the experimentally preferred $3\sigma$ ranges. For the smallest group $G_{f}=S_4$, one sees that $|m_{ee}|$ is determined to be around 0.015eV or 0.048eV in case of IO spectrum, which are accessible to the future experiments searching for $0\nu\beta\beta$ decay. In the case of NO, $|m_{ee}|$ could be smaller than $10^{-4}$ eV for certain values of the lightest neutrino mass, because cancellation between different terms in the expression of $|m_{ee}|$ can take place.

The residual symmetry enforces the second column of the PMNS to be trimaximal in this case. Therefore the $R$-matrix is of the form of $C_{13}$ given in Eq.~\eqref{eq:R_1st_row}. We can read out the CP invariants $I^{\alpha}_{13}$ relevant to leptogenesis as
\begin{equation}
\begin{split}
I^e_{13}=&\frac{1}{3}\sin\left(\varphi_1-\varphi_2\right)\,,\\
I^{\mu}_{13}=&-\frac{1}{3}\cos\left(\frac{\pi}{6}-\varphi_1+\varphi_2\right)\,,\\
I^{\tau}_{13}=&\frac{1}{3}\cos\left(\frac{\pi}{6}+\varphi_1-\varphi_2\right)\,.
\end{split}
\end{equation}
The numerical results of the baryon asymmetry for $(\varphi_1, \varphi_2)=(\pi, 0)$ are shown in figure~\ref{fig:leptogenesis_CaseII}. It is easy to see that the observed baryon asymmetry could be generated via leptogenesis except in the case of NO spectrum with $(K_1, K_2, K_3)=(-,\pm,+)$.

\begin{table}[t!]
\begin{center}
\begin{tabular}{|m{0.25\columnwidth}<{\centering}|m{0.7\columnwidth}<{\centering}|}
\hline\hline
Group Id & $(\varphi_1,\varphi_2)$ \\
\hline
$[24, 12]_{\vartriangle}$, $[48, 30]$ & $(\pi ,0)$\\ \hline $[96, 64]_{\vartriangle}$, $[192, 182]$ & $\left(-\frac{7 \pi }{12},\frac{\pi }{3}\right)$, $\left(-\frac{7 \pi }{12},\frac{\pi }{3}\right)$, $\left(-\frac{3 \pi }{4},\frac{\pi }{4}\right)$\\ \hline $[384, 568]_{\vartriangle}$, $[768, 1085335]$ & $\left(\frac{\pi }{24},-\frac{\pi }{24}\right)$, $\left(\frac{\pi }{24},-\frac{\pi }{24}\right)$, $\left(\frac{\pi }{6},-\frac{19 \pi }{24}\right)$, $\left(\frac{\pi }{6},-\frac{19 \pi }{24}\right)$, $\left(-\frac{7 \pi }{24},-\frac{5 \pi }{24}\right)$, $\left(-\frac{7 \pi }{24},-\frac{5 \pi }{24}\right)$, $\left(-\frac{5 \pi }{12},\frac{13 \pi }{24}\right)$, $\left(-\frac{5 \pi }{12},\frac{13 \pi }{24}\right)$, $\left(\frac{\pi }{8},-\frac{7 \pi }{8}\right)$\\ \hline $[600, 179]_{\vartriangle}$, $[1200, 682]$ & $\left(-\frac{\pi }{5},\frac{7 \pi }{10}\right)$, $\left(-\frac{\pi }{5},\frac{7 \pi }{10}\right)$, $\left(-\frac{\pi }{5},\frac{4 \pi }{5}\right)$, $\left(-\frac{\pi }{5},\frac{9 \pi }{10}\right)$, $\left(-\frac{\pi }{5},\frac{9 \pi }{10}\right)$, $\left(-\frac{7 \pi }{15},\frac{7 \pi }{15}\right)$, $\left(-\frac{7 \pi }{15},\frac{7 \pi }{15}\right)$, $\left(-\frac{\pi }{10},0\right)$, $\left(-\frac{\pi }{10},0\right)$, $\left(-\frac{23 \pi }{30},\frac{4 \pi }{15}\right)$, $\left(-\frac{23 \pi }{30},\frac{4 \pi }{15}\right)$, $\left(-\frac{3 \pi }{5},\frac{2 \pi }{5}\right)$, $\left(\frac{11 \pi }{15},\frac{2 \pi }{3}\right)$, $\left(\frac{11 \pi }{15},\frac{2 \pi }{3}\right)$, $\left(-\frac{2 \pi }{3},-\frac{19 \pi }{30}\right)$, $\left(-\frac{2 \pi }{3},-\frac{19 \pi }{30}\right)$, $\left(\frac{2 \pi }{15},\frac{\pi }{15}\right)$, $\left(\frac{2 \pi }{15},\frac{\pi }{15}\right)$, $\left(-\frac{8 \pi }{15},\frac{13 \pi }{30}\right)$, $\left(-\frac{8 \pi }{15},\frac{13 \pi }{30}\right)$\\ \hline $[648, 259]_{\vartriangle'}$, $[648, 260]$ & $\left(\frac{5 \pi }{9},-\frac{7 \pi }{18}\right)$, $\left(\frac{5 \pi }{9},-\frac{7 \pi }{18}\right)$, $\left(\frac{2 \pi }{3},-\frac{\pi }{3}\right)$, $\left(-\frac{7 \pi }{9},\frac{5 \pi }{18}\right)$, $\left(-\frac{7 \pi }{9},\frac{5 \pi }{18}\right)$, $\left(-\frac{4 \pi }{9},-\frac{5 \pi }{9}\right)$, $\left(-\frac{4 \pi }{9},-\frac{5 \pi }{9}\right)$, $\left(-\frac{2 \pi }{9},-\frac{\pi }{9}\right)$, $\left(-\frac{2 \pi }{9},-\frac{\pi }{9}\right)$\\ \hline $[1176, 243]_{\vartriangle}$ & $\left(-\frac{2 \pi }{7},\frac{5 \pi }{7}\right)$, $\left(\frac{20 \pi }{21},\frac{\pi }{21}\right)$, $\left(\frac{20 \pi }{21},\frac{\pi }{21}\right)$, $\left(\frac{3 \pi }{7},-\frac{4 \pi }{7}\right)$, $\left(\frac{19 \pi }{21},\frac{17 \pi }{21}\right)$, $\left(\frac{19 \pi }{21},\frac{17 \pi }{21}\right)$, $\left(\frac{5 \pi }{21},-\frac{2 \pi }{3}\right)$, $\left(\frac{5 \pi }{21},-\frac{2 \pi }{3}\right)$, $\left(\frac{19 \pi }{42},-\frac{2 \pi }{3}\right)$, $\left(\frac{19 \pi }{42},-\frac{2 \pi }{3}\right)$, $\left(-\frac{\pi }{6},\frac{17 \pi }{21}\right)$, $\left(-\frac{\pi }{6},\frac{17 \pi }{21}\right)$, $\left(-\frac{17 \pi }{21},-\frac{29 \pi }{42}\right)$, $\left(-\frac{17 \pi }{21},-\frac{29 \pi }{42}\right)$, $\left(-\frac{\pi }{7},\frac{6 \pi }{7}\right)$, $\left(\frac{11 \pi }{21},\frac{17 \pi }{42}\right)$, $\left(\frac{11 \pi }{21},\frac{17 \pi }{42}\right)$, $\left(-\frac{17 \pi }{21},\frac{5 \pi }{21}\right)$, $\left(-\frac{17 \pi }{21},\frac{5 \pi }{21}\right)$, $\left(-\frac{11 \pi }{21},\frac{8 \pi }{21}\right)$, $\left(-\frac{11 \pi }{21},\frac{8 \pi }{21}\right)$, $\left(-\frac{11 \pi }{21},\frac{11 \pi }{21}\right)$, $\left(-\frac{11 \pi }{21},\frac{11 \pi }{21}\right)$, $\left(\frac{\pi }{7},-\frac{11 \pi }{14}\right)$, $\left(\frac{\pi }{7},-\frac{11 \pi }{14}\right)$, $\left(-\frac{11 \pi }{21},-\frac{23 \pi }{42}\right)$, $\left(-\frac{11 \pi }{21},-\frac{23 \pi }{42}\right)$, $\left(\frac{2 \pi }{21},-\frac{20 \pi }{21}\right)$, $\left(\frac{2 \pi }{21},-\frac{20 \pi }{21}\right)$, $\left(0,-\frac{13 \pi }{14}\right)$, $\left(0,-\frac{13 \pi }{14}\right)$, $\left(-\frac{13 \pi }{21},\frac{11 \pi }{42}\right)$, $\left(-\frac{13 \pi }{21},\frac{11 \pi }{42}\right)$, $\left(-\frac{11 \pi }{14},-\frac{5 \pi }{7}\right)$, $\left(-\frac{11 \pi }{14},-\frac{5 \pi }{7}\right)$, $\left(-\frac{11 \pi }{42},\frac{16 \pi }{21}\right)$, $\left(-\frac{11 \pi }{42},\frac{16 \pi }{21}\right)$, $\left(-\frac{8 \pi }{21},\frac{2 \pi }{3}\right)$, $\left(-\frac{8 \pi }{21},\frac{2 \pi }{3}\right)$, $\left(-\frac{8 \pi }{21},-\frac{17 \pi }{42}\right)$, $\left(-\frac{8 \pi }{21},-\frac{17 \pi }{42}\right)$, $\left(\frac{4 \pi }{7},\frac{9 \pi }{14}\right)$, $\left(\frac{4 \pi }{7},\frac{9 \pi }{14}\right)$\\ \hline $[1536, 408544632]_{\vartriangle}$ & $\left(-\frac{47 \pi }{48},-\frac{23 \pi }{24}\right)$, $\left(-\frac{47 \pi }{48},-\frac{23 \pi }{24}\right)$, $\left(\frac{5 \pi }{16},\frac{5 \pi }{16}\right)$, $\left(-\frac{11 \pi }{48},-\frac{7 \pi }{48}\right)$, $\left(-\frac{11 \pi }{48},-\frac{7 \pi }{48}\right)$, $\left(-\frac{7 \pi }{16},\frac{9 \pi }{16}\right)$, $\left(\frac{23 \pi }{48},-\frac{29 \pi }{48}\right)$, $\left(\frac{23 \pi }{48},-\frac{29 \pi }{48}\right)$, $\left(-\frac{23 \pi }{48},-\frac{7 \pi }{12}\right)$, $\left(-\frac{23 \pi }{48},-\frac{7 \pi }{12}\right)$, $\left(-\frac{\pi }{8},-\frac{\pi }{16}\right)$, $\left(-\frac{\pi }{8},-\frac{\pi }{16}\right)$, $\left(\frac{9 \pi }{16},-\frac{\pi }{2}\right)$, $\left(\frac{9 \pi }{16},-\frac{\pi }{2}\right)$, $\left(-\frac{5 \pi }{48},-\frac{\pi }{12}\right)$, $\left(-\frac{5 \pi }{48},-\frac{\pi }{12}\right)$, $\left(\frac{37 \pi }{48},\frac{35 \pi }{48}\right)$, $\left(\frac{37 \pi }{48},\frac{35 \pi }{48}\right)$, $\left(-\frac{29 \pi }{48},\frac{17 \pi }{48}\right)$, $\left(-\frac{29 \pi }{48},\frac{17 \pi }{48}\right)$, $\left(\frac{35 \pi }{48},-\frac{\pi }{6}\right)$, $\left(\frac{35 \pi }{48},-\frac{\pi }{6}\right)$, $\left(\frac{11 \pi }{16},-\frac{\pi }{4}\right)$, $\left(\frac{11 \pi }{16},-\frac{\pi }{4}\right)$, $\left(-\frac{31 \pi }{48},\frac{11 \pi }{24}\right)$, $\left(-\frac{31 \pi }{48},\frac{11 \pi }{24}\right)$, $\left(\frac{\pi }{48},\frac{47 \pi }{48}\right)$, $\left(\frac{\pi }{48},\frac{47 \pi }{48}\right)$, $\left(\frac{3 \pi }{16},\frac{\pi }{8}\right)$, $\left(\frac{3 \pi }{16},\frac{\pi }{8}\right)$, $\left(-\frac{5 \pi }{24},-\frac{11 \pi }{48}\right)$, $\left(-\frac{5 \pi }{24},-\frac{11 \pi }{48}\right)$, $\left(\frac{\pi }{6},\frac{7 \pi }{48}\right)$, $\left(\frac{\pi }{6},\frac{7 \pi }{48}\right)$, $\left(\frac{17 \pi }{24},-\frac{19 \pi }{48}\right)$, $\left(\frac{17 \pi }{24},-\frac{19 \pi }{48}\right)$\\ \hline \hline
\end{tabular}
\caption{\label{tab:II}The predictions for PMNS matrix of the form $U^{II(a)}$ and $U^{II(b)}$, where the first column shows the group identification in \texttt{GAP} system, and the second column displays the achievable values of the parameters $\varphi_1$ and $\varphi_2$. We have shown at most two representatives flavor symmetry groups in the first column. If there is only one group predicting the corresponding values of $\varphi_1$ and $\varphi_2$ in the second column, this unique group would be listed. The full results of our analysis are provided at the website~\cite{webdata}. The subscripts $\Delta$ and $\Delta^{\prime}$ indicate that the corresponding groups belong to the type D group series $D_{n,n}^{(0)}\cong\Delta(6n^2)$ and $D_{9n^{\prime},3n^{\prime}}^{(1)}\cong(Z_{9n^{\prime}}\times Z_{3n^{\prime}})\rtimes S_3$, respectively. }
\end{center}
\end{table}

\begin{figure}[ht!]
\begin{center}
\includegraphics[width=0.5\linewidth]{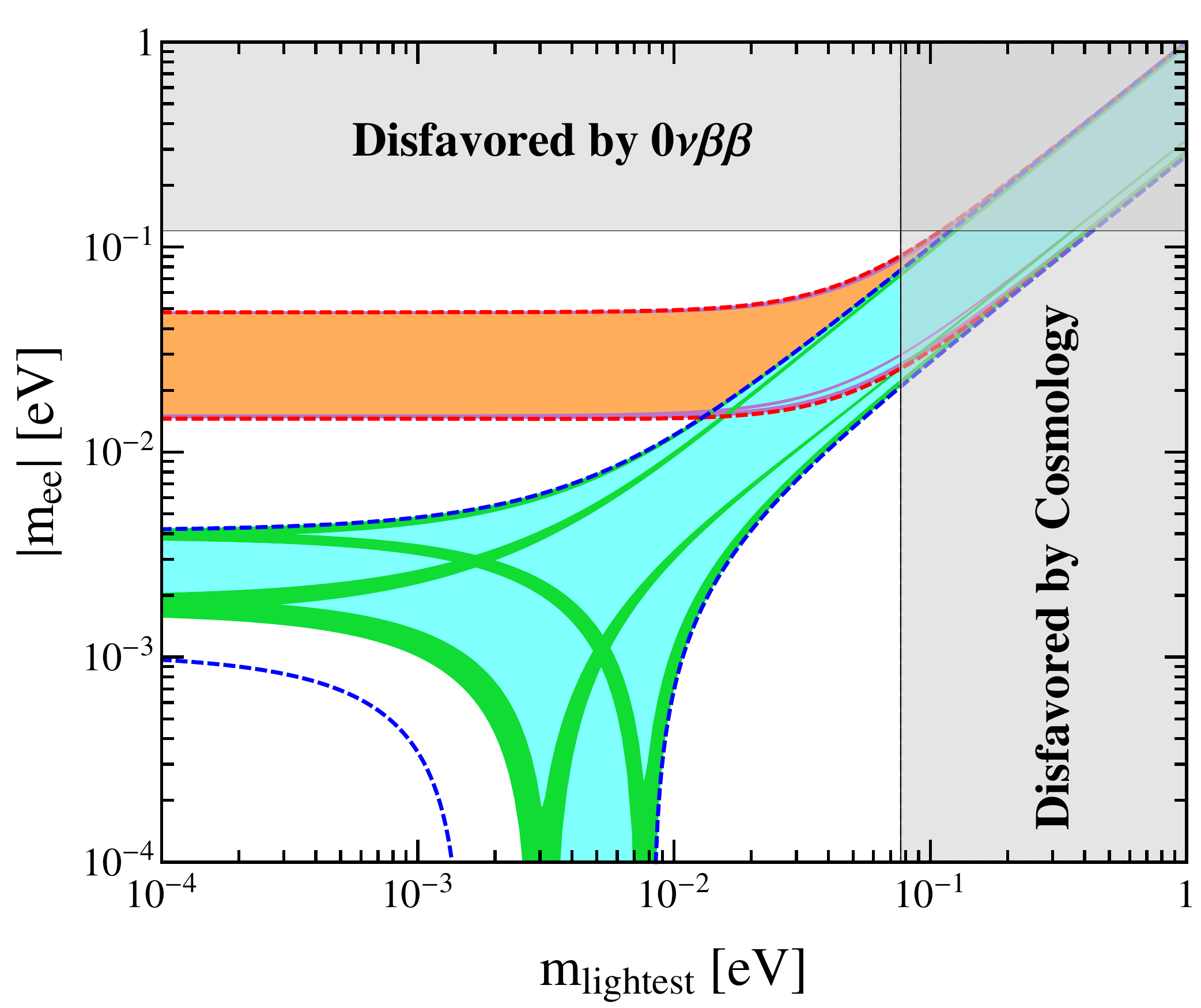}
\caption{\label{fig:mee_CaseII}Predictions of the $0\nu\beta\beta$ decay effective mass $|m_{ee}|$ with respect to the lightest neutrino mass $m_{\text{lightest}}$ for the mixing patterns $U^{II(a)}$ and $U^{II(b)}$. The red (blue) dashed lines indicate the most general allowed regions for IO (NO) spectrum obtained by varying the mixing parameters within their $3\sigma$ ranges~\cite{Gonzalez-Garcia:2014bfa}. The orange (cyan) areas denote the achievable values of $|m_{ee}|$ when $\varphi_1$ and $\varphi_2$ are taken to be free continuous parameters in the case of IO (NO). The purple and green regions are the theoretical predictions of the smallest flavor symmetry group which can generate this two mixing pattern. Note that the purple (green) region overlaps the orange (cyan) one. The present most stringent upper limits $|m_{ee}|<0.120$ eV from EXO-200~\cite{Auger:2012ar, Albert:2014awa} and KamLAND-ZEN~\cite{Gando:2012zm} is shown by horizontal grey band. The vertical grey exclusion band is the current limit on $m_{\text{lightest}}$ from the cosmological data of $\sum m_i<0.230$ eV by the Planck collaboration~\cite{Ade:2013zuv}.}
\end{center}
\end{figure}

\begin{figure}[ht!]
\begin{center}
\includegraphics[width=0.42\linewidth]{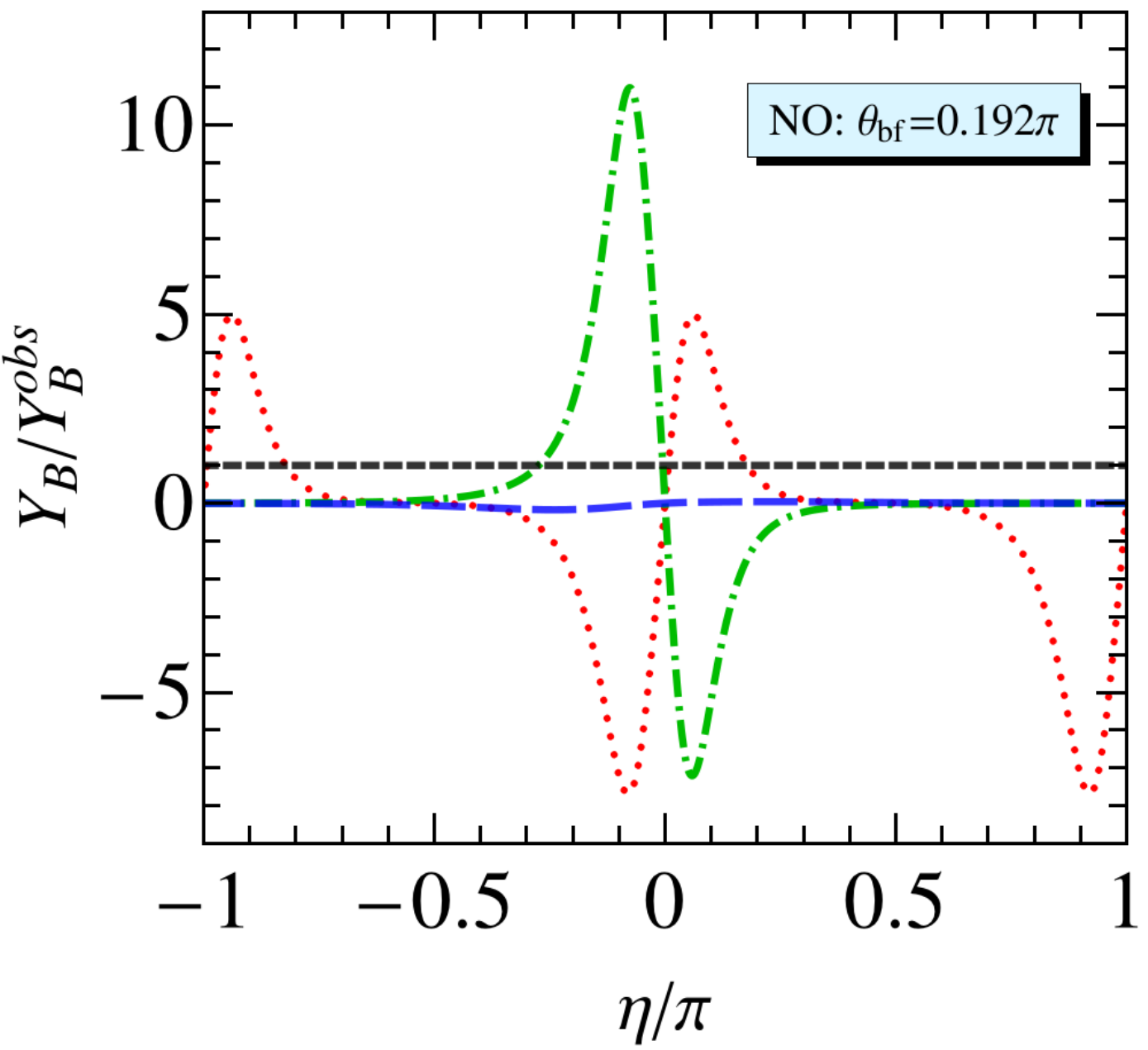}~~~
\includegraphics[width=0.42\linewidth]{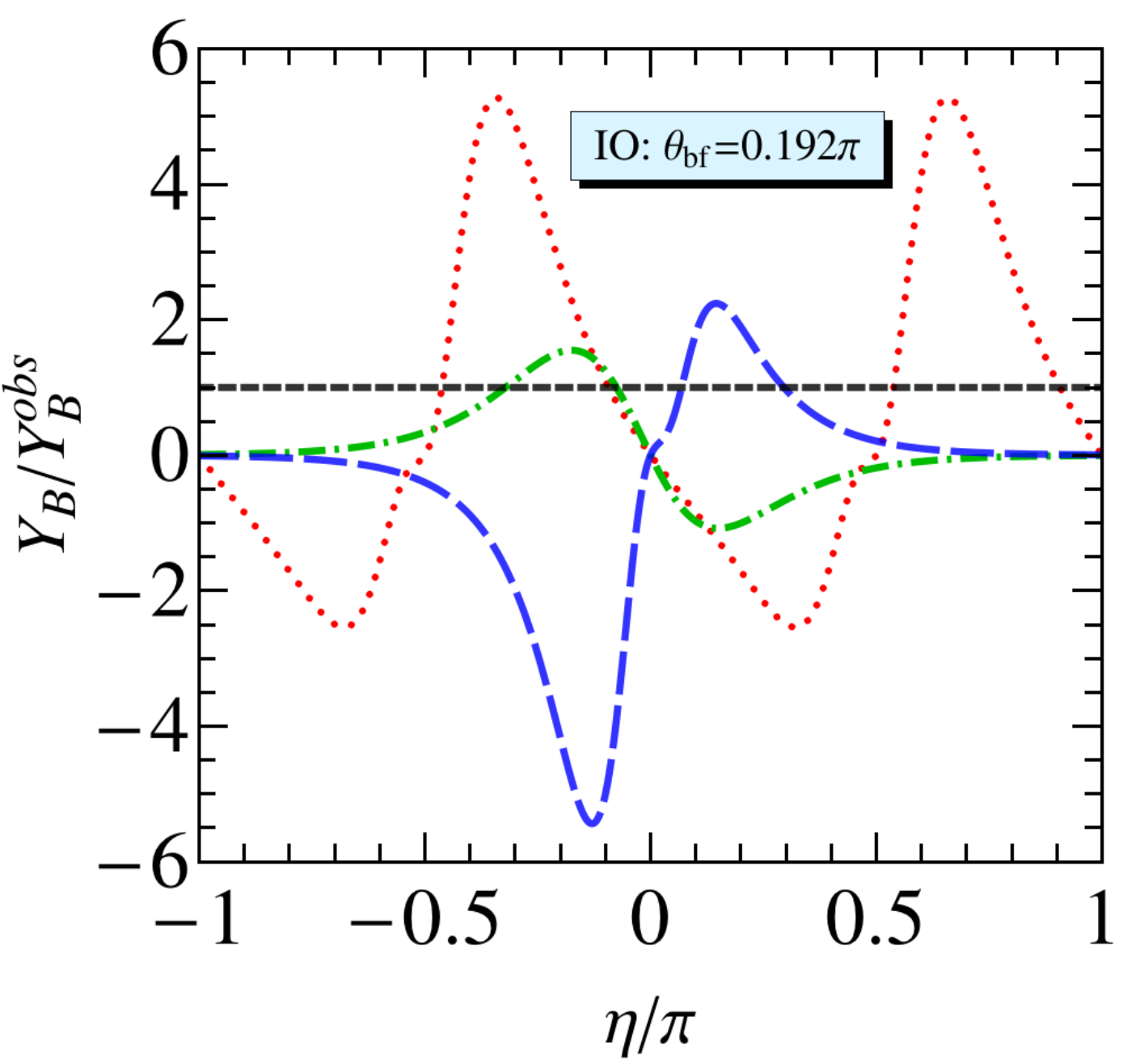}
\includegraphics[width=0.435\linewidth]{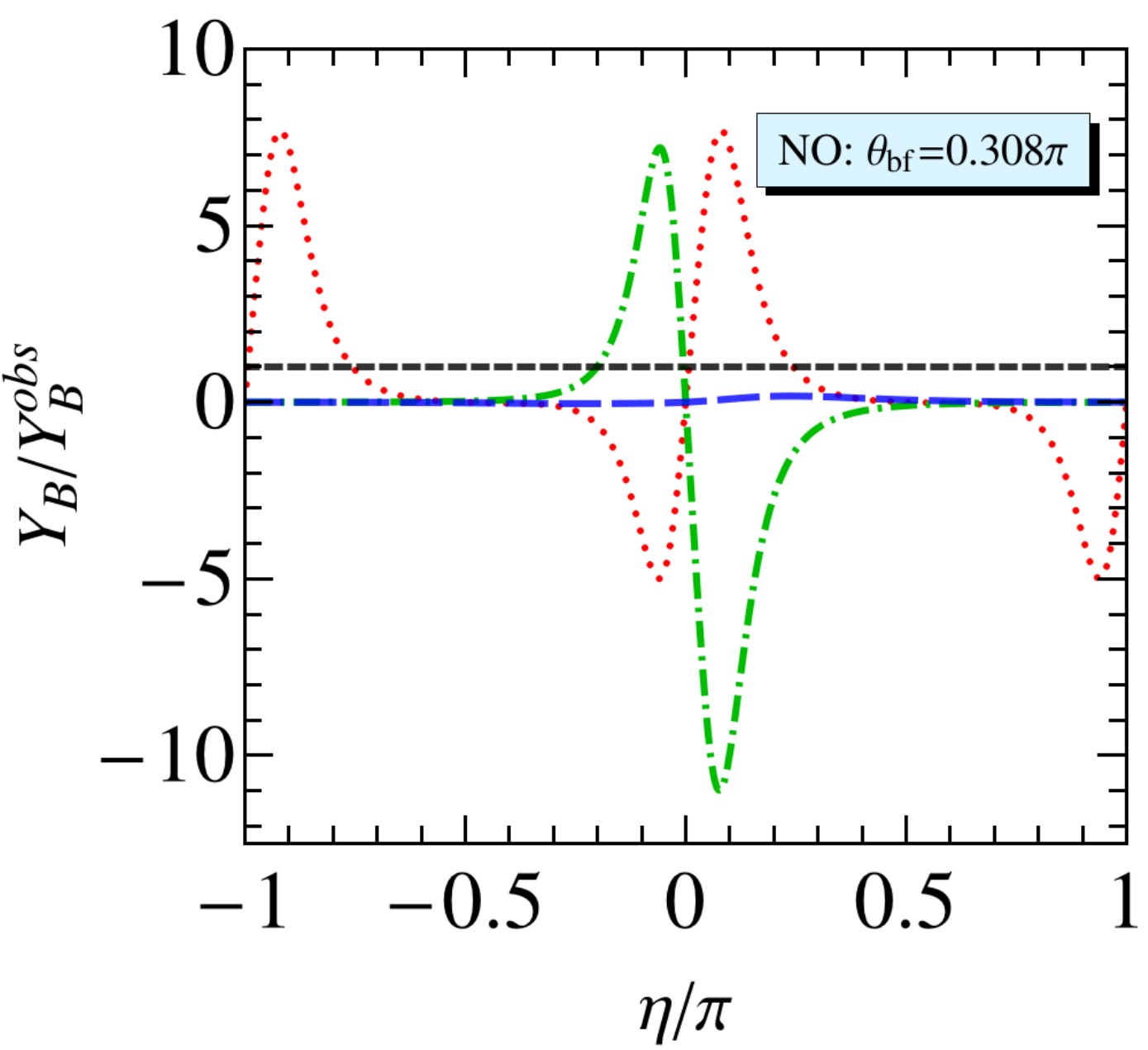}~~~
\includegraphics[width=0.42\linewidth]{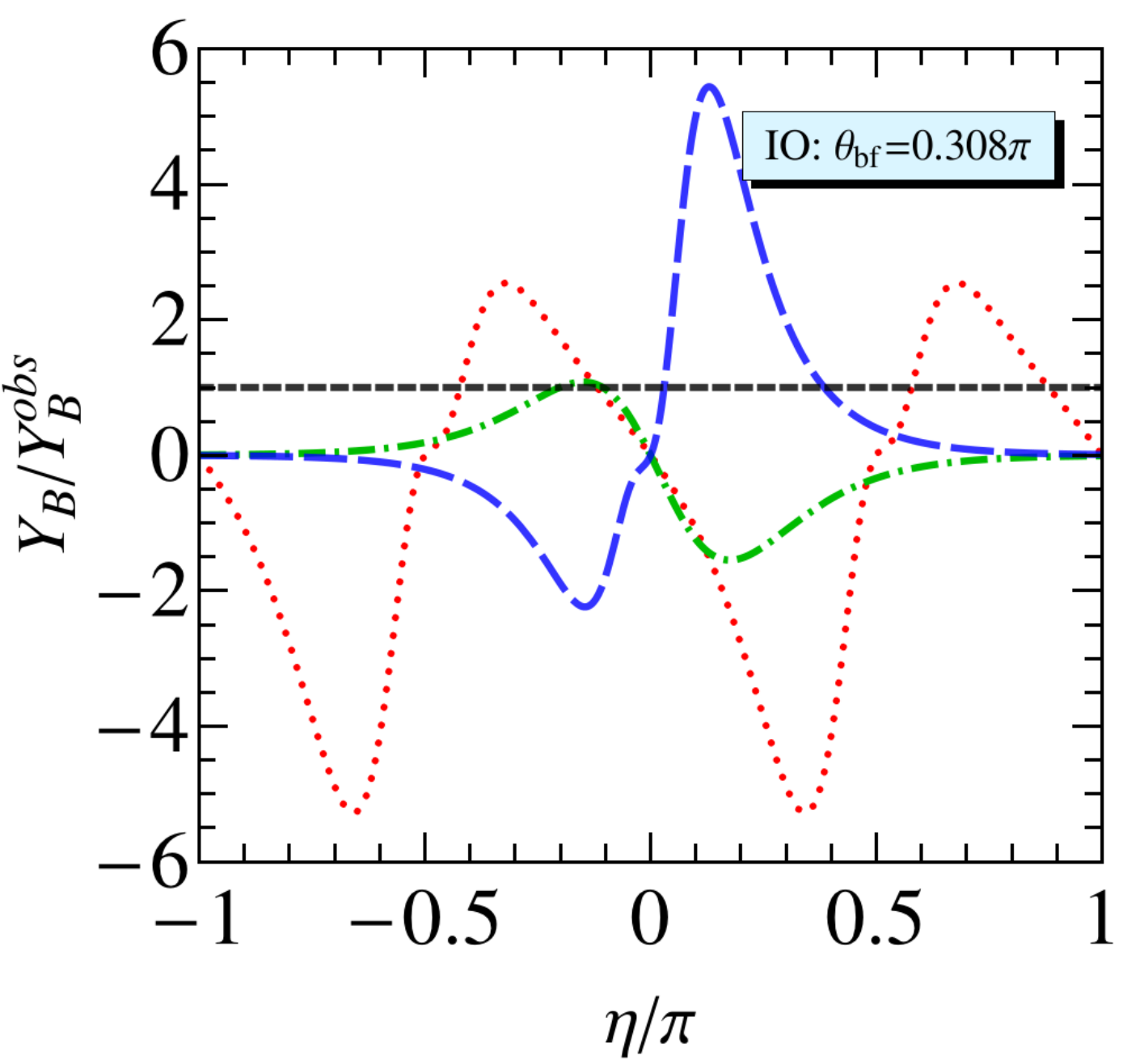}
\caption{\label{fig:leptogenesis_CaseII}
The prediction for $Y_B/Y_B^{obs}$ as a function of $\eta$ in case II with $(\varphi_1,\varphi_2)=(\pi,0)$, where $\theta_{bf}$ is the best fit value of $\theta$. We choose $M_1=5\times 10^{11}$ GeV and the lightest neutrino mass $m_1$ (or $m_3$) = 0.01eV. The red dotted, green dot-dashed, blue dashed lines correspond to $(K_1,K_2,K_3)=(+,\pm,+),(+,\pm,-)$ and $(-,\pm,+)$ respectively. The experimentally observed value $Y_B^{obs}$ is represented by the horizontal black dashed line.}
\end{center}
\end{figure}

\item[Case \uppercase\expandafter{\romannumeral3}]
\begin{equation}
\label{eq:PMNS_case_III}
U^{III}=\frac{1}{\sqrt{3}}
\left(\begin{array}{ccc}
 \sqrt{2} e^{i \varphi _1} \sin \varphi _2 & 1 & \sqrt{2} e^{i \varphi _1} \cos \varphi _2 \\
 \sqrt{2} e^{i \varphi _1} \cos \left(\varphi _2+\frac{\pi }{6}\right) & 1 & -\sqrt{2} e^{i \varphi _1} \sin \left(\varphi _2+\frac{\pi }{6}\right) \\
 -\sqrt{2} e^{i \varphi _1} \cos \left(\varphi _2-\frac{\pi }{6}\right) & 1 & \sqrt{2} e^{i \varphi _1} \sin \left(\varphi _2-\frac{\pi }{6}\right)
\end{array}
\right)S_{13}(\theta)Q^{\dagger}_{\nu}\,,
\end{equation}
where $\varphi_1$ and $\varphi_2$ are rational angles, and their values are determined by the residual symmetries. The admissible values of $\varphi_1$ and $\varphi_2$ and the representative flavor symmetry groups found from our group scan up to order 2000 are summarized in table~\ref{tab:III}. The full results are available at our website\cite{webdata}. Similar to case II, the second column of the mixing matrix is $(1, 1, 1)^{T}/\sqrt{3}$ as well. In particular, all the six row permutations lead to the same mixing pattern, if the freedom of redefining the parameters $\theta$, $\varphi_1$ and $\varphi_2$ is taken into account. For this mixing matrix $U^{(III)}$ in Eq.~\eqref{eq:PMNS_case_III}, the mixing angles read
\begin{equation}
\label{eq:mixing_angles_Case_III}\hskip-0.1in \sin^2\theta_{13}=\frac{2}{3} \cos ^2\left(\theta -\varphi _2\right),~~ \sin^2\theta_{12}=\frac{1}{3-2 \cos ^2\left(\theta -\varphi _2\right)},~~\sin^2\theta_{23}=\frac{\sin\left(2\theta-2\varphi_2+\frac{\pi}{6}\right)-1}{\cos\left(2\theta-2\varphi_2\right)-2}\,,
\end{equation}
which fulfill the following sum rules
\begin{equation}
3\cos^2\theta_{13}\sin^2\theta_{12}=1,\qquad
\sin^2\theta_{23}=\frac{1}{2}\pm\frac{1}{2}\tan\theta_{13}\sqrt{2-\tan^2\theta_{13}}\,.
\end{equation}
Inserting the best fit value $\sin^2\theta_{13}=0.0218$~\cite{Gonzalez-Garcia:2014bfa},we obtain
\begin{equation}
\sin^2\theta_{12}\simeq0.341,\qquad \sin^2\theta_{23}\simeq 0.395~~ \text{or}~~ 0.605\,,
\end{equation}
which are compatible with the present experimental data. By precisely measuring the solar and atmospheric mixing angles, the reactor neutrino experiment JUNO and long baseline neutrino oscillation experiments DUNE and Hyper-Kamiokande are able to exclude this mixing pattern or provide strong evidence for its relevance. Furthermore, the CP invariants are given by
\begin{equation}
J_{CP}=I_2=0,\qquad |I_1|=\frac{2}{9}\left|\sin2\varphi_1\right|\sin^2\left(\theta-\varphi_2\right)\,,
\end{equation}
which leads to
\begin{equation}
\delta_{CP}, \alpha_{31}=0~\text{or}~ \pi,\qquad \alpha_{21}~(\textrm{mod}~\pi)=\pm2\varphi_1\,.
\end{equation}
This indicates that both Dirac CP phase $\delta_{CP}$ and Majorana phase $\alpha_{31}$ are always trivial in this case. Subsequently we find for the effective Majorana mass $|m_{ee}|$ the following expression
\begin{equation}
|m_{ee}|=\frac{1}{3}\left|2m_1e^{2i\varphi_1}\sin^2\left(\theta-\varphi_2\right)+q_1m_2+2q_2m_3e^{2i\varphi_1}\cos^2\left(\theta-\varphi_2\right)\right|\,.
\end{equation}
We plot $|m_{ee}|$ as a function of the lightest neutrino mass $m_{\text{lightest}}$ in figure~\ref{fig:mee_CaseIII}. For the smallest flavor symmetry group $A_4$ which predicts $(\varphi_1, \varphi_2)=(\pi, 2\pi/3)$, all the three CP violation phases are conserved. As a result, the effective mass $|m_{ee}|$ is close to 0.027eV or 0.042eV in case of IO spectrum. It is notable that there is no cancellation in $|m_{ee}|$ for
any values of $m_{\text{lightest}}$ in the case of NO, and thus $|m_{ee}|$ has a lower bound $|m_{ee}|\geq 2.52\times10^{-3}$ eV.

As regards the leptogenesis, we find that both rephase invariant $I^{\alpha}_{13}$ and the CP asymmetry $\epsilon_{\alpha}$ are vanishing,
\begin{equation}
I^{e}_{13}=I^{\mu}_{13}=I^{\tau}_{13}=0,\qquad \epsilon_{e}=\epsilon_{\mu}=\epsilon_{\tau}=0\,.
\end{equation}
Hence the net baryon asymmetry can not be generated in this case, and appropriate subleading corrections are necessary in order to
to make the leptogenesis viable.

\begin{table}[hptb]
\begin{center}
\begin{tabular}{|m{0.25\columnwidth}<{\centering}|m{0.7\columnwidth}<{\centering}|}
\hline\hline
\texttt{Group Id} & $(\varphi_1,\varphi_2)$ \\
\hline
$[12, 3]$, $[24, 12]_{\vartriangle}$ & $\left(\pi ,\frac{2 \pi }{3}\right)$\\ \hline $[96, 64]_{\vartriangle}$, $[192, 182]$ & $\left(-\frac{3 \pi }{4},\frac{2 \pi }{3}\right)$\\ \hline $[384, 568]_{\vartriangle}$, $[768, 1085335]$ & $\left(-\frac{7 \pi }{8},0\right)$\\ \hline $[600, 179]_{\vartriangle}$, $[1200, 682]$ & $\left(-\frac{3 \pi }{5},\frac{\pi }{6}\right)$, $\left(\frac{4 \pi }{5},\frac{\pi }{6}\right)$\\ \hline $[648, 259]_{\vartriangle'}$, $[648, 260]$ & $\left(\frac{\pi }{3},\frac{2 \pi }{3}\right)$\\ \hline $[1176, 243]_{\vartriangle}$ & $\left(\frac{3 \pi }{7},\frac{\pi }{6}\right)$, $\left(\frac{5 \pi }{7},\frac{\pi }{6}\right)$, $\left(\frac{6 \pi }{7},\frac{2 \pi }{3}\right)$\\ \hline $[1536, 408544632]_{\vartriangle}$ & $\left(-\frac{7 \pi }{16},\frac{\pi }{6}\right)$, $\left(\frac{5 \pi }{16},\frac{\pi }{6}\right)$\\ \hline \hline
\end{tabular}
\caption{\label{tab:III}The predictions for PMNS matrix of the form $U^{III}$, where the first column shows the group identification in \texttt{GAP} system, and the second column displays the achievable values of the parameters $\varphi_1$ and $\varphi_2$. We have shown at most two representatives flavor symmetry groups in the first column. If there is only one group predicting the corresponding values of $\varphi_1$ and $\varphi_2$ in the second column, this unique group would be listed. The full results of our analysis are provided at the website~\cite{webdata}. The subscripts $\Delta$ and $\Delta^{\prime}$ indicate that the corresponding groups belong to the type D group series $D_{n,n}^{(0)}\cong\Delta(6n^2)$ and $D_{9n^{\prime},3n^{\prime}}^{(1)}\cong(Z_{9n^{\prime}}\times Z_{3n^{\prime}})\rtimes S_3$, respectively.}
\end{center}
\end{table}

\begin{figure}[ht!]
\begin{center}
\includegraphics[width=0.5\linewidth]{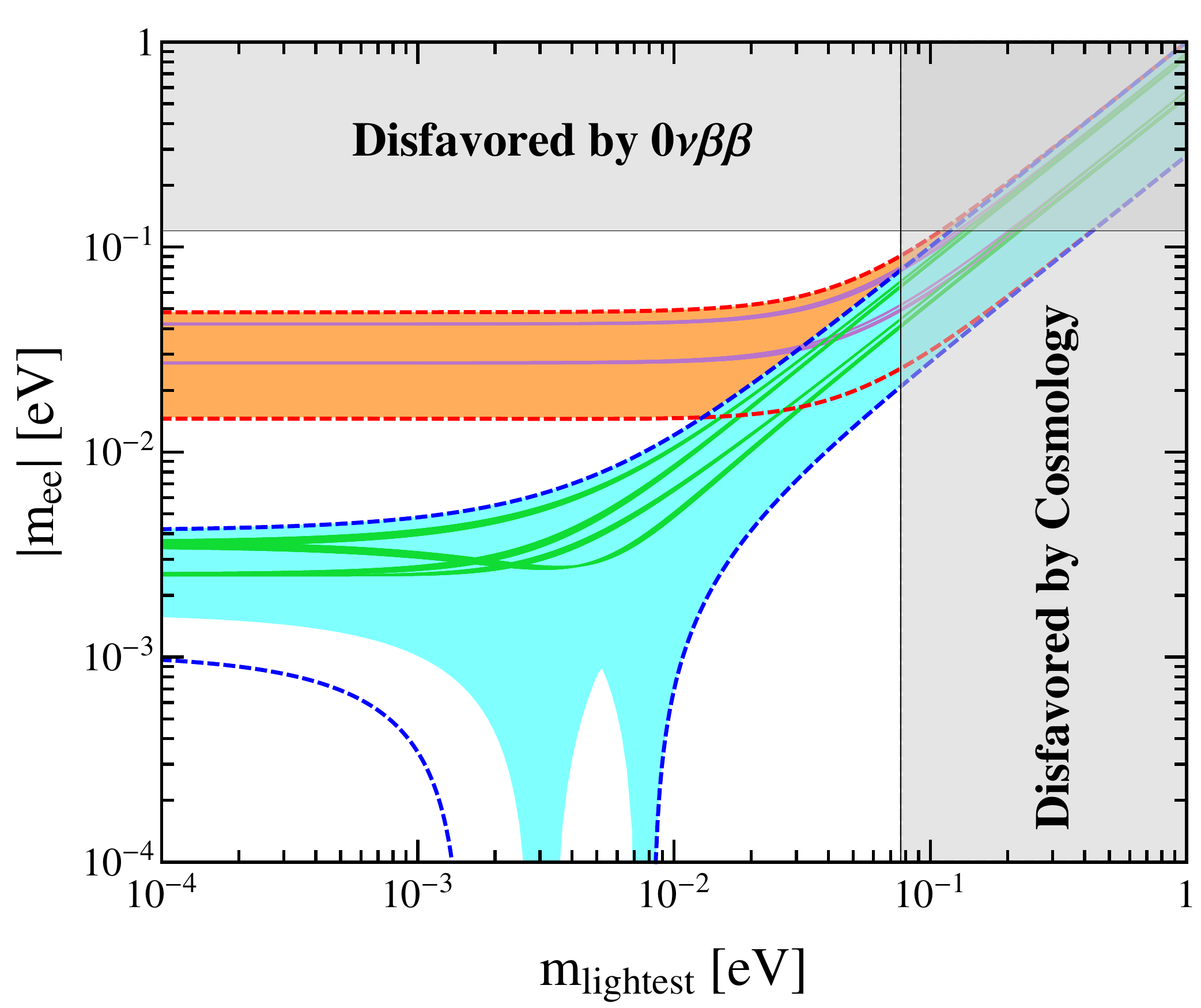}
\caption{\label{fig:mee_CaseIII}Predictions of the $0\nu\beta\beta$ decay effective mass $|m_{ee}|$ with respect to the lightest neutrino mass $m_{\text{lightest}}$ for the mixing pattern $U^{III}$. The red (blue) dashed lines indicate the most general allowed regions for IO (NO) spectrum obtained by varying the mixing parameters within their $3\sigma$ ranges~\cite{Gonzalez-Garcia:2014bfa}. The orange (cyan) areas denote the achievable values of $|m_{ee}|$ when $\varphi_1$ and $\varphi_2$ are taken to be free continuous parameters in the case of IO (NO). The purple and green regions are the theoretical predictions of the smallest flavor symmetry group which can generate this mixing pattern. Note that the purple (green) region overlaps the orange (cyan) one. The present most stringent upper limits $|m_{ee}|<0.120$ eV from EXO-200~\cite{Auger:2012ar, Albert:2014awa} and KamLAND-ZEN~\cite{Gando:2012zm} is shown by horizontal grey band. The vertical grey exclusion band is the current limit on $m_{\text{lightest}}$ from the cosmological data of $\sum m_i<0.230$ eV by the Planck collaboration~\cite{Ade:2013zuv}.}
\end{center}
\end{figure}

\item[Case \uppercase\expandafter{\romannumeral4}]

\vskip2cm

\begin{equation}
\begin{aligned}
U^{IV(a)}=&\left(
\begin{array}{ccc}
-\sqrt{\frac{\phi_g}{\sqrt{5}}} & \sqrt{\frac{1}{\sqrt{5}\phi_g}} & 0\\
\sqrt{\frac{1}{2\sqrt{5}\phi_g}} & \sqrt{\frac{\phi_g}{2\sqrt{5}}} & -\frac{1}{\sqrt{2}} \\
\sqrt{\frac{1}{2\sqrt{5}\phi_g}} & \sqrt{\frac{\phi_g}{2\sqrt{5}}} & \frac{1}{\sqrt{2}}
\end{array}
\right)S_{13}(\theta)Q^{\dagger}_{\nu}\,,\\
U^{IV(b)}=&\left(
\begin{array}{ccc}
-i\sqrt{\frac{\phi_g}{\sqrt{5}}} & \sqrt{\frac{1}{\sqrt{5}\phi_g}} & 0 \\
i\sqrt{\frac{1}{2\sqrt{5}\phi_g}} & \sqrt{\frac{\phi_g}{2\sqrt{5}}} & -\frac{1}{\sqrt{2}} \\
i\sqrt{\frac{1}{2\sqrt{5}\phi_g}} & \sqrt{\frac{\phi_g}{2\sqrt{5}}} & \frac{1}{\sqrt{2}}
\end{array}
\right)S_{13}(\theta)Q^{\dagger}_{\nu}\,,
\end{aligned}
\end{equation}
where $\phi_g=(\sqrt{5}+1)/2$ is the golden ratio. Notice that $U^{IV(b)}$ can be obtained from $U^{IV(a)}$ by multiplying the factor $i$ in its first column. Our group scanning reveals that these two mixing patterns can be obtained from the groups $[60,5]\cong A_5$, [120,35], [180,19] and many others shown in the website. Indeed, this case has been found in previous work on $A_5$ flavor symmetry and generalized CP~\cite{Li:2015jxa,DiIura:2015kfa,Ballett:2015wia}, and our results coincide with those. The PMNS mixing matrix $U^{IV(a)}$ leads to the following expressions for the mixing angles
\begin{equation}
\sin^2\theta_{13}=\frac{\phi_g}{\sqrt{5}} \sin ^2\theta,\quad
\sin^2\theta_{12}=\frac{4-2\phi_g}{5-2\phi_g+\cos 2\theta}\,,\quad\sin^2\theta_{23}=\frac{1}{2}-\frac{\sqrt{3-\phi_g}\sin 2\theta}{3\phi_g-2+\phi_g\cos 2\theta}\,.
\end{equation}
Obviously $U^{IV(a)}$ is a real matrix, therefore all the three CP invariants
vanish,
\begin{equation}
J_{CP}=I_1=I_2=0\,,
\end{equation}
which implies that each of the CP violation phases $\delta_{CP}, \alpha_{21}, \alpha_{31}$ is either 0 or $\pi$. Moreover, we see that the mixing angles fulfill the following sum rules
\begin{equation}
\label{eq:correlation_caseIV}\sin^{2}\theta_{12}\cos^2\theta_{13}=\frac{3-\phi_g}{5},\quad \sin^{2}\theta_{23}-\frac{1}{2}=\pm(\phi_g-1)\tan\theta_{13}\sqrt{1+(\phi_g-2)\tan^{2}\theta_{13}}\,,
\end{equation}
Using the $3\sigma$ range of the reactor mixing angle $0.0188\leq\sin^{2}\theta_{13}\leq0.0251$~\cite{Gonzalez-Garcia:2014bfa}, we get
\begin{equation}
0.282\leq\sin^{2}\theta_{12}\leq0.284,\qquad 0.401\leq\sin^{2}\theta_{23}\leq0.415~~\text{or}~~0.585\leq\sin^{2}\theta_{23}\leq0.599\,.
\end{equation}
These predictions for $\theta_{12}$ and $\theta_{23}$ will be testable at future neutrino facilities such as JUNO, DUNE, Hyper-Kamiokande and so on. For the mixing matrix $U^{IV(b)}$, the mixing angles read
\begin{equation}
\sin^2\theta_{13}=\frac{\phi_g}{\sqrt{5}} \sin ^2\theta,\quad
\sin^2\theta_{12}=\frac{4-2\phi_g}{5-2\phi_g+\cos 2\theta}\,,\quad \sin^2\theta_{23}=\frac{1}{2}\,.
\end{equation}
The solar and reactor mixing angles have the same form as that of  $U^{IV(a)}$, and consequently the correlation $\sin^{2}\theta_{12}\cos^2\theta_{13}=(3-\phi_g)/5$ given in Eq.~\eqref{eq:correlation_caseIV} still holds. The minimum value of $\chi^2$ is $\chi^2_{\text{min}}=4.045~(7.742)$ obtained at the best fitting values $\theta_{\text{bf}}=\pm 0.056\pi~(\pm 0.056\pi)$ for NO (IO) spectrum. For the CP violating phases, we find $\delta_{CP}$ is exactly maximal while both Majorana phases $\alpha_{21}$ and $\alpha_{31}$ are trivial with
\begin{equation}
|J_{CP}|=\frac{1}{4}\sqrt{\frac{\phi_g}{5\sqrt{5}}}\;|\sin 2\theta|\,,\qquad I_1=I_2=0\,.
\end{equation}
In this case, the general expression for the effective mass $|m_{ee}|$ is
\begin{equation}
\begin{split}
\left|m_{ee}\right|=\frac{1}{\sqrt{5}}\left|\phi_g m_1\cos^2\theta+\phi_g^{-1}q_1m_2+\phi_g q_2m_3\sin^2\theta\right|,\quad \text{for}~~U^{IV(a)}, \\
\left|m_{ee}\right|=\frac{1}{\sqrt{5}}\left|\phi_gm_1\cos^2\theta-\phi_g^{-1}q_1m_2+\phi_g q_2m_3\sin^2\theta\right|,\quad \text{for}~~U^{IV(b)}\,,
\end{split}
\end{equation}
where $q_1, q_2=\pm1$. Therefore the same values of $|m_{ee}|$ would be obtained if the parameter $q_1$ is of opposite sign for $U^{IV(a)}$ and $U^{IV(b)}$. After considering all possible values of $q_1$ and $q_2$, we display the allowed regions of $|m_{ee}|$ in figure~\ref{fig:mee_CaseIV}.
We see that $|m_{ee}|$ is close to 0.021eV or 0.048eV for IO while it is smaller than $10^{-4}$ eV for $0.0016~\mathrm{eV}\leq m_{\text{lightest}}\leq0.0024$~eV and $0.0051~\mathrm{eV}\leq m_{\text{lightest}}\leq0.0061$~eV in the case NO.

Then we come to study the resulting predictions for leptogenesis. All the rephasing invariants $I^{\alpha}_{13}$ are determined to be zero for $U^{IV(a)}$ so that the CP asymmetries $\epsilon_{\alpha}$ vanish and the matter-antimatter asymmetry of the universe can not
be generated without high order corrections. For the PMNS mixing matrix $U^{IV(b)}$, we find
\begin{equation}
I^e_{13}=0\,,\quad  I^{\mu}_{13}=-I^{\tau}_{13}=-\sqrt{\frac{1}{4 \sqrt{5}\phi_g }}\,.
\end{equation}
We plot the values of $Y_B $ versus $\eta$ in figure~\ref{fig:leptogenesis_CaseIV}. It is easy to see that the observed baryon asymmetry can be obtained via leptogenesis except in the case of NO with $(K_1, K_2, K_3)=(-,\pm,+)$.

\begin{figure}[ht!]
\begin{center}
\includegraphics[width=0.5\linewidth]{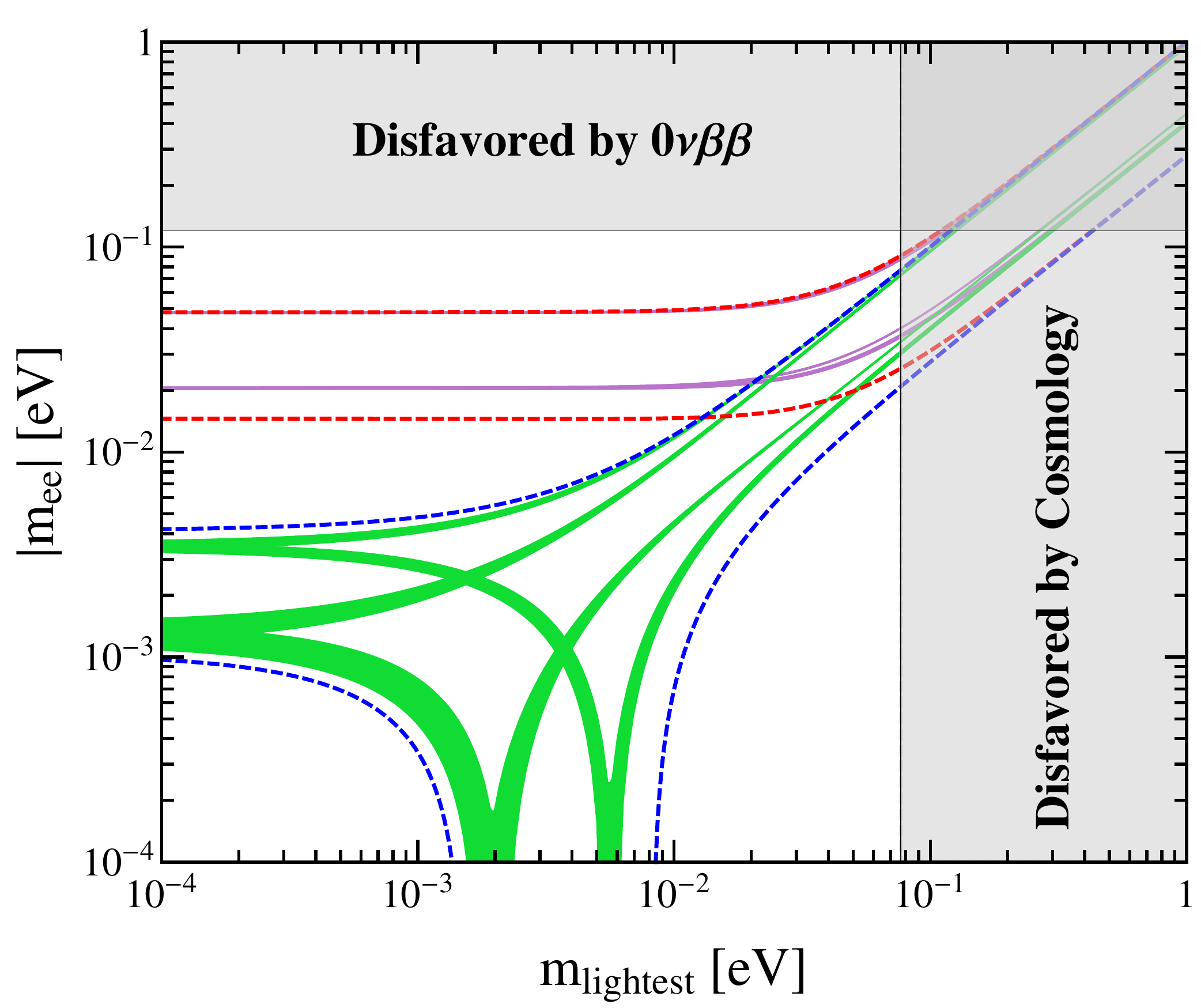}
\caption{\label{fig:mee_CaseIV}Predictions of the $0\nu\beta\beta$ decay effective mass $|m_{ee}|$ with respect to the lightest neutrino mass $m_{\text{lightest}}$ for the mixing patterns $U^{IV(a)}$ and $U^{IV(b)}$. The red (blue) dashed lines indicate the most general allowed regions for IO (NO) spectrum obtained by varying the mixing parameters within their $3\sigma$ ranges~\cite{Gonzalez-Garcia:2014bfa}. The purple and green regions are the theoretical predictions of these two mixing patterns. The present most stringent upper limits $|m_{ee}|<0.120$ eV from EXO-200~\cite{Auger:2012ar, Albert:2014awa} and KamLAND-ZEN~\cite{Gando:2012zm} is shown by horizontal grey band. The vertical grey exclusion band is the current limit on $m_{\text{lightest}}$ from the cosmological data of $\sum m_i<0.230$ eV by the Planck collaboration~\cite{Ade:2013zuv}.}
\end{center}
\end{figure}

\begin{figure}[ht!]
\begin{center}
\includegraphics[width=0.42\linewidth]{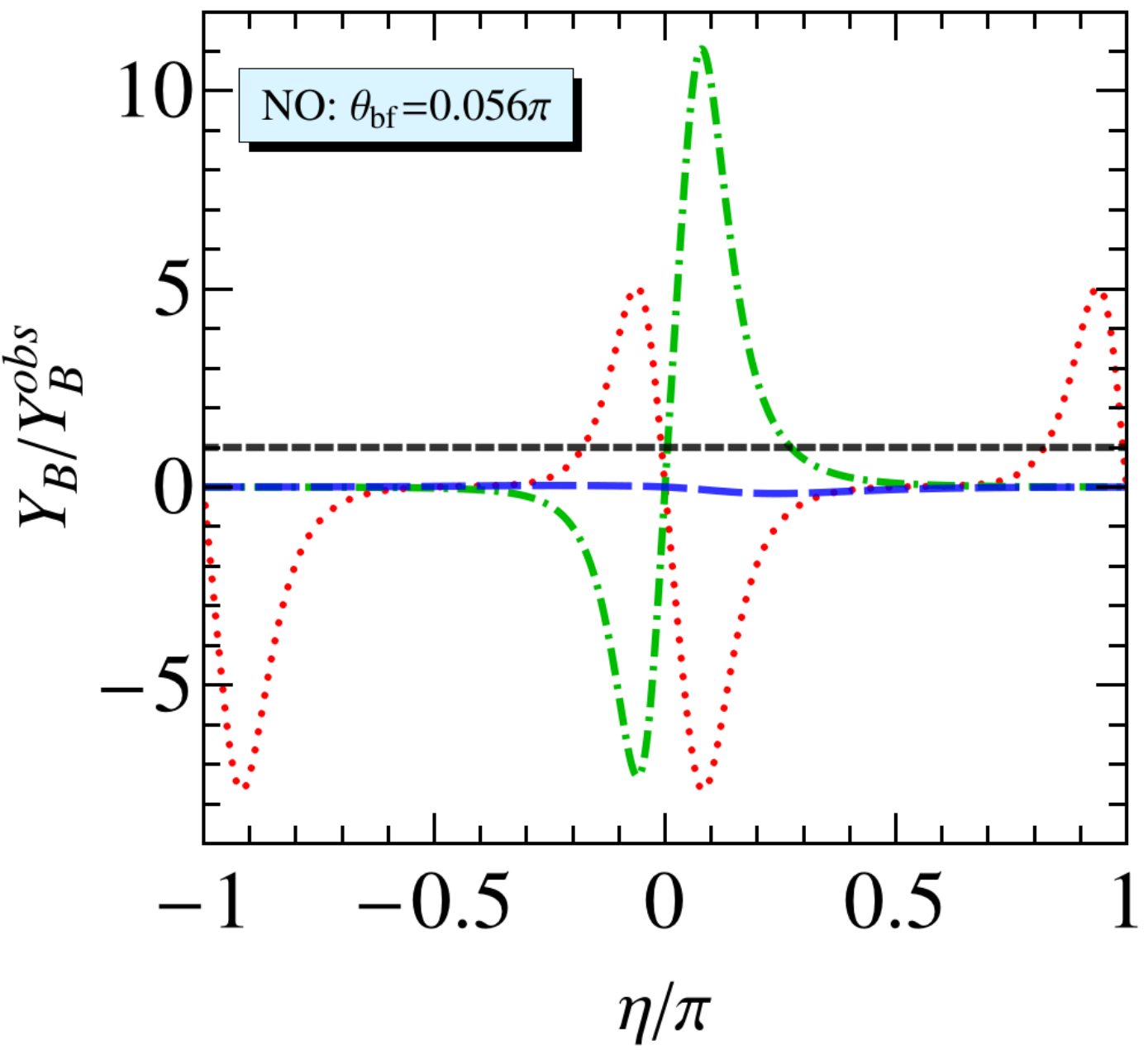}~~~
\includegraphics[width=0.42\linewidth]{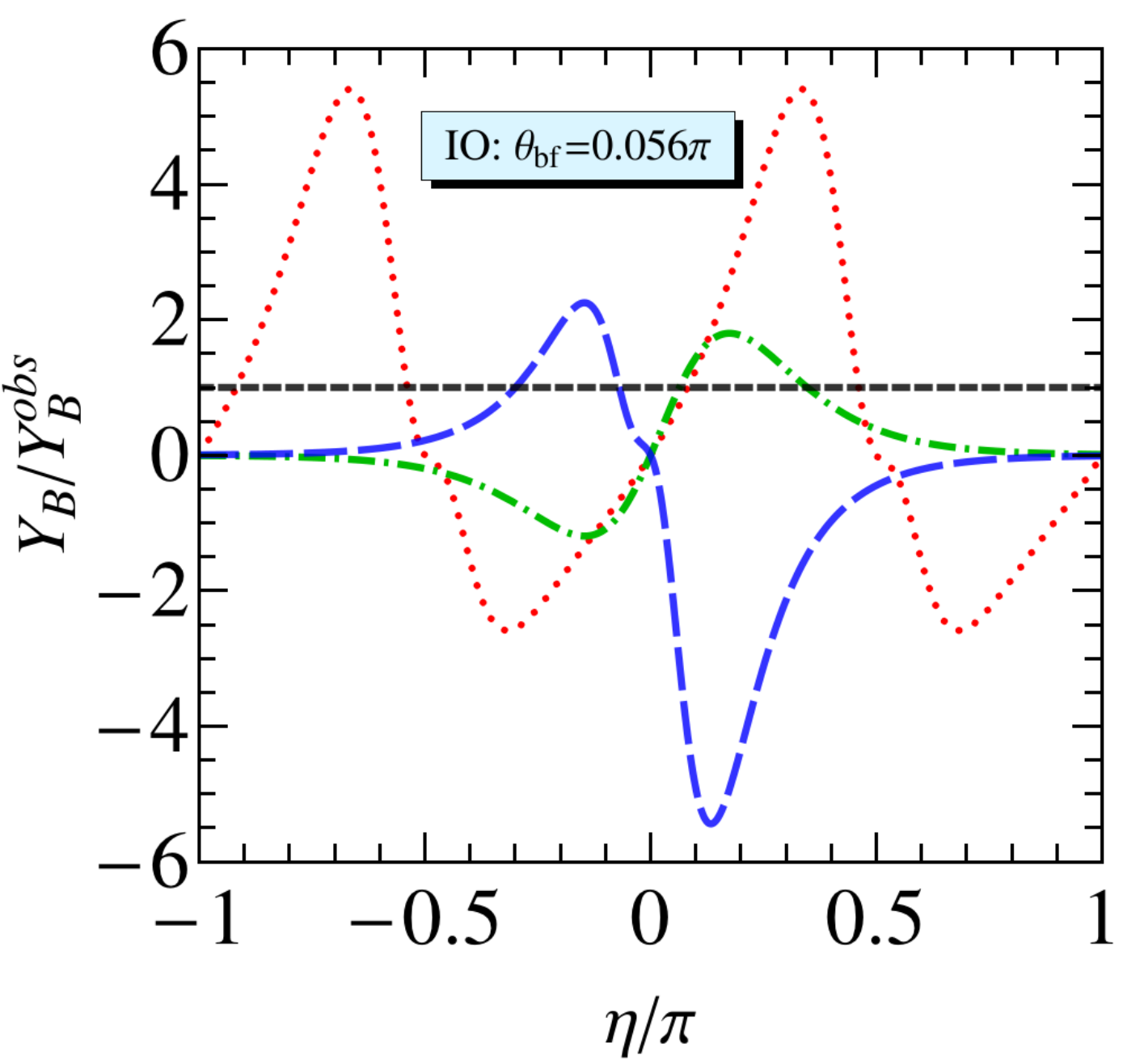}
\includegraphics[width=0.435\linewidth]{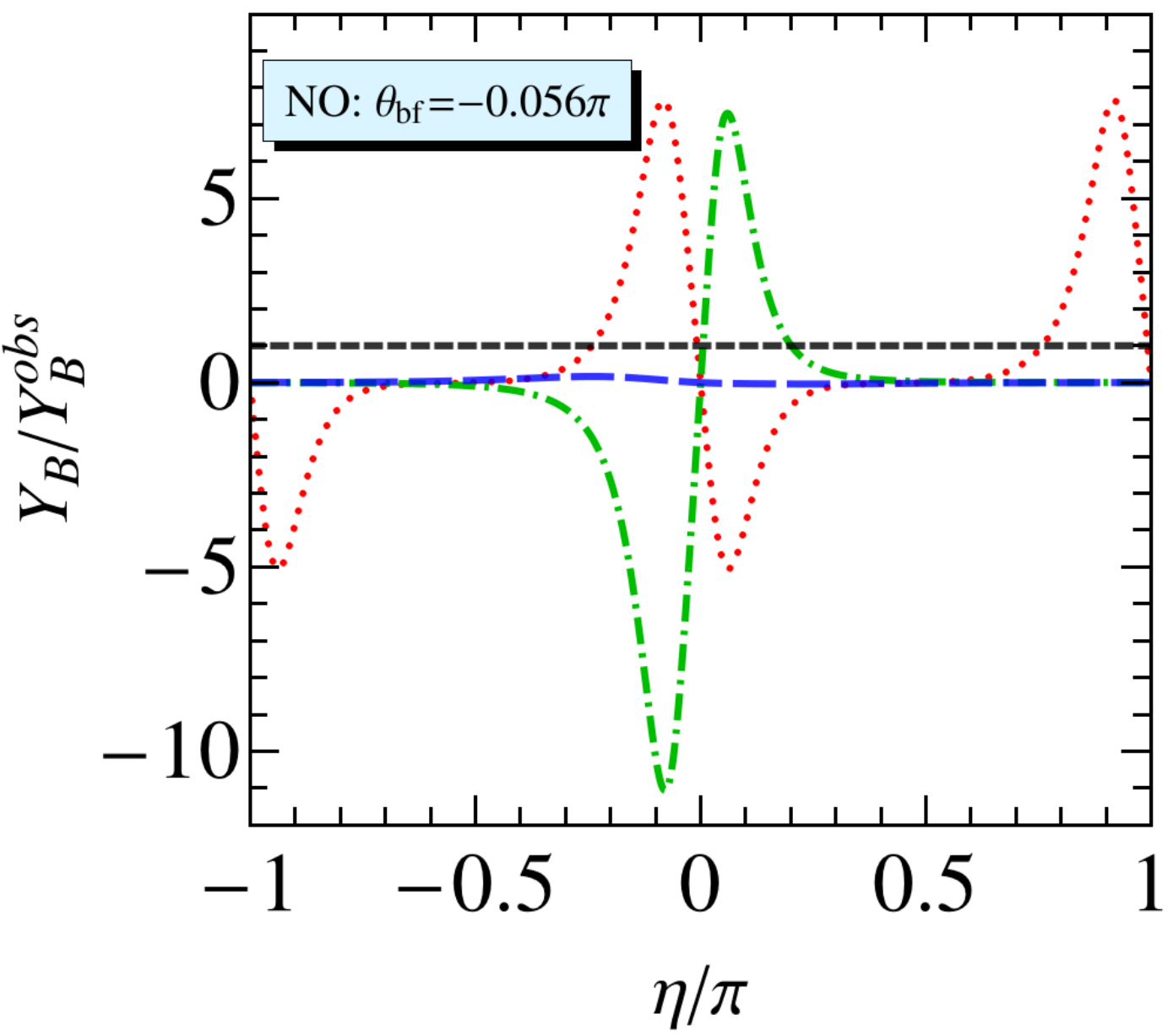}~~~
\includegraphics[width=0.435\linewidth]{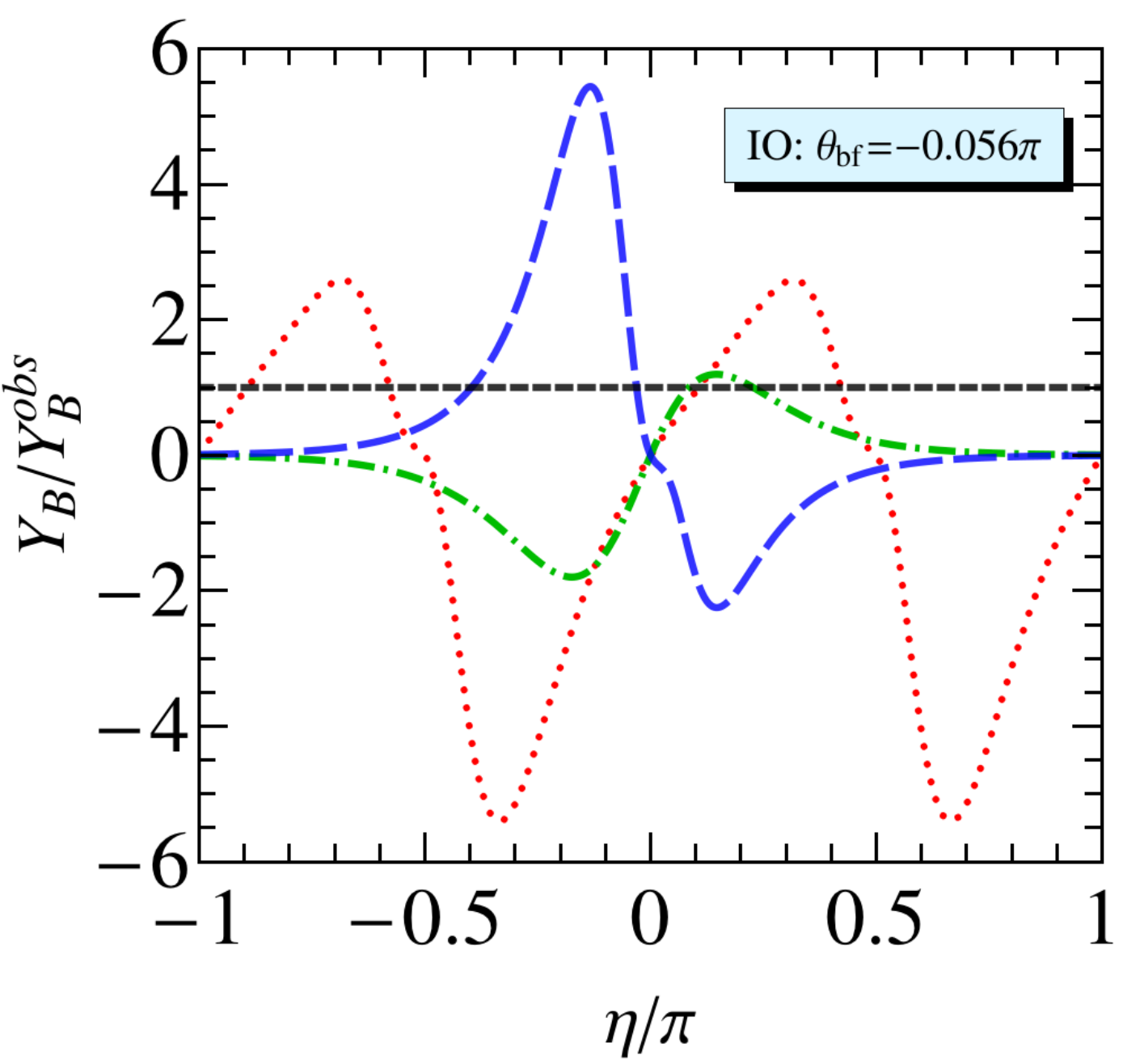}
\caption{\label{fig:leptogenesis_CaseIV}
The prediction for $Y_B/Y_B^{obs}$ as a function of $\eta$ in case IV(b), where $\theta_{bf}$ is the best fit value of $\theta$. We choose $M_1=5\times 10^{11}$ GeV and the lightest neutrino mass $m_1$ (or $m_3$) = 0.01eV. The red dotted, green dot-dashed, blue dashed lines correspond to $(K_1,K_2,K_3)=(+,\pm,+),(+,\pm,-)$ and $(-,\pm,+)$ respectively. The experimentally observed value $Y_B^{obs}$ is represented by the horizontal black dashed line.}
\end{center}
\end{figure}

\item[Case \uppercase\expandafter{\romannumeral5}]

\begin{equation}
\begin{aligned}
U^{V(a)}&=\frac{1}{2}\left(
\begin{array}{ccc}
 \phi_g  & 1 & \phi_g -1 \\
 \phi_g -1 & -\phi_g  & 1 \\
 1 & 1-\phi_g  & -\phi_g
\end{array}
\right)S_{23}(\theta)Q^{\dagger}_{\nu}\\
U^{V(b)}&=\frac{1}{2}\left(
\begin{array}{ccc}
 \phi_g  & 1 & \phi_g -1 \\
 1 & 1-\phi_g  & -\phi_g \\
 \phi_g -1 & -\phi_g  & 1
\end{array}
\right)S_{23}(\theta)Q^{\dagger}_{\nu}\,.
\end{aligned}
\end{equation}
Notice that these two mixing matrices are related through a exchange of
the second and third rows. Similar to case IV, this mixing pattern can be obtained from the flavor symmetry groups $[60,5]\cong A_5$, [120,35], [180,19] etc in combination with generalized CP. Earlier studies of this mixing pattern in the context of $A_5$ flavor symmetry and CP can be found in Refs.~\cite{Li:2015jxa,DiIura:2015kfa,Ballett:2015wia}. We can extract the following results for the mixing angles
\begin{equation}
\begin{split}
&\sin^2\theta_{13}=\frac{(\cos\theta-\phi_g\sin\theta)^2}{4\phi_g^2},\quad \sin^2\theta_{12}=\frac{(\phi_g\cos\theta+\sin\theta)^2}{4\phi_g^2-(\cos\theta-\phi_g\sin\theta)^2}\,,\\
&\sin^2\theta_{23}=\frac{\phi^2_{g}(\cos\theta+\phi_g\sin\theta)^2}{4\phi_g^2-(\cos\theta-\phi_g\sin\theta)^2}~~~\text{for}~~~U^{V(a)},\\
&\sin^2\theta_{23}=\frac{(\sin\theta-\phi^2_{g}\cos\theta)^2}{4\phi_g^2-(\cos\theta-\phi_g\sin\theta)^2}~~~\text{for}~~~U^{V(b)}\,.
\end{split}
\end{equation}
For the mixing pattern $U^{V(a)}$, the global minimum of $\chi^2$ is $\chi^2_{\text{min}}=6.190~(6.434)$ obtained at the best fitting values $\theta_{\text{bf}}=0.095\pi~(0.095\pi)$ for NO (IO) spectrum. Accordingly the mixing angels at $\theta=\theta_{\mathrm{bf}}$ are given by $\sin^2\theta_{12}=0.331$, $\sin^2\theta_{13}=0.022$ and $\sin^2\theta_{23}=0.524$ which are in excellent agreement with experimental data. For the PMNS matrix $U^{V(b)}$, $\chi^2$ is minimized at the best fitting point $\theta_{\text{bf}}=0.095\pi~(0.094\pi)$ with $\chi^2_{\text{min}}=4.477~(11.799)$, and the values obtained for the mixing angles are $\sin^2\theta_{12}=0.331$, $\sin^2\theta_{13}=0.022$ and $\sin^2\theta_{23}=0.476$. The CP invariants $J_{CP}$, $I_1$ and $I_2$ are found to vanish exactly so that both Dirac and Majorana CP phases take CP conserving values 0 and $\pi$. Similarly the bilinear invariants $I^{\alpha}_{23}$ are also zero. Hence a baryon asymmetry can not be obtained in this case unless the residual symmetries are further broken by higher order contributions. Furthermore, the two PMNS mixing matrices $U^{V(a)}$ and $U^{V(b)}$ yield the same expression for the effective Majorana mass $|m_{ee}|$
\begin{equation}
\left|m_{ee}\right|=\frac{1}{4}\left|\phi_g^2m_1+q_1m_2(\cos\theta+\phi^{-1}_g\sin\theta)^2+q_2m_3(\sin\theta-\phi^{-1}_g\cos\theta)^2\right|
\end{equation}
with $q_1, q_2=\pm1$. The predicted values of $|m_{ee}|$ from this mixing pattern are shown in figure~\ref{fig:mee_CaseV}. We find that $|m_{ee}|$ is around 0.016eV or 0.048eV in the case of IO spectrum, and it can be approximately vanishing for NO due to strong cancellations if the lightest neutrino mass is in the narrow range of $0.0023~\mathrm{eV}\leq m_{\text{lightest}}\leq0.0034$~eV and $0.0067~\mathrm{eV}\leq m_{\text{lightest}}\leq0.0078$~eV.

\begin{figure}[ht!]
\begin{center}
\includegraphics[width=0.5\linewidth]{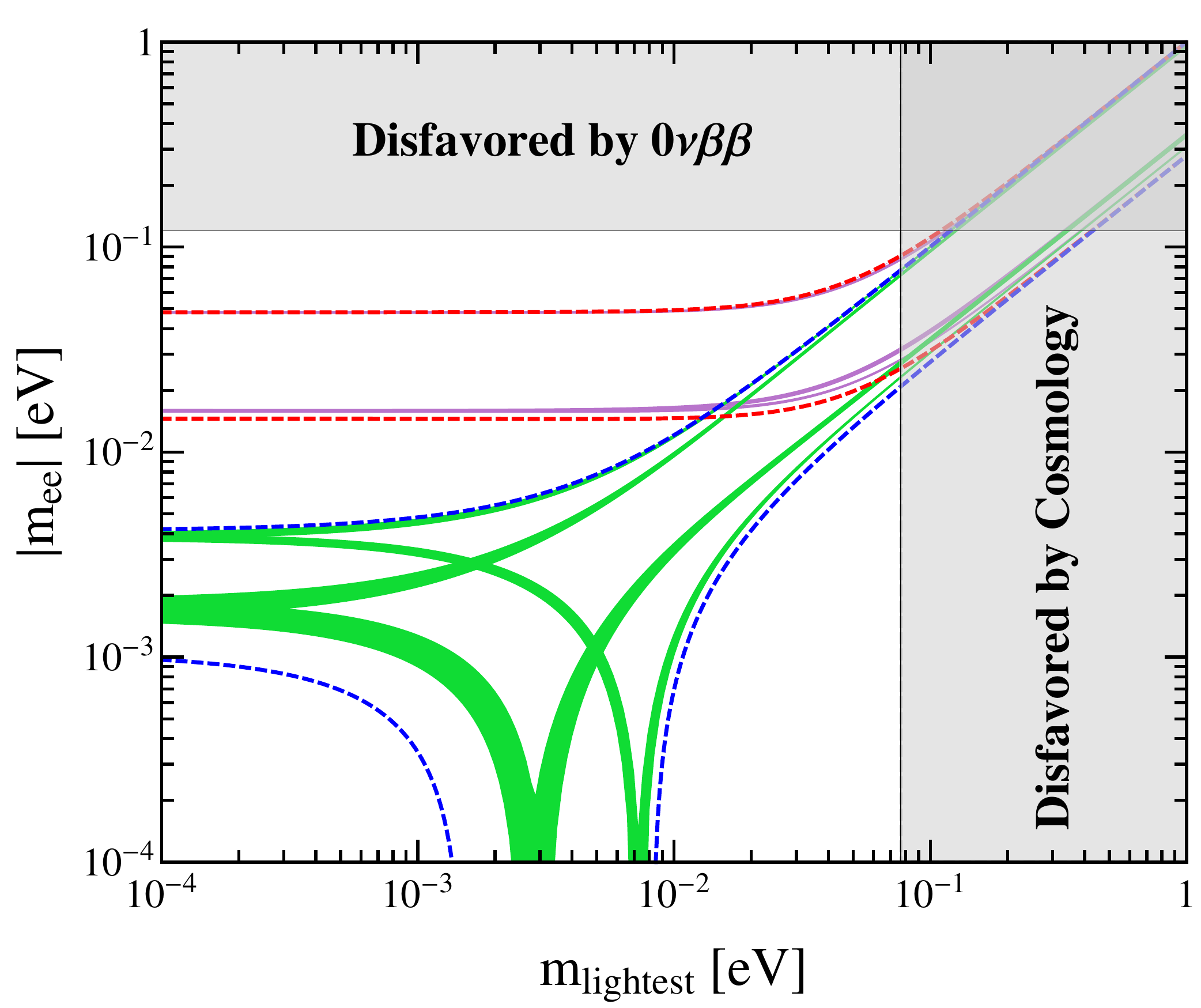}
\caption{\label{fig:mee_CaseV}Predictions of the $0\nu\beta\beta$ decay effective mass $|m_{ee}|$ with respect to the lightest neutrino mass $m_{\text{lightest}}$ for the mixing patterns $U^{V(a)}$ and $U^{V(b)}$. The red (blue) dashed lines indicate the most general allowed regions for IO (NO) spectrum obtained by varying the mixing parameters within their $3\sigma$ ranges~\cite{Gonzalez-Garcia:2014bfa}. The purple and green regions are the theoretical predictions of these two mixing patterns. The present most stringent upper limits $|m_{ee}|<0.120$ eV from EXO-200~\cite{Auger:2012ar, Albert:2014awa} and KamLAND-ZEN~\cite{Gando:2012zm} is shown by horizontal grey band. The vertical grey exclusion band is the current limit on $m_{\text{lightest}}$ from the cosmological data of $\sum m_i<0.230$ eV by the Planck collaboration~\cite{Ade:2013zuv}.}
\end{center}
\end{figure}

\item[Case \uppercase\expandafter{\romannumeral6}]

\begin{equation}
U^{VI}=\frac{1}{2\sqrt{3}}\left(
\begin{array}{ccc}
(\sqrt{3}-1)e^{i \varphi } &~ 2 &~ -(\sqrt{3}+1)e^{i\left(\varphi+\frac{3\pi}{4}\right)} \\
-(\sqrt{3}+1)e^{i\varphi} &~ 2 &~ (\sqrt{3}-1)e^{i\left(\varphi+\frac{3\pi}{4}\right)} \\
2e^{i\varphi} &~ 2 &~ 2e^{i\left(\varphi+\frac{3\pi}{4}\right)}
\end{array}
\right)S_{13}(\theta)Q^{\dagger}_{\nu}\,,
\end{equation}
where $\varphi=\arctan (2-\sqrt{7})$. This mixing pattern has not been discussed in the literature as far as we know. It can be achieved from the flavor symmetry groups [168,42], [336,209], [504,157] and others which are listed at the website~\cite{webdata}. The group [168,42] exactly is the known group $\Sigma(168)\cong PSL(2,7)$. It is the automorphism group of the Klein quartic as well as the symmetry group of the Fano plane. It is the second-smallest nonabelian simple group after the alternating group $A_5$. It has important applications in algebra, geometry, and number theory. $\Sigma(168)$ has also been recognized as quite interesting in discrete flavor symmetry theory~\cite{Luhn:2007yr}.  Notice that one column of the PMNS matrix is $(1, 1, 1)^{T}/\sqrt{3}$ in this case, and it should be identified as the second column in order to be compatible with the experimental data on lepton mixing angles. For the mixing matrices arising from the six possible row permutations of $U^{VI}$, four of them can accommodate the experimental data
\begin{equation}
\begin{aligned}
U^{VI(a)}=&U^{VI}_{PMNS}\,,\qquad U^{VI(b)}=P_{132}U^{VI}_{PMNS}\,,\\
U^{VI(c)}=&P_{213}U^{VI}_{PMNS}\,,\qquad U^{VI(d)}=P_{231}U^{VI}_{PMNS}\,.
\end{aligned}
\end{equation}
One see that $U^{VI(b)}$ and $U^{VI(d)}$ can be obtained from $U^{VI(a)}$ and $U^{VI(c)}$ respectively by exchanging the second and third rows. From the mixing matrices $U^{VI(a)}$ and $U^{VI(b)}$, the mixing angles and the three $CP$ rephasing invariants can be read out as
\begin{eqnarray}
\nonumber&&\sin^2\theta_{13}=\frac{1}{12}\left(4+2\sqrt{3}\cos2\theta+\sqrt{2}\sin2\theta\right),\\
\nonumber&&\sin^2\theta_{12}=\frac{4}{8-2\sqrt{3}\cos2\theta-\sqrt{2}\sin2\theta}\,,\\
\nonumber&&\sin^2\theta_{23}=\frac{4-2\sqrt{3}\cos2\theta+\sqrt{2}\sin2\theta}{8-2\sqrt{3}\cos2\theta-\sqrt{2}\sin2\theta}~~~\text{for}~~~U^{VI(a)}\,,\\
\nonumber&&\sin^2\theta_{23}=\frac{4-2\sqrt{2}\sin2\theta}{8-2\sqrt{3}\cos2\theta-\sqrt{2}\sin2\theta}~~~\text{for}~~~U^{VI(b)}\,,\\
\nonumber&& |J_{CP}|=\frac{1}{6\sqrt{6}}\left|\sin2\theta\right|\,,\quad |I_2|=\frac{1}{36}\left|\cos2\theta-\sqrt{6}\sin2\theta\right|\,,\\
&&|I_1|=\frac{1}{72}\left|2\sqrt{7}-\sqrt{3}+\left(2-\sqrt{21}\right)\cos2\theta-\sqrt{14}\sin2\theta\right|\,.
\end{eqnarray}
Then we can derive the following sum rules among the mixing angles
\begin{equation}
\label{eq:sum_rule_case_VI}\begin{aligned}
  \sin^2\theta_{12}\cos^2\theta_{13}&=\frac{1}{3}\,,\\
  \sin^2\theta_{23}\cos^2\theta _{13}&=\frac{1}{42} \left(9+15\cos2\theta_{13}\pm 2\sqrt{3}\sqrt{12 \cos 2 \theta _{13}-9 \cos 4 \theta _{13}-4}\right)~~\text{for $U^{VI(a)}$}~~\,,\\
  \sin^2\theta_{23}\cos^2\theta _{13}& =\frac{1}{21}\left(6+3\cos2\theta_{13}\pm\sqrt{3}\sqrt{12\cos2\theta_{13}-9\cos4\theta_{13}-4}\right)~~\text{for $U^{VI(b)}$}\,.
\end{aligned}
\end{equation}
Plugging in the best fitting value of the reactor angle $\sin^2\theta_{13}=0.0218$~\cite{Gonzalez-Garcia:2014bfa}, we have
\begin{equation}
 \begin{aligned}
 \label{eq:estimate_case_VI}\sin^2\theta_{12}=&0.341\,,\\
 \sin^2\theta_{23}=&0.559~~\text{or}~~0.578~~\text{for $U^{VI(a)}$}\,,\\
 \sin^2\theta_{23}=&0.441~~\text{or}~~0.422~~\text{for $U^{VI(b)}$}\,.
\end{aligned}
\end{equation}
\begin{table}[t!]\tabcolsep=0.11cm
\begin{center}
\scalebox{0.9}{\begin{tabular}{|c|c|c|c|c|c|c|c|c|c|}
\hline\hline
\multirow{2}{*}{} & \multirow{2}{*}{$\theta_{\text{bf}}/\pi$} & \multirow{2}{*}{$\chi^2_{\text{min}}$} & \multirow{2}{*}{$\sin^2\theta_{13}$} & \multirow{2}{*}{$\sin^2\theta_{12}$} & \multirow{2}{*}{$\sin^2\theta_{23}$} & \multirow{2}{*}{$\delta_{CP}/\pi$} &$\alpha_{21}/\pi$ & $\alpha'_{31}/\pi$ & \multirow{2}{*}{$(K_1,K_2,K_3)$} \\
&  &  &  &  &  &  & (mod 1) & (mod 1) &\\
\hline
\multirow{2}{*}{$U^{VI(a)}$} &  0.572 & 12.028 & 0.0222  &  0.341  &  0.554  &  0.667  &  0.839  &  0.106 & $(+,\pm,+),(+,\pm,-)$\\
&  [0.555] & [8.007] & [0.0218]  &  [0.341]  &  [0.578]  &  [0.763]  &  [0.845]  &  [0.926] & $[(+,\pm,+),(+,\pm,-)]$\\
\hline
\multirow{2}{*}{$U^{VI(b)}$} &  0.569 & 8.133 & 0.0219  &  0.341  &  0.443  &  1.680  &  0.839  &  0.082 & $(+,\pm,+),(+,\pm,-)$\\
&  [0.576] & [20.586] & [0.0227]  &  [0.341]  &  [0.452]  &  [1.646]  &  [0.837]  &  [0.146] & $[(+,\pm,+),(+,\pm,-)]$\\
\hline
\multirow{2}{*}{$U^{VI(c)}$} &  0.928 & 12.028 & 0.0222  &  0.341  &  0.554  &  1.333  &  0.392  &  0.894 & $(+,\pm,+),(+,\pm,-)$\\
&  [0.945] & [8.007] & [0.0218]  &  [0.341]  &  [0.578]  &  [1.237]  &  [0.385]  &  [0.074] & $[(-,\pm,+)]$\\
\hline
\multirow{2}{*}{$U^{VI(d)}$} &  0.931 & 8.133 & 0.0219  &  0.341  &  0.443  &  0.320  &  0.391  &  0.918 & $(+,\pm,+),(+,\pm,-)$\\
&  [0.924] & [20.586] & [0.0227]  &  [0.341]  &  [0.452]  &  [0.354]  &  [0.393]  &  [0.854] & $[(-,\pm,+)]$\\
\hline\hline
\end{tabular}}
\caption {Results of the $\chi^2$ analysis for case VI. We show the best fit value $\theta_{\text{bf}}$ of the parameter $\theta$, and $\chi^2_{\text{min}}$ is the global minimum of the $\chi^2$ function. The mixing angles and the CP violating phases for $\theta=\theta_{bf}$ are given as well. Note that the CP parity matrix $Q_{\nu}$ can shift the Majorana phases $\alpha_{21}$ and $\alpha'_{31}$ by $\pi$. In the last column we give the values of $K_{1,2,3}$ for which the observed baryon asymmetry can be generated via leptogenesis. The values in the square brackets are the corresponding results for the case of IO mass spectrum.\label{tab:CaseVIbparas}}
\end{center}
\end{table}
Obviously the atmospheric mixing angle $\theta_{23}$ is non-maximal in this case. The results of our $\chi^2$ analysis are summarized in table~\ref{tab:CaseVIbparas}. The mixing matrices $U^{VI(c)}$ and $U^{VI(d)}$ give rise to the following results for mixing angles and CP invariants
\begin{eqnarray}
\nonumber&&\sin^2\theta_{13}=\frac{1}{12}\left(4-2\sqrt{3}\cos2\theta+\sqrt{2}\sin2\theta\right)\,,\\
\nonumber&&\sin^2\theta_{12}=\frac{4}{8+2\sqrt{3}\cos2\theta-\sqrt{2}\sin2\theta}\,,\\
\nonumber&&\sin^2\theta_{23}=\frac{4+2\sqrt{3}\cos2\theta+\sqrt{2}\sin2\theta}{8+2\sqrt{3}\cos2\theta-\sqrt{2}\sin2\theta}~~~\text{for}~~~U^{VI(c)}\,,\\
\nonumber&&\sin^2\theta_{23}=\frac{4-2\sqrt{2}\sin2\theta}{8+2\sqrt{3}\cos2\theta-\sqrt{2}\sin2\theta}~~~\text{for}~~~U^{VI(d)}\,,\\
\nonumber&& |J_{CP}|=\frac{1}{6\sqrt{6}}\left|\sin2\theta\right|\,,\quad |I_2|=\frac{1}{36}\left|\cos2\theta+\sqrt{6}\sin2\theta\right|\,,\\
&&|I_1|=\frac{1}{72}\left|2\sqrt{7}+\sqrt{3}+\left(2+\sqrt{21}\right)\cos2\theta-\sqrt{14}\sin2\theta\right|\,.
\end{eqnarray}
We find the sum rules in Eq.~\eqref{eq:sum_rule_case_VI} and consequently the estimates given in Eq.~\eqref{eq:estimate_case_VI} are satisfied as well. Furthermore, the sum rule of Eq.~\eqref{eq:angle_phase_caseII} among the mixing angles and Dirac CP phase is fulfilled for all the above four permutations of the PMNS matrix. Consequently the comments below Eq.~\eqref{eq:angle_phase_caseII} also hold true here. As regards the neutrinoless double beta decay, the predictions for the effective mass $|m_{ee}|$ are given by
\begin{eqnarray}
\begin{aligned}
\left|m_{ee}\right|&=\frac{1}{12}\left|\left((\sqrt{3}-1)e^{i\varphi}\cos\theta+(1+\sqrt{3})e^{i\left(\frac{3\pi}{4}+\varphi\right)}\sin\theta\right)^2m_1+4q_1m_2\right.\\
&\left.+q_2m_3\left((1+\sqrt{3})e^{i\left(\frac{3\pi}{4}+\varphi\right)}\cos\theta-(\sqrt{3}-1)e^{i\varphi}\sin\theta\right)^2\right|~~\text{for}~~U^{VI(a)}~ \text{and}~ U^{VI(b)}\,,
\end{aligned}\\
\begin{aligned}
\left|m_{ee}\right|&=\frac{1}{12}\left|\left((1+\sqrt{3})e^{i\varphi}\cos\theta+(\sqrt{3}-1)e^{i\left(\frac{3\pi}{4}+\varphi\right)}\sin\theta\right)^2m_1+4q_1m_2\right.\\
&\left.+q_2m_3\left((\sqrt{3}-1)e^{i\left(\frac{3\pi}{4}+\varphi\right)}\cos\theta-(1+\sqrt{3})e^{i\varphi}\sin\theta\right)^2\right|~~\text{for}~~U^{VI(c)}~ \text{and}~ U^{VI(d)}\,.
\end{aligned}
\end{eqnarray}
The parameter $\theta$ freely varies in the range of $[0, \pi]$, and the observed values of the lepton mixing angles are required to be reproduced at $3\sigma$ level. The admissible regions of $|m_{ee}|$ as a function of $m_{\text{lightest}}$ are displayed in figure~\ref{fig:mee_CaseVI}. We can read off from this figure that $|m^{\text{IO}}_{ee}|\simeq0.019$ eV or 0.046eV and $|m^{\text{NO}}_{ee}|\geq0.00052$ eV for the mixing patterns $U^{VI(a)}$ and $U^{VI(b)}$ while $|m^{\text{IO}}_{ee}|\simeq0.030$ eV or 0.040eV and $|m^{\text{NO}}_{ee}|\geq0.0018$ eV for the mixing patterns $U^{VI(c)}$ and $U^{VI(d)}$, where $|m^{\text{IO}}_{ee}|$ and $|m^{\text{NO}}_{ee}|$ are the $0\nu\beta\beta$ decay effective masses corresponding to IO and NO mass orderings respectively.

Then we turn to study the implication for leptogenesis. One can read out the lepton asymmetry parameters $I^{\alpha}_{13}$ as follows
\begin{equation}
\begin{split}
&I^e_{13}=I^{\mu}_{13}=\frac{1}{6\sqrt{2}},\quad  I^{\tau}_{13}=-\frac{1}{3\sqrt{2}}~~~\text{for $U^{VI(a)}$ and $U^{VI(c)}$},\\
&I^e_{13}=I^{\tau}_{13}=\frac{1}{6\sqrt{2}},\quad  I^{\mu}_{13}=-\frac{1}{3\sqrt{2}}~~~\text{for $U^{VI(b)}$ and $U^{VI(d)}$}\,,
\end{split}
\end{equation}
which are constant values. The numerical results for $Y_{B}$ as a function of $\eta$ are plotted in figure~\ref{fig:leptogenesis_CaseVIa&b} and figure~\ref{fig:leptogenesis_CaseVIc&d}. We can see that the observed baryon asymmetry can be interpreted as an effect of leptogenesis for certain values of the parameters $K_{1,2,3}$, as listed in table~\ref{tab:CaseVIbparas}.

\begin{figure}[ht!]
\begin{center}
\includegraphics[width=0.45\linewidth]{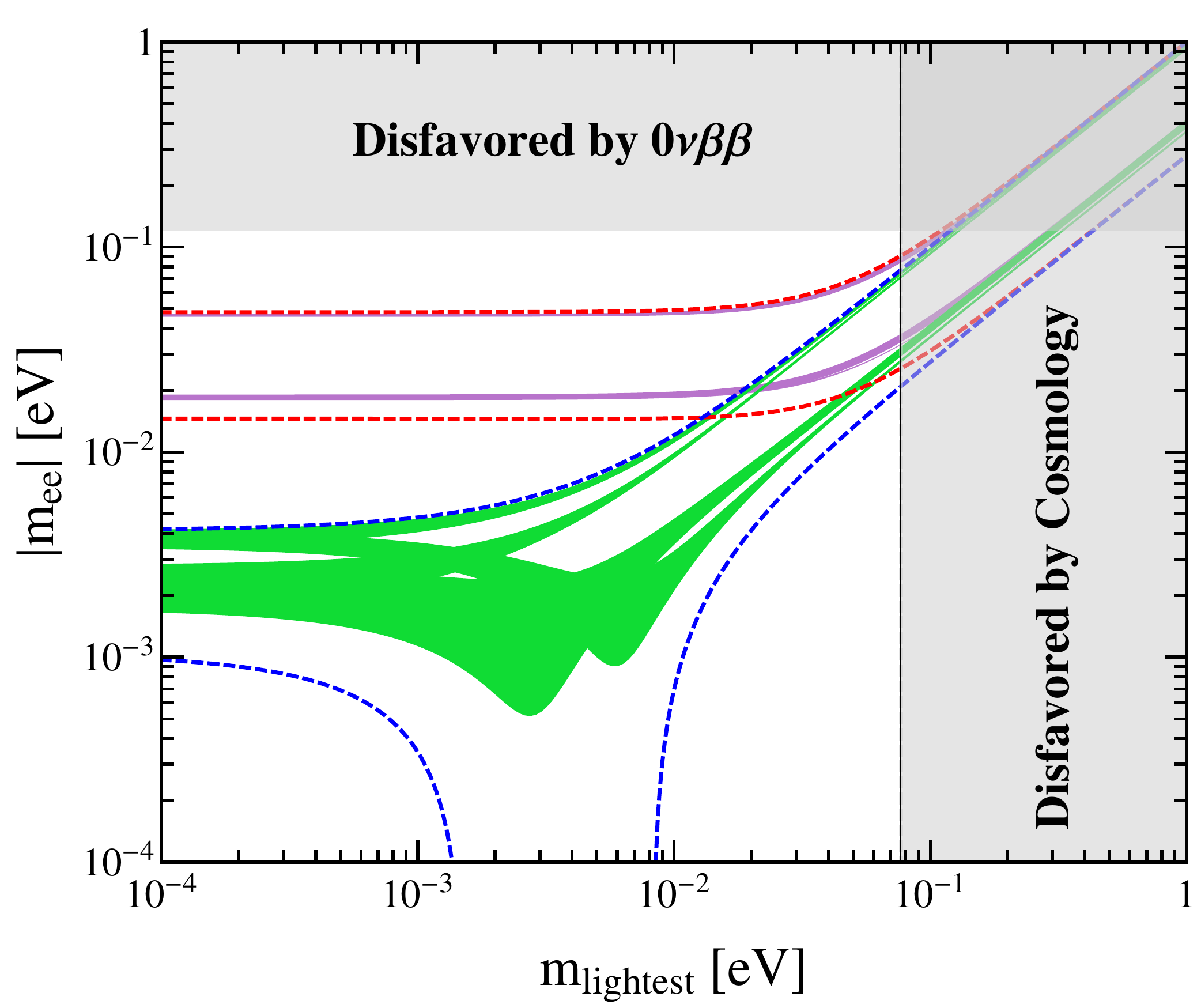}~~~
\includegraphics[width=0.45\linewidth]{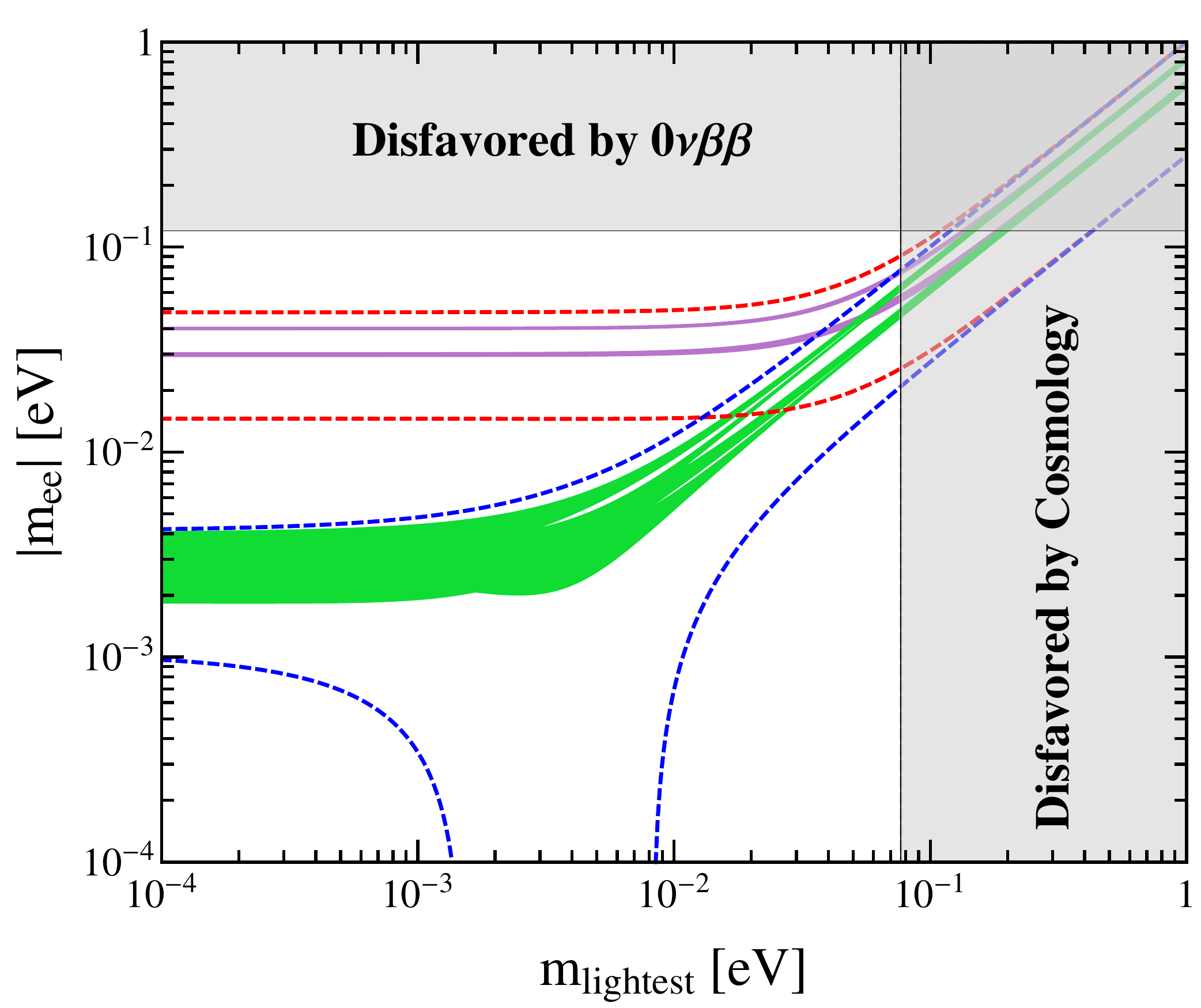}
\caption{\label{fig:mee_CaseVI}Predictions of the $0\nu\beta\beta$ decay effective mass $|m_{ee}|$ with respect to the lightest neutrino mass $m_{\text{lightest}}$ in the case VI. The left panel is the result for the mixing patterns $U^{I(a)}$ and $U^{I(b)}$, and the right panel is for $U^{I(c)}$ and $U^{I(d)}$. The red (blue) dashed lines indicate the most general allowed regions for IO (NO) spectrum obtained by varying the mixing parameters within their $3\sigma$ ranges~\cite{Gonzalez-Garcia:2014bfa}. The purple and green regions are the theoretical predictions of these two mixing patterns. The present most stringent upper limits $|m_{ee}|<0.120$ eV from EXO-200~\cite{Auger:2012ar, Albert:2014awa} and KamLAND-ZEN~\cite{Gando:2012zm} is shown by horizontal grey band. The vertical grey exclusion band is the current limit on $m_{\text{lightest}}$ from the cosmological data of $\sum m_i<0.230$ eV by the Planck collaboration~\cite{Ade:2013zuv}.}
\end{center}
\end{figure}

\begin{figure}[ht!]
\begin{center}
\begin{tabular}{cc}
\includegraphics[width=0.435\linewidth]{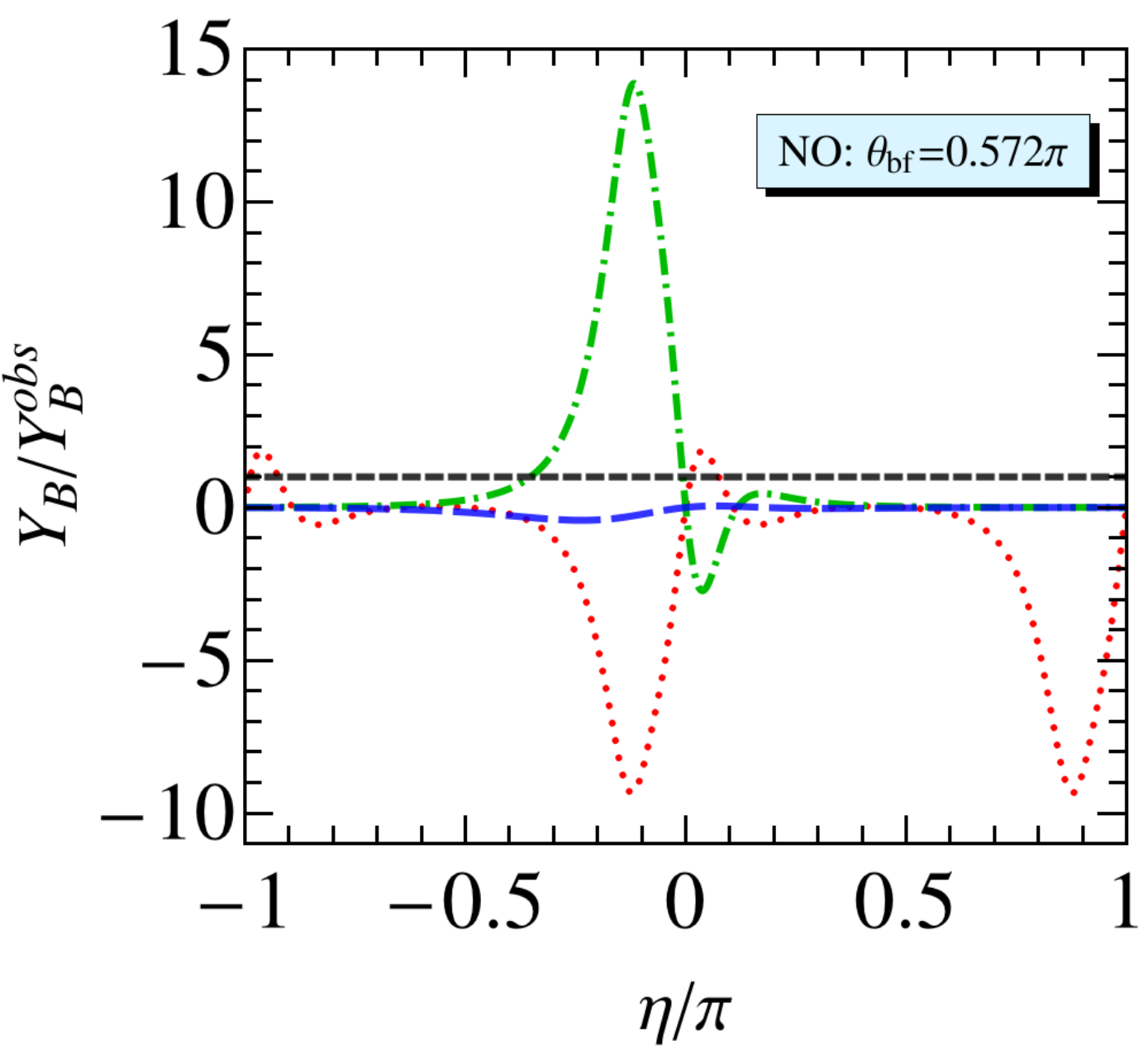}&
\includegraphics[width=0.435\linewidth]{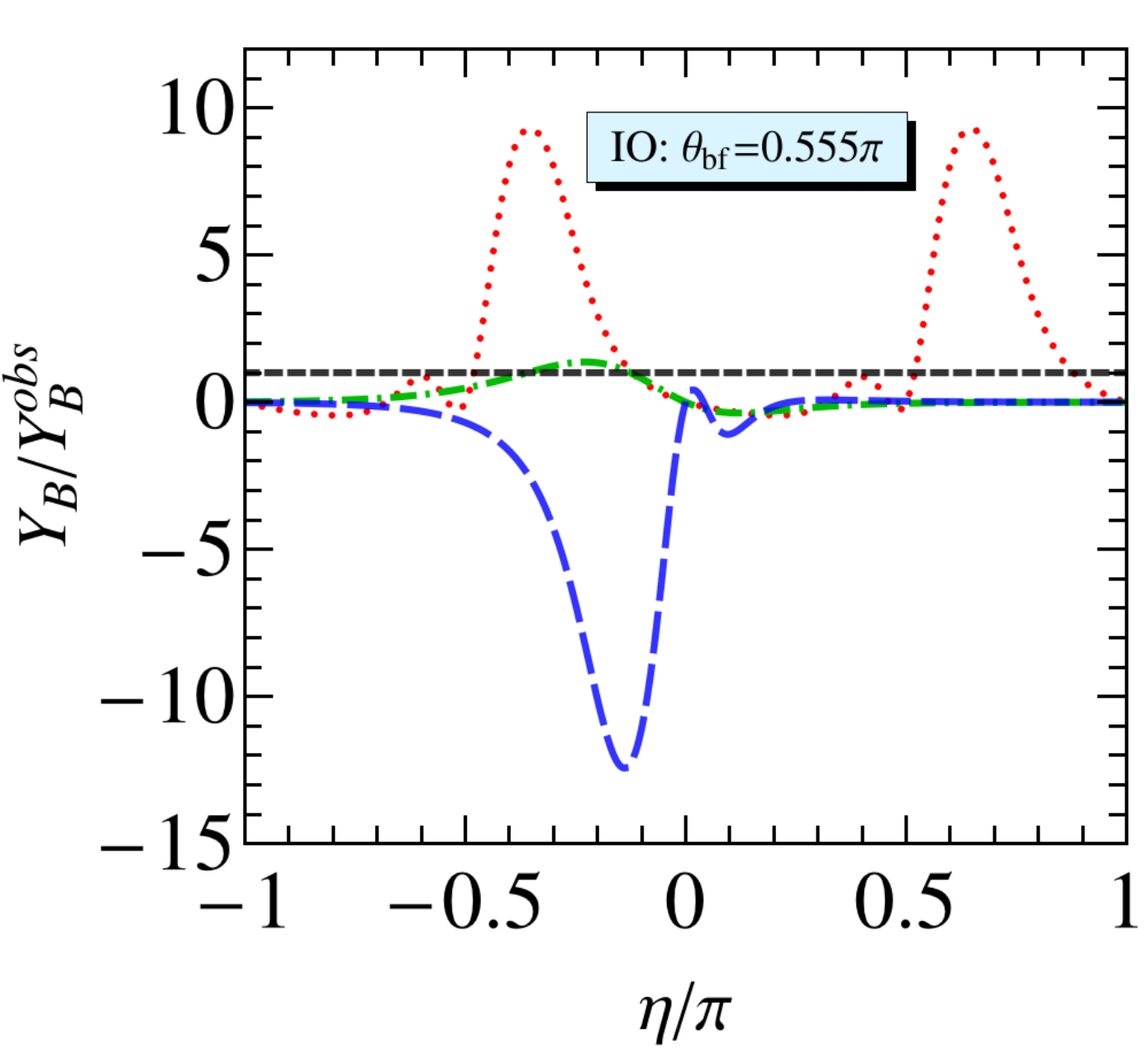}\\
~~~\includegraphics[width=0.42\linewidth]{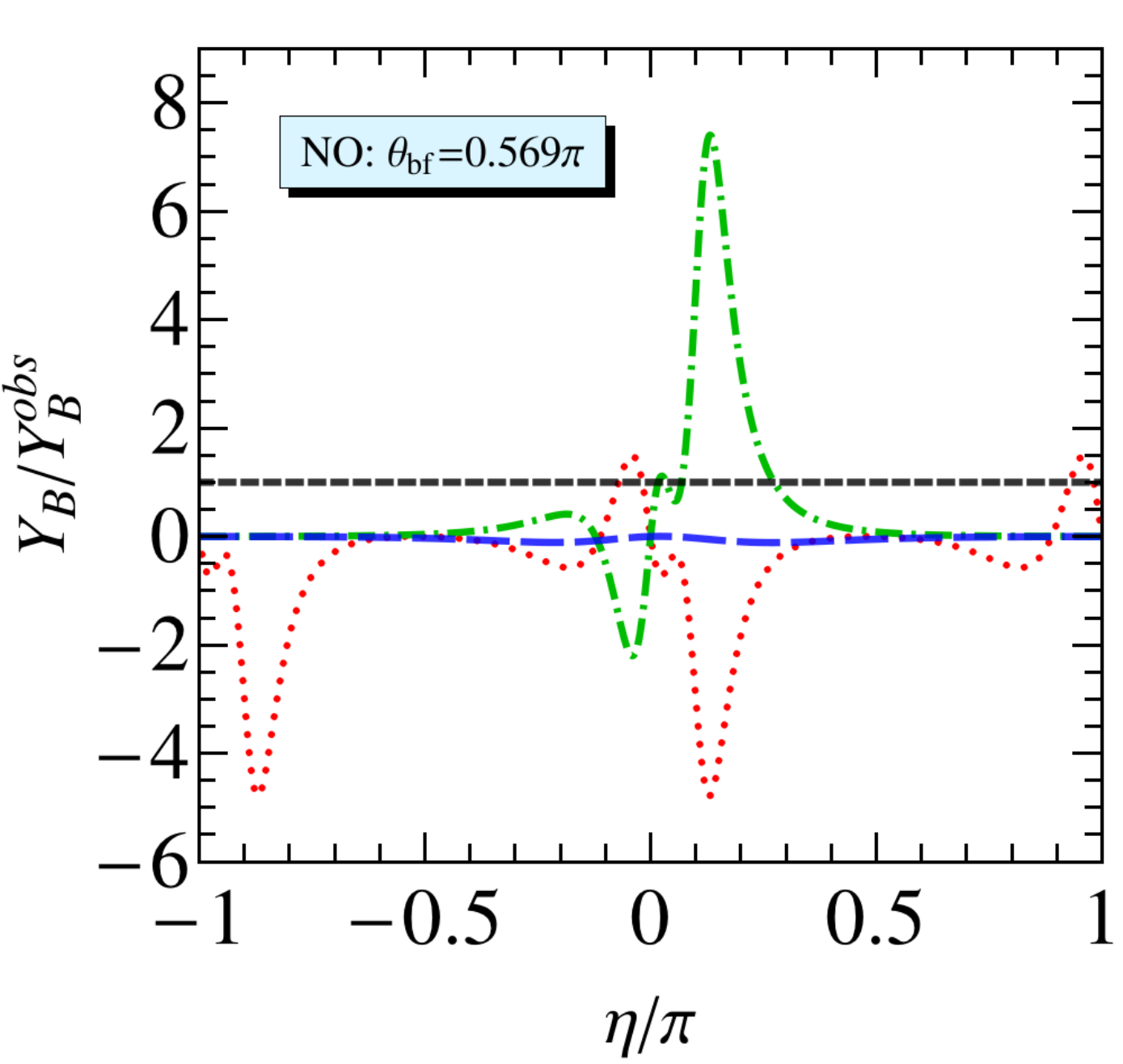}&~~~
\includegraphics[width=0.42\linewidth]{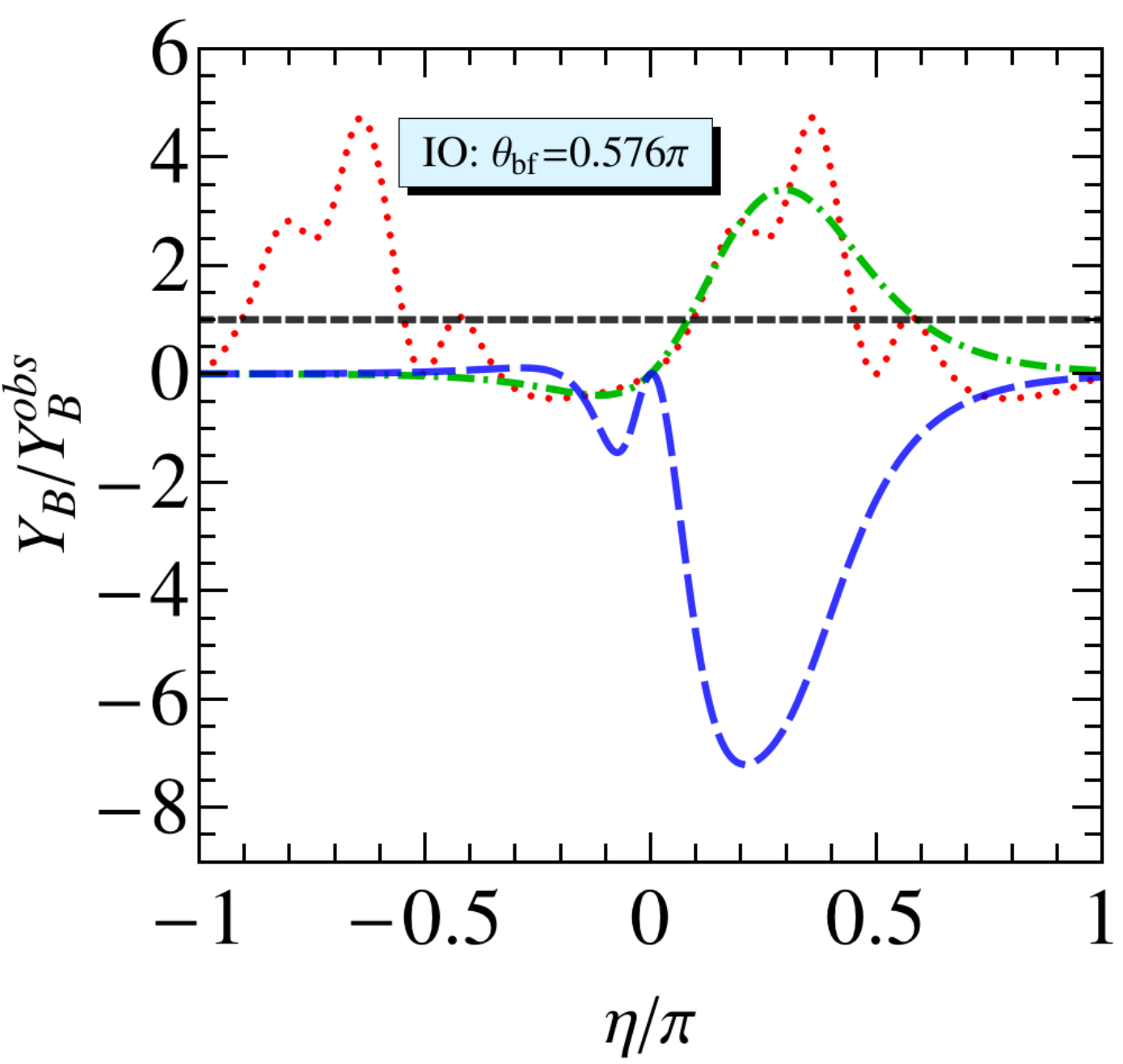}
\end{tabular}
\caption{\label{fig:leptogenesis_CaseVIa&b}
The prediction for $Y_B/Y_B^{obs}$ as a function of $\eta$ in case VI(a) and case VI(b) at the best fit value $\theta_{\mathrm{bf}}$, where the first and second rows correspond to the mixing patterns $U^{VI(a)}$ and $U^{VI(b)}$ respectively. We choose $M_1=5\times 10^{11}$ GeV and the lightest neutrino mass $m_1$ (or $m_3$) = 0.01eV. The red dotted, green dot-dashed, blue dashed lines correspond to $(K_1,K_2,K_3)=(+,\pm,+),(+,\pm,-)$ and $(-,\pm,+)$ respectively. The experimentally observed value $Y_B^{obs}$ is represented by the horizontal black dashed line.}
\end{center}
\end{figure}

\begin{figure}[ht!]
\begin{center}
\begin{tabular}{cc}
\includegraphics[width=0.435\linewidth]{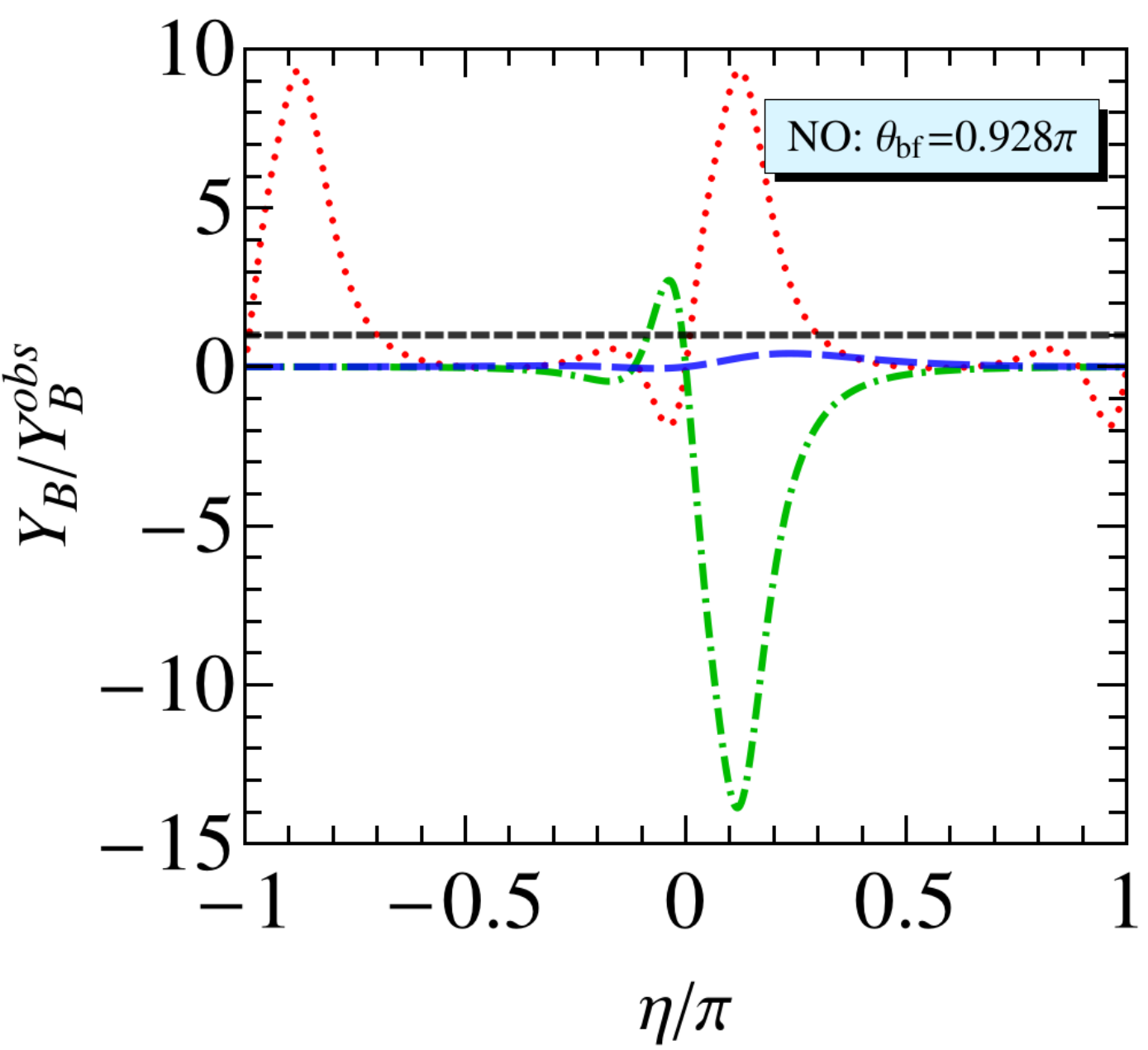} &
\includegraphics[width=0.435\linewidth]{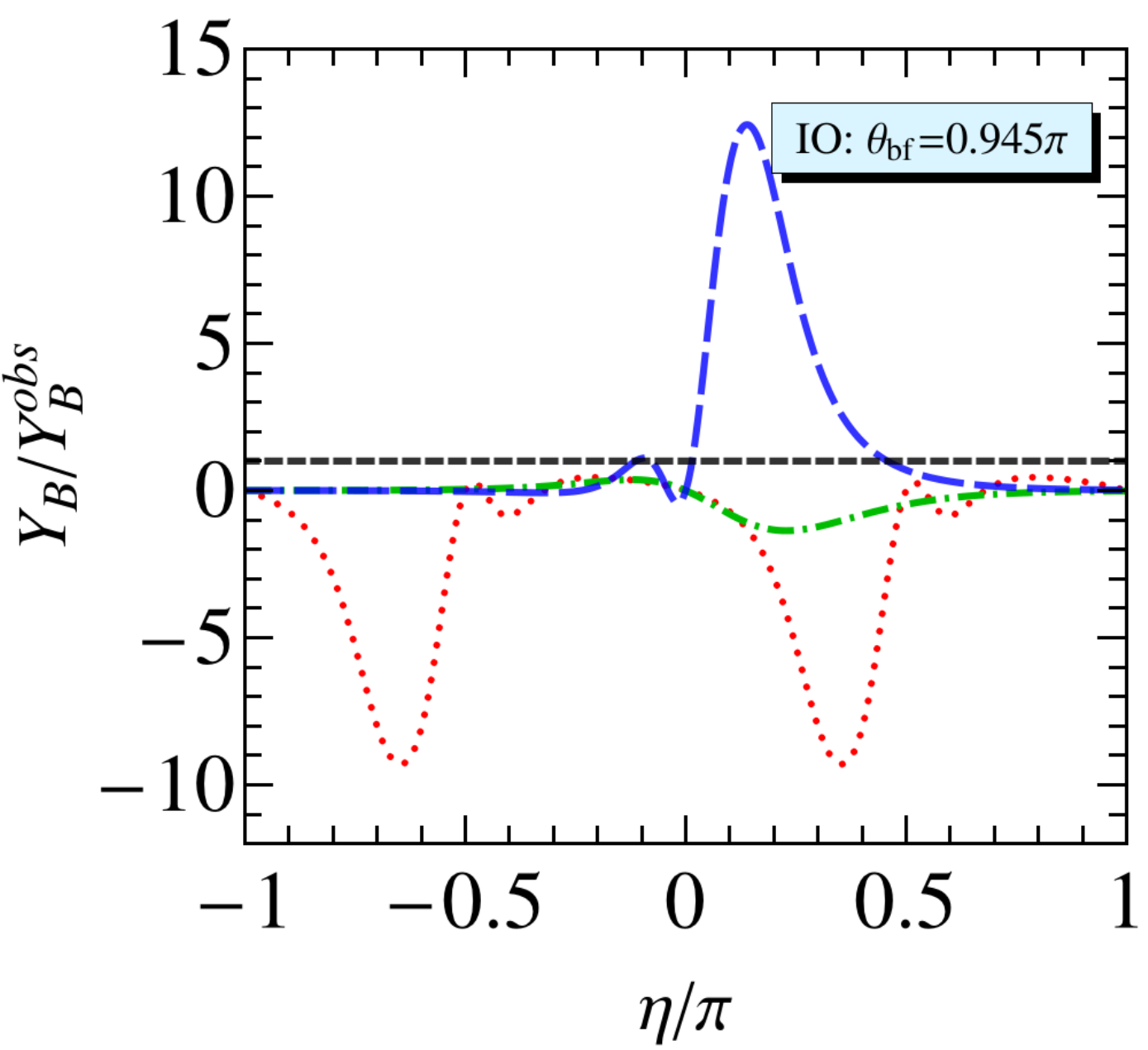}\\
~~~\includegraphics[width=0.42\linewidth]{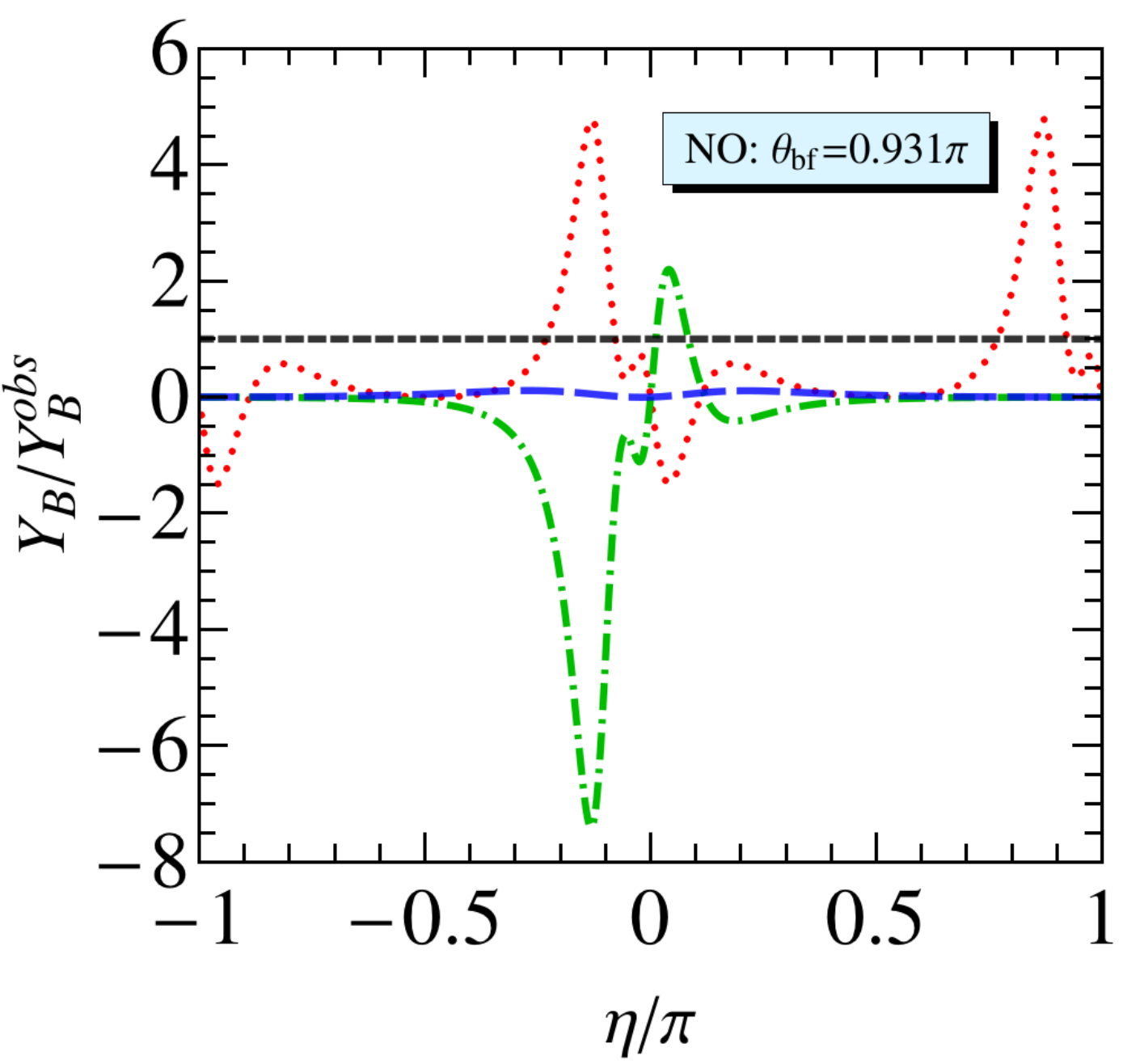} &~~~
\includegraphics[width=0.42\linewidth]{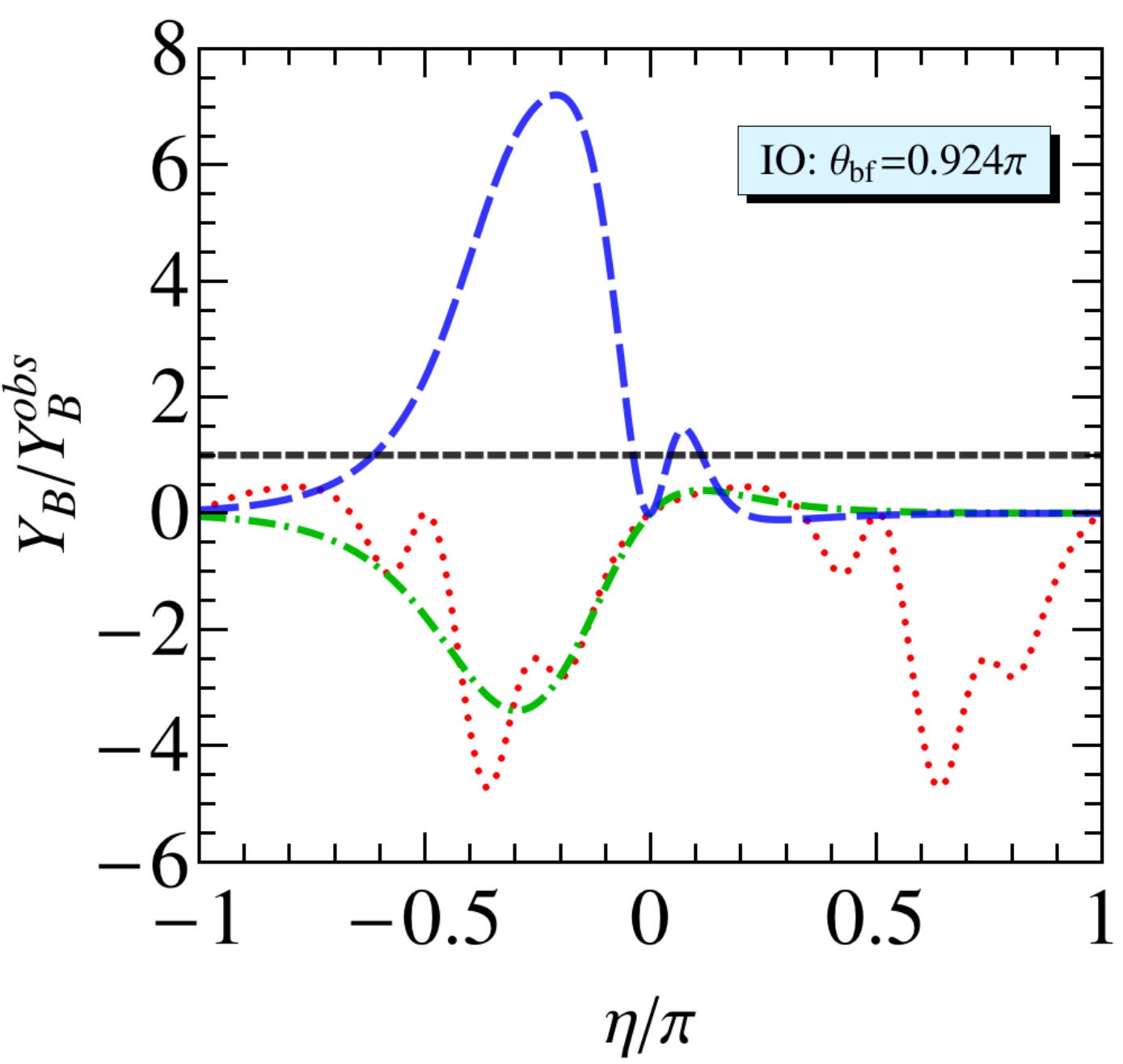}
\end{tabular}
\caption{\label{fig:leptogenesis_CaseVIc&d}
The prediction for $Y_B/Y_B^{obs}$ as a function of $\eta$ in case VI(c) and case VI(d) at the best fit value $\theta_{\mathrm{bf}}$, where the first and second rows correspond to the mixing patterns $U^{VI(c)}$ and $U^{VI(d)}$ respectively. We choose $M_1=5\times 10^{11}$ GeV and the lightest neutrino mass $m_1$ (or $m_3$) = 0.01eV. The red dotted, green dot-dashed, blue dashed lines correspond to $(K_1,K_2,K_3)=(+,\pm,+),(+,\pm,-)$ and $(-,\pm,+)$ respectively. The experimentally observed value $Y_B^{obs}$ is represented by the horizontal black dashed line.}
\end{center}
\end{figure}

\item[Case \uppercase\expandafter{\romannumeral7}]

\begin{equation}
\label{eq:PMNS_caseVII}
\begin{split}
U^{VII(a)}&=\frac{1}{2\sqrt{6}}\left(
\begin{array}{ccc}
  -\frac{\sqrt{3}}{s_3} &  2 \sqrt{2} &  \frac{s_2-s_1}{s_1 s_2} \\[1mm]
  \frac{\sqrt{3}}{s_2} &  2 \sqrt{2} & -\frac{s_1+s_3}{s_1 s_3} \\[1mm]
 \frac{\sqrt{3}}{s_1} &  2 \sqrt{2} & \frac{s_2+s_3}{s_2 s_3}
 \end{array}
\right)S_{23}(\theta)Q^{\dagger}_{\nu}\,,\\
U^{VII(b)}&=\frac{1}{2\sqrt{6}}\left(
\begin{array}{ccc}
  -\frac{\sqrt{3}}{s_3} &  2 \sqrt{2} &  \frac{s_2-s_1}{s_1 s_2} \\[1mm]
   \frac{\sqrt{3}}{s_1} &  2 \sqrt{2} & \frac{s_2+s_3}{s_2 s_3}
 \\[1mm]
  \frac{\sqrt{3}}{s_2} &  2 \sqrt{2} & -\frac{s_1+s_3}{s_1 s_3}
 \end{array}
\right)S_{23}(\theta)Q^{\dagger}_{\nu}\,,
\end{split}
\end{equation}
where $s_n\equiv\sin (2n\pi/7)$ with $n=1,2,3$. We note that that $U^{VII(a)}$ and $U^{VII(b)}$ are related by the exchange of the second and third rows. Similar to case VI, this mixing pattern can also be obtained from the flavor symmetry groups $[168,42]\cong\Sigma(168)$, [336,209], [504,157] and so forth in combination with generalized CP~\cite{webdata}. In this case, the column fixed by residual symmetry is
\begin{equation}
\frac{1}{2\sqrt{2}} \left(
\begin{array}{c}
 -1/s_3 \\ 1/s_2 \\ 1/s_1
\end{array}
\right) \approx \left(
\begin{array}{c}
 -0.815 \\  0.363\\ 0.452
\end{array}
\right),\quad \text{or} \quad \frac{1}{2\sqrt{2}} \left(
\begin{array}{c}
 -1/s_3  \\ 1/s_1 \\ 1/s_2
\end{array}
\right) \approx \left(
\begin{array}{c}
 -0.815 \\ 0.452 \\  0.363
\end{array}
\right)\,.
\end{equation}
It should be identified with the first column of the PMNS matrix to be in accordance with the experimental data. From the mixing matrices in Eq.~\eqref{eq:PMNS_caseVII}, we find the following results for the lepton mixing angles
\begin{equation}
\begin{split}
&\sin^2\theta_{13}=\frac{\left(2\sqrt{2}s_1s_2\sin\theta+\left(s_1-s_2\right)\cos\theta\right)^2}{24s_1^2s_2^2}\,,\\
&\sin^2\theta_{12}=\frac{\left(2\sqrt{2}s_1s_2\cos\theta+\left(s_2-s_1\right)\sin\theta\right)^2}{2\sqrt{2}s_1s_2\left(s_2-s_1\right)\sin2\theta-\left(s_2-s_1\right)^2\cos^2\theta+4s_1^2s_2^2(\cos2\theta+5)}\,,\\
&\sin^2\theta_{23}=\frac{s_2^2 \left(2\sqrt{2} s_1 s_3 \sin \theta + \left(s_1+s_3\right) \cos \theta \right)^2}{s_3^2 \left(2 \sqrt{2} s_1 s_2 \left(s_2-s_1\right) \sin 2 \theta-\left(s_2-s_1\right)^2 \cos ^2\theta +4 s_1^2 s_2^2 (\cos 2 \theta +5)\right)}~~\text{for}~~U^{VII(a)}\,,\\
&\sin^2\theta_{23}=\frac{s_1^2 \left(2\sqrt{2}s_2 s_3\sin\theta-(s_2+s_3)\cos\theta\right)^2}{s_3^2 \left(2 \sqrt{2} s_1 s_2 \left(s_2-s_1\right) \sin 2 \theta-\left(s_2-s_1\right)^2 \cos ^2\theta +4 s_1^2 s_2^2 (\cos 2 \theta +5)\right)}~~\text{for}~~U^{VII(b)}\,,
\end{split}
\end{equation}
and
\begin{equation}
J_{CP}=I_1=I_2=0\,,
\end{equation}
which implies that all the three CP violating phases $\delta_{CP}$, $\alpha_{21}$ and $\alpha_{31}$ are trivial. Expressing the parameter $\theta$ in terms of $\theta_{13}$, we can obtain the sum rules among the lepton mixing angles,
\begin{equation}
\begin{split}
  &8\cos^2\theta_{12}\cos^2\theta_{13}=\frac{1}{s_3^2}\,,\\
  &\sin^2\theta_{23}\cos^2\theta_{13}=\frac{\left(\sqrt{\left(8\cos^2\theta_{13}s_3^2-1\right)\left(s_2^2\left(8s_3^2-1\right)-s_3^2\right)}\pm s_3\sin\theta_{13}\right)^2}{s_2^2\left(8s_3^2-1\right)^2}~~~\text{for}~~~U^{VII(a)}\,,\\
 &\sin^2\theta_{23}\cos^2\theta_{13}=\frac{\left(\sqrt{\left(8\cos^2\theta_{13}s_3^2-1\right)\left(s_1^2\left(8s_3^2-1\right)-s_3^2\right)}\pm s_3\sin\theta_{13}\right)^2}{s_1^2\left(8s_3^2-1\right)^2}~~~\text{for}~~~U^{VII(b)}\,.
\end{split}
\end{equation}
Given the best fitting value of the reactor mixing angle $\sin^2\theta_{13}=0.0218$~\cite{Gonzalez-Garcia:2014bfa}, we obtain
\begin{equation}
\label{eq:estimate_case_VII}\sin^2\theta_{12}=0.321,\qquad \sin^2\theta_{23}=0.399~~\text{or}~~0.601\,.
\end{equation}
For this mixing pattern, the effective Majorana neutrino mass $|m_{ee}|$ is given by
\begin{equation}
\begin{split}
\left|m_{ee}\right|=\frac{1}{24}\left|\frac{3m_1}{s_3^2}+q_1m_2\left(2\sqrt{2}\cos\theta+\left(\frac{1}{s_1}-\frac{1}{s_2}\right)\sin\theta\right)^2\right.\\\left.+q_2m_3\left(-2\sqrt{2}\sin\theta+\left(\frac{1}{s_1}-\frac{1}{s_2}\right)\cos\theta\right)^2\right|\,,
\end{split}
\end{equation}
As shown in figure~\ref{fig:mee_CaseVII}, $|m_{ee}|$ is around 0.017eV or 0.048eV in the case of IO, while a noticeable cancellation occurs such that $|m_{ee}|$ can be smaller than $10^{-4}$ eV for NO if the lightest neutrino mass lies in the interval $[0.0022,0.0032]\,\mathrm{eV}$ or $[0.0064,0.0074]\,\mathrm{eV}$. Regarding the predictions for leptogenesis, all the relevant CP invariants $I^{\alpha}_{23}$ as well as the lepton asymmetries $\epsilon_{\alpha}$ are zero. Thus a model, realizing this pattern at leading order, should  receive moderate corrections to interpret the observed baryon asymmetry  as an effect of leptogenesis.

\begin{figure}[ht!]
\begin{center}
\includegraphics[width=0.5\linewidth]{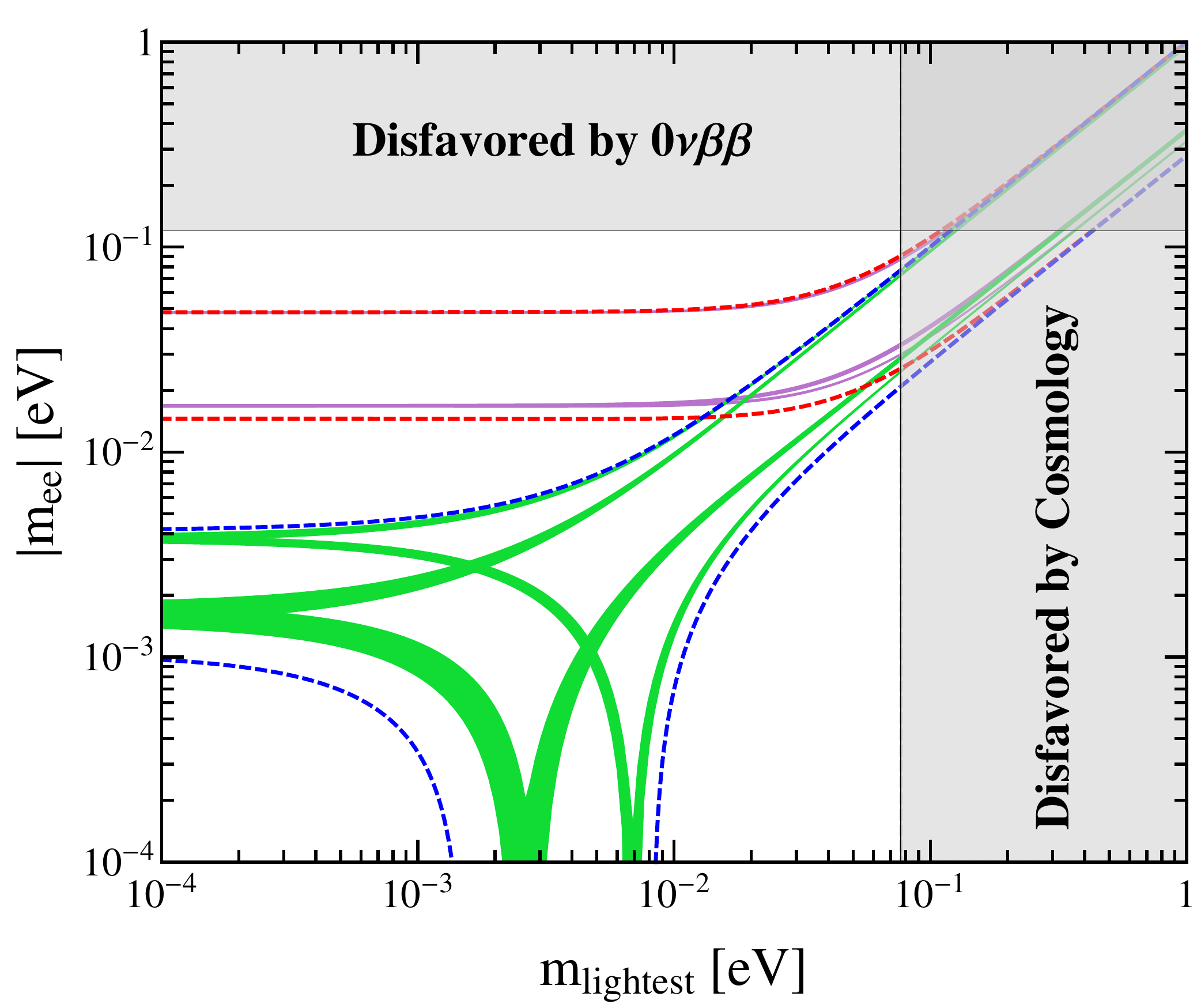}
\caption{\label{fig:mee_CaseVII}The Predictions of the $0\nu\beta\beta$ decay effective mass $|m_{ee}|$ with respect to the lightest neutrino mass $m_{\text{lightest}}$ for the mixing patterns $U^{VII(a)}$ and $U^{VII(b)}$. The red (blue) dashed lines indicate the most general allowed regions for IO (NO) spectrum obtained by varying the mixing parameters within their $3\sigma$ ranges~\cite{Gonzalez-Garcia:2014bfa}. The purple and green regions are the theoretical predictions of these two mixing patterns. The present most stringent upper limits $|m_{ee}|<0.120$ eV from EXO-200~\cite{Auger:2012ar, Albert:2014awa} and KamLAND-ZEN~\cite{Gando:2012zm} is shown by horizontal grey band. The vertical grey exclusion band is the current limit on $m_{\text{lightest}}$ from the cosmological data of $\sum m_i<0.230$ eV by the Planck collaboration~\cite{Ade:2013zuv}.}
\end{center}
\end{figure}

\end{description}

\subsection{\label{subsec:variant_semidirect_scan}Mixing patterns derived from the variant of semidirect approach}

In this approach, the residual flavor symmetries in the neutrino and charged lepton sectors are $K_4\times H^{\nu}_{CP}$ and $Z_2\times H^{l}_{CP}$ respectively. The prediction for the PMNS mixing matrix can be straightforwardly extracted from Eq.~\eqref{eq:Upmns2}. It is remarkable that the resulting mixing matrix has one column which is determined by the residual symmetries and which does not depend on the free parameter $\theta$. In exactly the same manner as the semidirect approach in section~\ref{subsec:semidirect_scan}, we perform a comprehensive scan over all possible finite discrete groups of the order less than 2000
with the help of \texttt{GAP}. We find only one type of mixing pattern which can accommodate the experimental data on lepton mixing angles for particular choices of the free parameter $\theta$
\begin{equation}
\label{eq:PMNS_Case_VIII}U^{VIII(a)}=\frac{1}{2}S^T_{13}(\theta)\left(
\begin{array}{ccc}
 \sqrt{2} e^{i \varphi _1} & -\sqrt{2} e^{i \varphi _1} & 0 \\
 1 & 1 & -\sqrt{2} e^{i \varphi _2} \\
 1 & 1 & \sqrt{2} e^{i \varphi _2}
\end{array}
\right)Q^{\dagger}_{\nu},\quad  U^{VIII(b)}=P_{132}U^{VIII(a)}_{PMNS}\,,
\end{equation}
where the viable values of $\varphi_1$, $\varphi_2$ and the representative flavor symmetry groups are summarized in table~\ref{tab:IV}. Notice that all these mixing patterns can be reproduced from the type D group series $\Delta(6n^2)$ and $D^{(1)}_{9n, 3n}$, and the small flavor symmetry groups $S_4$ and $\Delta(96)$ already allows a reasonable fit to the experimental data for this type of mixing pattern. This is consistent with the findings in Ref.~\cite{Li:2014eia}. Obvious $U^{VIII(b)}_{PMNS}$ is obtained from $U^{VIII(a)}_{PMNS}$ by exchanging the second and third rows. In this case, the row that is fixed by residual symmetry is $(1,1,-\sqrt{2}e^{i\varphi_2})/2$, and it could be the second or the third row of the PMNS mixing matrix. The predictions for the mixing angles read as
\begin{equation}
\label{eq:mixing_angles_case_XVII}
\begin{aligned}
&\qquad\qquad\sin^2\theta_{13}=\frac{1}{2}\sin^2\theta,\qquad
\sin^2\theta_{12}=\frac{1}{2}+\frac{\sqrt{2}\sin2\theta\cos\varphi_1}{3+\cos2\theta}\,,\\
&\sin^2\theta_{23}=\frac{2}{3+\cos2\theta}~~~\text{for}~~~U^{VIII(a)},\qquad
\sin^2\theta_{23}=\frac{1+\cos2\theta}{3+\cos2\theta}~~~\text{for}~~~U^{VIII(b)}\,,
\end{aligned}
\end{equation}
and the CP invariants take the form
\begin{equation}
\label{eq:CP_phases_case_XVII}
\begin{aligned}
&|J_{CP}|=\frac{1}{8\sqrt{2}}|\sin 2\theta \sin \varphi_1|\,,\qquad |I_1|=\frac{1}{8\sqrt{2}}|(1+3\cos 2\theta)\sin 2\theta\sin\varphi_1|\,,\\
&|I_2|=\frac{\sin^2\theta}{8}\left|\sqrt{2}\sin 2\theta\sin(2\varphi_2-\varphi_1)-2\cos^2\theta\sin 2(\varphi_2-\varphi_1)-\sin^2\theta\sin 2\varphi_2\right|\,.
\end{aligned}
\end{equation}
We easily see that the reactor and atmospheric mixing angles are related by
\begin{equation}
\sin^2\theta_{23}=\frac{1}{2\cos^2\theta_{13}}~~\text{for}~~U^{VIII(a)},\qquad
\sin^2\theta_{23}=\frac{\cos2\theta_{13}}{2\cos^2\theta_{13}}~~\text{for}~~U^{VIII(b)}\,.
\end{equation}
Given the $3\sigma$ range $0.0188\leq\sin^2\theta_{13}\leq0.0251$ of $\theta_{13}$~\cite{Gonzalez-Garcia:2014bfa}, the atmospheric mixing angle $\theta_{23}$ is determined to lie in the region of
\begin{equation}
0.510\leq\sin^2\theta_{23}\leq0.513~~\text{for}~~U^{VIII(a)},\qquad
0.487\leq\sin^2\theta_{23}\leq0.490~~\text{for}~~U^{VIII(b)}\,,
\end{equation}
which deviates from maximal mixing slightly. Similarly the sum rule among the reactor and solar mixing angles is given by
\begin{equation}
\sin^2\theta_{12}=\frac{1}{2}\pm\tan\theta_{13}\sqrt{1-\tan^2\theta_{13}}\,\cos\varphi_1\,,
\end{equation}
where the ``+'' and ``$-$'' signs are valid $0<\theta<\pi/2$ and $\pi/2<\theta<\pi$ respectively. For the experimentally favored $3\sigma$ interval of the reactor mixing angle, we get
\begin{equation}
0.342\leq\sin^2\theta_{12}\leq0.363\,.
\end{equation}

\begin{table}[t!]
\begin{center}
\begin{tabular}{|m{0.25\columnwidth}<{\centering}|m{0.7\columnwidth}<{\centering}|}
\hline\hline
\texttt{Group Id} & $(\varphi_1,\varphi_2)$ \\
\hline
$[24, 12]_{\vartriangle}$, $[48, 48]$ & $(\pi ,\pi )$\\ \hline $[96, 64]_{\vartriangle}$, $[192, 944]$ & $\left(0,\frac{3 \pi }{4}\right)$\\ \hline $[384, 568]_{\vartriangle}$, $[768, 1085727]$ & $\left(\frac{\pi }{8},-\frac{5 \pi }{8}\right)$, $\left(\frac{\pi }{8},\pi \right)$, $\left(0,\frac{7 \pi }{8}\right)$, $\left(\frac{\pi }{8},-\frac{7 \pi }{8}\right)$, $\left(-\frac{7 \pi }{8},\frac{3 \pi }{4}\right)$\\ \hline $[600, 179]_{\vartriangle}$, $[1200, 1011]$ & $\left(0,-\frac{4 \pi }{5}\right)$, $\left(0,-\frac{9 \pi }{10}\right)$, $\left(-\frac{\pi }{10},\frac{9 \pi }{10}\right)$, $\left(-\frac{\pi }{5},-\frac{4 \pi }{5}\right)$, $\left(-\frac{\pi }{10},\frac{7 \pi }{10}\right)$, $\left(-\frac{\pi }{5},\pi \right)$, $\left(-\frac{\pi }{5},\frac{9 \pi }{10}\right)$, $\left(-\frac{\pi }{10},-\frac{9 \pi }{10}\right)$, $\left(-\frac{\pi }{5},\frac{4 \pi }{5}\right)$, $\left(-\frac{\pi }{10},\pi \right)$, $\left(-\frac{\pi }{10},-\frac{7 \pi }{10}\right)$, $\left(-\frac{\pi }{5},-\frac{9 \pi }{10}\right)$\\ \hline $[648, 259]_{\vartriangle'}$, $[648, 260]$ & $\left(-\frac{5 \pi }{6},\frac{2 \pi }{3}\right)$, $\left(-\frac{5 \pi }{6},\pi \right)$, $\left(-\frac{5 \pi }{6},-\frac{2 \pi }{3}\right)$, $\left(-\pi ,-\frac{5 \pi }{6}\right)$\\ \hline $[1176, 243]_{\vartriangle}$ & $\left(0,-\frac{5 \pi }{7}\right)$, $\left(0,-\frac{13 \pi }{14}\right)$, $\left(\frac{13 \pi }{14},-\frac{13 \pi }{14}\right)$, $\left(\frac{13 \pi }{14},-\frac{9 \pi }{14}\right)$, $\left(\frac{13 \pi }{14},-\frac{6 \pi }{7}\right)$, $\left(-\frac{6 \pi }{7},\frac{13 \pi }{14}\right)$, $\left(0,-\frac{6 \pi }{7}\right)$, $\left(-\frac{\pi }{14},\frac{13 \pi }{14}\right)$, $\left(\frac{13 \pi }{14},\frac{5 \pi }{7}\right)$, $\left(\frac{3 \pi }{14},\pi \right)$, $\left(\frac{13 \pi }{14},\pi \right)$, $\left(\frac{\pi }{7},\frac{4 \pi }{7}\right)$, $\left(\frac{3 \pi }{14},\frac{5 \pi }{7}\right)$, $\left(-\frac{\pi }{14},\frac{11 \pi }{14}\right)$, $\left(-\frac{11 \pi }{14},\frac{13 \pi }{14}\right)$, $\left(\frac{\pi }{7},\frac{11 \pi }{14}\right)$, $\left(\frac{3 \pi }{14},-\frac{6 \pi }{7}\right)$, $\left(\frac{\pi }{7},\pi \right)$, $\left(\frac{3 \pi }{14},\frac{6 \pi }{7}\right)$, $\left(\frac{\pi }{7},-\frac{11 \pi }{14}\right)$, $\left(\frac{3 \pi }{14},-\frac{13 \pi }{14}\right)$, $\left(\frac{3 \pi }{14},-\frac{5 \pi }{7}\right)$, $\left(\frac{\pi }{7},-\frac{6 \pi }{7}\right)$, $\left(\frac{\pi }{7},-\frac{9 \pi }{14}\right)$\\ \hline $[1536, 408544632]_{\vartriangle}$ & $\left(-\frac{13 \pi }{16},\frac{7 \pi }{8}\right)$, $\left(\frac{\pi }{16},\frac{13 \pi }{16}\right)$, $\left(\frac{\pi }{16},\pi \right)$, $\left(\frac{\pi }{16},-\frac{13 \pi }{16}\right)$, $\left(\frac{\pi }{16},\frac{7 \pi }{8}\right)$, $\left(-\pi ,\frac{15 \pi }{16}\right)$, $\left(\frac{\pi }{16},-\frac{15 \pi }{16}\right)$, $\left(\frac{\pi }{16},-\frac{3 \pi }{4}\right)$, $\left(\frac{\pi }{16},-\frac{9 \pi }{16}\right)$, $\left(\frac{3 \pi }{16},-\frac{11 \pi }{16}\right)$, $\left(\frac{\pi }{8},-\frac{13 \pi }{16}\right)$, $\left(\frac{3 \pi }{16},-\frac{7 \pi }{8}\right)$, $\left(\frac{\pi }{8},\frac{13 \pi }{16}\right)$, $\left(\frac{3 \pi }{16},\frac{15 \pi }{16}\right)$, $\left(\frac{3 \pi }{16},-\frac{3 \pi }{4}\right)$, $\left(\frac{3 \pi }{16},\pi \right)$, $\left(0,\frac{11 \pi }{16}\right)$, $\left(\frac{\pi }{8},\frac{15 \pi }{16}\right)$, $\left(\frac{3 \pi }{16},-\frac{15 \pi }{16}\right)$, $\left(\frac{\pi }{8},\frac{9 \pi }{16}\right)$, $\left(\frac{3 \pi }{16},\frac{11 \pi }{16}\right)$, $\left(-\frac{15 \pi }{16},\frac{5 \pi }{8}\right)$\\ \hline\hline
\end{tabular}
\caption{\label{tab:IV}The predictions for PMNS matrix of the form $U^{VIII(a)}$ and $U^{VIII(b)}$, where the first column shows the group identification in \texttt{GAP} system, and the second column displays the achievable values of the parameters $\varphi_1$ and $\varphi_2$. We have shown at most two representatives flavor symmetry groups in the first column. If there is only one group predicting the corresponding values of $\varphi_1$ and $\varphi_2$ in the second column, this unique group would be listed. The full results of our analysis are provided at the website~\cite{webdata}. The subscripts $\Delta$ and $\Delta^{\prime}$ indicate that the corresponding groups belong to the type D group series $D_{n,n}^{(0)}\cong\Delta(6n^2)$ and $D_{9n^{\prime},3n^{\prime}}^{(1)}\cong(Z_{9n^{\prime}}\times Z_{3n^{\prime}})\rtimes S_3$, respectively. }
\end{center}
\end{table}

As a example, for $\sin^2\theta_{13}=0.0251$ ($\theta\simeq0.072\pi$ or $\theta\simeq1.928\pi$) and $\varphi_1=\pi$ (or $0$), we find the value of the solar mixing angle $\sin^2\theta_{12}\simeq0.342$
which is within the $3\sigma$ range. Therefore $\sin^2\theta_{12}$ is generically predicted to be close to its $3\sigma$ upper limit in this case\footnote{The $3\sigma$ ranges of $\sin^2\theta_{12}$ obtained by distinct global fitting groups have some minor difference: $0.270\leq\sin^2\theta_{12}\leq0.344$ from the NuFIT group~\cite{Gonzalez-Garcia:2014bfa}, $0.278\leq\sin^2\theta_{12}\leq0.375$ from the Valencia group~\cite{Forero:2014bxa} and  $0.250\leq\sin^2\theta_{12}\leq0.354$ given by the Italian group~\cite{Capozzi:2016rtj}.}. Notice that better agreement of the predicted values of $\sin^2\theta_{12}$ with the experimental results could be achieved in a concrete model with small corrections.

Moreover, we find that the Dirac CP phase is correlated with the mixing angles as follows
\begin{equation}
\cos\delta_{CP}=\pm\frac{\left(3\cos2\theta_{13}-1\right)\cot2\theta_{12}}{4\sqrt{\cos2\theta_{13}}\,\sin\theta_{13}}\,,
\end{equation}
where the ``+'' and ``$-$'' correspond to $U^{VIII(a)}$ and $U^{VIII(b)}$ respectively. If the reactor and solar mixing angles vary within the $3\sigma$ intervals $0.0188\leq\sin^2\theta_{13}\leq0.0251$ and $0.270\leq\sin^2\theta_{12}\leq0.344$~\cite{Gonzalez-Garcia:2014bfa}, we obtain
\begin{equation}
\cos\delta_{CP}\in\pm [0.983,1]\,.
\end{equation}
Hence $\delta_{CP}$ is predicted to be around $0$ or $\pi$ in this case. This mixing pattern would be ruled out if large CP violation effect is discovered in planned long baseline experiments.

\begin{figure}[hptb]
\begin{center}
\includegraphics[width=0.5\linewidth]{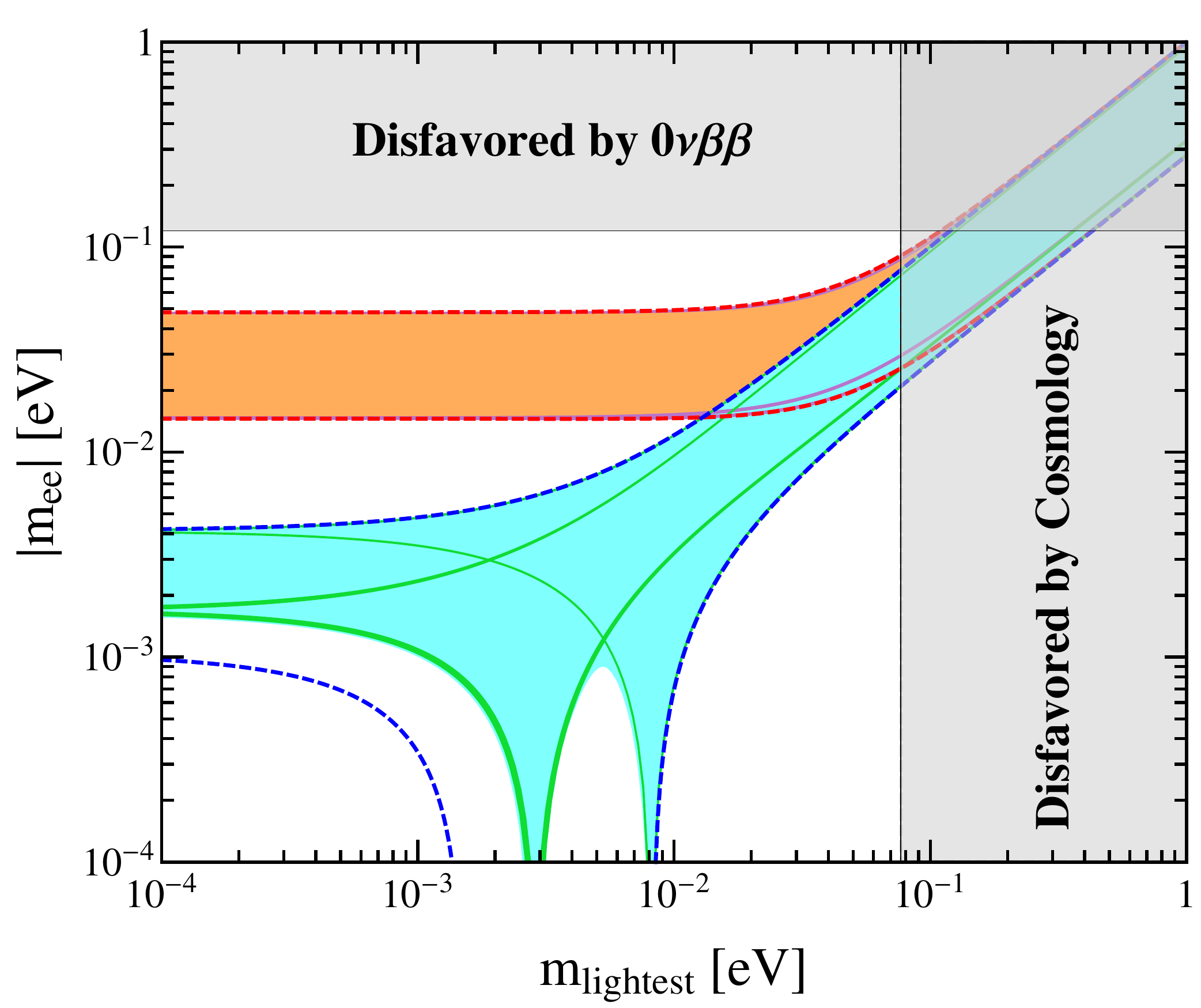}
\caption{\label{fig:mee_CaseRI}Predictions of the $0\nu\beta\beta$ decay effective mass $|m_{ee}|$ with respect to the lightest neutrino mass $m_{\text{lightest}}$ for the mixing patterns $U^{VIII(a)}$ and $U^{VIII(b)}$. The red (blue) dashed lines indicate the most general allowed regions for IO (NO) spectrum obtained by varying the mixing parameters within their $3\sigma$ ranges~\cite{Gonzalez-Garcia:2014bfa}. The orange (cyan) areas denote the achievable values of $|m_{ee}|$ when $\varphi_1$ and $\varphi_2$ are taken to be free continuous parameters in the case of IO (NO). The purple and green regions are the theoretical predictions of the smallest flavor symmetry group which can generate these two mixing patterns. Note that the purple (green) region overlaps the orange (cyan) one. The present most stringent upper limits $|m_{ee}|<0.120$ eV from EXO-200~\cite{Auger:2012ar, Albert:2014awa} and KamLAND-ZEN~\cite{Gando:2012zm} is shown by horizontal grey band. The vertical grey exclusion band is the current limit on $m_{\text{lightest}}$ from the cosmological data of $\sum m_i<0.230$ eV by the Planck collaboration~\cite{Ade:2013zuv}.}
\end{center}
\end{figure}

From the mixing matrix shown in Eq.~\eqref{eq:PMNS_Case_VIII}, we can extract the expression for the effective Majorana mass $|m_{ee}|$,
\begin{equation}
\left|m_{ee}\right|=\frac{1}{4}\left|m_1\left(\sqrt{2}e^{i\varphi_1}\cos\theta-\sin\theta\right)^2+q_1m_2\left(\sin\theta+\sqrt{2}e^{i\varphi_1}\cos\theta\right)^2+2q_2m_3e^{2i\varphi_2}\sin^2\theta\right|\,,
\end{equation}
with $q_{1,2}=\pm1$. We plot the possible region of $|m_{ee}|$ as a function of the lightest neutrino mass $m_{\text{lightest}}$ in figure~\ref{fig:mee_CaseRI}. In the limit of $|G_{f}|\rightarrow\infty$, we see that the entire $3\sigma$ region for IO and a sizable part for NO can be reproduced. For the particular value of $(\varphi_1, \varphi_2)=(\pi,\pi)$ which can be achieved from $S_4$ flavor symmetry combined with CP symmetry, we can read off from this figure $|m^{\text{IO}}_{ee}|\simeq0.015$ eV or $|m^{\text{IO}}_{ee}|\simeq0.048$ eV and $|m^{\text{NO}}_{ee}|$ is highly suppressed for $0.0026~\text{eV}\leq m_{\text{lightest}}\leq0.0031$ eV and $0.0079~\text{eV}\leq m_{\text{lightest}}\leq0.0084$ eV.

As has been shown in Ref.~\cite{Chen:2016ptr}, if a Klein four flavor symmetry is preserved by the neutrino mass matrix, all the leptogenesis CP asymmetries $\epsilon_{\alpha}$ would vanish and this result is independent of the concrete form of the residual Klein flavor symmetry transformation. Since the residual flavor symmetry of the neutrino sector is $K_4$ in the variant of the semidirect approach, a net baryon asymmetry can not be generated, and appropriate higher order corrections are necessary to have successful leptogenesis.

\section{\label{sec:conclusion}Conclusions}

Flavor and CP symmetries have been widely used to predict leptonic mixing parameters. In the present work, we take into account the generalized CP symmetry and perform an exhaustive scan of the lepton mixing patterns which can be obtained from the discrete finite groups up to order 2000 with the help of computer program~\texttt{GAP}. The generalized CP transformations are required to correspond to class-inverting automorphisms of the flavor symmetry group $G_f$, so that the consistency conditions between flavor and CP symmetry can be fulfilled. If $G_f$ doesn't possess a class-inverting automorphism, a CP symmetry could possibly be consistently defined in a model which contains only a subset of irreducible representations of $G_f$.

The flavor and CP symmetries have to be broken at low energy. The PMNS mixing matrix is fully fixed by the residual symmetries of the neutrino and charged lepton mass matrices, and we
do not need to consider how the residual symmetries are dynamically realized. In this work, we have considered two scenarios: the semidirect approach and the variant of the semidirect approach. In the semidirect approach, the residual symmetries of the charged lepton and neutrino mass matrices are $G_{l}\rtimes H^{l}_{CP}$ and $Z_2\times H^{\nu}_{CP}$ respectively,
where $G_{l}$ can be any abelian subgroup of $G_{f}$ capable of distinguishing the three generations. In the variant of the semidirect approach, the flavor and CP symmetries are assumed to be broken down to $Z_2\times H^{l}_{CP}$ and  $K_4\times H^{\nu}_{CP}$ in the charged lepton and neutrino sectors respectively. The PMNS matrix can be determined from the representation matrix of the residual symmetry without reconstructing the neutrino and charged lepton mass matrices, and the master formula is given by Eq.~\eqref{eq:Upmns1} and Eq.~\eqref{eq:Upmns2} respectively. We see that the PMNS matrix depends on only a free parameter $\theta$ which can take values in the range of $\left[0, \pi\right)$ in both approaches. Nevertheless, one column of the PMNS matrix is fixed to certain constant value by the residual symmetry in the semidirect approach while one row is fixed in its variant.

For each discrete flavor group which has a faithful three-dimensional irreducible representation and a class-inverting outer automorphism, all the possible remnant symmetries and the resulting predictions for lepton flavor mixing are studied. All these results are available at our website~\cite{webdata}. We find that all the mixing patterns which can accommodate the experimental data on the mixing angles can be  organized into eight different cases up to possible permutations of rows and columns. It is remarkable that the mixing matrices of case I, case II and case III can be reproduced from the $\Delta(6n^2)$ or $D^{(1)}_{9n, 3n}$ groups combined with the CP symmetry. The list of the mixing matrices associated with $\Delta(6n^2)$ and $D^{(1)}_{9n, 3n}$ agrees exactly with those given in Refs.~\cite{Hagedorn:2014wha,Ding:2014ora,Li:2016ppt}. The smallest group which can produce the mixing patterns of case IV and case V is the alternating group $A_5$. These two mixing patterns have really been found in the literature of $A_5$ flavor symmetry with generalised CP~\cite{Li:2015jxa,DiIura:2015kfa,Ballett:2015wia}. The mixing patterns of case VI and case VII are completely new as far as we know. They can be achieved from the flavor symmetry group $\Sigma(168)\cong PSL(2,7)$ and CP symmetry. The second column of the resulting PMNS mixing matrix is trimaximal in case II, case III and case VI, and therefore the sum rule $3\sin^2\theta_{12}\cos^2\theta_{13}=1$ is satisfied and the solar mixing angle is bounded from below $\sin^2\theta_{12}\geq1/3$. In the variant of the semidirect approach, only one type of mixing matrix denoted as case VIII can yield a good fit to the experimental data, and one row of the PMNS matrix is $(1, 1, -\sqrt{2}e^{i\varphi_2})/2$. The solar mixing angle $\theta_{12}$ is predicted to the close to its $3\sigma$ upper bound, and the atmospheric mixing angle is around $\sin^2\theta_{23}\simeq0.49$ or $\sin^2\theta_{23}\simeq0.51$. As a result, the paradigm of the generalized CP symemtry should be testable by precisely measuring $\theta_{12}$ and $\theta_{23}$ at future reactor neutrino experiments such as JUNO and long baseline experiments DUNE and Hyper-K.

Furthermore, the implications of residual symmetry in $0\nu\beta\beta$ decay and flavored thermal leptogenesis are studied. The predicted values of the effective Majorana mass $|m_{ee}|$ are within the sensitivity of planned experiments for IO neutrino mass spectrum, the known cancellation of the different terms in $|m_{ee}|$ may occur in the case of NO although $|m_{ee}|$ could have a non-trivial lower limit for a certain finite group. As regards the leptogenesis, the $R$-matrix in the Casas-Ibarra parametrization only depends on one single parameter $\eta$ because of the constraint imposed by remnant symmetry. The total lepton asymmetry $\epsilon_1\equiv\epsilon_e+\epsilon_{\mu}+\epsilon_{\tau}$ is determined to be zero such that the unflavored leptogenesis does not work. One the other hand, all the lepton charge asymmetries $\epsilon_{\alpha}$ ($\alpha=e$, $\mu$, $\tau$) are vanishing in case III, case V, case VII and case VIII, consequently the matter-antimatter asymmetry of the universe can not be explained via leptogenesis unless the postulated residual symmetry is further broken at the subleading level. For the remaining case I, case II, case IV and case VI,  the measured value of the baryon asymmetry can be generated for certain values of the parameters $\eta$ and $K_{1,2,3}$ which are determined by the CP parity of the neutrino states.

Many interesting mixing patterns and the associated residual symmetry provide new opportunity for model building. It would be interesting to construct concrete models in which the breaking of the symmetry group to the residual symmetry is achieved dynamically. Inspired by the above promising results obtained for lepton mixing, it is appealing to investigate whether the quark mixing angles and the precisely measured CP violating phase can be obtained as a result of mismatched remnant symmetries in the down quark and up quark sectors if the generalized CP symmetry is considered.

\section*{Acknowledgements}

We are grateful to Peng Chen for stimulating discussions on leptogenesis. This work is supported by the National Natural Science Foundation of China under Grant Nos. 11275188, 11179007 and 11522546.

\appendix

\section{\label{sec:equivalence}Equivalent conditions of distinct mixing patterns }

In both the semidirect approach and the variant of the semidirect approach discussed in section~\ref{sec:approach}, two distinct residual symmetries could lead to the same PMNS mixing matrix up to permutations of rows and columns and redefinition of the free parameter $\theta$ and the CP parity matrix $Q_{\nu}$.
Then the lepton mixing matrices following from these two residual symmetries would be called equivalent. For example, the mixing matrices predicted by two residual symmetries conjugate under a group element are equivalent, as shown in the end of section~\ref{sec:approach}. In the following, we shall derive the most general equivalent conditions for both approaches.

\subsection{\label{subsec:equiv_con_semidirect}Equivalence in semidirect approach}

Let us consider two generic residual symmetries in the semidirect approach, their predictions for the lepton mixing matrix can be written as
\begin{equation}
\label{eq:PMNS_equ}
\begin{aligned}
U_1=Q_{l1}^{\dagger}P_{l1}^T\Sigma_1 S_{23}(\theta_1) P_{\nu 1}Q^{\dagger}_{\nu 1}\,,\\
U_2=Q_{l2}^{\dagger}P_{l2}^T\Sigma_2 S_{23}(\theta_2) P_{\nu 2}Q^{\dagger}_{\nu 2}\,,
\end{aligned}
\end{equation}
where $\Sigma=\Sigma_{l}^{\dagger}\Sigma_{\nu}$, $\Sigma_1$ and $\Sigma_2$
are the corresponding results of $\Sigma$ for the two postulated residual symmetries. $Q_{l1,2}$ are arbitrary diagonal phase matrices and $Q_{\nu1,2}$ are unitary diagonal matrices with nonvanishing entries $\pm1$ and $\pm i$. $P_{l1,2}$ and $P_{\nu 1,2}$ are permutation matrices, and they can take the six possible forms in Eq.~\eqref{eq:per_matrix}. Moreover, $\theta_1$ and $\theta_2$ are free continuous parameters within the fundamental interval of $\left[0, \pi\right)$. For any given values of $\theta_1$ and the matrices $Q_{l1}, P_{l1}, Q_{\nu1}, P_{\nu1}$, if the corresponding values of $\theta_{2}$ as well as $Q_{l2}, P_{l2}, Q_{\nu2}, P_{\nu2}$ can be found such the equality $U_1=U_2$ is fulfilled, these two residual symmetries would be equivalent, i.e.,
\begin{equation}
Q_{l1}^{\dagger}P_{l1}^T\Sigma_1 S_{23}(\theta_1) P_{\nu 1}Q^{\dagger}_{\nu1}=Q_{l2}^{\dagger}P_{l2}^T\Sigma_2S_{23}(\theta_2)P_{\nu 2} Q^{\dagger}_{\nu2}\,,
\end{equation}
from which we can define a matrix $\Xi$ which is independent of $\theta_1$ and $\theta_2$ as follows
\begin{equation}
\Xi\equiv\Sigma_1^{\dagger}P_{l1}Q_{l1}Q_{l2}^{\dagger}P_{l2}^T\Sigma_2=S_{23}(\theta_1)P_{\nu 1}Q^{\dagger}_{\nu 1}Q_{\nu 2}P_{\nu 2}^TS_{23}^T(\theta_2)\,.
\end{equation}
For convenience, introducing the notations $P_l=P_{l1}P_{l2}^T$, $Q_l=P_{l2}Q_{l1}Q_{l2}^{\dagger}P^T_{l2}$, $P_{\nu}=P_{\nu 1}P_{\nu 2}^T$ and $Q_\nu =P_{\nu2}Q_{\nu 1}Q_{\nu 2}^{\dagger}P^T_{\nu2}$, then we have
\begin{equation}
\label{eq:M0}\Xi=\Sigma_1^{\dagger}P_{l}Q_{l}\Sigma_2=S_{23}(\theta_1)P_{\nu}Q_{\nu}S_{23}^T(\theta_2)\,,
\end{equation}
which implies
\begin{equation}
\Xi\Xi^T=S_{23}(\theta_1)Q'^2_{\nu}S_{23}^T(\theta_1)\,,
\end{equation}
where $Q'_{\nu}=P_{\nu}Q_{\nu}P_{\nu}^T$. Since $\Xi$ doesn't depend on the parameters $\theta_1$ and $\theta_2$, the right hand side of the above equation has to be independent of $\theta_1$. This requires $Q'_{\nu}$ should be of the form
\begin{equation}
Q'^2_{\nu}=\pm\text{diag}(1,\pm \mathbb{1}_{2\times 2})\,.
\end{equation}
Therefore the $(22)$ and $(33)$ elements of $Q'_{\nu}$ are either $\pm1$ or $\pm i$ simultaneously while the (11) element denoted as $q_{\nu}$ is independently $\pm1$ and $\pm i$. Without loss of generality, we assume that the fixed column by residual symmetries is the first column of the PMNS matrix, thus the permutation matrices $P_{\nu1}$ and $P_{\nu2}$ as well as $P_{\nu}$ can be either $P_{123}$ or $P_{132}$. Using the properties $S^{T}_{23}(\theta)=S_{23}(-\theta)$, $P_{132}S_{23}(\theta)=S_{23}(-\theta)P_{132}$ and $\text{diag}(1, 1, -1)S_{23}(\theta)=S_{23}(-\theta)\text{diag}(1, 1, -1)$, we can obtain
\begin{equation}
\label{eq:m1m2}
\Xi=\Sigma_1^{\dagger}P_lQ_l\Sigma_2=S_{23}(\theta_1)Q'_{\nu}P_{\nu}S^T_{23}(\theta_2)=S_{23}(\theta_0)Q'_{\nu}P_{\nu}\,,
\end{equation}
where $\theta_0=\theta_1\pm\theta_2$, ``+'' and ``$-$'' depend on the values of $Q'_{\nu}$ and $P_{\nu}$.  Assuming the common first column of $\Sigma_1$ and $\Sigma_2$ is $v_1$, the (11) entry of  then the $(11)$ entry of the $\Xi$ matrix is
\begin{equation}
\label{eq:v1} v_1^{\dagger}P_lQ_lv_1=q_{\nu}\,.
\end{equation}
We parameterize $v_1$  and $Q_l$ as $v_1=(a,b,c)^T$ and $Q_l=\text{diag}(e^{i\alpha_1},e^{i\alpha_2},e^{i\alpha_3})$, where $a$, $b$, $c$ can be set to be positive real numbers by redefining the charged lepton fields with the property $a^2+b^2+c^2=1$. In the following we shall discuss the constraints of Eqs.~(\ref{eq:m1m2}, \ref{eq:v1}) for the six possible forms of $P_{l}$ one by one.

Firstly, in the case of $P_l=P_{123}=\mathbb{1}_{3\times3}$, Eq.~\eqref{eq:v1} becomes
\begin{equation}
e^{i\alpha_1}a^2+ e^{i\alpha_2}b^2+e^{i\alpha_3}c^2=q_{\nu}\,.
\end{equation}
Taking the absolute value of the both sides of this equation, we obtain
\begin{equation}
\left|e^{i\alpha_1}a^2+ e^{i\alpha_2}b^2+e^{i\alpha_3}c^2\right|\leq a^2+b^2+c^2=1=|q_{\nu}|\,.
\end{equation}
This equality is fulfilled if and only if
\begin{equation}
e^{i\alpha_1}=e^{i\alpha_2}=e^{i\alpha_3}=q_{\nu}\,.
\end{equation}
Thus $Q_l=q_{\nu}\mathbb{1}_{3\times3}$, and Eq.~\eqref{eq:m1m2} reduces to
\begin{equation}
\label{eq:m1m2v2_nu}
\Omega\equiv\Sigma_1^{\dagger}P_l\Sigma_2=q^*_{\nu}S_{23}(\theta_0)Q'_{\nu}P_{\nu}\,,
\end{equation}
which can be written into a equivalent and more compact form
\begin{equation}
\label{eq:MMT_nu}
\Omega\Omega^T=q^{*2}_{\nu}Q'^2_{\nu}=\text{diag}(1,\pm \mathbb{1}_{2\times 2})\,,
\end{equation}
Conversely, if the condition of Eq.~\eqref{eq:m1m2v2_nu} or Eq.~\eqref{eq:MMT_nu} is satisfied, one can easily see that the two PMNS mixing matrices $U_1$ and $U_2$ in Eq.~\eqref{eq:PMNS_equ} would be equivalent.

For the case of $P_{l}=P_{132}$, then Eq.~\eqref{eq:v1} becomes
\begin{equation}
e^{i\alpha_1}a^2+ e^{i\alpha_2}bc+e^{i\alpha_3}bc=q_{\nu}\,.
\end{equation}
Taking the absolute value on both sides of this equation, we get
\begin{equation}
\begin{aligned}
\left|e^{i\alpha_1}a^2+e^{i\alpha_2}bc+e^{i\alpha_3}bc\right|\leq&a^2+2bc\leq a^2+b^2+c^2=1\,,
\end{aligned}
\end{equation}
which requires
\begin{equation}
e^{i\alpha_1}=e^{i\alpha_2}=e^{i\alpha_3},\qquad b=c\,.
\end{equation}
Consequently the equivalent condition in Eq.~\eqref{eq:m1m2v2_nu} and Eq.~\eqref{eq:MMT_nu} is also fulfilled with $P_{l}=P_{132}$. In other words, if the second and third elements $b$ and $c$ of the fixed column are the same, we should further consider the equivalent condition of Eq.~\eqref{eq:MMT_nu} with $P_l=P_{132}$. In the same manner, we can analyze the remaining cases of $P_{l}=P_{213}$, $P_{321}$, $P_{231}$ and $P_{312}$. The resulting constraints on the phases $\alpha_{1,2,3}$ and the constraints on the elements $a$, $b$ and $c$ are summarized in table~\ref{tab:equiv_semidirect}. One can see that $e^{i\alpha_1}=e^{i\alpha_2}=e^{i\alpha_3}=q_{\nu}$ always needs to be satisfied. As a consequence, we summarize that the most general equivalent condition of two mixing pattern is given by Eq.~\eqref{eq:MMT_nu} in the semidirect approach, and $P_l$ is the permutation matrix under which the fixed column $v_1$ is invariant $P_{l}v_{1}=v_1$.

\begin{table}[hptb]
\begin{center}
\begin{tabular}{|c|c|c|}
\hline\hline
$P_l$ & \texttt{Constraint on $\alpha_{1,2,3}$} & \texttt{Constraint on $a$, $b$ and $c$} \\[0.5mm]\hline
$P_{123}$ & \multirow{6}{*}{$e^{i\alpha_1}=e^{i\alpha_2}=e^{i\alpha_3}=q_{\nu}$}  &  \rule{0.04\textwidth}{.4pt}\\\cline{1-1}\cline{3-3}

$P_{132}$ &  &  $b=c$ \\ \cline{1-1} \cline{3-3}

$P_{213}$ & &  $a=b$ \\ \cline{1-1} \cline{3-3}

$P_{321}$ &   &  $a=c$ \\ \cline{1-1} \cline{3-3}

$P_{231}$ & &  $a=b=c$ \\ \cline{1-1} \cline{3-3}

$P_{312}$ & &  $a=b=c$ \\  \hline \hline
\end{tabular}
\caption{\label{tab:equiv_semidirect}Constraints on the fixed column $v_1=\left(a, b, c\right)^{T}$ and the phase matrix $Q_{l}=\text{diag}(e^{i\alpha_1},e^{i\alpha_2},e^{i\alpha_3})$ imposed by the equivalent condition in the semidirect approach.}
\end{center}
\end{table}

\subsection{\label{subsec:equiv_con_variant}Equivalence in variant of the semidirect approach}

Given two distinct set of residual symmetries in this approach, as shown in section~\ref{subsec:variant_semi_app}, the lepton mixing matrices read as
\begin{equation}
\begin{aligned}
U_1=Q_{l1}P_{l1}^TS^T_{23}(\theta_1)\Sigma_1P_{\nu 1}Q^{\dagger}_{\nu 1}\,,\\
U_2=Q_{l2}P_{l2}^TS^T_{23}(\theta_2)\Sigma_2P_{\nu 2}Q^{\dagger}_{\nu 2}\,,
\end{aligned}
\end{equation}
where $\Sigma=\Sigma_{l}^{\dagger}\Sigma_{\nu}$. In the following, we shall derive the criteria to determine whether the above two PMNS matrices $U_1$ and $U_2$ are essentially the same up to rows and columns permutations and the redefinition of the parameter $\theta$. In other words, if the solution(s) for $\theta_2$ and the $P_{l1,2}, Q_{l1,2}, P_{\nu 1,2}, Q_{\nu 1,2}$ matrices can be found for any given value of $\theta_1$, so that the equality $U_1=U_2$ is fulfilled, and then $U_1$ and $U_2$ would be equivalent, i.e.
\begin{equation}
Q_{l1}P_{l1}^TS^T_{23}(\theta_1)\Sigma_1P_{\nu 1}Q^{\dagger}_{\nu1}=Q_{l2}P_{l2}^TS^T_{23}(\theta_2)\Sigma_2P_{\nu 2}Q^{\dagger}_{\nu2}\,,
\end{equation}
which leads to
\begin{equation}
\label{eq:2M0}
\Xi\equiv \Sigma_1P_{\nu}Q_{\nu}\Sigma^{\dagger}_2=S_{23}(\theta_1)P_lQ_lS_{23}^T(\theta_2)\,,
\end{equation}
with $P_{\nu}=P_{\nu 1}P_{\nu 2}^T$, $Q_\nu=P_{\nu2}Q^{\dagger}_{\nu1}Q_{\nu 2}P^T_{\nu 2}$, $P_l=P_{l1}P_{l2}^T$ and $Q_l=P_{l2}Q^{\dagger}_{l1}Q_{l2}P^T_{l2}$. Thus the product of $\Xi$ and its transpose is
\begin{equation}
\Xi\Xi^T=S_{23}(\theta_1)Q'^2_lS_{23}^T(\theta_1)\,,
\end{equation}
where $Q'_l=P_lQ_lP_l^T$ is a diagonal phase matrix. Since $\Xi$
is a constant matrix and it doesn't depend on $\theta_1$, we have
\begin{equation}
Q'_{l}=\text{diag}(\pm e^{i\gamma/2}, \pm e^{i\alpha/2},\pm e^{i\alpha/2})\,,
\end{equation}
where $\alpha$ and $\gamma$ are real, and ``$\pm$'' can be chosen independently. In the variant of the semidirect approach, one row of the PMNS matrix is fixed by the postulated residual symmetry. Without loss of generality, we assume that the fixed row is the first row of the PMNS matrix. As a result, the permutation matrices $P_{l1}$, $P_{l2}$ and $P_l$ can be either $P_{123}$ or $P_{132}$, thus we obtain the equivalent condition
\begin{equation}
\label{eq:Xi_variant_semi}\Xi=\Sigma_1P_{\nu}Q_{\nu}\Sigma^{\dagger}_2=S_{23}(\theta_1)P_lQ_lS^T_{23}(\theta_2)=S_{23}(\theta_1)Q'_lP_lS^T_{23}(\theta_2)=Q'_lP_lS^T_{23}(\theta_0)\,,
\end{equation}
with $\theta_0=\theta_2\pm\theta_1$. If the two mixing patterns $U_1$ and $U_2$ are equivalent, the first row of $\Sigma_1$ and $\Sigma_2$ must be equal, and it is denoted as  $u_1=(c_1,c_2,c_3)=(|c_1|e^{i\delta_1}, |c_2|e^{i\delta_2},|c_3|e^{i\delta_3})$ with $|c_1|^2+|c_2|^2+|c_3|^2=1$.
Notice that we can set the phases $\delta_1=0$ and $\delta_{2,3}\in [0,\frac{\pi}{2})$ by redefining the matrices $Q_{l}$ and $Q_{\nu}$.
The (11) element of $\Xi$ can be read from Eq.~\eqref{eq:Xi_variant_semi} as
\begin{equation}
\label{eq:u1}u_1P_{\nu}Q_{\nu}u^{\dagger}_1=\pm e^{i\gamma/2}\equiv q_l\,.
\end{equation}
We parameterize $Q_{\nu}=\text{diag}(q_{\nu 1},q_{\nu 2}, q_{\nu 3})$ and $q_{\nu 1,2,3}=\pm 1,\pm i$. In the following, we shall analyze the equivalent condition of Eq.~\eqref{eq:Xi_variant_semi} and the constraint of Eq.~\eqref{eq:u1} for the six possible values of $P_{\nu}$.

If $P_{\nu}=P_{123}=\mathbb{1}_{3\times3}$, Eq.~\eqref{eq:u1} reduces to
\begin{equation}
q_{\nu 1}|c_1|^2+q_{\nu 2}|c_2|^2+q_{\nu3}|c_3|^2=q_l\,,
\end{equation}
Subsequently taking absolute value of the both sides of this equation, we obtain
\begin{equation}
\Big|q_{\nu 1}|c_1|^2+q_{\nu 2}|c_2|^2+q_{\nu 3}|c_3|^2\Big|=1\,,
\end{equation}
which requires
\begin{equation}
q_{\nu 1}=q_{\nu 2}=q_{\nu 3}=q_{l}.
\end{equation}
Therefore the equivalent condition of Eq.~\eqref{eq:Xi_variant_semi} becomes
\begin{equation}
\label{eq:2m1m2v2}
\Omega\equiv\Sigma_1P_{\nu}\Sigma^{\dagger}_2=q^*_lQ'_lP_lS^T_{23}(\theta_0)\,,
\end{equation}
or equivalently
\begin{equation}
\label{eq:2MMT}
\Omega\Omega^T=q^{*2}_lQ'^2_l=\text{diag}(1,e^{i \alpha'},e^{i \alpha'})\,,
\end{equation}
where $\alpha'=\alpha-\gamma$.

For the case of $P_{\nu}=P_{132}$, Eq.~\eqref{eq:u1} takes the form
\begin{equation}
\label{eq:P132_variant}q_{\nu 1}|c_1|^2+q_{\nu 2}c_3c_2^*+q_{\nu 3}c_2c_3^*=q_l\,,
\end{equation}
from which we obtain
\begin{equation}
\Big|q_{\nu 1}|c_1|^2+q_{\nu 2}c_2^*c_3+q_{\nu 3}c_2c_3^*\Big|\leq |c_1|^2+2|c_2||c_3|\leq |c_1|^2+|c_2|^2+|c_3|^2=1=|q_{l}|\,.
\end{equation}
Thus Eq.~\eqref{eq:P132_variant} is satisfied if and only if
\begin{equation}
q_{\nu 1}=e^{i(\delta_3-\delta_2)}q_{\nu 2}=e^{-i(\delta_3-\delta_2)}q_{\nu 3},\qquad |c_2|=|c_3|\,,
\end{equation}
which leads to $e^{i(\delta_3-\delta_2)}=\pm 1,\pm i$. Considering $\delta_3-\delta_2\in (-\frac{\pi}{2},\frac{\pi}{2})$, we have
\begin{equation}
\delta_2=\delta_3, \qquad q_{\nu 1}=q_{\nu 2}=q_{\nu 3}=q_l\,.
\end{equation}
Therefore the equivalent condition is still $\Omega\Omega^T=\text{diag}(1,e^{i \alpha'},e^{i \alpha'})$ given by Eq.~\eqref{eq:2MMT} with $\Omega=\Sigma_1P_{\nu}\Sigma^{\dagger}_2$ and $P_{\nu}=P_{132}$.

For all the six possible values of $P_{\nu}$, the corresponding constraints on the fixed row $u_1=(|c_1|, |c_2|e^{i\delta_2},|c_3|e^{i\delta_3})$ and the phase matrix $Q_{\nu}=\text{diag}(q_{\nu 1},q_{\nu 2}, q_{\nu 3})$ are summarized in table~\ref{tab:equiv__variant_semidirect}. We see that the equivalent condition can be written as $\Omega\Omega^T=\text{diag}(1,e^{i \alpha'},e^{i \alpha'})$ with $\Omega=\Sigma_1P_{\nu}Q'_{\nu}\Sigma^{\dagger}_2$. The matrix $Q'_{\nu}$ is an identity matrix $Q'_{\nu}=\mathbb{1}_{3\times3}$ in the case of $P_{\nu}=P_{123}$, $P_{132}$, $P_{213}$ and $P_{321}$. Nevertheless, depending on the values of $\delta_2$ and $\delta_3$, we have $Q'_{\nu}=\mathbb{1}_{3\times3}$, $e^{-i\pi/6}\text{diag}\left(1, i, 1\right)$, $e^{-i\pi/6}\text{diag}\left(1, 1, i\right)$ or $e^{-i\pi/3}\text{diag}\left(1, i, i\right)$ for $P_{\nu}=P_{231}$, $P_{312}$. Using this simple criteria, one can easily determine whether two residual symmetries give rise to the same lepton mixing patter.

\begin{table}[hptb]
\begin{center}
\begin{tabular}{|c|c|c|c|}
\hline\hline
$P_{\nu}$ & \texttt{Constraint on $q_{\nu1,2,3}$} &
\texttt{Constraint on $|c_{1,2,3}|$} &  \texttt{Constraint on $\delta_{2,3}$} \\[0.5mm]\hline
$P_{123}$ & $q_{\nu1}=q_{\nu2}=q_{\nu3}=q_{l}$  & --- & --- \\\hline

$P_{132}$ & $q_{\nu1}=q_{\nu2}=q_{\nu3}=q_{l}$  &  $|c_2|=|c_3|$  &  $\delta_{2}=\delta_3$ \\ \hline

$P_{213}$ &  $q_{\nu1}=q_{\nu2}=q_{\nu3}=q_{l}$   &  $|c_1|=|c_2|$  & $\delta_{2}=0$ \\ \hline

$P_{321}$ & $q_{\nu1}=q_{\nu2}=q_{\nu3}=q_{l}$  &  $|c_1|=|c_3|$ & $\delta_{3}=0$ \\ \hline

\multirow{3}{*}[-4pt]{$P_{231}$} &  $q_{\nu1}=q_{\nu2}=q_{\nu3}=q_{l}$ & \multirow{3}{*}[-4pt]{$|c_1|=|c_2|=|c_3|=\frac{1}{\sqrt{3}}$} & $\delta_2=\delta_3=0$\\ \cline{2-2}\cline{4-4}

  &   $q_{\nu1}=-iq_{\nu2}=q_{\nu3}=e^{-i\pi/6}q_{l}$ & & $\delta_2=\pi/3$, $\delta_3=\pi/6$\\ \cline{2-2}\cline{4-4}

  &   $q_{\nu1}=-iq_{\nu2}=-iq_{\nu3}=e^{-i\pi/3}q_{l}$ &  & $\delta_2=\pi/6$, $\delta_3=\pi/3$\\\hline

\multirow{3}{*}[-4pt]{$P_{312}$} & $q_{\nu1}=q_{\nu2}=q_{\nu3}=q_{l}$ &
\multirow{3}{*}[-4pt]{$|c_1|=|c_2|=|c_3|=\frac{1}{\sqrt{3}}$} & $\delta_2=\delta_3=0$\\ \cline{2-2}\cline{4-4}

  &  $q_{\nu1}=-iq_{\nu2}=-iq_{\nu3}=e^{-i\pi/3}q_{l}$ & & $\delta_2=\pi/3$, $\delta_3=\pi/6$\\ \cline{2-2}\cline{4-4}

  &   $q_{\nu1}=q_{\nu2}=-iq_{\nu3}=e^{-i\pi/6}q_{l}$ &  & $\delta_2=\pi/6$, $\delta_3=\pi/3$ \\  \hline \hline
\end{tabular}
\caption{\label{tab:equiv__variant_semidirect}Constraints on the fixed row $u_1=(|c_1|, |c_2|e^{i\delta_2},|c_3|e^{i\delta_3})$ and the phase matrix $Q_{\nu}=\text{diag}(q_{\nu 1},q_{\nu 2}, q_{\nu 3})$ imposed by the equivalent condition in the variant of the semidirect approach.}
\end{center}
\end{table}

\end{document}